\documentclass[12pt]{amsart}
\usepackage{amssymb}
\usepackage{amsmath}
\usepackage{bm}
\usepackage{epsfig,subfigure}
\usepackage[norelsize,lined,algoruled]{algorithm2e}
\usepackage{color}
\usepackage{float}
\usepackage{fullpage}
\usepackage{hyperref}
\usepackage{textgreek}
\usepackage[title]{appendix}
\usepackage{url}
\usepackage[square]{natbib}

\SetKwInput{KwInput}{Input}
\SetAlCapSkip{1em}
\SetKwFunction{defined}{defined}
\SetKwFunction{notdefined}{not defined}
\SetKwFunction{mean}{mean}
\SetKwProg{Fn}{Function}{}{}
\SetKwFunction{FnGetLowerBound}{GetAlphaLowerBound}%
\SetKwFunction{FnGetAlpha}{GetAlphaEstimate}%

\newcommand{\dotstar}{\cdot \ast}

\newcommand{\bfa}{\mathbf{a}}
\newcommand{\bfb}{\mathbf{b}}
\newcommand{\bfc}{\mathbf{c}}

\newcommand{\bfd}{\mathbf d}
\newcommand{\bfe}{\mathbf e}
\newcommand{\dx}{\Delta_x}
\newcommand{\dy}{\Delta_y}
\newcommand{\dz}{\Delta_z}

\newcommand{\Gr}{G^{(r)}}

\newcommand{\bfh}{\mathbf{h}}

\newcommand{\Kmax}{K_{\mathrm{max}}}
\newcommand{\bfm}{\mathbf m}
\newcommand{\bfmmin}{\bfm_{\mathrm{min}}}
\newcommand{\bfmmax}{\bfm_{\mathrm{max}}}
\newcommand{\nbr}{n_r}
\newcommand{\nbx}{n_x}
\newcommand{\nby}{n_y}
\newcommand{\nbz}{n_z}
\newcommand{\nsx}{s_x}
\newcommand{\nsy}{s_y}
\newcommand{\pad}{\mathrm{pad}}

\newcommand{\padxl}{p_{x_{\texttt{L}}}}
\newcommand{\padxr}{p_{x_{\texttt{R}}}}
\newcommand{\padyl}{p_{y_{\texttt{L}}}}
\newcommand{\padyr}{p_{y_{\texttt{R}}}}

\newcommand{\bfr}{\mathbf{r}}

\newcommand{\Rmn}[2]{\mathcal{R}^{#1\times #2}}
\newcommand{\Rm}[1]{\mathcal{R}^{#1}}

\newcommand{\bfw}{\mathbf{w}}
\newcommand{\bfx}{\mathbf{x}}
\newcommand{\bfy}{\mathbf{y}}
\newcommand{\bfz}{\mathbf{z}}
\newcommand{\bfeta}{\boldsymbol{\eta}}

\newcommand{\Phid}{ \Phi_{\mathrm{d}}}
\newcommand{\Phis}{ \Phi_{\mathrm{S}}}
\newcommand{\Phia}{ \Phi_{\mathrm{\alpha}}}
\newcommand{\Wd}{\mathbf{W_{\bfd}}}
\newcommand{\Wz}{\mathbf{W_{\mathrm{z}}}}
\newcommand{\WL}{\mathbf{W_{{\mathrm{L}}}}}
\newcommand{\WLk}{\mathbf{W}^{(k)}_{{\mathrm{L}}}}
\newcommand{\Wh}{\mathbf{W_{\mathrm{h}}}}
\newcommand{\W}{\mathbf{W}}

\newcommand{\bfdo}{\mathbf{d}_{\mathrm{obs}}}

\newcommand{\bfde}{\mathbf{d}_{\mathrm{exact}}}
\newcommand{\bfme}{\bfm_{\mathrm{exact}}}
\newcommand{\bfma}{\mathbf{m}_{\mathrm{apr}}}

\newcommand{\tildeGk}{\tilde{G}^{(k)}}

\newcommand{\tilder}{\tilde{\bfr}}
\newcommand{\tildeG}{\tilde{G}}
\newcommand{\argmin}[1]{\underset{#1}{\arg \min}\,}

\newcommand{\blockdiag}{\texttt{blockdiag}}
\newcommand{\irls}{\texttt{IRLS}~}

\newcommand{\svd}{\texttt{SVD}~}

\newcommand{\floor}{\texttt{floor}}
\newcommand{\round}{\texttt{round}}
\newcommand{\flops}{\texttt{flops}}

\newcommand{\fftt}{\texttt{fft2}}
\newcommand{\ifftt}{\texttt{ifft2}}

\newcommand{\zeros}{\texttt{zeros}}

\newcommand{\bttb}{\texttt{BTTB}}
\newcommand{\sbttb}{\texttt{SBTTB}}
\newcommand{\bccb}{\texttt{BCCB}}

\newcommand{\oneDFFT}{\texttt{1DFFT}}
\newcommand{\gravity}{\texttt{gravity}}
\newcommand{\magnetic}{\texttt{magnetic}}

\title[Fast Focusing Inversion]{A fast methodology for large-scale focusing inversion of gravity and magnetic data using the structured model matrix and the $2D$ fast Fourier transform\footnote{{\today}}}


\begin{document}
\maketitle
\smallskip
\centerline{\scshape Rosemary A. Renaut}
{\footnotesize
 \centerline{School of Mathematical and Statistical Sciences}
   \centerline{Arizona State University}
   \centerline{Tempe, AZ 85287, USA}
   \centerline{\email{renaut@asu.edu}}
}
\smallskip
\centerline{\scshape Jarom D. Hogue}
{\footnotesize
 \centerline{School of Mathematical and Statistical Sciences, Arizona State University}
   \centerline{Tempe, AZ 85287, USA}
  \centerline{\email{jdhogue@asu.edu}}
}
\smallskip
\centerline{\scshape Saeed Vatankhah}

{\footnotesize
 \centerline{Institute of Geophysics, University of Tehran, Tehran, Iran}
 \centerline{Hubei Subsurface Multi-scale Imaging Key Laboratory, Institute of Geophysics and Geomatics}
 \centerline{China University of Geosciences, Wuhan, China.}
\centerline{\email{svatan@ut.ac.ir}}
}

\begin{abstract}
Focusing inversion of potential field data for the recovery of sparse subsurface structures from surface measurement data on a uniform grid is discussed. For the uniform grid the model sensitivity matrices  exhibit block Toeplitz Toeplitz block structure, by blocks for each depth layer of the subsurface. Then, through embedding in circulant matrices, all forward operations with the sensitivity matrix, or its transpose, are realized using the fast two dimensional Fourier transform. Simulations demonstrate that  this fast inversion algorithm can be implemented on standard desktop computers with sufficient memory for storage of volumes up to size $n \approx 1M$. The linear systems of equations arising in the focusing inversion algorithm are solved using  either Golub-Kahan-bidiagonalization or randomized singular value decomposition algorithms in which all matrix operations with the sensitivity matrix are implemented using the fast Fourier transform. These two algorithms are contrasted for efficiency for large-scale problems with respect to  the sizes of the projected subspaces adopted for the solutions of the linear systems.  The presented results confirm earlier studies that the randomized algorithms are to be preferred  for the inversion of gravity data, and that it is sufficient to use projected spaces of size approximately $m/8$, for data sets of size $m$. In contrast, the Golub-Kahan-bidiagonalization leads to more efficient implementations for the inversion of magnetic data sets, and it is again sufficient to use projected spaces of size approximately $m/8$. Moreover, it is sufficient to use projected spaces of size $m/20$ when $m$ is large, $m \approx  50000$,  to reconstruct volumes with  $n \approx 1M$.  Simulations support the presented conclusions and are verified on the inversion of a practical magnetic data set that is obtained over the Wuskwatim Lake region in Manitoba, Canada.
 \end{abstract}

\section{Introduction}\label{sec:introduction}
The determination of the subsurface structures from measured potential field data is important for many practical applications concerned with oil and gas exploration, mining, and regional investigations, \cite{blakely_1995,Nabighian:05}. There are many approaches that can be considered for the inversion of potential field data sets. These range from techniques that directly use the inversion of a forward model described by a sensitivity matrix for gravity and magnetic potential field data, as in, for example, \cite{BoCh:2001,Farquharson:2008, LeOl:06, LiOl:96, LiOl:98, Pilkington:97,SiBa:06} and \cite{PoZh:99}. Other approaches avoid the problem with the storage and generation of a large sensitivity matrix  by employing alternative approaches, as in \cite{CoWiZh:10,UiBa:12,VANRK:19}. Of those that do handle the sensitivity matrix, some techniques to avoid the large scale challenge, include wavelet and  compression techniques, \cite{LiOl:03, PoZh:02} and \cite{Voronin2015}. Of interest here is the development of an approach that takes advantage of the structure that can be realized for the sensitivity matrix, and then enables the use of the fast Fourier transform for fast 
matrix operations, and  avoids the high storage overhead of the matrix.

The efficient inversion of three dimensional gravity data using the $2$D fast Fourier transform (\texttt{2DFFT}) was presented in the Master's thesis of \citep{bruun2007}. There, it was observed that the sensitivity matrix exhibits a block Toeplitz Toeplitz block (\texttt{BTTB}) structure provided that the data  measurement positions are uniform and carefully related to the grid defining the volume discretization. It is this structure which facilitates the use of the \texttt{2DFFT} via the embedding of the required kernel entries that define the sensitivity matrix within a block Circulant Circulant block (\texttt{BCCB}) matrix, and which is explained in \cite{ChanFuJin:2007,Vogel:2002}.  Then, \cite{ZhangWong:15} used the \texttt{BTTB} structure for fast computations with the sensitivity matrix, and employed this within an algorithm for the inversion of gravity data using a smoothing regularization, allowing for  variable heights of the individual depth layers in the domain. They also applied  optimal preconditioning for the \texttt{BTTB} matrices using the approach of \cite{ChanFuJin:2007}.  Their approach was then optimized by \cite{ChenLiu:18} but only for efficient forward gravity modeling  and with a slight modification in the way that the matrices for each depth layer of the domain are defined using the approximation of the forward integral equation.  In particular, \cite{ZhangWong:15}  use a multilayer approximation of the gravity kernel, rather than the derivation of the kernel integral in \cite{Li1998}.  They noted, however,  that their  approach is  subject to greater potential for error on coarse-grained domains because it does not use the exact kernel integral developed by \cite{Li1998}. \cite{bruun2007} also developed  an algorithm that is even  more efficient in memory and computation than the use of the \texttt{BTTB} for each depth layer by using an upward continuation method to deal with the issue that measured data are only provided at the surface of the domain. They concluded that this was not suitable for practical problems. Finally, they also considered the interpolation of data not on the uniform grid to the uniform grid, hence removing the restriction on the uniform placement of measurement stations on the surface, but potentially introducing some error due to the interpolation. On the other hand, their study  did not include Tikhonov stabilization for the solution of the linear systems, and hence did not implement state-of-the-art approaches for resolving complex structures with general L$_p$ norm regularizers ($0\le p\le 2$). Moreover, standard techniques for inclusion of depth weighting, and imposition of constraint conditions were not considered. The focus of this work is, therefore, a demonstration and validation of efficient solvers that are more general and can be effectively employed for the independent focusing inversion of both large scale gravity and magnetic potential field data sets. It should be noted, moreover, that the approach can be applied also for domains with padding, which is of potential benefit for structure identification near the boundaries of the analyzed volume. 

First, we note that the fast computation of geophysics kernel models using the fast Fourier transform (\texttt{FFT}) has already been considered in a number of different contexts. These include calculation using the Fourier domain as in \cite{Li2018,ZHAO2018294}, and  also by  \cite{Pilkington:97}  in conjunction with the conjugate gradient method for solving the magnetic susceptibility inverse problem. Fast forward modeling of the magnetic kernel on an undulated surface, combined with spline interpolation of the surface data was also suggested by \cite{Li2018}  using an implementation of the model in the wave number domain. Further, fast forward and high accuracy modeling of the gravity kernel using the Gauss \texttt{2DFFT} was  discussed by \cite{ZHAO2018294}. Moreover, the derivation of the forward modeling operators that yield the \texttt{BTTB} structure for the magnetic and gravity kernels in combination with domain padding and the staggered placement of measurement stations with respect to the domain prisms at the surface was carefully presented in \cite{hogue2019tutorial}. Hence, here, we only present necessary details concerning
the development of the forward modeling approach

Associated with the development of a focusing inversion algorithm, is the choice of solver within the inversion algorithm,  the choice of regularizer 
for focusing the subsurface structures, and a decision on determination of suitable regularization parameters. With respect to the solver, small scale problems can be solved using the full singular value decomposition (\texttt{SVD}) of the sensitivity matrix, which is not feasible for the large scale. Moreover, the use of the \texttt{SVD} for focusing inversion has been well-investigated in the literature, see for example \citep{vatan:2014,VAR:2014b}, while choices and implementation details  for focusing inversion are reviewed in \citep{VaReSh:20}. Furthermore, methods that yield useful approximations of the \texttt{SVD}, hence enabling automatic but efficient techniques for  choice of the regularization parameters have also been discussed in \citep{RVA:15,VRA:2017} when considered with iterative Krylov methods based on the Golub-Kahan Bidiagonalization (\texttt{GKB}) algorithm, \citep{PaigeSau2}, and in \cite{saeed6, vatankhah2019improving} when adopted using the randomized singular value decomposition (\texttt{RSVD}), \citep{HMT:11}. Recommendations for the application of the \texttt{RSVD} with power iteration, and the sizes of the projected spaces to be used for both \texttt{GKB} and \texttt{RSVD} were presented, but only within the context of problems that can be solved without the use of the \texttt{2DFFT}. Thus, a complete validation of these
 algorithms for the solution of the large scale focusing inversion problem, 
 with considerations contrasting the effectiveness of these algorithms in the large scale, is still important, and is addressed here. 

We comment, further, that there is an alternative approach for the comparison of \texttt{RSVD} and \texttt{GKB} algorithms, which was discussed by \cite{LUIKEN2020}. The focus there, on the other hand, was on the effective determination of both  the size of the projected space and the determination of the optimal regularization parameter, using these algorithms. Their \texttt{RSVD} algorithm used the range finder suggested in \citep[Algorithm 4]{HMT:11}, rather than the power iteration. They concluded with their one rather small example for an under-determined sensitivity matrix of size $400$ by $2500$ that this was not successful. The test for the \texttt{GKB} approach was successful for this problem, but it is still rather small scale as compared to the problems considered here. Instead as stated, we return to the problem of assessing a suitable size of the projected space to be used for large scale inversion of magnetic and gravity data, using the techniques that provide an approximate \texttt{SVD} and hence efficient and automatic estimation of the regularization parameter concurrently with solving large scale problems. We use the method of Unbiased Predictive Risk Estimation (\texttt{UPRE}) for automatically estimating the regularization parameters, as extensively discussed elsewhere, \cite{Vogel:2002}.

\textit{Overview of main scientific contributions.} This work provides a comprehensive study of the application of the \texttt{2DFFT} in focusing inversion algorithms for gravity and magnetic potential field data sets.  Specifically, our main contributions are as follows.
(i) A detailed review of the mechanics for the inversion of potential field data using focusing inversion algorithms based on the iteratively regularized least squares algorithm in conjunction with the solution of linear systems using \texttt{GKB} or \texttt{RSVD} algorithms; (ii)  The extension of these approaches for the use of the \texttt{2DFFT} for all forward multiplications with the sensitivity matrix, or its transpose; (iii) Comparison of the computational cost when using the \texttt{2DFFT} as compared to the sensitivity matrix, or its transpose, directly, when implemented within the inversion algorithm, and dependent on the sizes of the projected spaces adopted for the inversion; (iv) Presentation of numerical experiments that confirm that the \texttt{RSVD} algorithm is more efficient than the \texttt{GKB} for the inversion of gravity data sets, for larger problems than previously considered; (v) A new comparison the use of \texttt{GKB} as compared for \texttt{RSVD} for the inversion of magnetic data sets, showing that \texttt{GKB} is to be preferred; (vi) Finally, all conclusions are confirmed by application on a practical data set, demonstrating that the methodology is suitable for focusing inversion of large scale data sets and can provide parameter reconstructions with more than $1$M variables using a laptop computer. 

The paper is organized as follows. In Section~\ref{sec:methodology} we present the general methodology used for the independent inversion of gravity and magnetic potential field data. The \texttt{BTTB} details are reviewed in Section~\ref{sec:forward} and stabilized inversion is reviewed in Section~\ref{sec:inversion}. Details for the numerical solution of the inversion formulation are provided in Section~\ref{sec:solution} and the algorithms are in Section~\ref{sec:algorithm}. The estimated computational cost of each algorithm, in terms of the number of floating point operations \texttt{flops} is given in Section~\ref{sec:compcost}. Numerical results applying the presented algorithms to synthetic and practical data are described in Section~\ref{sec:numerics}, with the details that apply to all computational implementations given in Section~\ref{sec:algdetails} and the generation of the  synthetic data used in the simulations provided in Section~\ref{sec:volsimulation}. Results assessing comparison of computational costs for one iteration of the algorithm for use with, and without, the \texttt{2DFFT} are discussed in Section~\ref{sec:costperiteration}. The  convergence  of the \texttt{2DFFT}-based algorithms for problems of increasing size is discussed in Section~\ref{sec:methodconvergence}. Validating results for the inversion of real magnetic data obtained over a portion of
the Wuskwatim Lake region in Manitoba, Canada are provided in Section~\ref{sec:real} and conclusions in Section~\ref{sec:conclusions}. \ref{Appendix:BTTB}  provides  brief details on the implementation of the computations using the embedding of the \texttt{BTTB} matrix in the \texttt{BCCB} matrix and the \texttt{2DFFT}, and supporting numerical evidence of the figures illustrating the results are provided in a number of tables in \ref{app:table}.

\section{Methodology}\label{sec:methodology}
\subsection{Forward Model and \texttt{BTTB} Structure}\label{sec:forward}
We consider the inversion of measured potential field data $\bfdo$ that describes the response at the surface 
due to unknown subsurface model parameters $\bfm$. The data and model parameters are connected   via the forward model
\begin{equation}\label{linearequation}
\bfdo=G \bfm,
\end{equation}
where $G$ is the sensitivity, or model, matrix. 
This linear relationship is obtained via the discretization of a Fredholm integral equation of the first kind, 
\begin{equation}\label{continuousforwardmodel}
d(a,b,c)=\int \int \int h(a,b,c,x,y,z) \zeta(x,y,z) dx \,dy\, dz,
\end{equation}
where exact values $\bfd$ and $\bfm$ are the discretizations of continuous functions $d$ and $\zeta$, respectively, and  $G$ in \eqref{linearequation} provides the discrete approximation of the integrals of the kernel function $h$ over the volume cells. For the specific kernels  associated with gravity and magnetic data, assuming for magnetic data that there is no remanence magnetization or self-magnetization,  $h$ is spatially invariant  in all dimensions, $h(a,b,c,x,y,z)=h(x-a,y-b,z-c)$ and \eqref{continuousforwardmodel} describes a convolution operation. 

\begin{figure}
\begin{center}
\includegraphics[width=.55\textwidth]{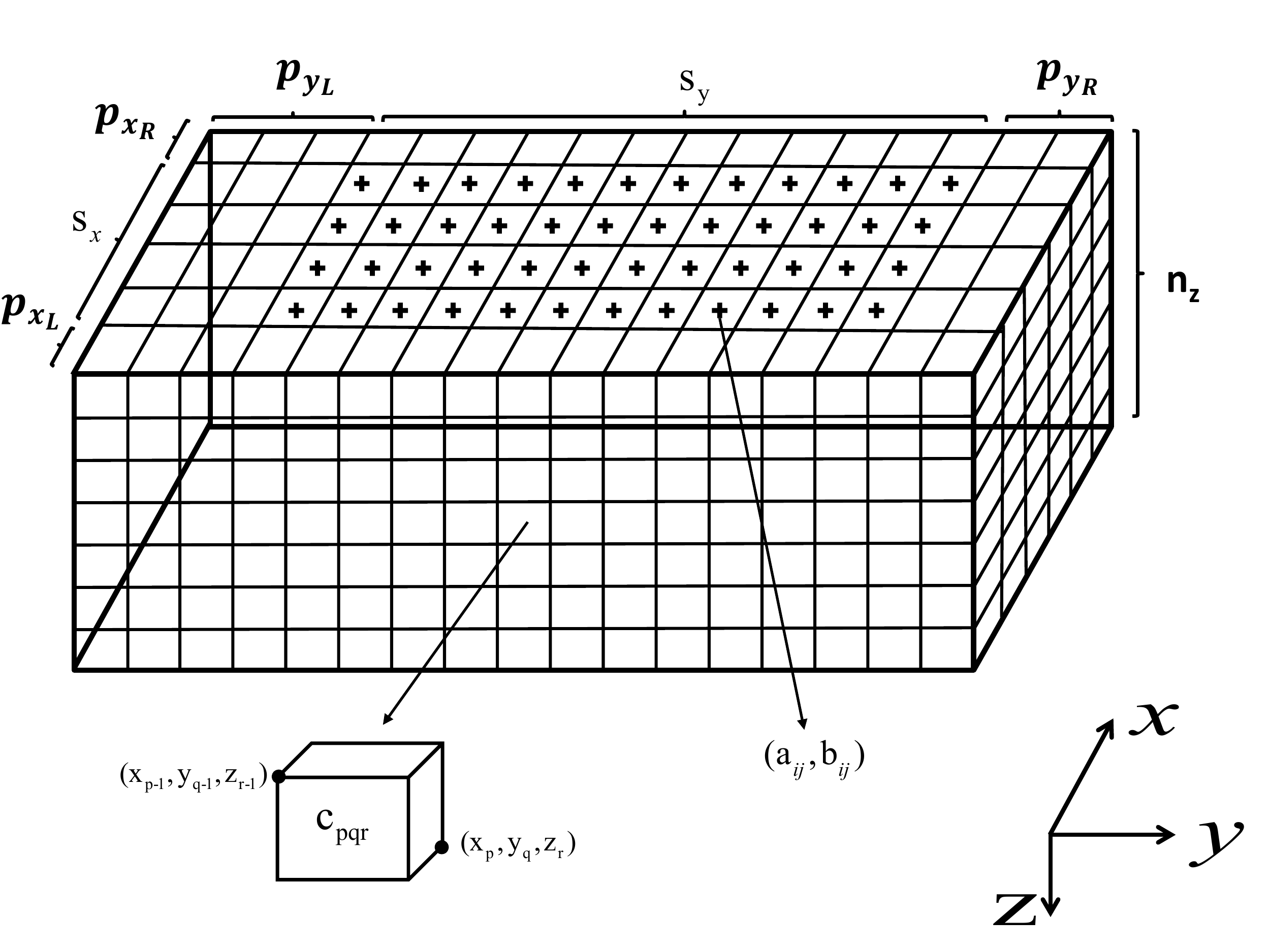}
\caption{The configuration of prism $c_{pqr}$, $1\le p\le \nsx+\padxl+\padxr=\nbx$, $1\le q \le \nsy+\padyl+\padyr=\nby$, $1\le r\le \nbz$, in the volume relative to a station on the surface at location $s_{ij}=(a_{ij}, b_{ij})$, $1\le i \le \nsx$, $1\le j \le \nsy$. Here the stations are shown as  located at the centers of the cells on the surface of the domain and that there are no measurements taken in the padded portion of the domain. \label{figure1}} \end{center}
\end{figure}

Using the formulation of the integral of the kernel as derived by \cite{haaz1953,Li1998} for the gravity kernel, and by  \cite{RaoBabu:91} for the magnetic kernel, sensitivity matrix $G$ decomposes by column blocks as 
\begin{equation}\label{blockG}
G=[G^{(1)}, \dots, G^{(\nbz)}],
\end{equation}
where block $\Gr$ is for the $r^{\mathrm{th}}$ depth layer. The individual entries in $G$ correspond to the projections of the contributions from prisms $c_{pqr}$ in the volume to measurement stations, denoted by $s_{ij}$, at or near the surface. The configurations of the volume and measurement domains are illustrated in Figure~\ref{figure1}. Here it is assumed that the measurement stations are all on the surface with coordinates $(a_i,b_j,0)$ in $(x,y,z)$. 
Prism $c_{pqr}$ of the domain has dimensions  $\dx$, $\dy$ and $\dz$  in $x$, $y$ and $z$ directions with coordinates that are integer multiples of $\dx$, $\dy$ and $\dz$, and is indexed by $1\le p\le \nsx+\padxl+\padxr=\nbx$, $1\le q \le \nsy+\padyl+\padyr=\nby$, and $1\le r\le \nbz$. This indexing assumes  that there is padding around the domain in $x$ and $y$ directions  by additional borders 
of $\padxl$, $\padxr$, $\padyl$ and $\padyr$ cells.  The distinction between the padded and unpadded portions of the domain is that there are no measurement stations in the padded regions. This yields $G\in \Rmn{m}{n}$  where $m=\nsx  \nsy$, and $n=\nbx \nby \nbz$, and  each $\Gr \in \Rmn{m}{\nbr}$, where $\nbr=\nbx \nby$. 

In \eqref{blockG}, $m\le \nbr \ll n$ and the system is drastically underdetermined for any reasonable discretization of the depth  ($z$) dimension  of the volume. Moreover, when $n$ is large the use of the matrix $G$ requires both significant computational cost for evaluation of matrix-matrix operations and significant storage. Without taking account of structure in $G$, and assuming that a dot product of real vectors of length $n$ requires $2n$ floating point operations ($\flops$), calculating $G H$, for $H\in\Rmn{n}{p}$, takes $\mathcal{O}(2nmp)$ $\flops$ and storage of matrix $G$  uses approximately $8mn\times 1e^{-9}$GB\footnote{We assume one double floating point number requires $8$ bytes and note $1$ byte is $10^{-9}$GB.}. For example, suppose $p=m=n/8$ and $n=10^6$, then storage of $G$ requires approximately $1000$GB, and the single matrix multiplication uses $\approx 10^{18}/32~ \flops$ or $10^7 \mathrm{G}\flops$, without any consideration of additional software and system overheads.  These observations limit the ability to do large scale stabilized inversion of potential field data in real time using  current desktop computers, or laptops, without taking into account any further information on the structure of $G$. This is the topic of the further discussion here. 

\cite{bruun2007} observed that the configuration of the locations of the stations in relation to the domain discretization is significant in generating $\Gr$ with structure that can be effectively utilized to improve the efficiency of operations with $G$ and to reduce the storage requirements. Assuming that the stations are always placed uniformly with respect to the domain prisms, and provided that the distances between stations are fixed in $x$ and $y$, then matrix $\Gr$ for the gravity kernel has symmetric \texttt{BTTB} structure ($\sbttb$). Then, it is possible to embed $\Gr$ in a $\bccb$  matrix and matrix operations can be efficiently performed using  the \texttt{2DFFT}, as explained in \citep{Vogel:2002}. This structure was also discussed and then utilized for efficient forward operations with $G$ in  \cite{ChenLiu:18}. There it was assumed that the stations are placed symmetrically with respect to the domain coordinates, as illustrated for the staggered configuration in Figure~\ref{figure1} with the stations at the center of the cells on the surface.  With respect to the magnetic kernel, 
\cite{bruun2007} demonstrated $\Gr$ can also exhibit $\bttb$ structure, but they did not use the standard computation of the magnetic kernel integral as described in \cite{RaoBabu:91}. On the other hand, a thorough derivation of the $\bttb$ structure for $\Gr$ using the   approach of \cite{RaoBabu:91} has been given in \cite{hogue2019tutorial}. That analysis also considered for the first time the use of the padding for the domain and the modifications required in the generation of the required entries in the matrix $\Gr$. It should be noted, as shown in \cite{hogue2019tutorial}, that regardless of whether operations with $G$ are implemented using the \texttt{2DFFT} or by direct multiplication, it is far faster to generate $G$ taking advantage of the $\bttb$ structure.  Here, we are concerned with efficient stabilized inversion of potential field data using this $\bttb$ structure, and thus refer to \ref{Appendix:BTTB} for a brief discussion of the implementation of the needed operations using $G$ when 
implemented using the \texttt{2DFFT},  and point to \cite{hogue2019tutorial} for the details.  

\subsection{Stabilized Inversion}\label{sec:inversion}
The solution of \eqref{linearequation} is an ill-posed problem; even if $G$ is well-conditioned the problem is underdetermined because $m \ll n$. There is a considerable literature on the solution of this ill-posed problem and we refer in particular to \cite{VaReSh:20} for a relevant overview, and specifically the use of the unifying framework for determining an acceptable solution of \eqref{linearequation} by stabilization. Briefly, here we estimate $\bfm^*$ as the minimizer of the nonlinear objective function $\Phia(\bfm)$ subject to bound constraints $\bfmmin\le \bfm \le \bfmmax$
\begin{equation}\label{globalfunction}
\bfm^*=\argmin{\bfmmin\le \bfm \le \bfmmax} \{\Phia(\bfm) \}= \argmin{\bfmmin\le \bfm \le \bfmmax}\{\Phid(\bfm)+\alpha^2 \Phis(\bfm)\}. 
\end{equation}
Here $\alpha$ is a regularization parameter which trades off the relative weighting of the two terms $\Phid(\bfm)$ and $\Phis(\bfm)$, which are  respectively the weighted data misfit  and stabilizer, given by
\begin{equation}\label{dfandstab}
\Phid(\bfm) = \|\Wd (G\bfm-\bfdo)\|_2^2, \text{ and } \Phis(\bfm)=\|\Wh \Wz \WL D (\bfm-\bfma)\|_2^2.
\end{equation}
The weighting matrices $\Wd$, $\Wh$, $ \Wz$ and $\WL$ are all diagonal, with dimensions that depend on the size of $D$,  which can be used to yield an approximation for a derivative. Here, while  we assume throughout that $D=I_{n \times n}$\footnote{We use $I_{n \times n}$ to denote the identity matrix of size $n \times n$.} and refer to \citep[Eq. (5)]{VaReSh:20}  for the modification in the weighting matrices that is required for derivative approximations using $D$, we present this general formulation in order to place the work in context of  generalized Tikhonov inversion. We also use $\bfma=\mathbf{0}$, but when initial estimates for the parameter are available, perhaps from physical measurements, note that these can be incorporated into $\bfma$ as  an initial estimate for $\bfm$. The weighting matrix $\Wd$ has entries  $(\Wd)_{ii}=1/\sigma_i$ where we suppose that the measured data can be given by $\bfdo=\bfde+\bfeta$, where $\bfde$ is the exact but unknown data, and $\bfeta$ is a noise vector drawn from uncorrelated Gaussian data with variance components $\sigma_i^2$.   

Whereas stabilizer matrix $\WL$ in $\W=\Wh \Wz\WL$  depends on $\bfm$, $\Wh$ and $\Wz$ are constant  hard constraint and constant depth weighting matrices. Although $\Wh$ can be used to impose specific known values for entries of $\bfm$, as discussed in \citep{BoCh:2001}, we will use $\Wh=I_{n \times n}$. Depth weighting $\Wz $ is routinely used in the context of potential field inversion and is imposed to counteract the natural decay of the kernel with depth. With the same column structure as for $G$, $\Wz=\blockdiag(\Wz^{(1)}, \dots, \Wz^{(\nbz)})$ where 
$\Wz^{(r)}=(.5(z_r+z_{r-1}))^{-\beta} I_{\nbr \times \nbr}$, $.5(z_r+z_{r-1})$ is the average depth for depth level $r$, and $\beta$ is a parameter that depends on the data set, \citep{LiOl:96}.  Now, diagonal matrix $\WL$ depends on the parameter vector $\bfm$ via  $i^{\mathrm{th}}$ entry given by
 \begin{equation}\label{WLdefn}
(\WL)_{ii} = \left((\bfm_i-(\bfma)_i)^2+\epsilon^2\right)^{\frac{\lambda-2}{4}}, \quad i=1\dots n,
\end{equation}
where parameter $\lambda$  determines the form of the stabilization, and  focusing parameter $0<\epsilon \ll 1$  is chosen to avoid division by zero. We use $\lambda=1$ which yields an approximation to the $L_1$ norm  as described in \citep{WoRo:07}, and is preferred for inversion of potential field data, although we note that the implementation makes it easy to switch to $\lambda=0$, yielding a solution which is compact, or $\lambda=2$ for a smooth solution.  Based on prior studies we use $\epsilon^2=1e-9$, \citep{VRA:2017}. 

\subsection{Numerical Solution}\label{sec:solution}
We first reiterate that  \eqref{globalfunction} is only nonlinear in $\bfm$ through the definition of $\WL$. Supposing that $\WL$ is constant and that  $\mathrm{null}(\Wd G) \cap \mathrm{null}(\W)=\emptyset$,  then the solution $\bfm^*$ of \eqref{globalfunction} without the bound constraints is given analytically by 
\begin{equation}\label{constantWL}
\bfm = \bfma+(G^T\Wd^T\Wd G +\alpha^2 \W^T\W)^{-1}G^T\Wd^T \Wd(\bfdo- G\bfma). 
\end{equation}
Equivalently, assuming that $\W$ is invertible, and defining $\tildeG=\Wd G \W^{-1}$, $\tilder=\Wd(\bfdo-G\bfma)$ and  $\bfy=\bfm-\bfma$, then $\bfy$ solves the normal equations 
\begin{equation}\label{normalequations}
\bfy=\W^{-1}(\tildeG^T\tildeG+\alpha^2 I)^{-1} \tildeG^T\tilder,
\end{equation}
and $\bfm^*$ can be found by restricting $\bfy+\bfma$ to lie within the bound constraints. 

Now \eqref{normalequations} can be used to obtain the iterative solution for \eqref{globalfunction} using the iteratively reweighted least squares (\texttt{IRLS})  as described in \cite{VaReSh:20}. Specifically, we use superscript $k$ to indicate a variable at an iteration $k$, and replace $\alpha$ by $\alpha^{(k)}$, $\WL$ by matrix $\WLk$  with entries $(\WLk)_{ii}=\left((\bfm^{(k-1)}_i-\bfm^{(k-2)}_i)^2+\epsilon^2\right)^{\frac{\lambda-2}{4}}$ and $\bfm-\bfma$   by $\bfm-\bfm^{(k-1)}$,  initialized with $\WL^{(1)}=I$, and $\bfm^{(0)}=\bfma$ respectively. Then  $\bfy^{(k)}$ is found as the solution of the normal equations \eqref{normalequations}, and $\bfm^{(k)}$ is the restriction of $\bfy^{(k)}+\bfm^{(k-1)}$ to the bound constraints.

This use of the \irls algorithm for the incorporation of the stabilization term $\Phis$
contrasts the implementation discussed in \citep{ZhangWong:15} for the inversion of potential field gravity data. 
In their presentation, they considered the solution of the general smoothing Tikhonov formulation described by 
\eqref{normalequations} for general fixed smoothing operator $D$ replacing $\WL$. For the solver, they used the re-weighted regularized conjugate solver for iterations to improve $\bfm^{(k)}$ from $\bfma$. They also included  a penalty function to impose positivity in $\bfm^{(k)}$, depth weighting to prevent the accumulation of the solution at the surface, and adjustment of $\alpha$ with iteration $k$ to encourage decrease in the data fit term.   Moreover, they showed that it is possible to pick approximations $D$ which also exhibit \texttt{BTTB} structure for each depth layer, so that $D\bfx$ can also be implemented by layer using the 2DFFT. Although we do not consider the unifying stabilization framework here with general operator $D$ as described in \cite{VaReSh:20}, it is a topic for future study, and a further extension of the work, therefore, of \cite{ZhangWong:15} for the more general stabilizers.  In the earlier work of the use of the  \texttt{BTTB}  structure arising in potential field inversion,  \cite{bruun2007} investigated the use of a truncated  \texttt{SVD} and the conjugate gradient least squares method for the minimization of the data fit term without regularization. They also considered the direct solution of the constant Tikhonov function with  $\WL=I$ and a fixed regularization parameter $\alpha$, for which the solution uses the filtered \svd in the small scale. Here, not only do we use the unifying stabilization framework, but we also estimate $\alpha$ at each iteration of the \irls algorithm. 
The \irls algorithm is implemented with two different solvers that yield effective approximations of a truncated SVD. One is based on the randomized singular value decomposition  (\texttt{RSVD}), and the second uses the Golub Kahan Bidiagonalization  (\texttt{GKB}).

\subsection{Algorithmic Details}\label{sec:algorithm}
The \irls algorithm relies on the use of an appropriate solver for finding $\bfy^{(k)}$ as the solution of the normal equations \eqref{normalequations} for each update $k$, and a method for estimating the regularization parameter $\alpha^{(k)}$. While any suitable computational scheme can be used to update $\bfm^{(k)}$, the determination of $\alpha^{(k)}$ automatically can be challenging. But if the solution technique generates the \svd for $\tildeGk$, or an approximation to the  \texttt{SVD}, such as by use of the  \texttt{RSVD} or \texttt{GKB}  factorization for $\tildeGk$, then there are many efficient techniques that can be used such as the unbiased predictive risk estimator (\texttt{UPRE}) or  generalized cross validation (\texttt{GCV}). 
The obtained estimate for  $\alpha^{(k)}$ depends on the estimator used and there is extensive literature on the subject, e.g. \cite{hansenbook}. Thus, here, consistent with earlier studies on the use of the \texttt{GKB} and \texttt{RSVD} for stabilized inversion,  we use the \texttt{UPRE}, denoted by $U(\alpha)$,  for all iterations $k>1$, and refer to \cite{RVA:15} and \cite{saeed6,vatankhah2019improving} for the details on the \texttt{UPRE}. The \texttt{GKB} and \texttt{RSVD} algorithms play, however, a larger role in the discussion and thus for clarity are given here as Algorithms~\ref{Alg:GKB} and \ref{Alg:RSVDQ}, respectively.

For the use of the \texttt{GKB} we note that Algorithm~\ref{Alg:GKB} uses the factorization $\tildeG A_{t_p}=H_{t_p+1}B_{t_p}$, where $A_{t_p} \in \Rmn{n}{t_p}$ and $H_{t_p+1}\in\Rmn{m}{t_p+1}$.   Steps~\ref{MGS} and \ref{MGS2} of Algorithm~\ref{Alg:GKB} apply the modified Gram-Schmidt re-orthogonalization to the columns of $A_{t_p}$ and $H_{t_p+1}$, as is required to avoid the loss of column orthogonality. This factorization is then used in Step~\ref{BtSVD} to obtain the rank $t_p$ approximate \svd given by $\tildeG=(H_{t_p+1}U_{t_p})\Sigma_{t_p}(A_{t_p}V_{t_p})^T$. The quality of this approximation depends on the conditioning of $\tildeG$, \citep{PaigeSau2}. In particular,  the projected system of the \texttt{GKB} algorithm inherits the ill-conditioning of the original system, rather than just the dominant terms of the full \svd expansion. Thus, the approximate singular values include dominant terms that are good approximations to the dominant singular values of the original system, as well as very small singular values that approximate the tail of the singular spectrum of the original system.  The accuracy of the dominant terms increases quickly with increasing $t_p$, \cite{PaigeSau2}. Therefore,  to effectively  regularize the dominant spectral terms from the rank $t_p$ approximation, in Step~\ref{tupre}  we use the truncated \texttt{UPRE}  that was discussed and introduced in \cite{VRA:2017}.  Specifically, a suitable choice for $\alpha^{(k)}$ is found 
using the truncated \svd of $B_{t_p}$ with $t$ terms. Then, in Step~\ref{yupdategkb}, $\bfy^{(k)}$ is found using   all terms in the expansion of $B_{t_p}$. The matrix $\Gamma(\alpha, \Sigma)$ in Step~\ref{yupdategkb} is the diagonal matrix with entries $\sigma_i/(\sigma_i^2+\alpha^2)$.  In our simulations we  use $t_p = \floor(1.05\,t)$  corresponding to  $5\%$ increase in the space obtained. This contrasts to using just $t$ terms and will include terms from the tail of the spectrum. Note, furthermore, that the top $t$ terms, from the projected space of size $t_p>t$ will 
be more accurate estimates of the true dominant $t$ terms than if obtained with $t_p=t$. Effectively, by using a $5\%$ increase of $t$  in the calculation of $t_p$, we assume that the first $t$ terms from the $t_p$ approximation  provide good approximations of the dominant $t$ spectral components of the original matrix $\tildeG$.  We reiterate that the presented algorithm depends on parameters $t_p$ and $t$. At Step~\ref{tupre} in Algorithm~\ref{Alg:GKB} $\alpha^{(k)}$ is found using the projected space of size $t$ but  the update for $\bfy$ in Step~\ref{yupdategkb}  uses the oversampled projected space of size $t_p$.  The results presented 
for the synthetic tests will demonstrate that this uniform choice for $t_p$ is a suitable compromise between taking $t_p$ too small and contaminating
 the solutions by components from the less accurate approximations of the small components, and a reliable, 
 but larger, choice for $t_p$ that provides a good approximation of the dominant terms within reasonable computational cost.

\begin{center}
\begin{algorithm}[htb!] 
\caption{Use \texttt{GKB} algorithm  for factorization  $\tildeG A_{t_p}=H_{t_p+1}B_{t_p}$ and obtain solution $\bfy$ of \eqref{normalequations}. \label{Alg:GKB}}
\SetAlgoLined \LinesNumbered
  \KwIn{$\tilder \in  \Rm{m}$, 
$\tildeG \in \Rmn{m}{n}$, a target rank $t$ and  size of oversampled projected problem $t_p$, $t<t_p \ll m$.}
\KwOut {\  $\alpha$ and $\bfy$.\label{gkbfact}}
Set $\bfa=\zeros(n,1)$, $B=\texttt{sparse}(\zeros(t_p+1,t_p))$, $H=\zeros(m,t_p+1)$, $A=\zeros(n,t_p)$\;
Set $\beta=\|\tilder\|_2$, $\bfh=\tilder/\beta$, $H(:,1)=\bfh$\;
\For{$i=1:t_p$ }{
$\bfb=\tildeG^T \bfh - \beta \bfa$ \label{gkbGT}\;
\For{$j=1:i-1$} {$\bfb=\bfb-(A(:,j)^T\bfb)A(:,j)$ (modified Gram-Schmidt (\texttt{MGS}))\label{MGS}}
$\gamma=\|\bfb\|_2$, $\bfa=\bfb/\gamma$, $B(i,i)=\gamma$, $A(:,i)=\bfa$\;
$\bfc=\tildeG \bfa- \gamma \bfh$ \label{gkbG}\;
\For{$j=1:i$} {$\bfc=\bfc-(H(:,j)^T\bfc)H(:,j)$ (\texttt{MGS}) \label{MGS2}}
$\beta=\|\bfc\|_2$, $\bfh=\bfc/\beta$, $B(i+1,i)=\beta$, $H(:,i+1)=\bfh$\;}
\texttt{SVD} for sparse matrix: $U_{t_p} \Sigma_{t_p} V_{t_p}^T = \texttt{svds}(B,t_p)$\label{BtSVD}\;
Apply \texttt{UPRE} to find $\alpha$  using $U_{t_p}(:,1:t)$ and $\Sigma_{t_p}(1:t,1:t)$\label{tupre}\;
Solution $\bfy=\|\tilder\|_2A_{t_p}V_{t_p} \Gamma(\alpha,\Sigma_{t_p}) U_{t_p}(1,:)^T $\label{yupdategkb}\; 
\end{algorithm}
\end{center}

\begin{center}
\begin{algorithm}[htb!] 
\caption{Use \texttt{RSVD}  with one power iteration  to compute an approximate \svd of $\tildeG$ and obtain solution $\bfy$ of \eqref{normalequations}\label{Alg:RSVDQ}}
\SetAlgoLined \LinesNumbered
  \KwIn{$\tilder \in  \Rm{m}$,
$\tildeG \in \Rmn{m}{n}$, a target matrix rank $t$ and size of oversampled projected problem $t_p$,  $t<t_p \ll m$.}
\KwOut {\ $\alpha$ and $\bfy$.}
 Generate a Gaussian random matrix $\Omega \in \Rmn{t_p}{m}$ \label{omega}\;
 $Y=\Omega \tildeG \in \Rmn{t_p}{n}$  \label{Y}\;
$[Q,\sim]=\texttt{qr}(Y^{T},0)$, $Q \in \Rmn{n}{t_p}$. \label{Q}(economic QR decomposition) \;
$Y=\tildeG Q \in \Rmn{m}{t_p}$ \label{Y1}\;
 $[Q,\sim]=\texttt{qr}(Y,0)$, $Q \in \Rmn{m}{t_p}$ \label{Q1}\;
 $Y=Q^T \tildeG $, $Y \in \Rmn{t_p}{n}$ \label{Y2}\;
$[Q,\sim]=\texttt{qr}(Y^T,0)$, $Q \in \Rmn{n}{t_p}$ \label{Q2}\;
$ B=\tildeG Q \in \Rmn{m}{t_p}$ \label{B}\;
 Compute  $Y=B^TB \in \Rmn{t_p}{t_p}$ \label{BTB}\;
 Eigen-decomposition of $B^TB$: $[\tilde{V} , D ]=\texttt{eig}((Y+Y^T)/2)$  \label{eigen}\;
 $S=\texttt{diag}(\sqrt{|\texttt{real}(D)|})$, $[S, \texttt{indsort}]=\texttt{sort}(S,\mathrm{'}descend')$\;
$\tilde{\Sigma}_t=\texttt{diag}(S(1:t))$, $\tilde{V}=\tilde{V}(:,\texttt{indsort}(1:t))$, $\tilde{U}=\tilde{V}./(S(1:t)^T)$\;
 Apply \texttt{UPRE} to find $\alpha$  using $\tilde{U}$, $\tilde{\Sigma}_t$, and $B^T\tilder$\;
 Solution $\bfy=Q\tilde{V} \Gamma(\alpha,\tilde{\Sigma}_t )\tilde{U}^T (B^T\tilder)$\;
Note if we form $ \tilde{V}_t=Q \tilde{V}$; and $ \tilde{U}_t=B \tilde{U}$ $  \tilde{\Sigma}_{t}^{-1}$,\label{SVDcomponent} then 
$\tilde{U}_t \tilde{\Sigma}_t  \tilde{V}_t^T$ is a $t$-rank approximation of matrix $\tildeG$ \;
\end{algorithm}
\end{center}
 
The algorithm presented in Algorithm~\ref{Alg:RSVDQ}, denoted as \texttt{RSVD},  includes a  single power iteration  in Steps~\ref{Q} to \ref{Y2}.  Without the use of the power iteration in the \texttt{RSVD} it is necessary to use larger projected systems in order to obtain a good approximation of the singular space of the original system, \cite{HMT:11}. Further, it was shown in \citep{vatankhah2019improving}, that when using \texttt{RSVD} for potential field inversion, it is better to apply a power iteration. 
Skipping the power iteration steps leads to  a less accurate approximation of the dominant singular space. Moreover, the gain from taking more than one power iteration is insignificant as compared to the  increased computational time required. As with the \texttt{GKB}, the \texttt{RSVD}, with and without power iteration, depends on two parameters $t$ and $t_p$, where here $t$ is the target rank and $t_p$ is size of the oversampled system, $t_p>t$. For given $t$ and $t_p$ the algorithm uses an eigen decomposition  with $t_p$ terms to find the \svd approximation of $\tildeG$ with $t_p$ terms. Hence, the total projected space is of size $t_p$, the size of the oversampled system, which is then restricted to size $t$ for estimating the approximation of $\tildeG$.  \footnote{We note that using $(Y+Y^T)/2$ in Step~\ref{eigen} of Algorithm~\ref{Alg:RSVDQ}, rather than $Y$, assures that the matrix is symmetric which is important for the efficiency of $\texttt{eig}$.}

It is clear that the \texttt{RSVD} and \texttt{GKB} algorithms provide approximations for the spectral expansion of $\tildeG$, with the  quality of this approximation dependent on both $t$ and $t_p$, and hence the quality of the obtained solutions $\bfy^{(k)}$ at a given iteration is dependent on these choices for $t$ and $t_p$. As noted, the \texttt{GKB} algorithm inherits the ill-conditioning of $\tildeG$ but the \texttt{RSVD} approach provides the dominant terms, and is not impacted by the tail of the spectrum. Thus, we may not expect to use the same choices for the pairs $t$ and $t_p$ for these algorithms. \cite{vatankhah2019improving} investigated the choices for $t$ and $t_p$ for both gravity and magnetic kernels. When using \texttt{RSVD} with the single power iteration they showed that suitable choices for $t$, when $t_p=t+10$, are $t \gtrsim m/s$, where $s\approx 8$ for the gravity problem and $s\approx 4$ for magnetic data inversion. This  contrasts using $s \approx 6$ and $s\approx 2$ without power iteration, for gravity and magnetic data inversion, respectively. On the other hand, results presented in \citep{VRA:2017} suggest using  $t_p \gtrsim m/s$ where $s \lesssim 20$ for the inversion of gravity data using the \texttt{GKB}~algorithm. This leads to the range of $t$ used in the simulations to be discussed in Section~\ref{sec:numerics}. We use the choices $s=40$, $25$, $20$, $8$, $6$, $4$ and $ 3$. This permits  a viable comparison of cost and accuracy for \texttt{GKB} 
and \texttt{RSVD}.  Observe that, for the large scale cases considered here, we chose to test with least $s= 3$ rather than $s= 2$.
 Indeed, using $s=2$ generates a    large overhead of testing for a wide range of parameter choices, and suggests that we would need relatively large subspaces defined by $t=m/2$, offering limited gain in speed and computational cost. 

\subsection{Computational Costs}\label{sec:compcost}
Of interest is the computational cost of (i) the practical  implementations of  the \texttt{GKB} or \texttt{RSVD} algorithms for finding the parameter vector $\bfy^{(k)}$ when operations with matrix $G$ are implemented using the \texttt{2DFFT}, and (ii) the associated impact of the choices of $t_p$ on the comparative costs of these algorithms with increasing $m$ and $n$.  In the estimates we focus on the dominant costs in terms of $\flops$, recalling that the underlying cost of a dot product of two vectors of length $m$ is assumed to be $2m$. Further, the costs ignore  any overheads of data movement and data access. 

First, we address the evaluation of matrix products with $\tildeG$ or $\tildeG^T$ required at Steps~\ref{gkbGT} and \ref{gkbG}  of Algorithm~\ref{Alg:GKB} and \ref{Y}, \ref{Y1}, \ref{Y2} and \ref{B} of Algorithm~\ref{Alg:RSVDQ}.  Matrix operations with $G$, rather than $\tildeG$, use the \texttt{2DFFT}, as described in \ref{Appendix:BTTB} for $G\bfx$, $G^T\bfy$ and $\bfy^TG$, based on the discussion in \citep{Vogel:2002}. The cost of a single matrix vector operation in each case is $4\nbx\nby \nbz \log_2(4\nbx\nby)= 4n \log_2(4\nbr)$. This  includes the operation of the \texttt{2DFFT} on the reshaped components of $\bfx_r \in \Rm{\nbx\nby}$ and the inverse \texttt{2DFFT} of the component-wise product  of $\hat{\bfx}_r$ with $\hat{G}^{(r)}$, for $r=1:\nbz$, but ignores the lower cost of forming the component-wise products and summations over vectors of size $\nbr$. Thus, multiplication with a matrix of size $n\times t_p$ has dominant cost
\begin{equation}\label{2dfftmultcostd}
4nt_p \log_2(4\nbr),
\end{equation}
in place of $2mnt_p$. 
In the \irls algorithm we need to use operations with $\tildeG=\Wd G \W^{-1}$ rather than $G$. But this is handled   immediately by using suitable component-wise multiplications of the diagonal matrices and vectors. Specifically, 
\begin{equation}\label{forwardmult}
\tildeG\bfx=\Wd(G(\W^{-1} \bfx))
\end{equation}
and the \texttt{2DFFT} is applied for the evaluation of $G\bfw$ where  $\bfw=\W^{-1}\bfx$. Then, given $\bfz=G\bfw$, a second  component-wise multiplication, $\Wd \bfz$,  is applied   to complete the process. Within the algorithms, matrix-matrix operations are also required but, clearly, operations $\tildeGk X$, $(\tildeGk)^TZ$, $Z^T\tildeGk$ are just loops over the relevant columns (or rows) of the matrices $X$ and $Z$, with the appropriate weighting matrices provided before and after application of the \texttt{2DFFT}. The details are provided in \ref{Appendix:BTTB}.

Now, to determine the impact of the choices for $t$ (and $t_p$) we estimate the dominant costs for finding the solution of \eqref{normalequations} using the \texttt{GKB} and \texttt{RSVD} algorithms. This is the major cost of the \irls algorithm. The assumptions for the dominant costs of standard algorithms, given in Table~\ref{table2},   are quoted from \cite{golub2013matrix}. But note that the cost for $\texttt{eig}$ depends significantly on problem size and symmetry. Here $t$ can be quite large, when $m$ is large, but the matrix is symmetric, hence we use the estimate $9t^3$, \cite[Algorithm 8.3.3]{golub2013matrix}. To be complete we note that $\texttt{svds}$ for the sparse bidiagonal matrix $B$ is achieved at cost which is at most quadratic in the variables. A comment on the cost of the $\texttt{qr}$ operation is also required. Generally, in forming the $QR$ factorization of a matrix we would maintain the information on the Householder reflectors that are used in the reduction of the matrix to upper triangular form, rather than accumulating the matrix $Q$. The cost is reduced significantly if $Q$ is not accumulated. But, as we can see from Steps~\ref{Y}, \ref{Y1}, \ref{Y2} and \ref{B} of Algorithm~\ref{Alg:RSVDQ}, we will need to evaluate products of $Q$ with $\tildeG$ or its transpose. To take advantage of the \texttt{2DFFT} we then need to first evaluate a product of $Q$ with a diagonal scaling matrix, which amounts to accumulation of matrix $Q$. Experiments, that are not reported here, show that it is more efficient to accumulate $Q$ as given in Algorithm~\ref{Alg:RSVDQ}, rather than to  to first evaluate the product of $Q$ with a diagonal scaling matrix without pre accumulation. Then,  the cost for accumulating $Q$  is $2t^2(m-t/3)$ for a matrix of size $m\times t$, \cite[page 255]{golub2013matrix} yielding a total cost for the \texttt{qr} step of $4t^2(m-t/3)$, as also reported in \cite{XiangZou:13}.
\begin{table}[htb!]
\begin{center}
\begin{tabular}{c|c|c|c|c|c}
\hline
 $GX$& $G^TY$&$\texttt{svds}(B)$
 &$\texttt{MGS}(C)$&$\texttt{eig}(A^TA)$&$[Q,\sim]=\texttt{qr}(Z)$\\ \hline
 $2mnt$&$2mnt$&$6t(m+t)$
 &$2mt^2$&$9t^3$&$4t^2(m-t/3)$\\
 \hline 
\end{tabular}
\caption{Computational costs for standard operations. Matrix $G\in\Rmn{m}{n}$, $X\in\Rmn{n}{t}$, $Y\in\Rmn{m}{t}$, sparse bidiagonal $B\in\Rmn{t+1}{t}$, $A^TA\in\Rmn{t}{t}$, and $Z\in\Rmn{m}{t}$.  The modified Gram-Schmidt for $C\in\Rmn{m}{i}$ is repeated for $i=1:t$, yielding the given estimate. These costs use the basic unit that the inner product $\bfx^T\bfx$ for $\bfx$ of length $n$ requires $2n$ operations. \label{Table1}}
\end{center}
\end{table}

Using the results in Table~\ref{Table1} we can estimate the dominant costs of Algorithms~\ref{Alg:GKB} and \ref{Alg:RSVDQ}. In the estimates we do not distinguish between costs based on $t_p$ or $t$, noting  $t_p=\floor(1.05\,t)$ and $t=m/s$. We also ignore the distinction between $m$ and $\nbr$, where $\nbr>m$ for padded domains. Moreover, the cost of finding $\alpha^{(k)}$ and then evaluating $\bfy^{(k)}$ is of lower order than the dominant costs involved with finding the needed factorizations. Using $\mathit{LOT}$ to indicate the  lower order terms that are ignored, and assuming the calculation without the use of the \texttt{2DFFT},  the most significant terms yield
\begin{eqnarray}\label{costgkbwithg}
\texttt{CostG}_{\texttt{GKB}} &=&4nmt  + 2t^2(n+m)  +\mathit{LOT}\\ 
\texttt{CostG}_{\texttt{RSVD}} &=& 8nmt +4t^2(2n+m-t) +2mt^2+9t^3+\mathit{LOT} \nonumber  \\
&=& 8nmt +4t^2(2n+3/2m)+5t^3+ \mathit{LOT}. \label{costrsvdwithg}
\end{eqnarray}
When using the \texttt{2DFFT},  the first two entries $2mnt$ in Table~\ref{Table1} are replaced  by $4nt \log_2(4\nbr)$. Then, using $m\approx \nbr$, it is just the first term in each estimate that is replaced leading to the costs with the \texttt{2DFFT} as   
\begin{eqnarray}\label{costgkb}
\texttt{Cost}_{\texttt{GKB}} &=&8nt \log_2(4m) + 2t^2(n+m)  +\mathit{LOT}\\ 
\texttt{Cost}_{\texttt{RSVD}}&=& 16nt \log_2(4m) +4t^2(2n+3/2m)+5t^3+ \mathit{LOT}. \label{costrsvd}
\end{eqnarray}
Both pairs of equations suggest, just in terms of \texttt{flop} count,  that $\texttt{Cost}_{\texttt{RSVD}}  > 2\,\texttt{Cost}_{\texttt{GKB}}$. Thus, we would hope to use a smaller $t$ for the \texttt{RSVD}, than for the \texttt{GKB}, in order to obtain a comparable cost. This expectation contradicts earlier experiments contrasting these algorithms for the inversion of gravity data, using the \texttt{RSVD} without power iteration, as discussed in \cite{saeed6}. Alternatively, it would be desired that the \texttt{RSVD} should converge in the \irls far faster than the \texttt{GKB}. Further, theoretically, the gain of using the \texttt{2DFFT} 
is that the major terms are $8t^2n$ and $2t^2n$ for the \texttt{RSVD} and \texttt{GKB}, respectively. as compared to $8nmt>8t^2n$ and $4mnt>2t^2n$, noting $t<m$.  Specifically, even though the costs should  go up with order $nt^2$ eventually with the \texttt{2DFFT}, this is still far slower than the increase  $mnt$ that arises without taking advantage of the structure. 

Now, as discussed in \cite{XiangZou:13}, measuring the computational cost just in terms of the \texttt{flop} count can be misleading. It was noted by  \cite{XiangZou:13} that a distinction between the \texttt{GKB} and \texttt{RSVD} algorithms, where the latter is without the power iteration, is  that the operations required in the \texttt{GKB} involve many \texttt{BLAS2} (matrix-vector) operations, requiring repeated access to the matrix or its transpose, as compared to \texttt{BLAS3}  (matrix-matrix) operations for \texttt{RSVD} implementations. On the other hand, within the \texttt{qr} algorithm, the Householder operations also involve \texttt{BLAS2} operations. Hence, when using \textsc{Matlab}, the major distinction should be  between the use of functions that are \texttt{builtin} and compiled, or are not compiled. In particular, the functions \texttt{qr} and \texttt{eig} are  \texttt{builtin} and hence optimized, but all other operations that are used in the two algorithms do not use any compiled code. Specifically, there is no compiled option for the \texttt{MGS} used in steps~\ref{MGS} and \ref{MGS2} of Algorithm~\ref{Alg:GKB}, while almost all operations in Algorithm~\ref{Alg:RSVDQ} use \texttt{builtin} functions or \texttt{BLAS3} operations for matrix products that do not involve the matrices with \texttt{BTTB} structure.  Thus, in the evaluation of the two algorithms in the \textsc{Matlab} environment, we will consider computational costs directly, rather than just the estimates given by \eqref{costgkb} -\eqref{costrsvd}. On the other hand, the estimates of the \texttt{flop} counts should be relevant for higher-level programming environments, and are thus relevant more broadly. We also note that in all implementations none of the results quoted will use multiple cores or GPUs.

\section{Numerical Experiments}\label{sec:numerics}
We now validate the fast and efficient methods for inversion of potential field data using the \texttt{BTTB}  structure of the gravity and magnetic kernel matrices. 
\subsection{Implementation parameter choices}\label{sec:algdetails} 
Diagonal depth weighting matrix $\Wz$ uses $\beta=0.8$ for the gravity problem, and $\beta=1.4$ for the magnetic problem, consistent with recommendations in  
\cite{LiOl:98} and \cite{Pilkington:97}, respectively. Diagonal $\Wd$ is determined by the noise in the  data, and hard constraint matrix $\Wh$ is taken to be the identity. Moreover, we use $\bfma=0$, indicating no imposition of prior information on the parameters. Regularization parameter $\alpha^{(k)}$ is found using the \texttt{UPRE} method for $k>1$, but initialized with  appropriately large $\alpha^{(1)}$ given by 
\begin{equation}\label{eq:alpha1}
\alpha^{(1)}= \left(\frac{n}{m}\right)^{3.5} \frac{\sigma_1}{\texttt{mean}(\sigma_i)}.
\end{equation}
Here $\sigma_i$ are the estimates of the ordered singular values for $\Wd G \W^{-1}$ given by the use of the \texttt{RSVD} or \texttt{GKB}~algorithm, and the mean value is taken only over $\sigma_i>0$. This follows the practice  implemented in \cite{saeed6,RVA:15} for studies using the \texttt{RSVD} and \texttt{GKB}, and which was based on the recommendation to use a large value for $\alpha^{(1)}$,  \citep{FaOl:04}. 
In order to contrast the performance and computational cost of the \texttt{RSVD} and \texttt{GKB}~algorithms with increasing problem size $m$, different sizes $t$ of the projected space  for the solution are obtained using $t=\floor(m/s)$. Generally, the \texttt{GKB}~is successful with larger values for $s$ (smaller $t$) as compared to that needed for the \texttt{RSVD} algorithm. Hence, following recommendations for both algorithms, as discussed in Section~\ref{sec:algorithm}, we use the range of $s$ 
 from $40$ to $3$, given by $s=40$, $25$, $20$, $8$, $6$, $4$ and $ 3$,  corresponding to increasing $t$, but also limited by $5000$. 

For all simulations, the \irls algorithm is iterated to convergence as determined by the $\chi^2$ test for the predicted data, 
\begin{equation}\label{chi2}
\|\Wd(G\bfm^{(k)}-\bfdo)\|_2^2 \le m + \sqrt{2m}, 
\end{equation}
or 
\begin{equation}\label{scaledchi2}
\frac{\|\Wd(G\bfm^{(k)}-\bfdo)\|_2^2}{m+\sqrt{2m}} \le 1. 
\end{equation}
If this is not attained for $k\le \Kmax$, the iteration is terminated. Noisy data are generated for observed data $\bfdo=\bfde+\bfeta$ using 
\begin{equation}\label{noisydata}
\bfeta_i= (\tau_1|(\bfde)_i|+\tau_2\|\bfde\|_\infty)\bfe_i
\end{equation}
where $\bfe$ is drawn from a Gaussian normal distribution with mean $0$ and variance $1$. The pairs $(\tau_1,\tau_2)$ are chosen to provide a signal to noise ratio (\texttt{SNR}), as calculated by 
\begin{equation}\label{SNR}
\texttt{SNR}=20 \log_{10} \frac{\|\bfde\|_2}{\|\bfdo-\bfde\|_2},
\end{equation}
that is approximately constant across the increasing resolutions of the problem. Recorded for all simulations are  (i) the values of the  relative error $\texttt{RE}^{(k)}$, as defined by
\begin{equation}\label{RE}
\texttt{RE}= \frac{\|\bfme-\bfm^{(k)}\|_2}{\|\bfme\|_2},
\end{equation}
(ii) the number of iterations to convergence $K$ which is limited to $25$ in all cases, (iii) the scaled $\chi^2$ estimate given by \eqref{scaledchi2} at the final iteration, and (iv) the time to convergence measured in seconds, or to iteration $25$ when convergence is not achieved. 

\subsection{Synthetic data}\label{sec:volsimulation}
\begin{figure}[ht!]\begin{center}
\subfigure[Iso-surface of the volume structure. \label{figure2a}]{\includegraphics[width=0.49\textwidth]{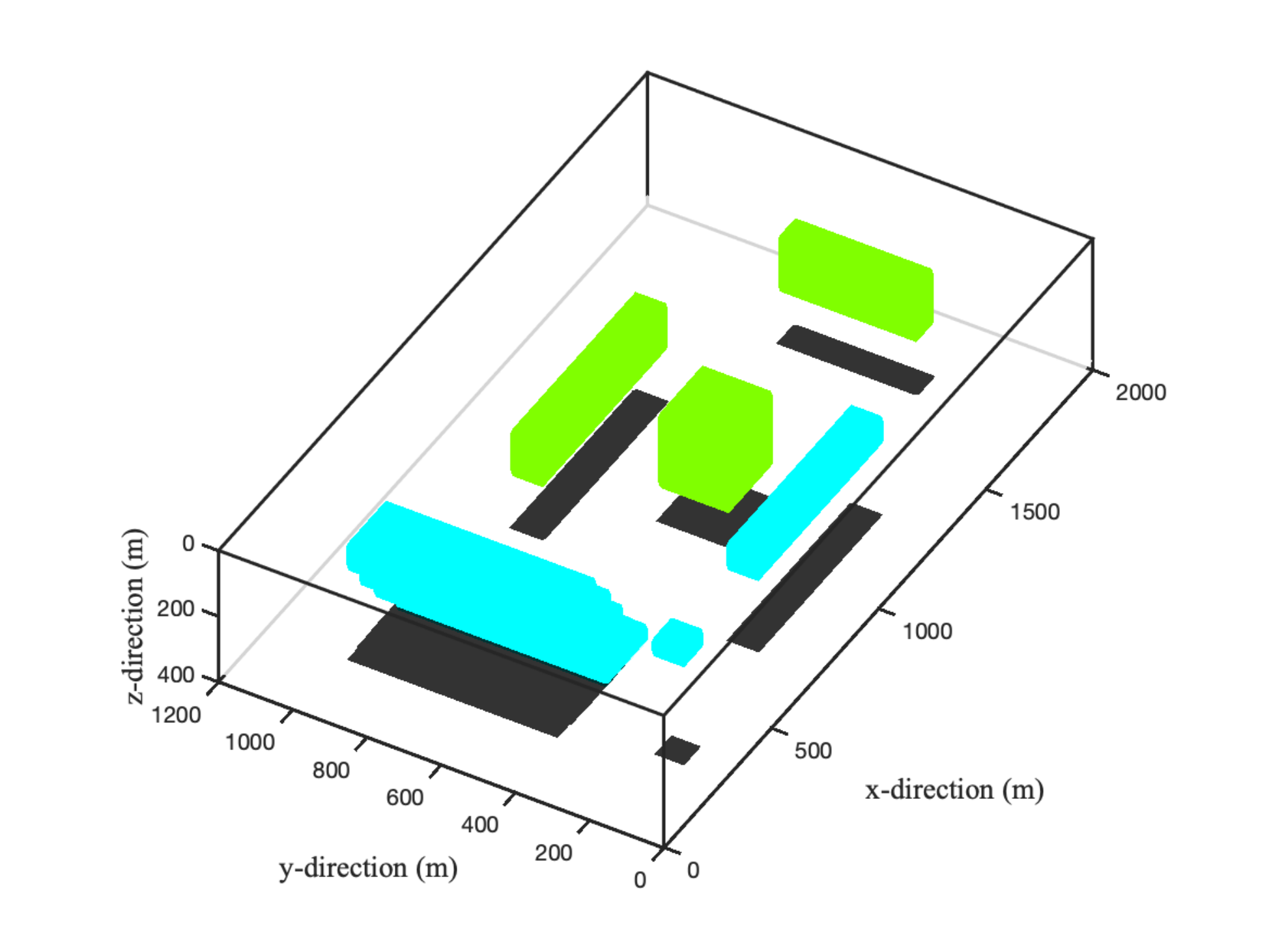}}
\subfigure[Cross-section of the volume structure.  \label{figure2b}]{\includegraphics[width=0.49\textwidth]{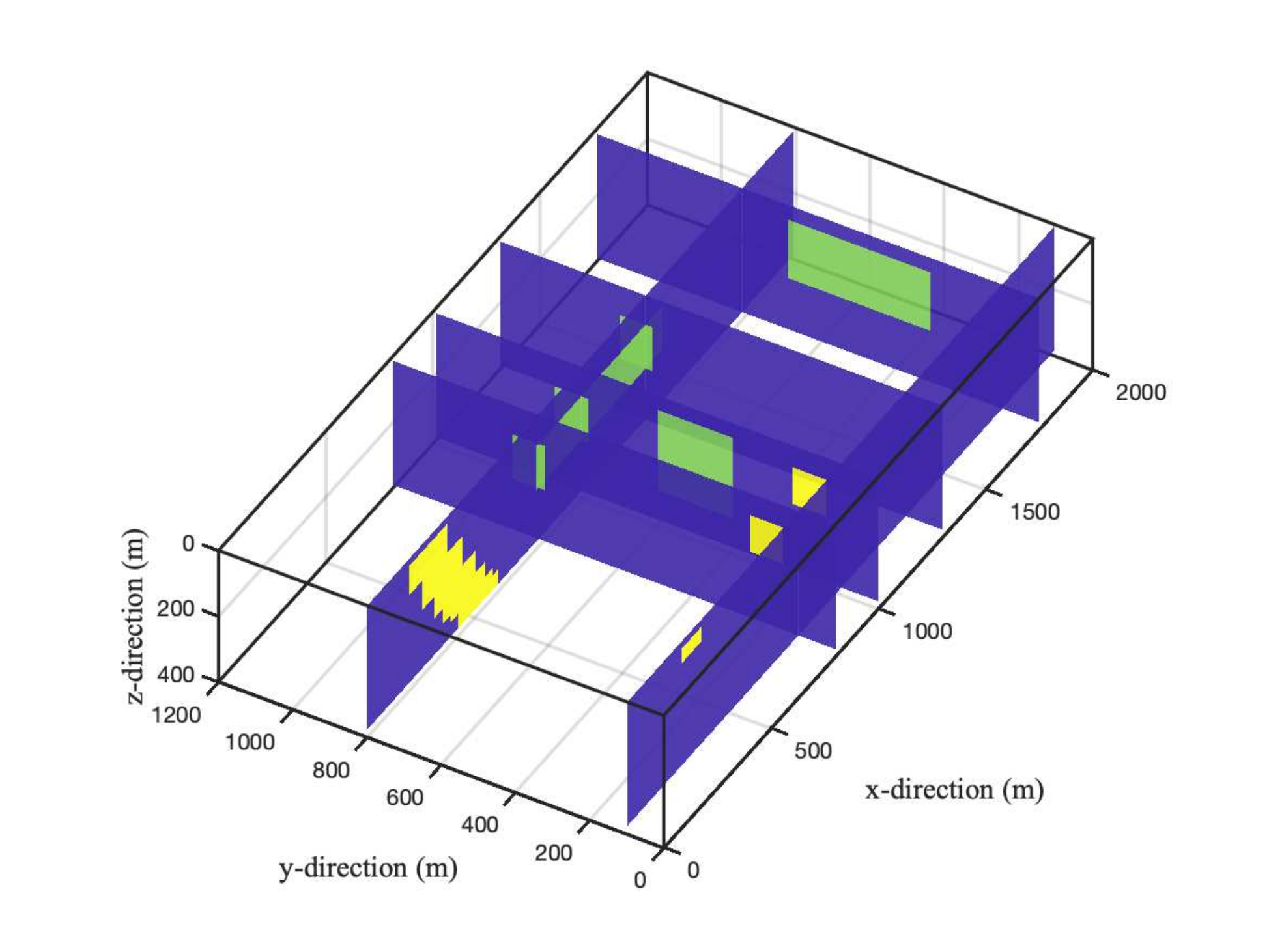}}
\caption{The basic volume structure within the domain of size $2000\times 1200\times 400$. The extent of each structure is shown by the shadow on the base of the volume. The same structure is used for the results using the padded domain.\label{figure2}}
\end{center}
\end{figure}

For the validation of the algorithms, we pick a volume structure with a number of boxes of different dimensions, and a six-layer dipping dike. The same structure is used for generation of the gravity and magnetic potential field data. For gravity data the densities of all aspects of the structure are set to $1$, with the homogeneous background set to $0$. For the magnetic data, the dipping dike,  one extended well and one very small well have susceptibilities $.06$. The three other structures have  susceptibilities set to $.04$. The distinction between these structures with different susceptibilities  is illustrated in the illustration of the iso-structure in Figure~\ref{figure2a} and the cross-section in Figure~\ref{figure2b}. 
The domain volume is discretized   in $x$, $y$ and $z$ into the number of blocks as indicated by triples $(\nsx,\nsy,\nbz)$ with increasing resolution for increasing values of these triples.  They are generated by taking  $(\nsx,\nsy,\nbz) = (25,15,2)$, and then scaling each dimension  by scaling factor $\ell\ge4$ for the test cases, correspondingly, $\nsx\nsy = 375$ is scaled by $\ell^2$ with increasing $\ell$, yielding a minimum problem size with $m=6000$ and $n=48000$. The grid sizes are thus given by the triples 
$(\dx,\dy,\dz)=(2000/\nsx, 1200/\nsy, 400/\nbz)$.  The   problem sizes considered for each simulation are  detailed in Table~\ref{table2}. For padding we compare the case with $\pad=0\%$ and $\pad=5\%$ padding across $x$ and $y$ dimensions. These are rounded to the nearest integer yielding $\padxl=\padxr=\round(\pad \, \nsx)$, and $\nbx=\nsx+2\,\round (
\pad\,\nsx)$.  $\nby$ is calculated in the same way, yielding $n =(\nsx+2\,\round (
\pad\,\nsx))(\nsy+2\,\round (
\pad\,\nsy)) \nbz$. Certainly, the choice to use $\pad=5\%$ is quite large, but  is chosen to demonstrate that the solutions obtained 
using the \texttt{2DFFT} are robust to boundary conditions, and thus not impacted by the restriction due to lack of padding or very small padding.  

\begin{figure}[ht!]\begin{center}
\subfigure[True gravity anomaly\label{figure3a}]{\includegraphics[width=0.45\textwidth]{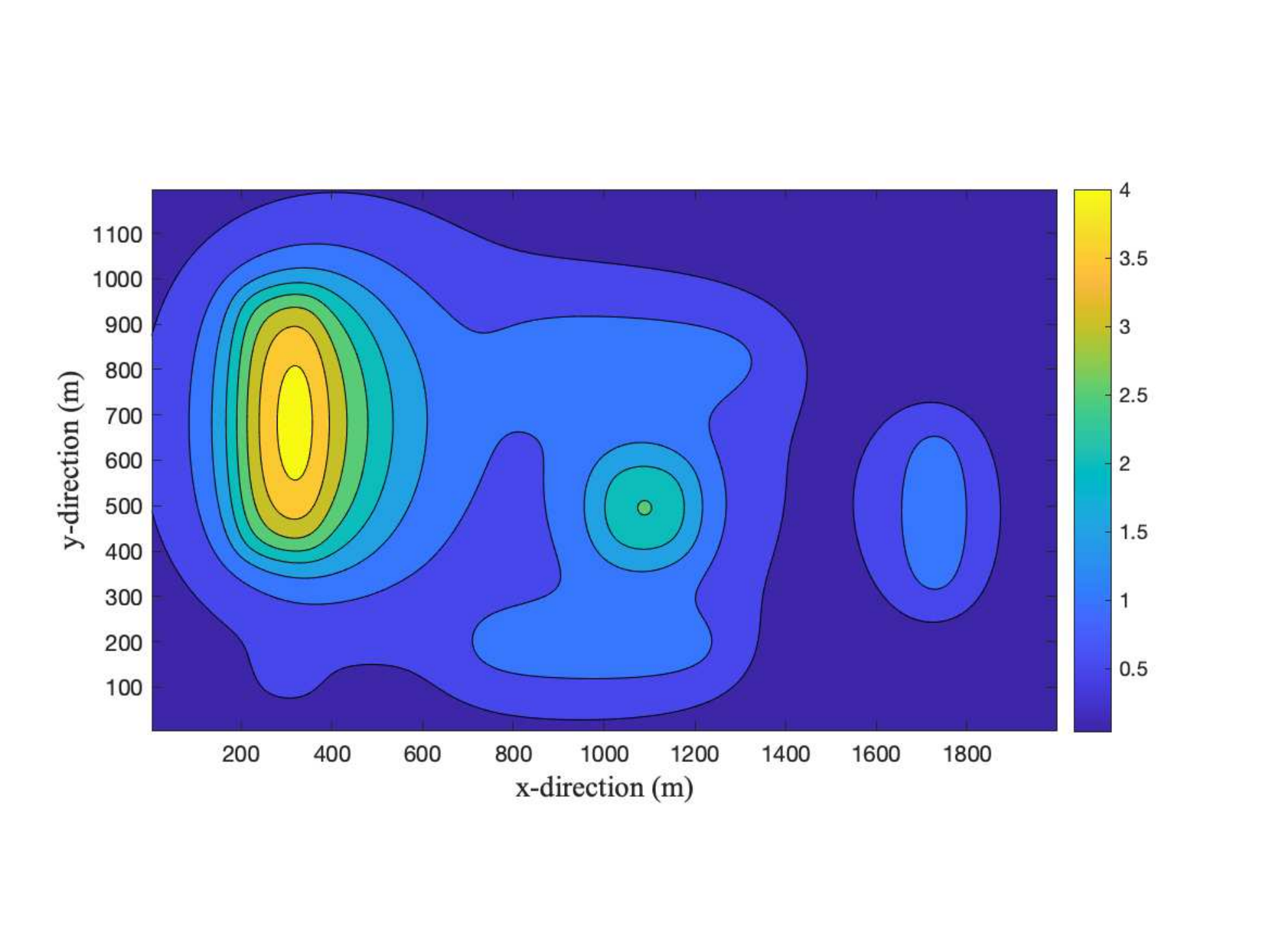}}
\subfigure[Noisy gravity anomaly\label{figure3b}]{\includegraphics[width=0.45\textwidth]{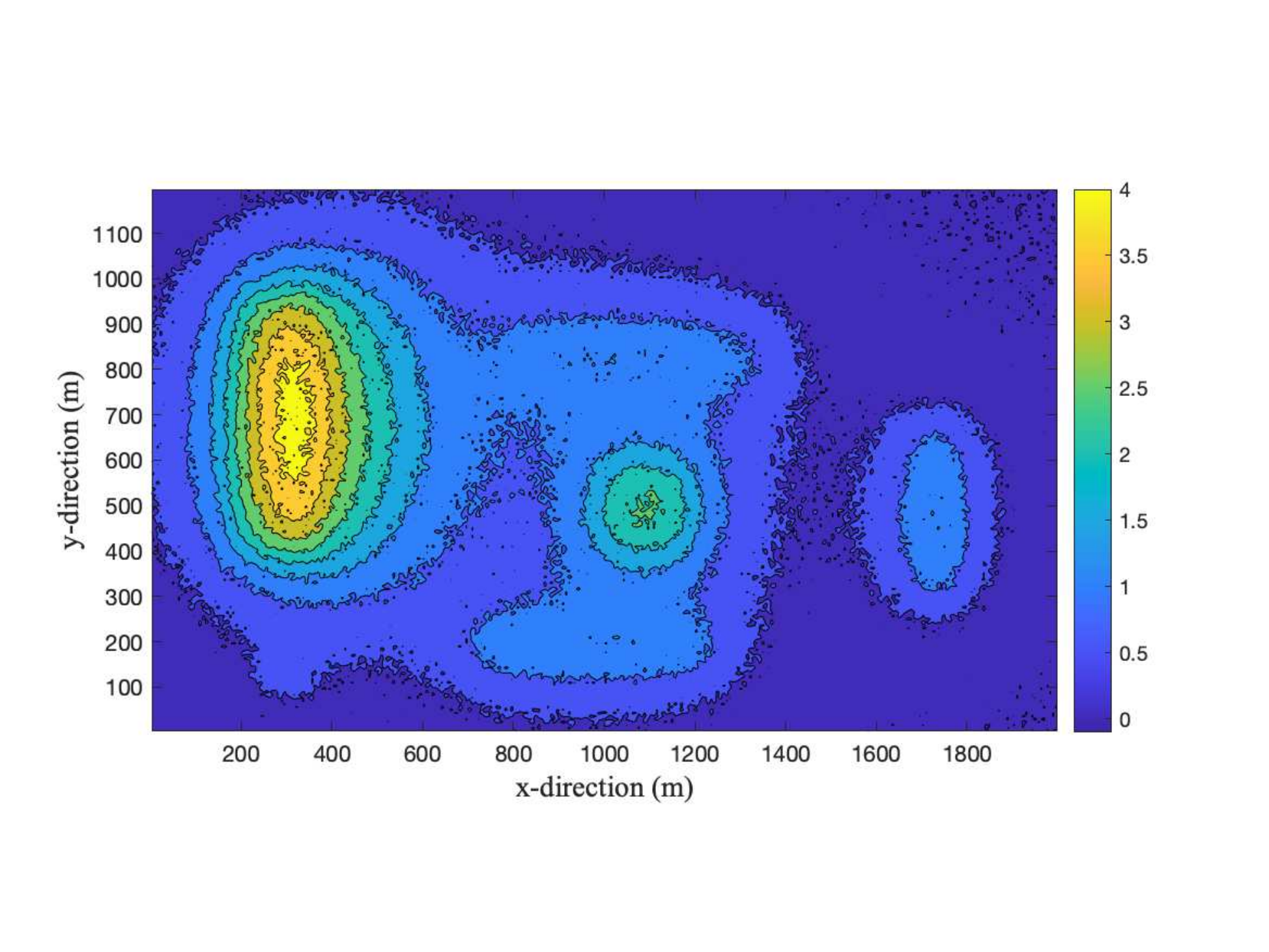}}
\subfigure[True magnetic anomaly\label{figure3c}]{\includegraphics[width=0.45\textwidth]{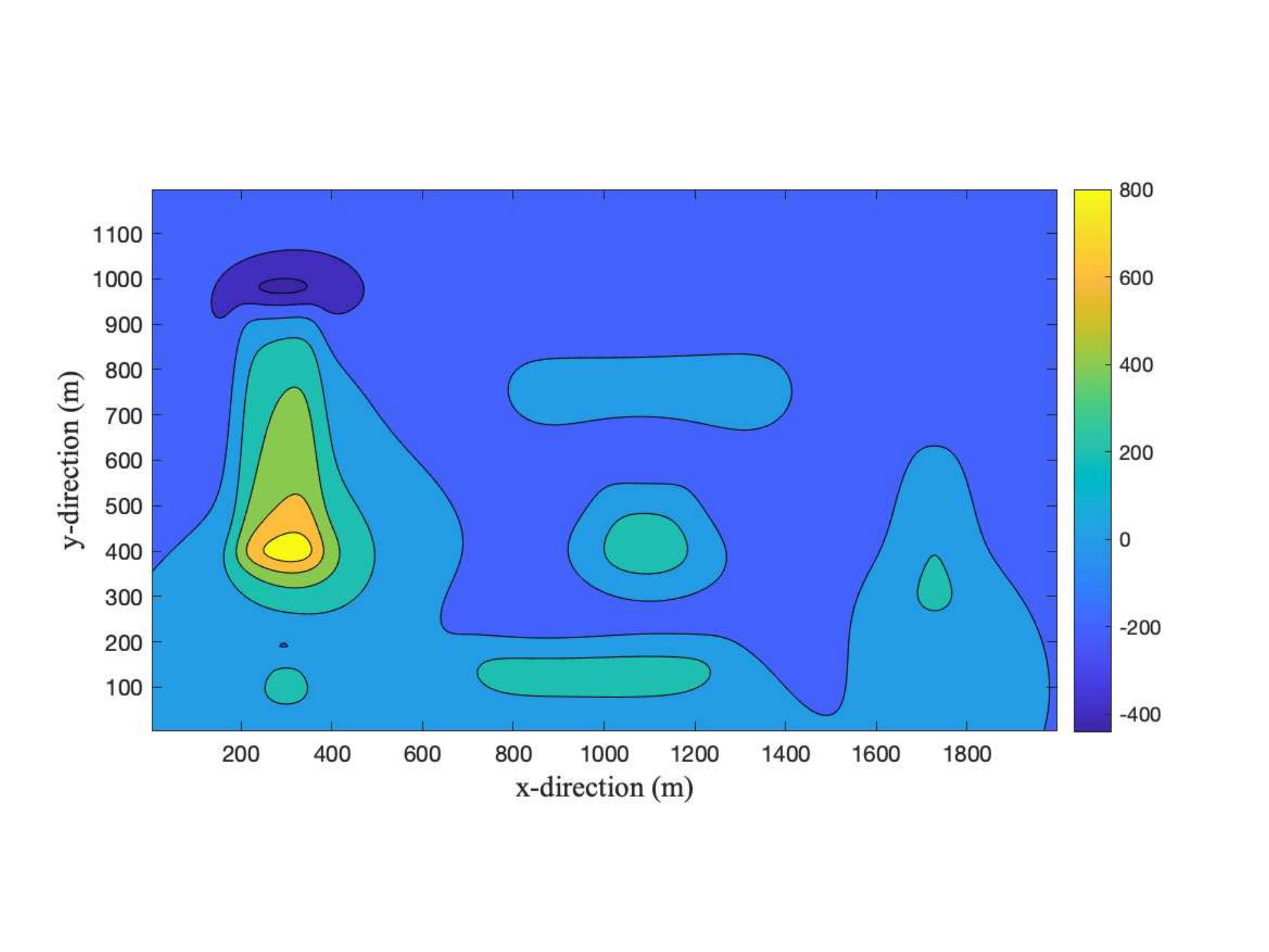}}
\subfigure[Noisy magnetic  anomaly\label{figure3d}]{\includegraphics[width=0.45\textwidth]{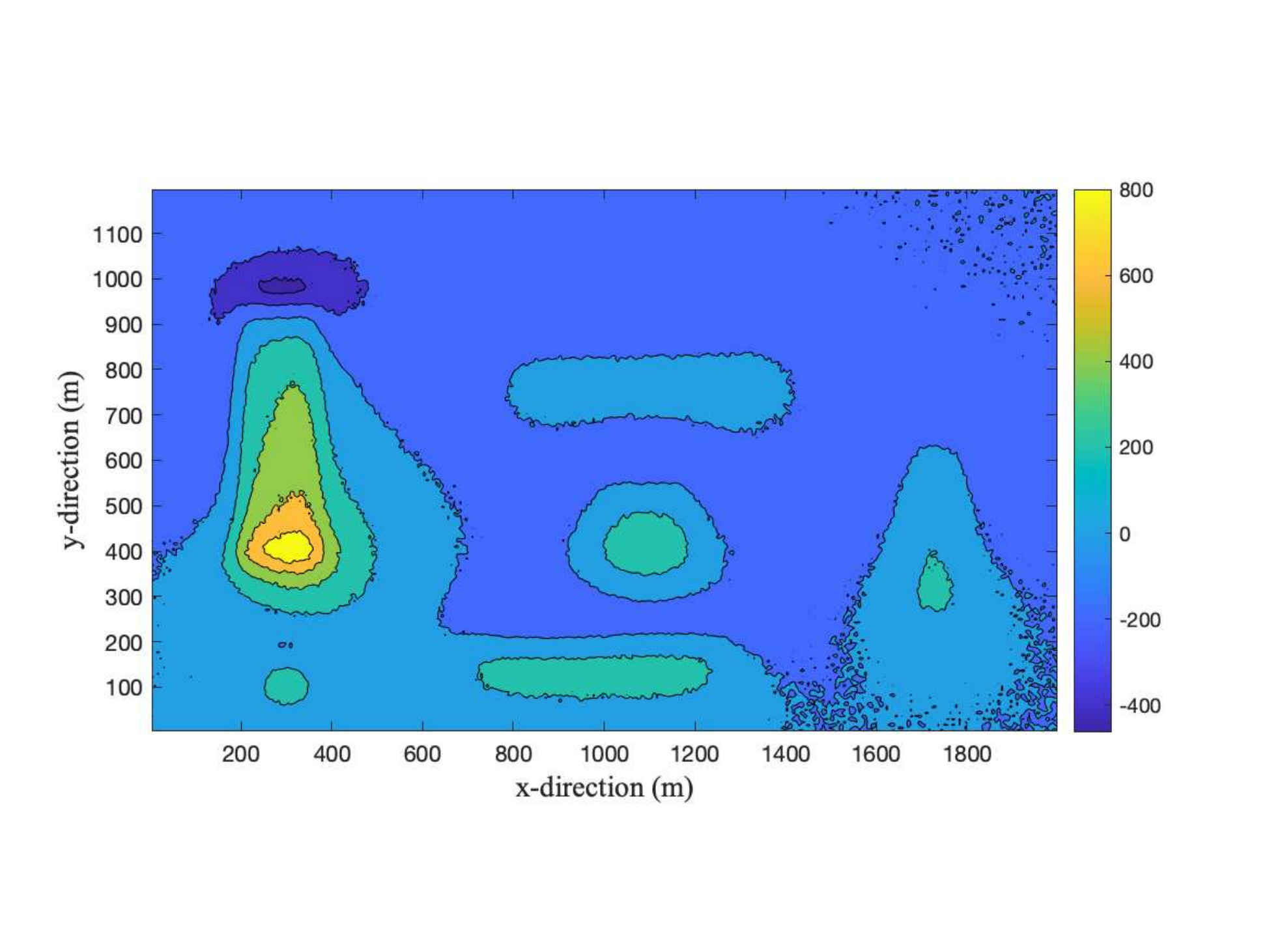}}
\caption{The calculated true and noisy anomalies for the volume structure given in Figure~\ref{figure2a}, where the units are mGal and nT for gravity and magnetic data, respectively. The anomalies used for the inversion using the padded domain are exactly the same as given here.  \label{figure3}}
\end{center}
\end{figure}

For these structures and resolutions, noisy data are generated as given in \eqref{noisydata} to yield an \texttt{SNR} of approximately $24$ across all scales as calculated using \eqref{SNR}. This results 
in different choices of  $\tau_1$ and $\tau_2$ for each problem size and dependent on the gravity or magnetic data case, denoted by $(\tau_1^{\mathrm{g}},\tau_2^{\mathrm{g}})$ and $(\tau_1^{\mathrm{m}},\tau_2^{\mathrm{m}})$, respectively. In all cases we use $\tau_1^{\mathrm{g}}=\tau_1^{\mathrm{m}}=.02$ and adjust $\tau_2$. The simulations for the choices of $\tau_2^{\mathrm{g}}$ and $\tau_2^{\mathrm{m}}$ for increasing problem sizes are detailed in Table~\ref{Table1}. As an example we illustrate the true and noisy data for gravity and magnetic data, when $\ell=12$, in Figure~\ref{figure3}. 

\begin{table}[htb!]\begin{center}
\begin{tabular}{|c|c|c|c|c|c|c|c|c|}
\hline
$\ell$& $(\nsx,\nsy,\nbz)$ & $m$ & $n$   &$n_{\pad}$&$\tau_2^{\mathrm{g}}$&$\tau_2^{\mathrm{m}}$&\texttt{SNR}$^{\mathrm{g}}$&\texttt{SNR}$^{\mathrm{m}}$\\
\hline
$4$&   $(100, 60,    8)  $&$ 6000   $&  $ 48000 $&    $58080$& $.0138$&$.0081$&$24.0$&$24.0$\\ \hline
$5$&   $(125, 75,   10) $&$ 9375   $&  $ 93750 $&  $113710$& $.0147$&$.0083$&$24.0$&$24.0$\\ \hline
$6$&   $(150, 90,   12) $&$ 13500 $& $ 162000$& $199200$& $.0133$&$.0074$&$24.0$&$24.0$\\ \hline
$7$&   $(175, 105, 14) $&$ 18375 $& $ 257250$& $310730$& $.0133$&$.0070$&$24.0$&$24.0$\\ \hline
$8$&   $(200, 120, 16) $&$ 24000 $& $ 384000$& $464640$& $.0133$&$.0071$&$24.0$&$24.1$\\ \hline
$9$&   $(225, 135, 18) $&$ 30375 $& $ 546750$& $662450$& $.0133$&$.0069$&$24.0$&$24.0$\\ \hline
$10$& $(250, 150, 20) $&$ 37500 $& $ 750000$& $916320$& $.0132$&$.0070$&$24.0$&$24.0$\\ \hline
$11$& $(275, 165, 22) $&$ 45375 $& $ 998250$&$1206500$& $.0135$&$.0075$&$24.0$&$24.0$\\ \hline
$12$& $(300, 180, 24) $&$ 54000 $& $1296000$&$1568160$& $.0135$&$.0075$&$24.0$&$24.0$\\ \hline
\end{tabular}
\caption{Dimensions of the volume used in the experiments with scaling of the small problem size $(25,15,2)$ by  scale factor $\ell$ in each dimension. $m$ and $n$ are the dimensions of the measurement vector and the volume domain, respectively, $G\in \Rmn{m}{n}$. Here $m=\nsx\nsy=375\ell^2$ and 
$n=m\nbz$ where $\nbx=\nsx$ and $\nby=\nsy$ without padding. Here,  we use $n_{\pad}=\nbx\nby\nbz$ to denote the volume dimension $n$ with $5\%$ padding, 
using $\nbx=\nsx+2\,\round (\pad\,\nsx )$ and $\nby=\nsy+2\,\round (\pad\,\nsy)$ for padding obtained using a percentage, $\pad$, on each side of the domain so that $\padxl=\padxr=\round(\pad \, \nsx)$, and similarly for $\nsy$. . \label{table2}}
\end{center}\end{table}

\subsection{Numerical Results}\label{sec:results}
The validation and analysis of the algorithms for the inversion of the potential field data is presented in terms of  (i) the cost per iteration of the algorithm (Section~\ref{sec:costperiteration}), (ii) the total cost to convergence of the algorithm  (Section~\ref{sec:methodconvergence}), and (iii) the quality of the obtained solutions,  (Section~\ref{sec:methodsolutions}). Supporting quantitative data that summarize the illustrated results are presented as Tables in \ref{app:table}. 
\subsubsection{Comparative cost of \texttt{RSVD} and \texttt{GKB} algorithms per \texttt{IRLS} iteration}\label{sec:costperiteration}
\begin{figure}[ht!]\begin{center}
\subfigure[Running time \texttt{magnetic}: \texttt{GKB}.\label{figure4a}]{\includegraphics[width=0.45\textwidth]{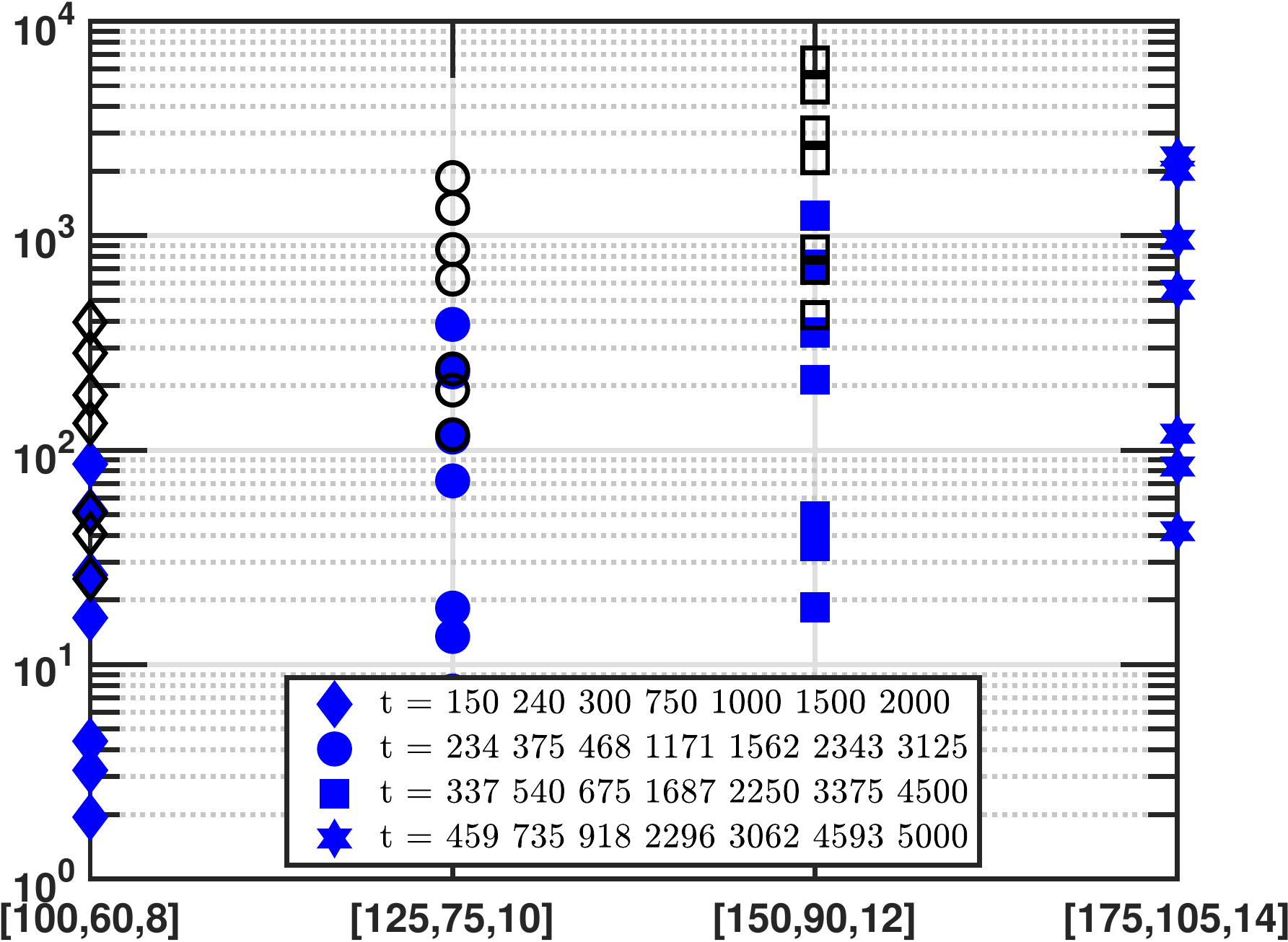}}
\subfigure[Running time \texttt{magnetic}: \texttt{RSVD}.\label{figure4b}]{\includegraphics[width=0.45\textwidth]{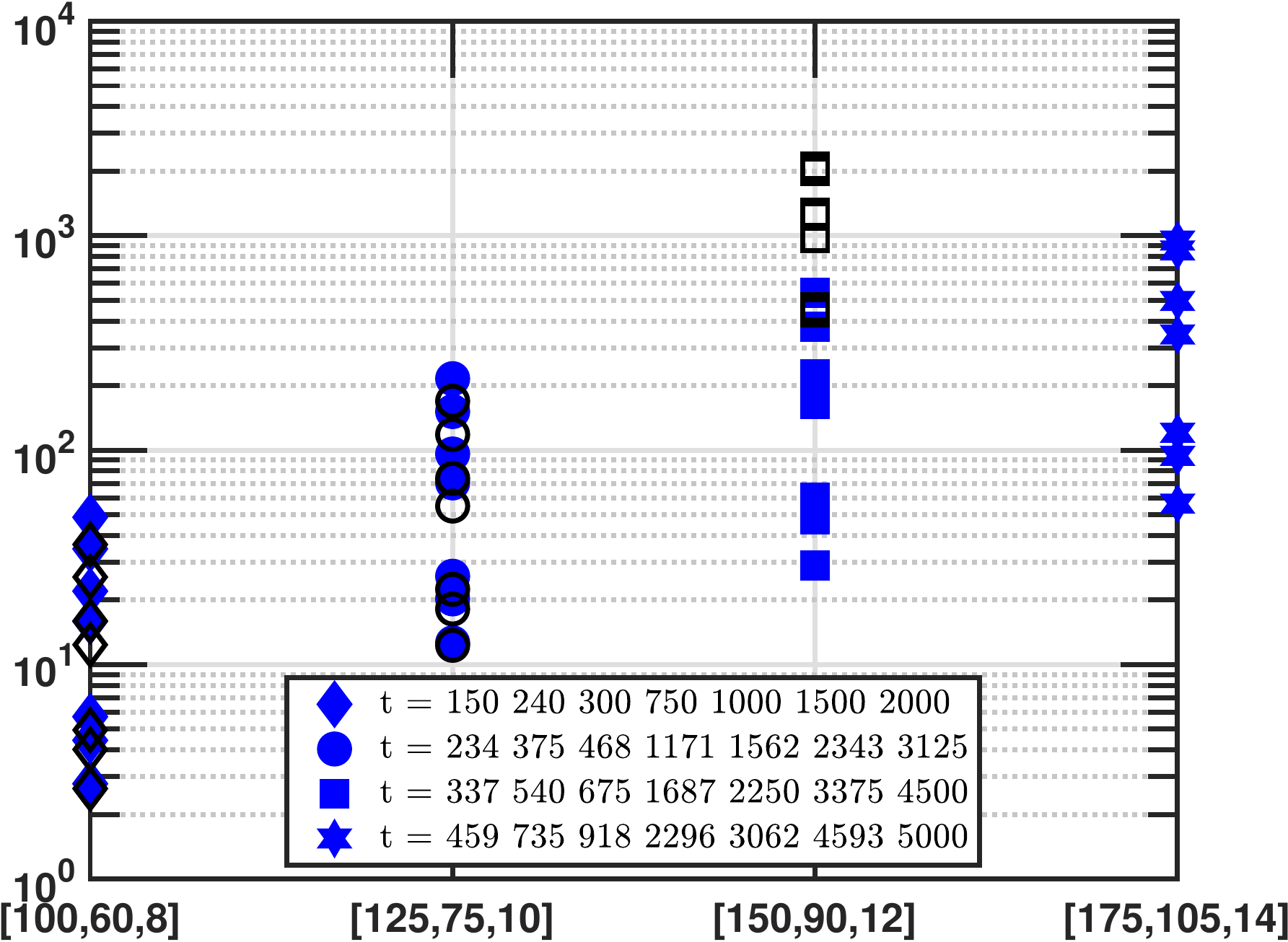}}
\subfigure[Running time \texttt{gravity}: \texttt{GKB}.\label{figure4c}]{\includegraphics[width=0.45\textwidth]{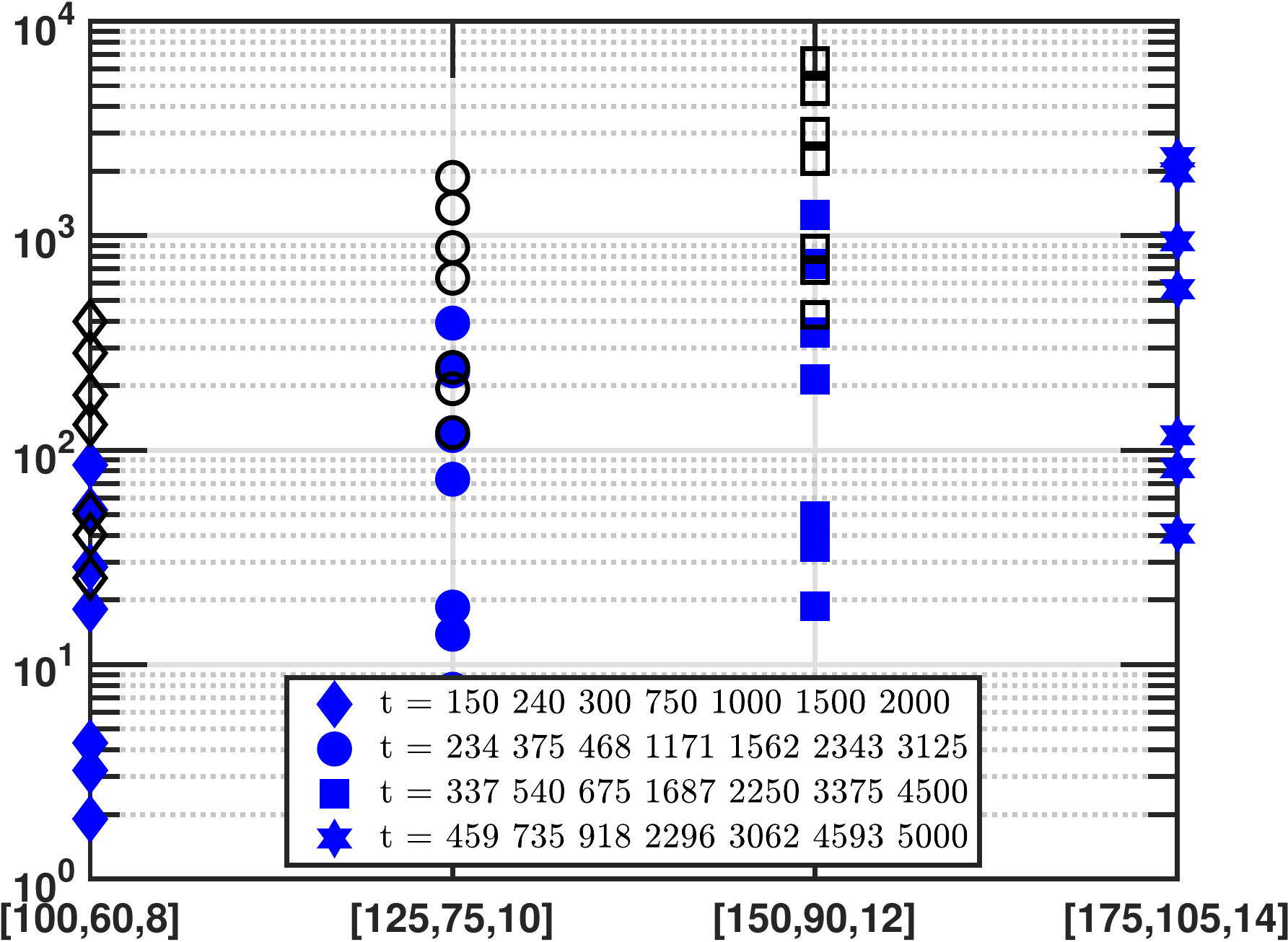}}
\subfigure[Running time \texttt{gravity}: \texttt{RSVD}. \label{figure4d}]{\includegraphics[width=0.45\textwidth]{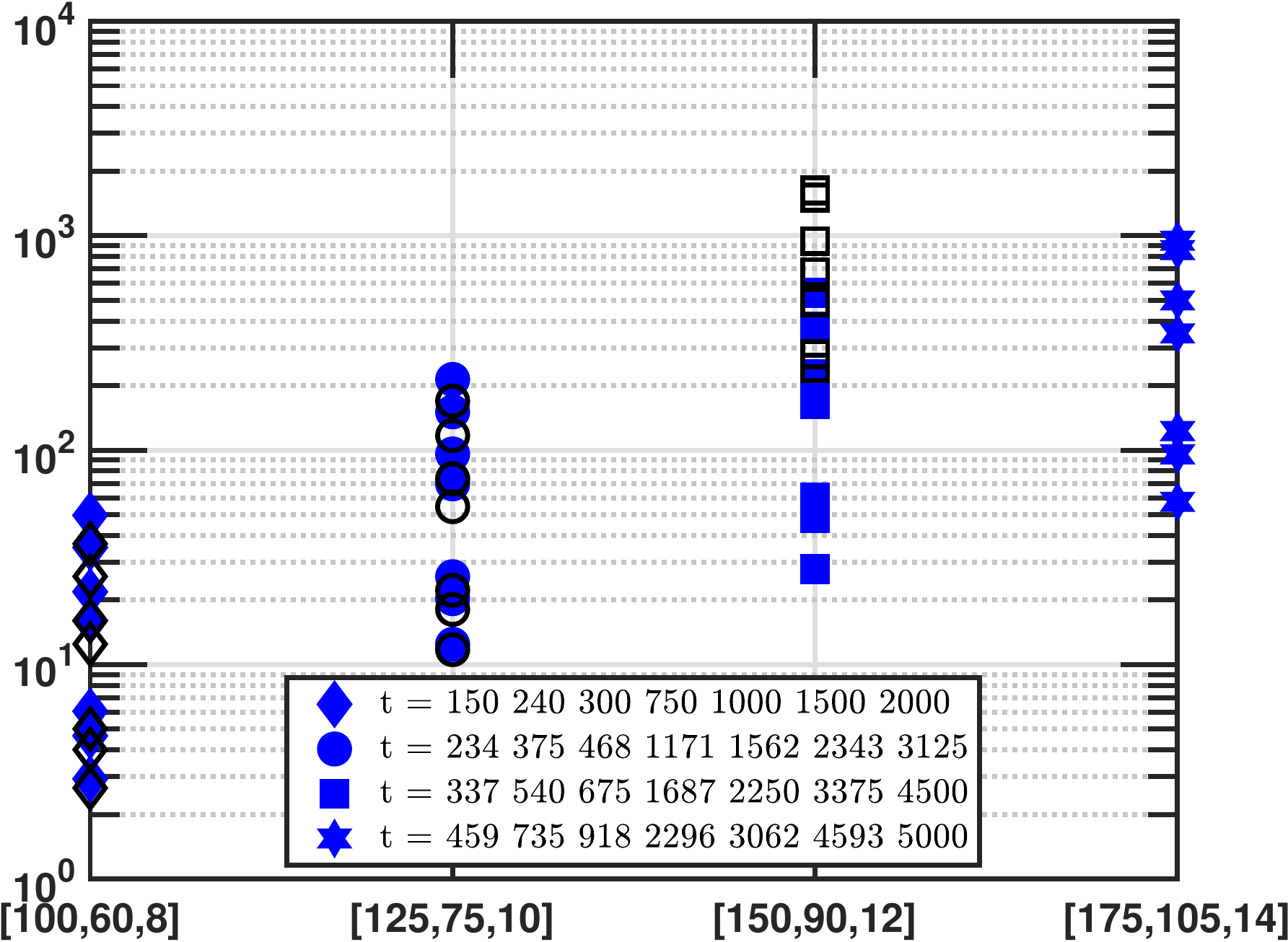}}
\caption{Running time in seconds for one iteration of the inversion algorithm for the inversion of  \texttt{magnetic} and \texttt{gravity} data, without padding the volume domain. Problems are of increasing size, as indicated by the $x-$axis for triples $[\nbx, \nby, \nbz]$ and increasing projection size $t$ ($y-$axis using $\log$ scale) determined by fractions of $m=\nsx\nsy$. In Figures~\ref{figure4a} and \ref{figure4c} the running time for the \texttt{GKB} algorithm using $t_p=\floor(1.05t)$ (an oversampling percentage $5\%$),  for the \texttt{magnetic} and \texttt{gravity} problems respectively. In Figures~\ref{figure4b} and \ref{figure4d} the equivalent running times using the \texttt{RSVD} algorithm with one power iteration.   In these plots the solid symbols represent the timing for one iteration of the algorithm using the \texttt{2DFFT} and the open symbols represent the timing for the same simulation using $G$ directly.  Matrix $G$ for problem size $\ell=7$, which corresponds to triple $[175, 105, 14]$, requires too much memory for implementation in the specific computing environment. \label{figure4}}
\end{center}\end{figure}

\begin{figure}[ht!]\begin{center}
\subfigure[\texttt{magnetic} data: $\texttt{Cost}_{\texttt{G}}/\texttt{Cost}_{\texttt{2DFFT}} $.\label{figure5a}]{\includegraphics[width=0.45\textwidth]{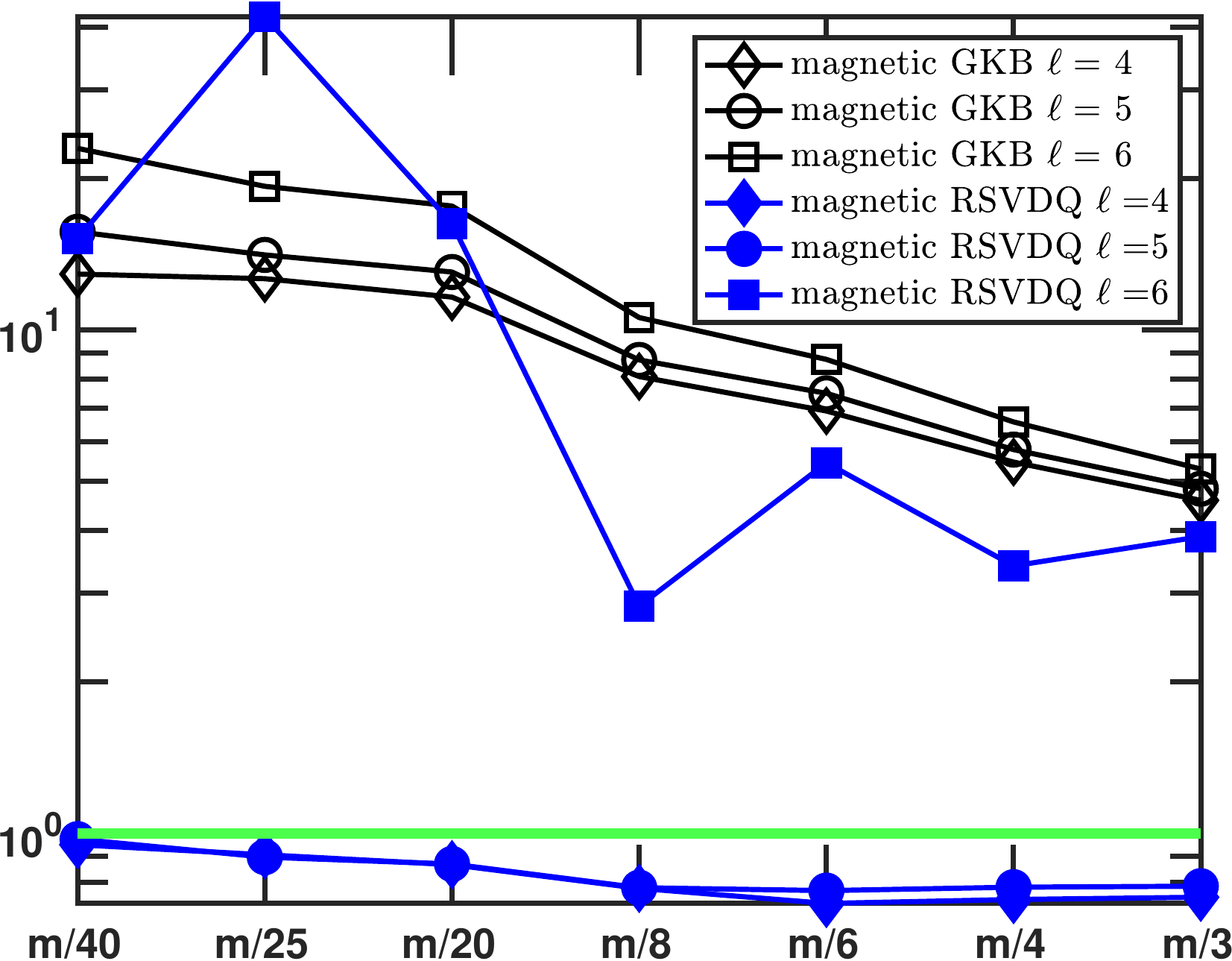}}
\subfigure[\texttt{gravity} data: $\texttt{Cost}_{\texttt{G}}/\texttt{Cost}_{\texttt{2DFFT}} $.\label{figure5b}]{\includegraphics[width=0.45\textwidth]{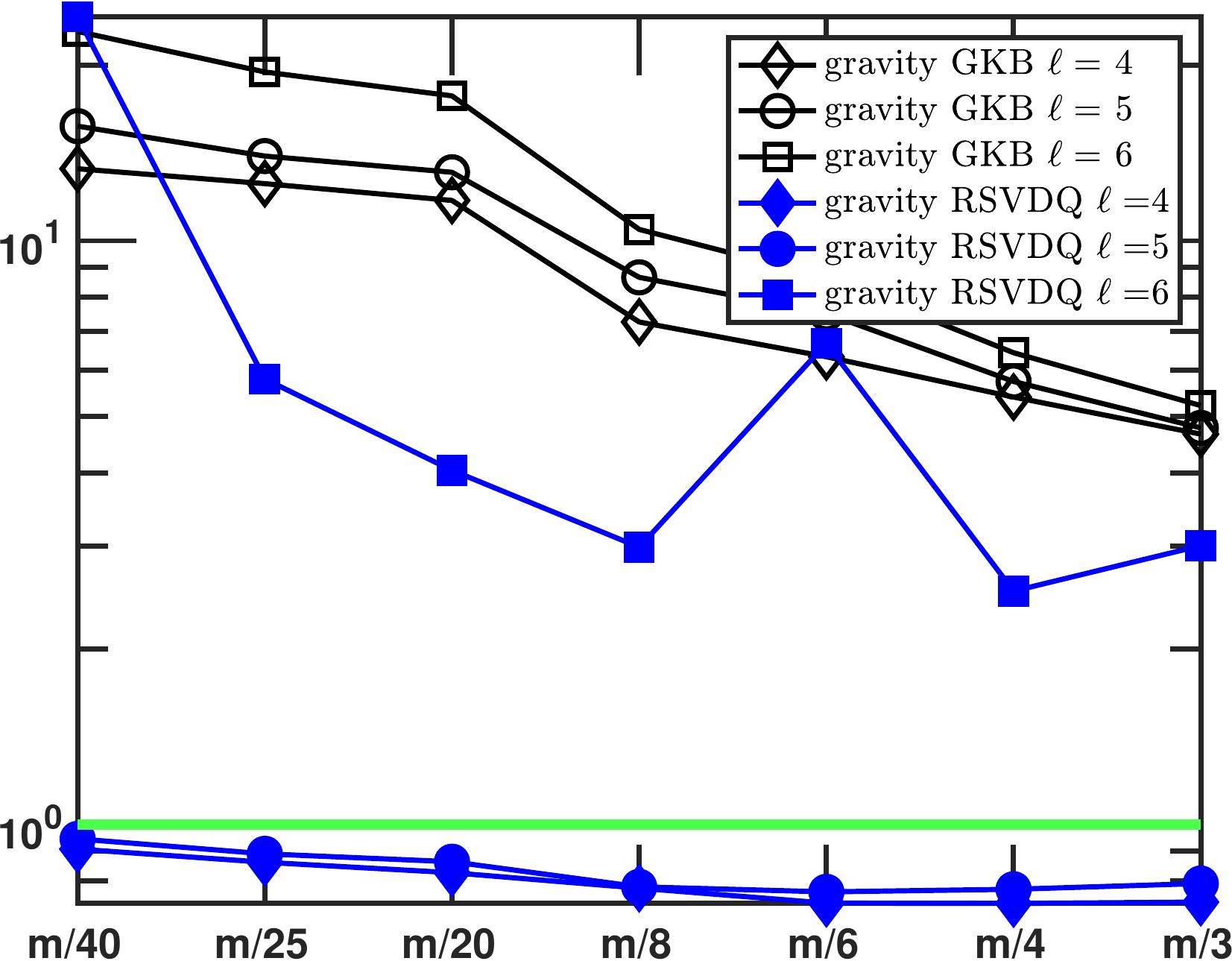}}
\caption{Relative computational cost for one iteration of the \texttt{IRLS}  algorithm using $G$ directly as compared to the  \texttt{2DFFT}, as indicated by $\texttt{Cost}_{\texttt{G}}/\texttt{Cost}_{\texttt{2DFFT}}$, for the data presented in Figure~\ref{figure4}, for the \texttt{magnetic} and \texttt{gravity}  problems, Figures~\ref{figure5a}-\ref{figure5b}. Here, the values for the relative cost that are less than $1$, below the horizontal line at $y=1$, indicate that it is more efficient to use $G$ directly. Values that are greater than $1$ indicate that it is more efficient to use the \texttt{2DFFT}.  Open symbols indicate the \texttt{GKB} algorithm and solid symbols the \texttt{RSVD} algorithm.   In each case the given plots for a fixed $\ell$ are for increasing projection size $t$ as given by $m/s$ for the selections of $t$ as used in Figure~\ref{figure4}.\label{figure5}}
\end{center}\end{figure}

\begin{figure}[ht!]\begin{center}
\subfigure[\texttt{magnetic} data: $\texttt{Cost}_{\texttt{GKB}}/\texttt{Cost}_{\texttt{RSVD}} $.\label{figure6a}]{\includegraphics[width=0.45\textwidth]{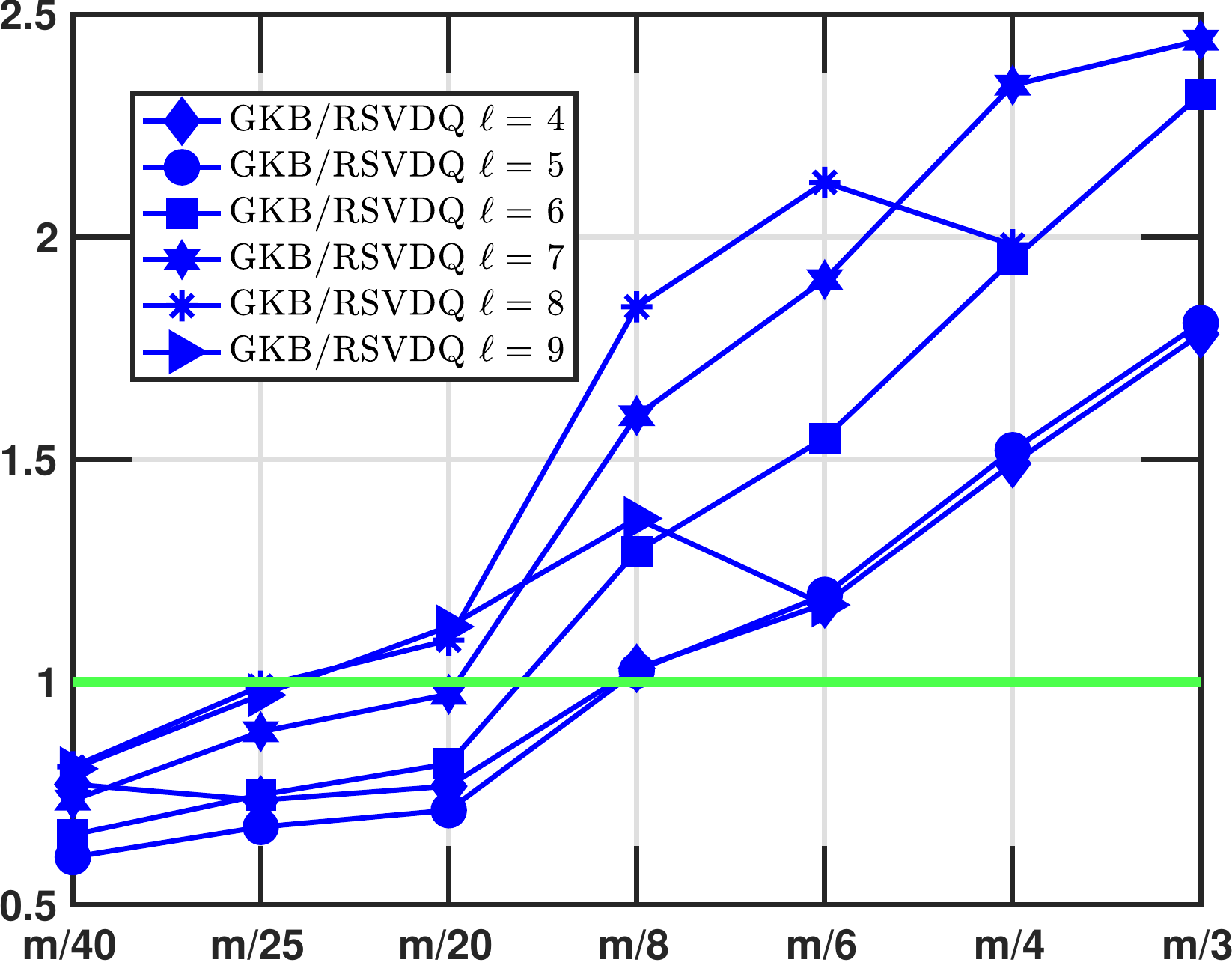}}
\subfigure[\texttt{gravity} data: $\texttt{Cost}_{\texttt{GKB}}/\texttt{Cost}_{\texttt{RSVD}}. $\label{figure6b}]{\includegraphics[width=0.45\textwidth]{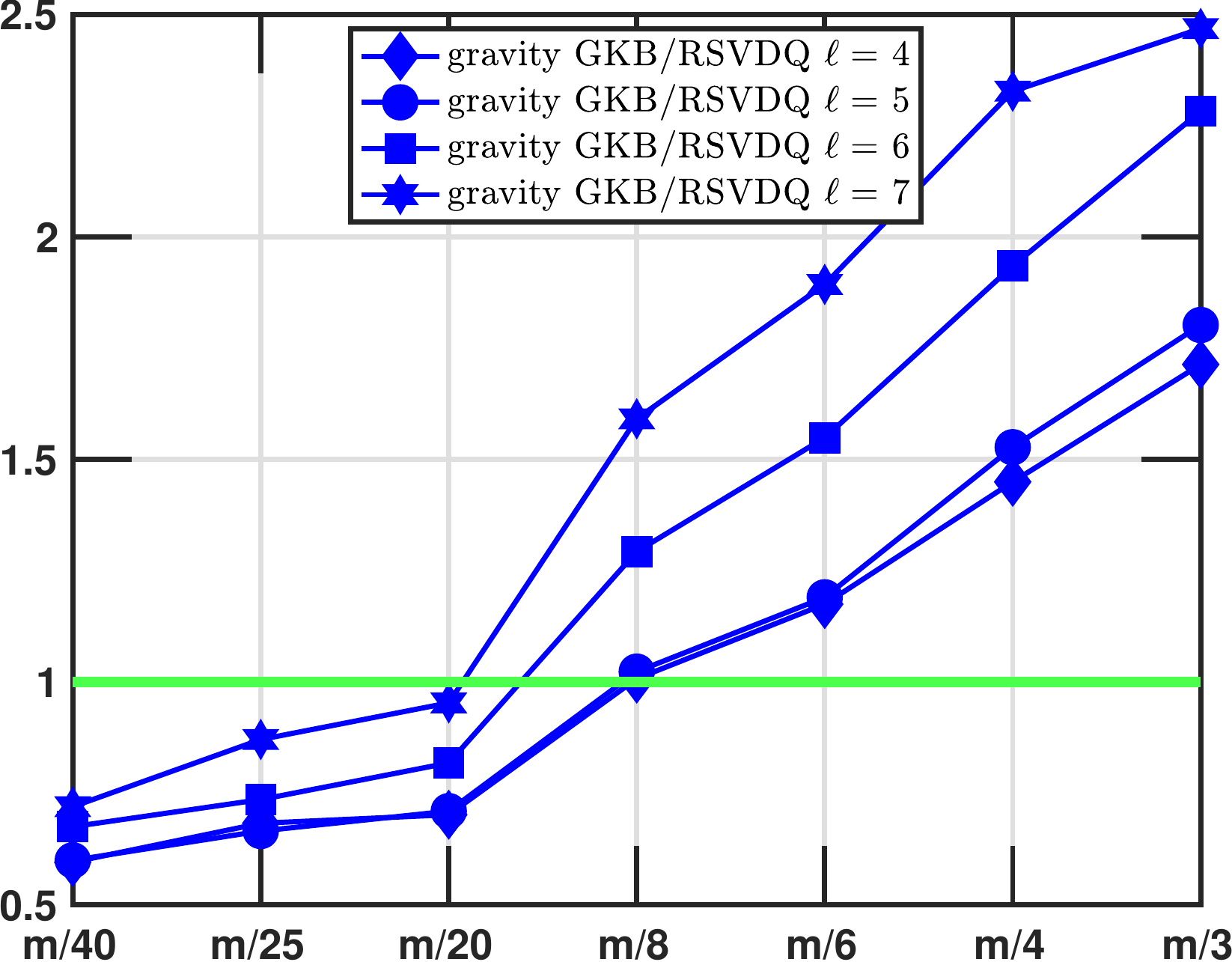}}
\caption{The relative computational cost for one iteration of the \irls algorithm for inversion using the \texttt{GKB} as compared to the \texttt{RSVD} algorithm ($\texttt{Cost}_{\texttt{GKB}}/\texttt{Cost}_{\texttt{RSVD}} $), for given $\ell$ and projected size $t$. In each case the given plots for a fixed $\ell$ are for increasing projection size $t$ as given by $m/s$ as in Figure~\ref{figure5}. The   horizontal line at $y=1$ represents the data for which the costs are the same, independent of whether using the \texttt{RSVD} or \texttt{GKB} algorithms. The \texttt{GKB} is more efficient when $t$ is maintained small, $s=40$, $25$ and $20$. The gain in using the \texttt{GKB} decreases, however, as $\ell$ increases. For small $\ell$ and $t$, the estimates confirm the computational cost estimates in \eqref{costgkb}-\eqref{costrsvd}, but for larger projection sizes $t$, the \texttt{RSVD} is more efficient. In Figure~\ref{figure6a} the relative costs are also included for $\ell=8$ and $\ell=9$, where $t\le 5000$. \label{figure6}}
\end{center}\end{figure}

\begin{figure}[ht!]\begin{center}
\subfigure[Running time (padded): \texttt{GKB}.  \label{figure7a}]{\includegraphics[width=0.45\textwidth]{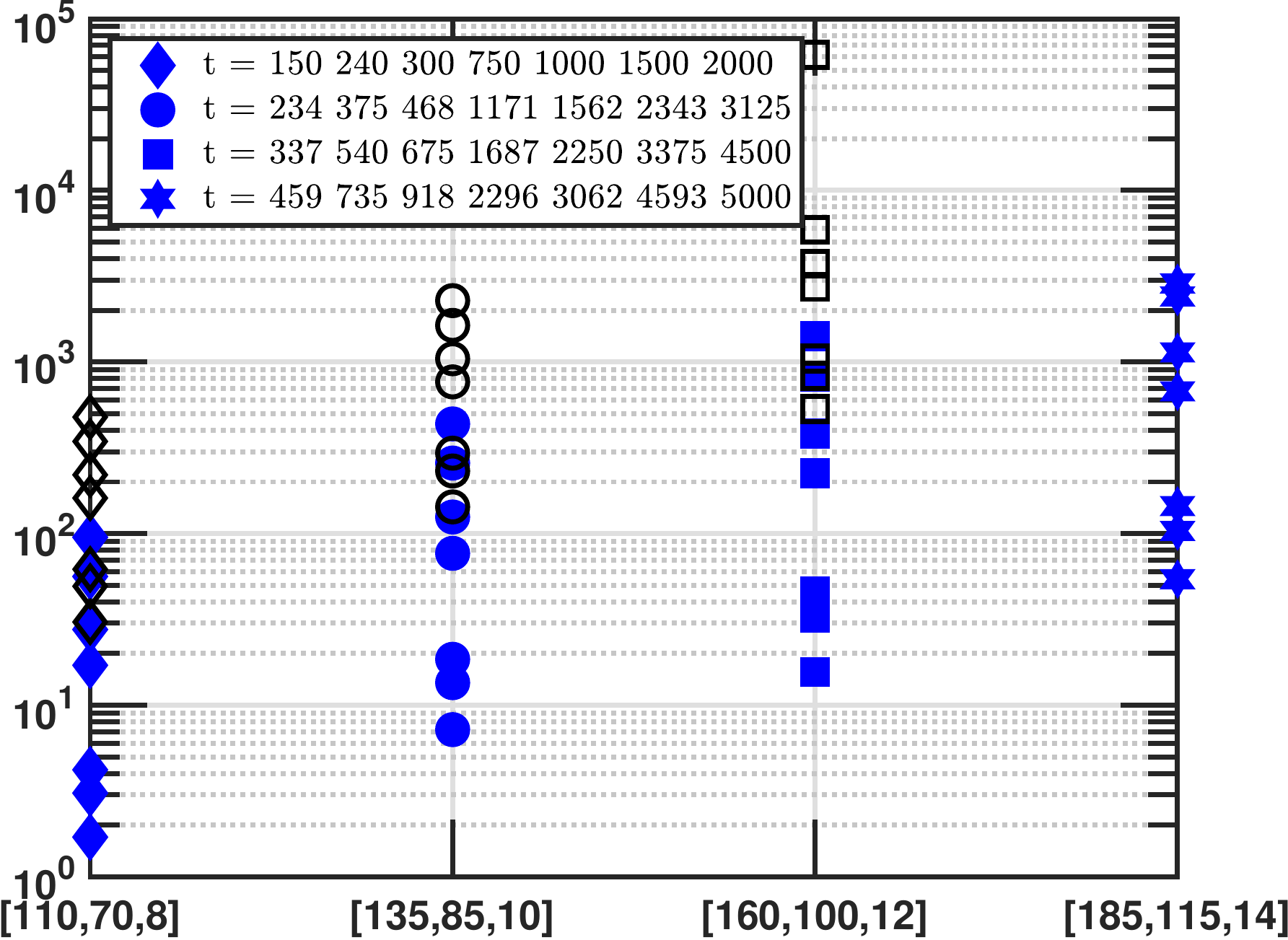}}
\subfigure[Running time (padded): \texttt{RSVD}.      \label{figure7b}]{\includegraphics[width=0.45\textwidth]{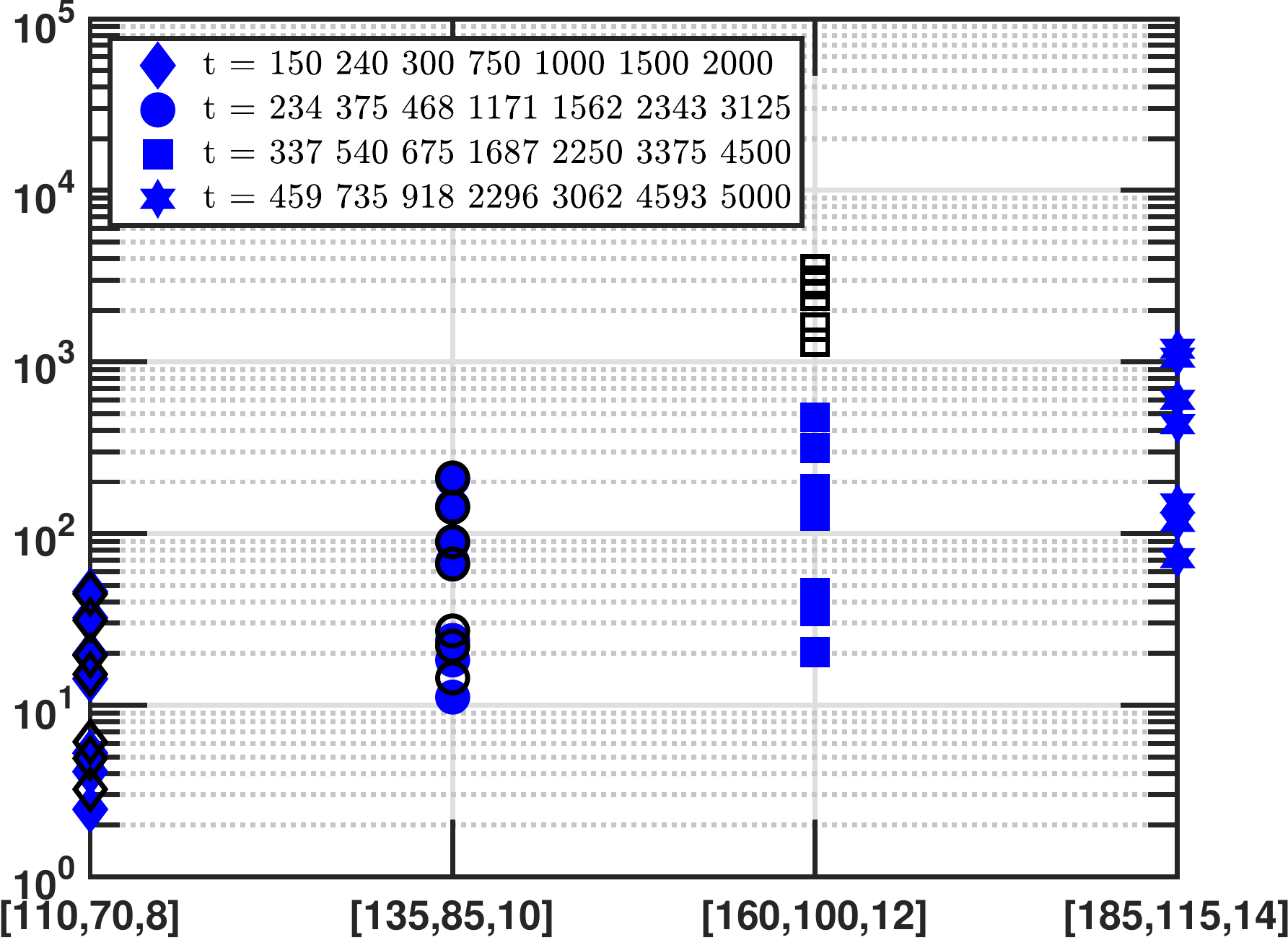}}
\subfigure[$\texttt{Cost}_{\texttt{G}}/\texttt{Cost}_{\texttt{2DFFT}} $ (Padded).\label{figure7c}]{\includegraphics[width=0.43\textwidth]{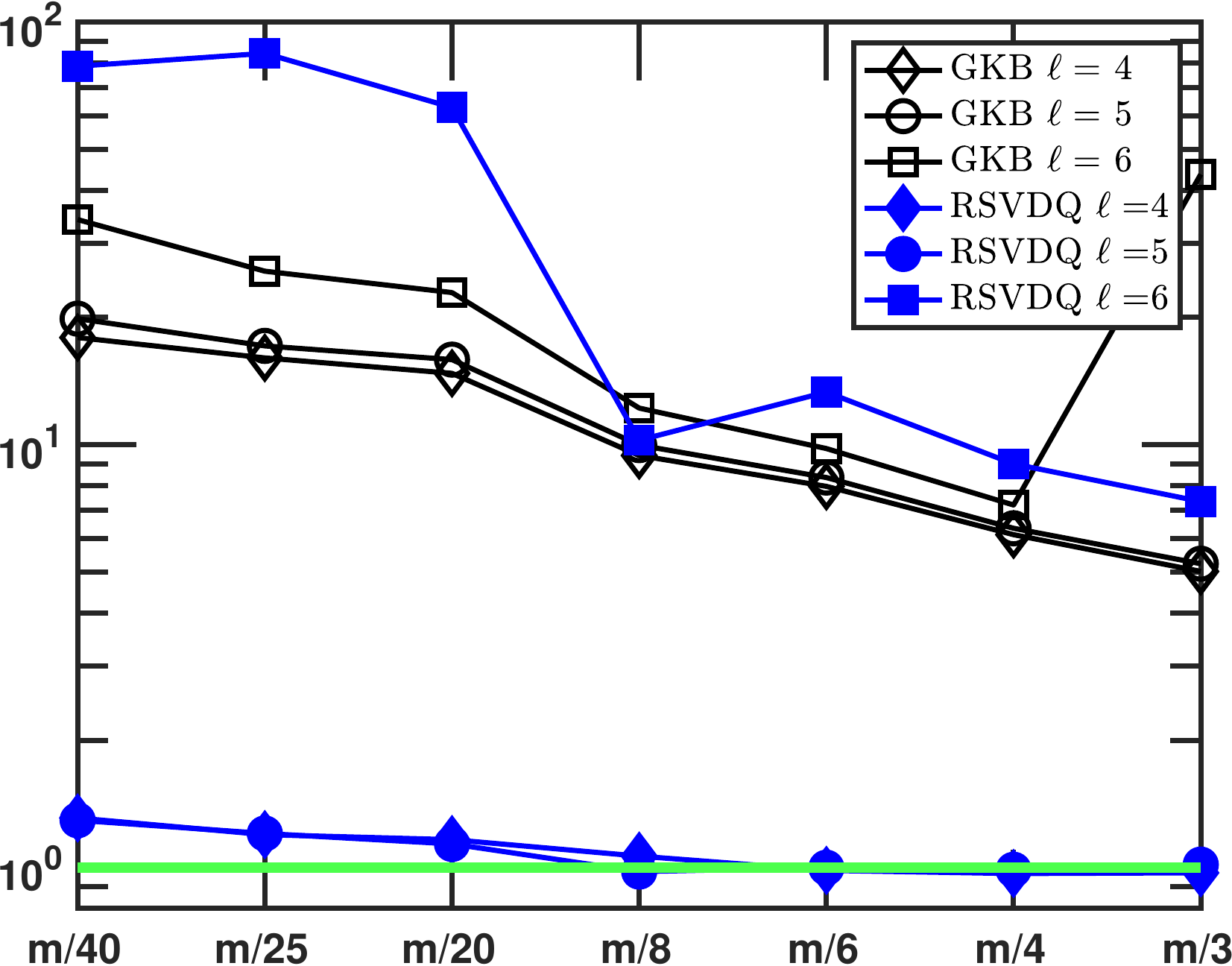}}
\subfigure[$\texttt{Cost}_{\texttt{GKB}}/\texttt{Cost}_{\texttt{RSVD}}$ (Padded).\label{figure7d}]{\includegraphics[width=0.43\textwidth]{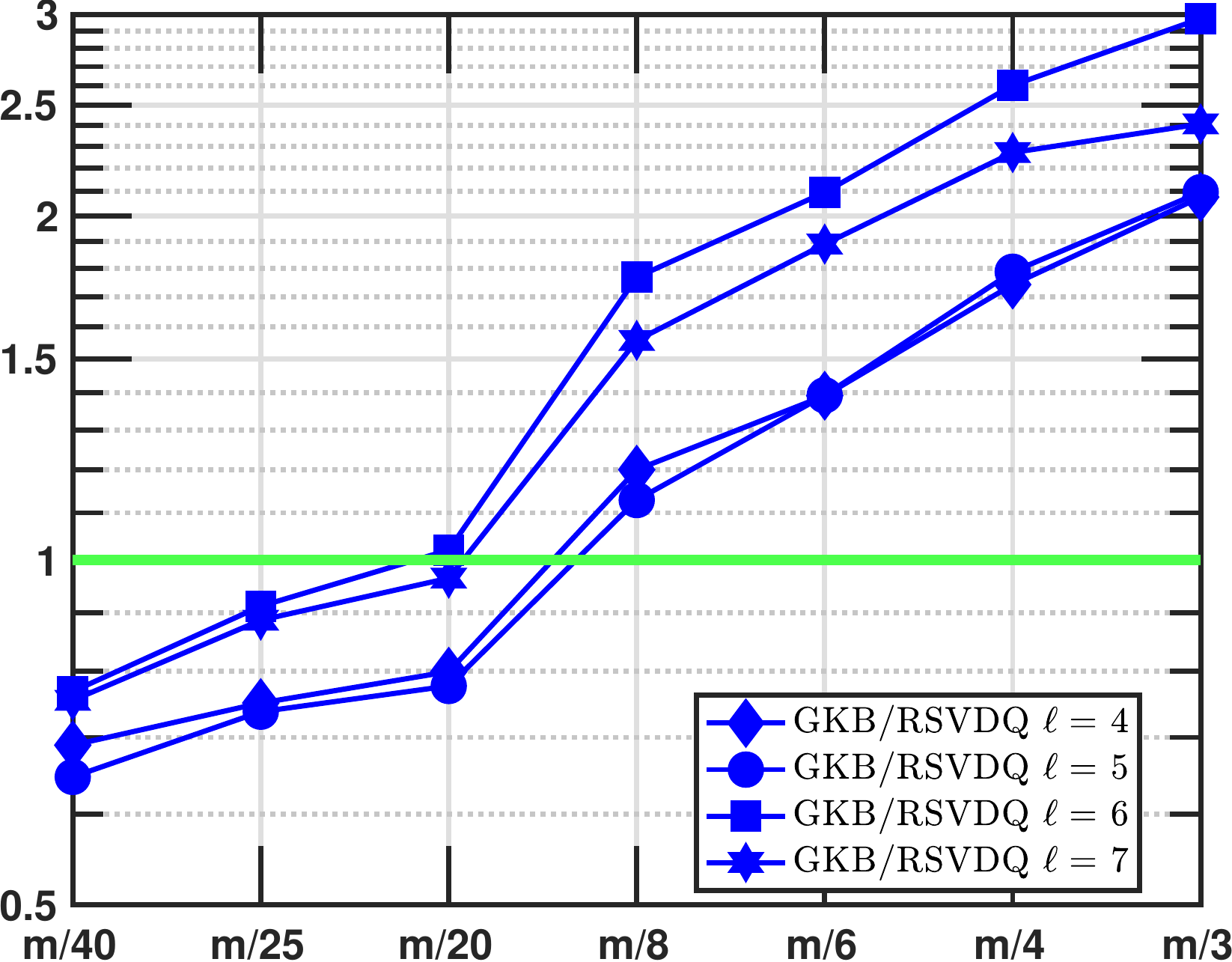}}
\caption{In Figures~\ref{figure7a}-\ref{figure7b} the running time in seconds for one iteration of the inversion algorithm for the inversion of  \texttt{magnetic}  data, for the same problems as in Figure~\ref{figure4a} and \ref{figure4b} but with padding, $\pad=5\%$, added to  the volume domain. Problems are of increasing size, as indicated by the $x-$axis for triples $[\nbx, \nby, \nbz]$ and increasing projection size $t$ ($y-$axis using $\log$ scale) determined by fractions of $m=\nsx\nsy$.  In these plots the solid symbols represent the timing for one iteration of the algorithm using the \texttt{2DFFT} and the open symbols represent the timing for the same simulation without using the \texttt{2DFFT} for the kernel operations.   In Figure~\ref{figure7c} the relative costs for these results, as also provided in Figure~\ref{figure5a} for the case without padding, and in Figure~\ref{figure7d} the relative costs of the two algorithms with the \texttt{2DFFT}, as in Figure~\ref{figure6a} without padding.  \label{figure7}}
\end{center}\end{figure}

We investigate the computational cost, as measured in seconds, for one iteration of the inversion algorithm using both the direct multiplications using matrix $G$, respectively, $G^T$,  and the circulant embedding, for the resolutions up to $\ell=6$ that are indicated in Table~\ref{table2}, using both the \texttt{RSVD} and \texttt{GKB} algorithms, and for both gravity and magnetic data.  For fair comparison, all the timing results that are reported use \textsc{Matlab} release  2019b implemented on the same iMac $4.2$GHz Quad-Core Intel Core i7 with  $32$GB RAM. In this environment, the size of the matrix $G$ is too large for effective memory usage when $\ell>6$. The details of the timing results for one step of the \texttt{IRLS} algorithm  are illustrated in Figures~\ref{figure4}-\ref{figure7}, with the specific values for the magnetic data case, given in Table~\ref{tableB.3}. 
 
Figure~\ref{figure4} provides an overview of the computational cost with increasing projection size $t$, for a given $m$, when the algorithm is implemented using $G$ directly, or using the \texttt{2DFFT}. These costs exclude the cost of generating $G$. In these plots, we use the open symbols for calculations using $G$ and solid symbols when using the \texttt{2DFFT}. The same symbols are used for each choice of $t$ and $\ell$. An initial observation, confirming expectation, is that the timings for equivalent problems and methods, are almost independent of whether the potential field data are gravity or magnetic, comparing Figures~\ref{figure4a}-\ref{figure4b} with Figures~\ref{figure4c} and \ref{figure4d}. The lack of entries for triple $[175,105,14]$ indicates that the matrix $G$ is too large for the operations, $\ell=7$. With increasing $\ell$, (increasing values of the triples along the $x-$axis), it can also be observed that the open symbols are more spread out vertically, confirming that the algorithms using $G$ directly are more expensive for problems at these resolutions. 

In  Figure~\ref{figure5} we plot the relative computational costs  for one iteration of the \irls algorithm using the matrix $G$  as compared to the algorithm using the \texttt{2DFFT}, as indicated by $\texttt{Cost}_{\texttt{G}}/\texttt{Cost}_{\texttt{2DFFT}}$, for the data presented in Figure~\ref{figure4}. Along the $x-$axis we give the size $t$ used for the projected problem in terms of the ratio $m/s$. The lines with solid blue symbols are for results using the \texttt{RSVD} algorithm, and the open black symbols are for the \texttt{GKB} algorithm. Here, the values for the relative cost that are less than $1$, below the horizontal green line at $y=1$, indicate that for the specific algorithm it is more efficient to use $G$ directly. Values that are greater than $1$ indicate that  it is more efficient to use the \texttt{2DFFT} for the given algorithm and problem size.
It is apparent that it is not beneficial to use the \texttt{2DFFT} for the smaller scale implementation of the \texttt{RSVD} algorithm, when $\ell=4$ or $5$. But the situation is completely reversed using  the \texttt{GKB} algorithm for all choices of $\ell$ and the \texttt{RSVD} algorithm for $\ell\ge 6$. Thus, the relative gain in reduced computational cost, by using the \texttt{2DFFT} depends on the algorithm used within the \texttt{IRLS} inversion algorithm. The decrease in efficiency for a given size problem, fixed $\ell$ but increasing size $t$ (in the $x-$axis), is explained by the theoretical discussion relating to equations \eqref{costgkb}-\eqref{costrsvd}. As $t$ increases the impact of the efficient matrix multiplication using the \texttt{2DFFT} is reduced. Again the gravity and magnetic data results are comparable.

Figure~\ref{figure4} provides no information on the relative costs of the \texttt{GKB} and \texttt{RSVD} algorithms with increasing $\ell$, independent of the use of the \texttt{2DFFT}. Figure~\ref{figure6}  shows the relative computational costs, $\texttt{Cost}_{\texttt{GKB}}/\texttt{Cost}_{\texttt{RSVD}}$. Note that Figure~\ref{figure6a} also includes results for larger problems. These plots demonstrate that the relative costs for a single iteration are not constant across all $t$ with the \texttt{GKB} generally cheaper for smaller $t$, and the \texttt{RSVD} cheaper for larger $t$. These results confirm the analysis of the computational cost in terms of $\flops$ provided in \eqref{costgkb}-\eqref{costrsvd} for small $t$.  The relative computational costs increase from roughly $0.6$ to $2.5$, increasing with both $\ell$ and $t$.  Still, this improved relative performance of  \texttt{RSVD}  with increasing $\ell$ and $t$ appears to violate the \texttt{flop} count analysis in \eqref{costgkb}-\eqref{costrsvd}. As discussed in Section~\ref{sec:compcost}, this  is a feature of the implementation.  While \texttt{RSVD}  is implemented using the \textsc{Matlab} \texttt{builtin} function $\texttt{qr}$ which uses compiled code for faster implementation,   \texttt{GKB}  only uses \texttt{builtin} operations for performing the \texttt{MGS} reorthogonalization of the basis matrices $A_{t_p}$ and $H_{t_p}$. 
Once again results are comparable for inversion of both gravity and magnetic data sets.

Figure~\ref{figure7} summarizes magnetic data timing results from Table~\ref{tableB.3} for domains which are padded with   $5\%$ padding in $x$ and $y$ directions. Data illustrated in Figures~\ref{figure7a}-\ref{figure7b} are equivalent to the results presented in Figures~\ref{figure4a}-\ref{figure4b}, but with padded volume domains. Again these results show  the open symbols are more spread out vertically, for increasing $\ell$, confirming that the algorithms using $G$ directly are more expensive for problems at these resolutions, with greater impact when using the \texttt{GKB} algorithm for small $\ell$.  This is further confirmed in Figure~\ref{figure7c}, equivalent to Figure~\ref{figure5a}, showing that the computational cost of performing one step of the \texttt{IRLS} algorithm using matrix $G$ directly, is always greater than that using the \texttt{2DFFT}. This is more emphasized for the \texttt{GKB} algorithm. The relative costs  shown in Figure~\ref{figure7d},  equivalent to Figure~\ref{figure6a},  again shows that the \texttt{GKB} algorithm is cheaper for small $t$ when $\ell$ is small. But as the problem size increases and the projected problem size also increases, it is more efficient to use the \texttt{RSVD} algorithm,  consistent with the observations for the unpadded domains.

\subsubsection{Comparative cost of \texttt{RSVD} and \texttt{GKB} algorithms to convergence}\label{sec:methodconvergence}
\begin{figure}[ht!]\begin{center}
\subfigure[\texttt{magnetic} $\texttt{Cost}_{\texttt{GKB}}/\texttt{Cost}_{\texttt{RSVD}}. $\label{figure8a}]{\includegraphics[width=0.44\textwidth]{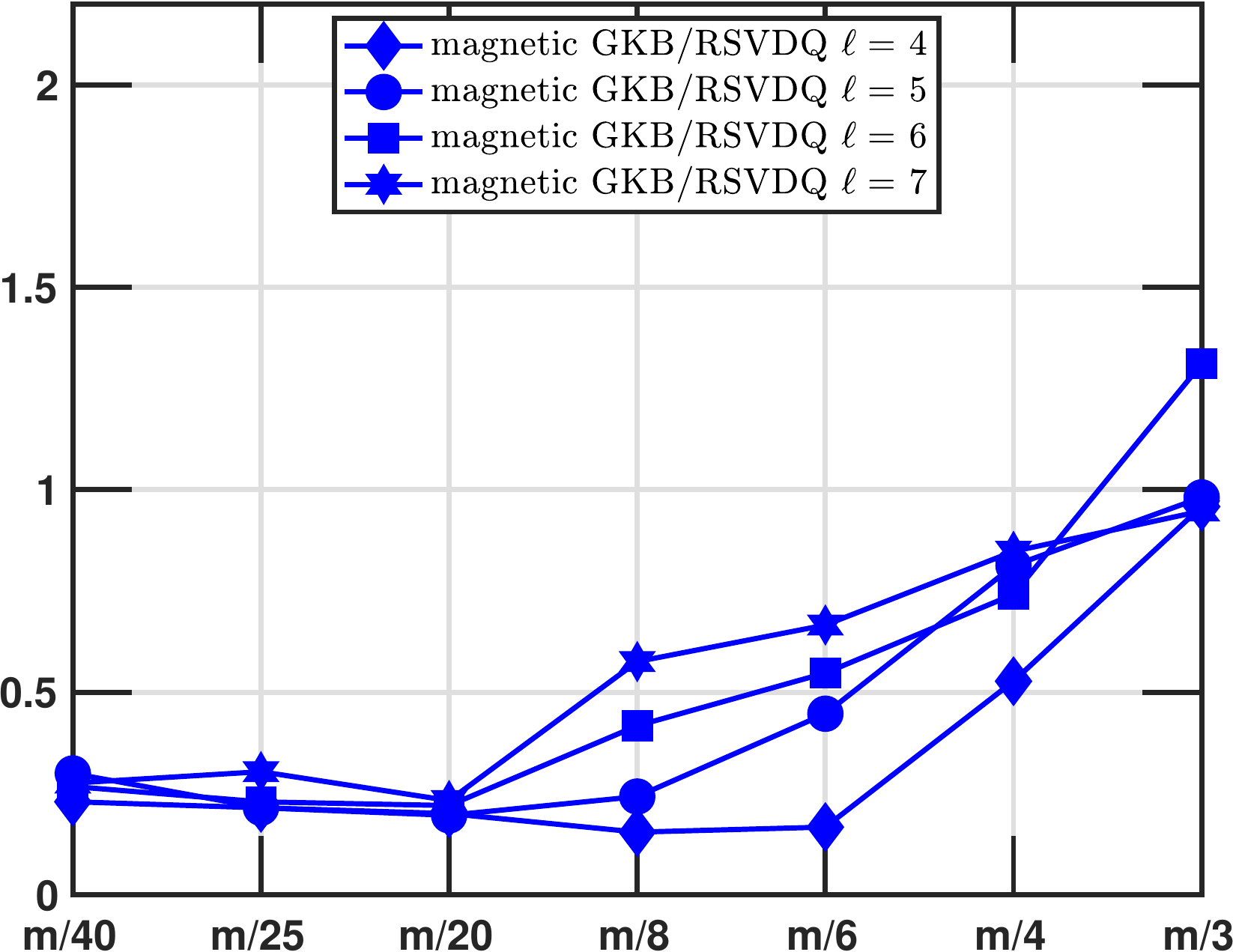}}
\subfigure[\texttt{gravity} $\texttt{Cost}_{\texttt{GKB}}/\texttt{Cost}_{\texttt{RSVD}}. $ \label{figure8b}]{\includegraphics[width=0.44\textwidth]{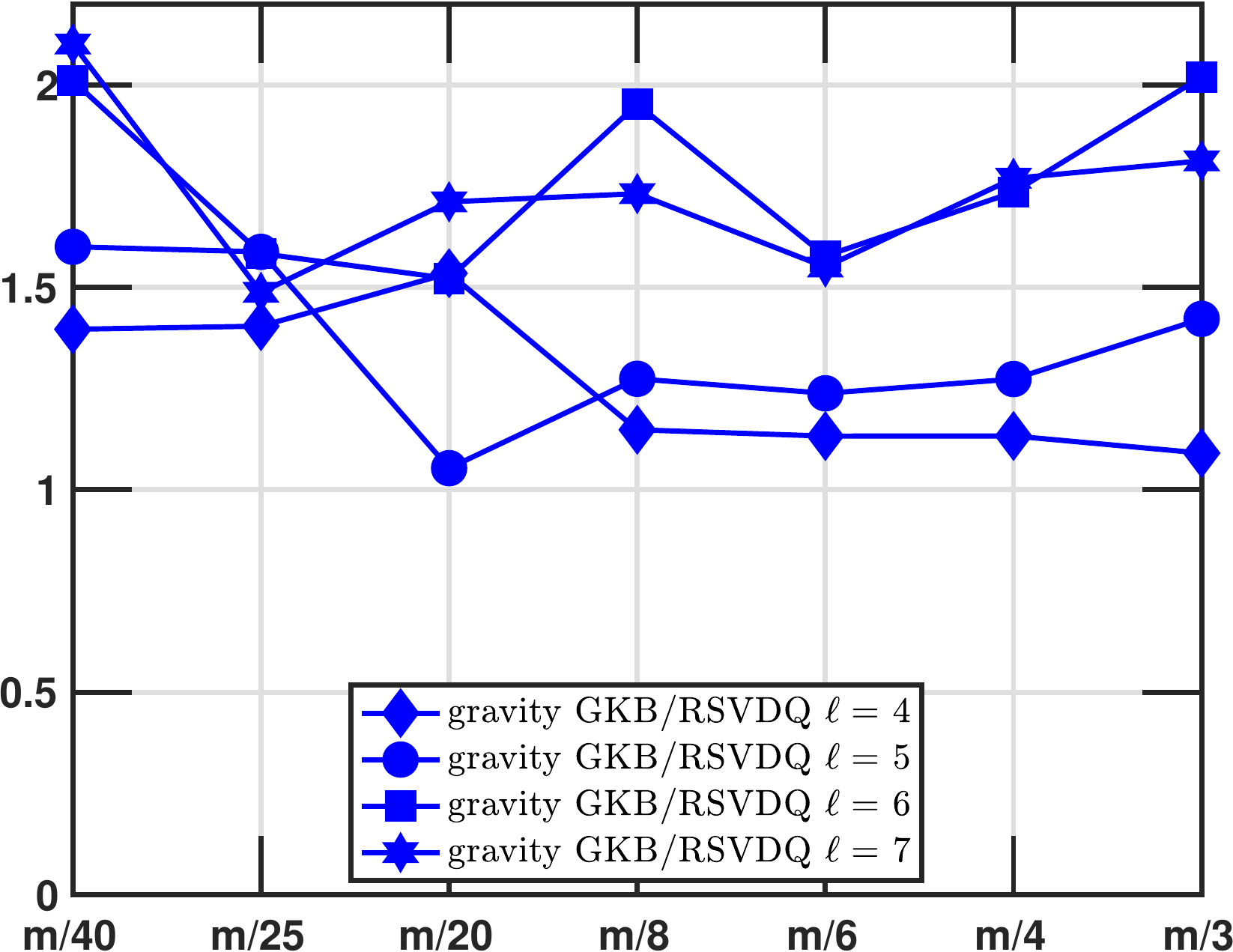}}
\caption{Computational cost to convergence of the $\irls$ algorithm for inversion using the \texttt{GKB} as compared to the \texttt{RSVD} algorithm, $\texttt{Cost}_{\texttt{GKB}}/\texttt{Cost}_{\texttt{RSVD}} $, for the \texttt{magnetic} and \texttt{gravity} problems respectively, in Figures~\ref{figure8a}-\ref{figure8b}. \label{figure8}}
\end{center}\end{figure}

The computational cost of the \texttt{IRLS} algorithm for solving the inversion problem to convergence depends on the choice of $t$, the choice of \texttt{GKB} or \texttt{RSVD} algorithms, and whether solving the magnetic or the gravity problem.   In Table~\ref{tableB.4} we report the timing results for the inversion of gravity and magnetic data for problems of increasing size $\ell$ and projected spaces of sizes $t_p$. The relative total computational costs to convergence, $\texttt{Cost}_{\texttt{GKB}}/\texttt{Cost}_{\texttt{RSVD}} $, (the last two columns in Table~\ref{tableB.4})  are illustrated via Figures~\ref{figure8a}-\ref{figure8b}, for the magnetic and gravity results, respectively. There is a distinct difference between the two problems. The results in Figure~\ref{figure8a} for the magnetic problem demonstrate a strong preference for the use of the \texttt{GKB} algorithm, except for large $t$, $t=\floor(m/3)$.   In contrast, the \texttt{RSVD} algorithm is always most efficient for the solution of the gravity problem, which is consistent with the conclusion presented in \cite{saeed6} for \texttt{RSVD} without  power iteration. Moreover, the data presented in Table~\ref{tableB.6} for the gravity problem, indicate that the \texttt{RSVD} algorithm generally converges more quickly and yields a smaller relative error. Furthermore, if based entirely on the calculated \texttt{RE}, the results suggest that good results can be achieved for relatively small $t$ as compared to $m$, certainly $s\gtrsim 8$ leads to generally acceptable error estimates, and in contrast to the case without the power iteration, here with power iteration, the errors using the \texttt{GKB} are generally larger for comparable choices of $t$. 

For the magnetic data,  the results in Table~\ref{tableB.5} demonstrate that the \texttt{RSVD} algorithm generally requires more iterations than the \texttt{GKB} algorithm, and that the obtained relative errors are then comparable, or slightly larger. This is then reflected  in Figure~\ref{figure8a} that the \texttt{GKB} algorithm is most efficient. Referring back to Table~\ref{tableB.5}, it is the case that the \texttt{RSVD} algorithm often reaches the maximum number of iterations, $K=25$, without convergence, when \texttt{GKB} has converged in less than half the number of iterations, when $t$ is small relative to $m$, $t=\floor(m/s)$ with $s=40$, $25$ and $20$. This verifies that the \texttt{RSVD} needs to take a larger projected subspace $t$ in order to capture the required dominant spectral space when solving the magnetic problem, as compared to the gravity problem, and confirms the conclusions presented in \cite{vatankhah2019improving}.  On the other hand, the use of the \texttt{GKB} as compared to the \texttt{RSVD} was not discussed in \cite{vatankhah2019improving}. Our results now lead to a new conclusion concerning these two algorithms for solving the magnetic data inversion problem. In particular, the results suggest  that the \texttt{GKB} algorithm be adopted for inversion of magnetic data. Further, the results suggest that the relative error obtained using the \texttt{GKB} generally decreases with increasing $t$, and that it is necessary to use subspaces with $t$ at least as large as $\floor(m/8)$. It remains to verify these assertions by illustrating the results of the inversions and the predicted anomalies for a selection of cases.

\subsubsection{Illustrating Solutions with Increasing $\ell$ and $t$}\label{sec:methodsolutions}
\begin{figure}[ht!]\begin{center}
\subfigure[ \texttt{GKB}:   $\ell=4$, $\quad \quad $ $\quad t=750$,    $(9,248\textrm{s})$.    \label{figure9a}]{\includegraphics[width=0.24\textwidth]{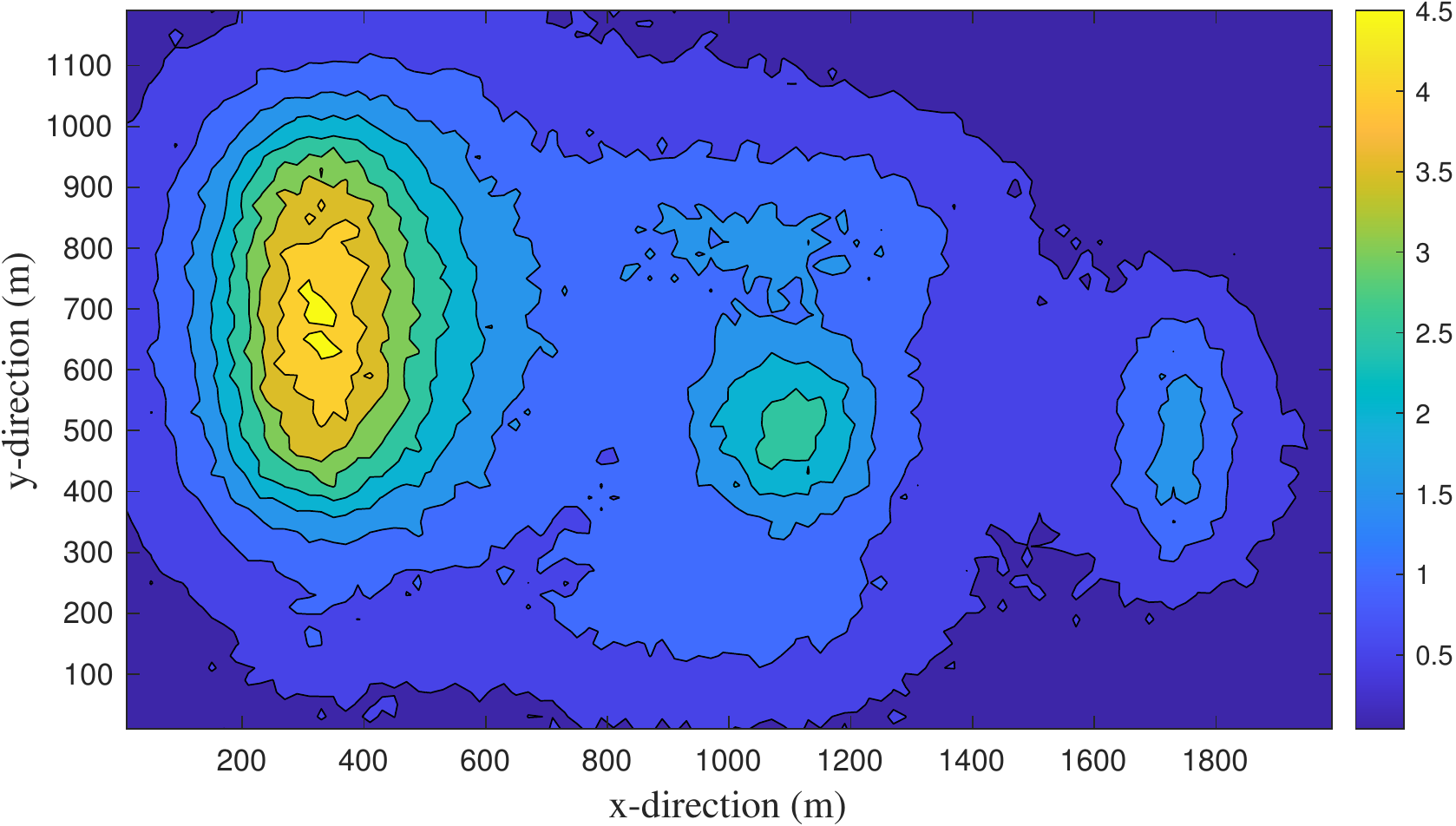}}
\subfigure[$\ell=4$, $t=1500$,     $(7,494\textrm{s})$.    \label{figure9b}]{\includegraphics[width=0.24\textwidth]{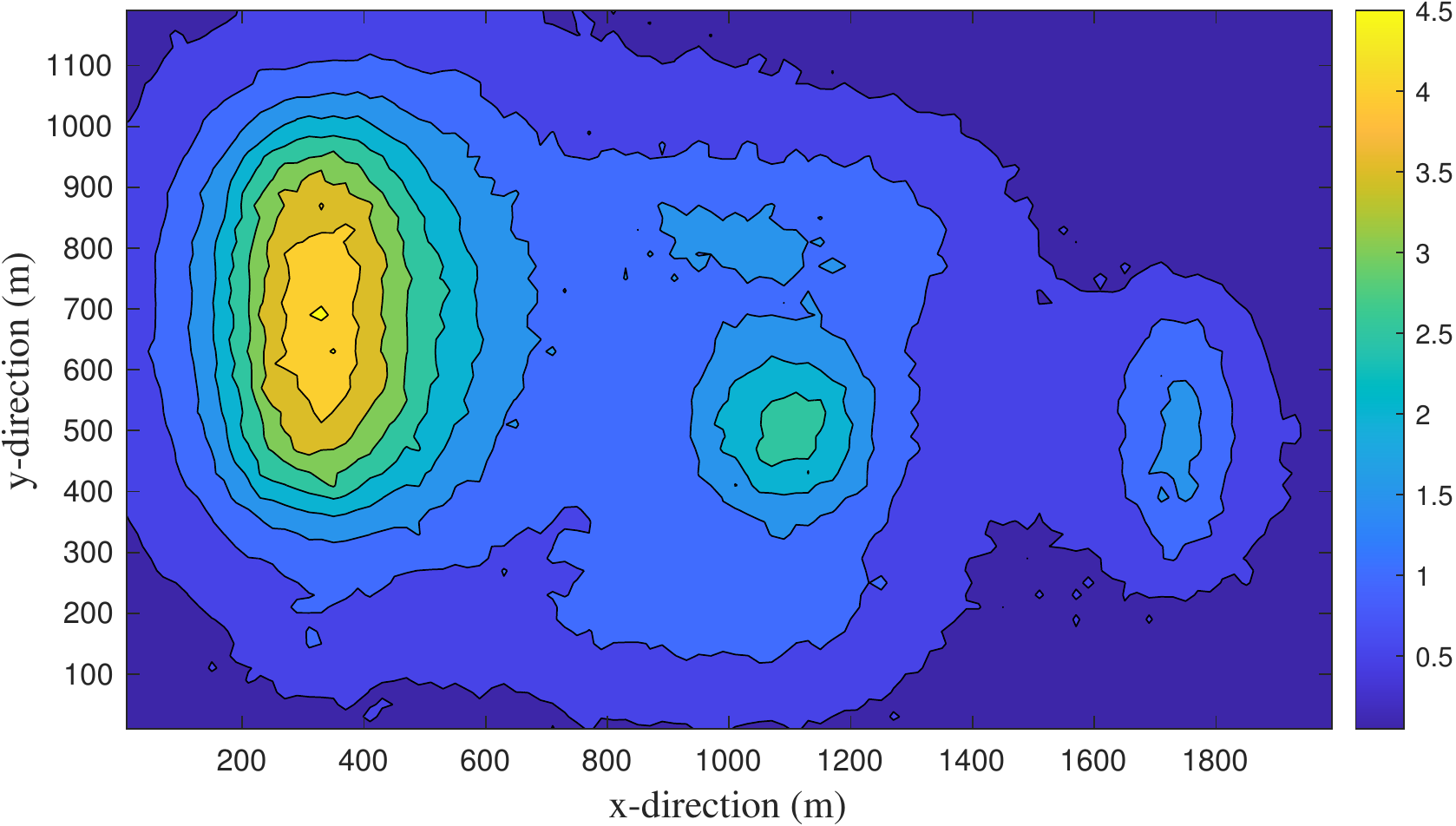}}
\subfigure[$\ell=7$, $t=2296$,     $(11,5732\textrm{s})$. \label{figure9c}]{\includegraphics[width=0.24\textwidth]{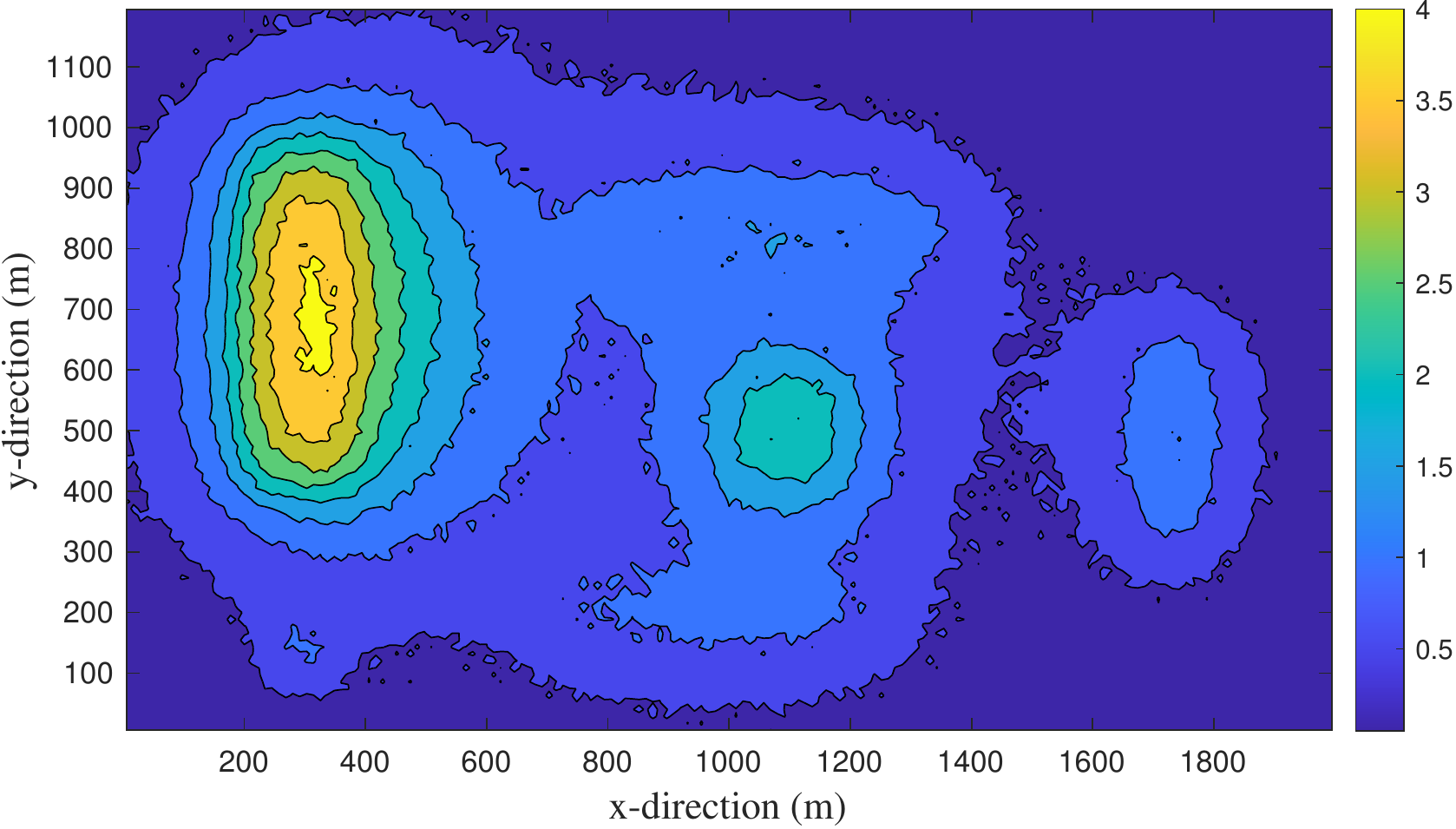}}
\subfigure[$\ell=7$, $t=4593$,     $(8,12347\textrm{s})$.\label{figure9d}]{\includegraphics[width=0.24\textwidth]{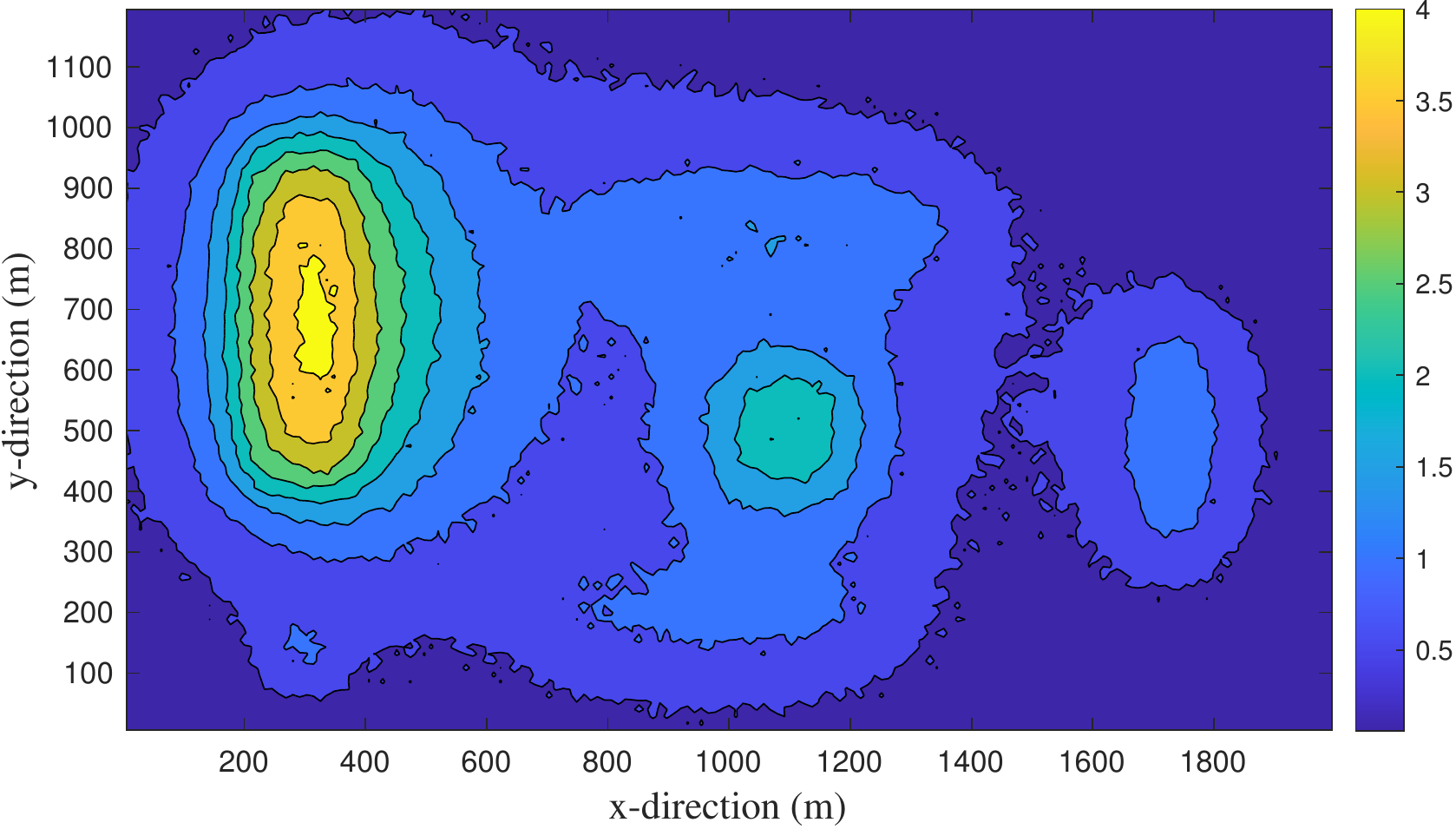}}
\subfigure[ \texttt{RSVD}: $(6,216\textrm{s})$.    \label{figure9e}]{\includegraphics[width=0.24\textwidth]{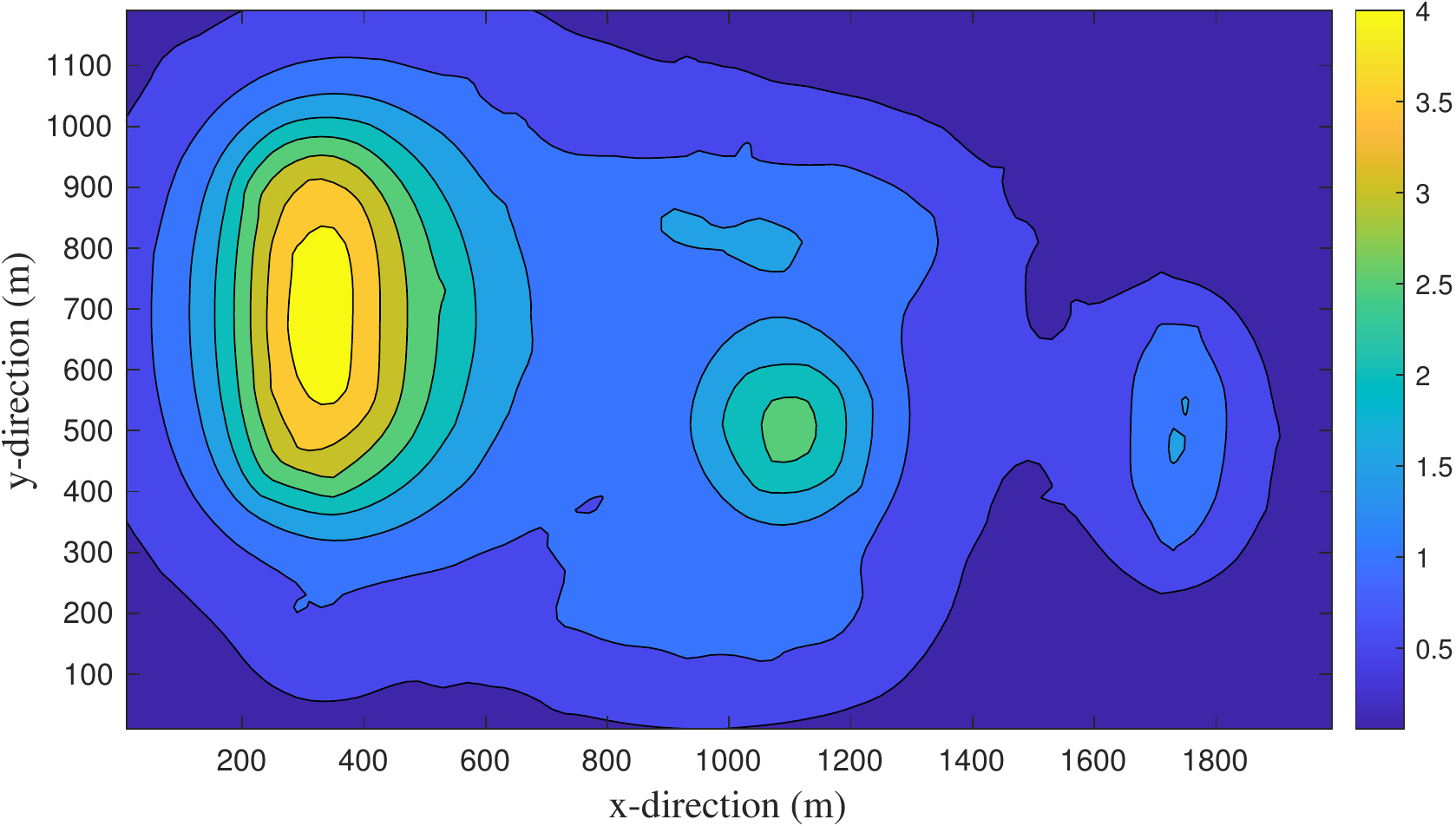}}
\subfigure[  $(6,436\textrm{s})$.    \label{figure9f}]{\includegraphics[width=0.24\textwidth]{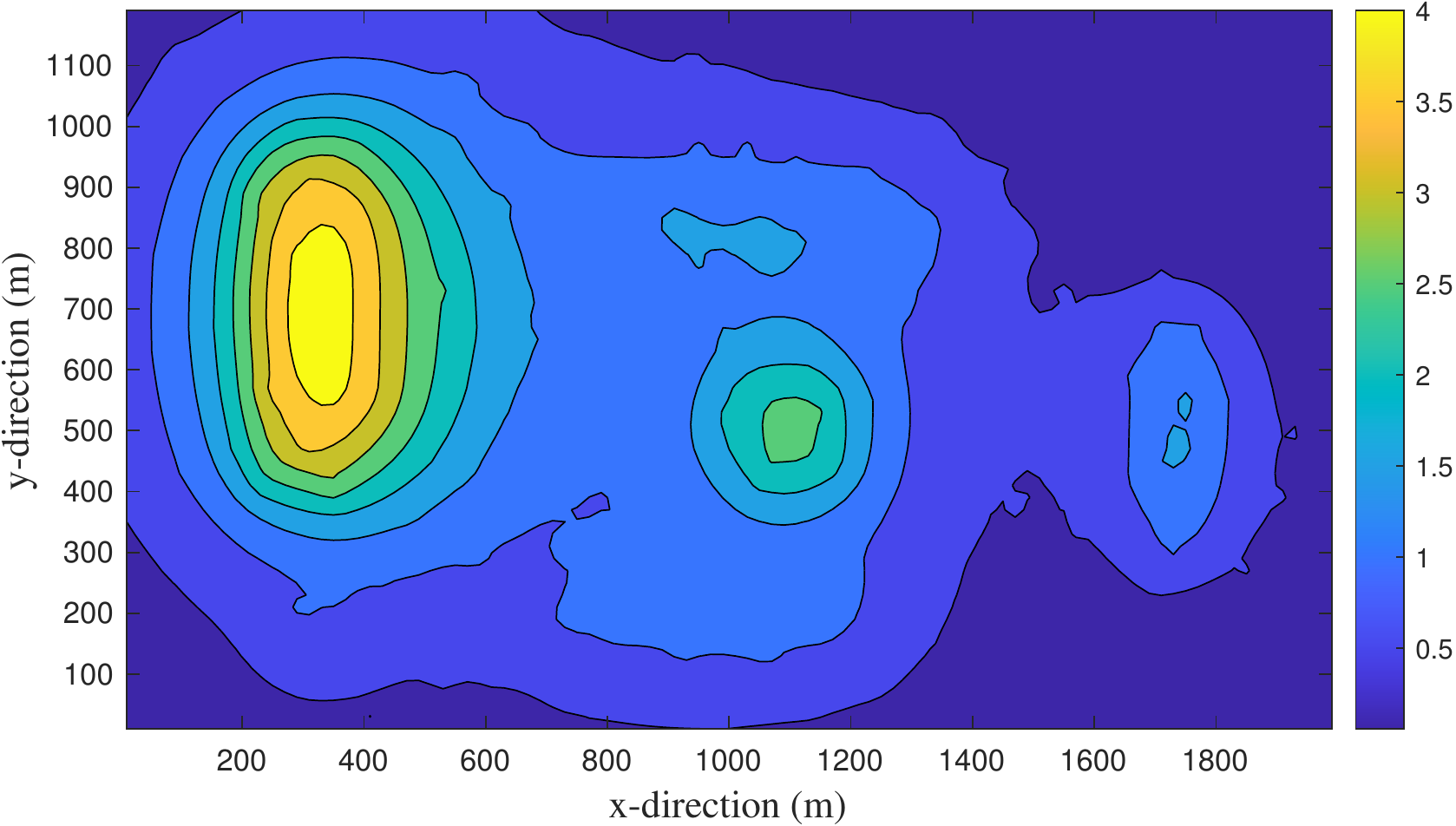}}
\subfigure[  $(7,3311\textrm{s})$. \label{figure9g}]{\includegraphics[width=0.24\textwidth]{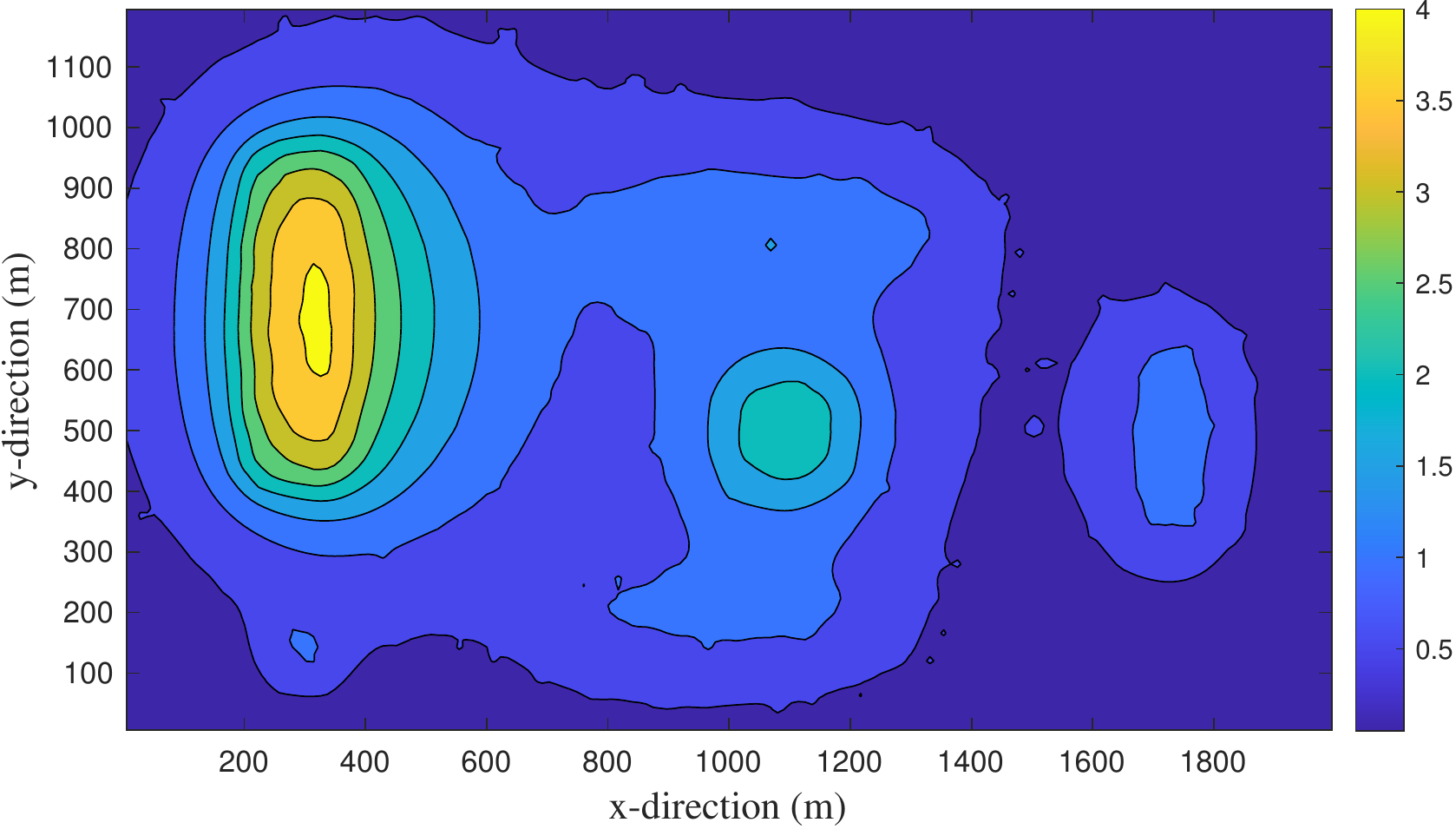}}
\subfigure[   $(7,6979\textrm{s})$.  \label{figure9h}]{\includegraphics[width=0.24\textwidth]{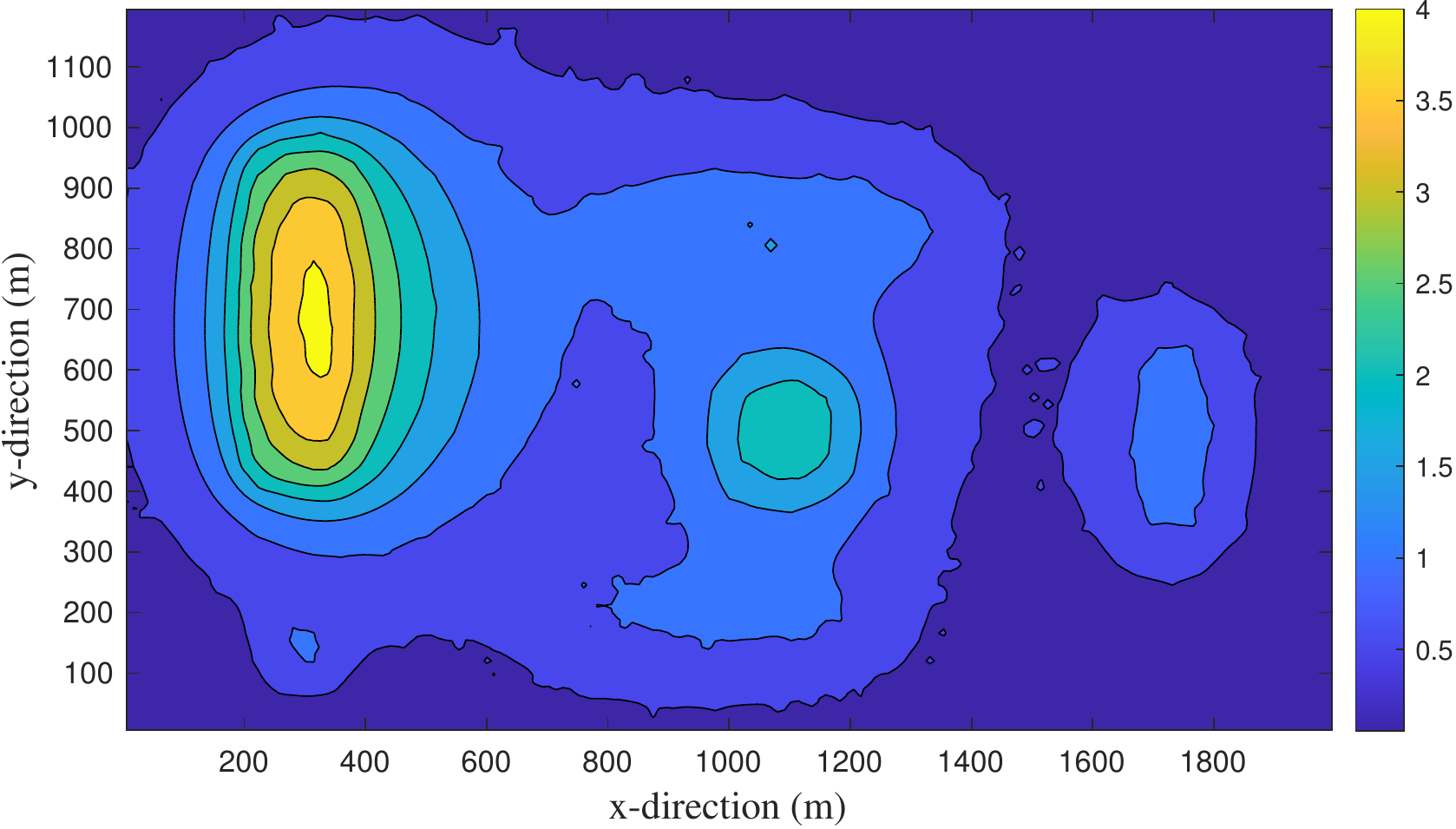}}
\subfigure[ \texttt{GKB}:  $(.76, .90)$.\label{figure9i}]{\includegraphics[width=0.24\textwidth]{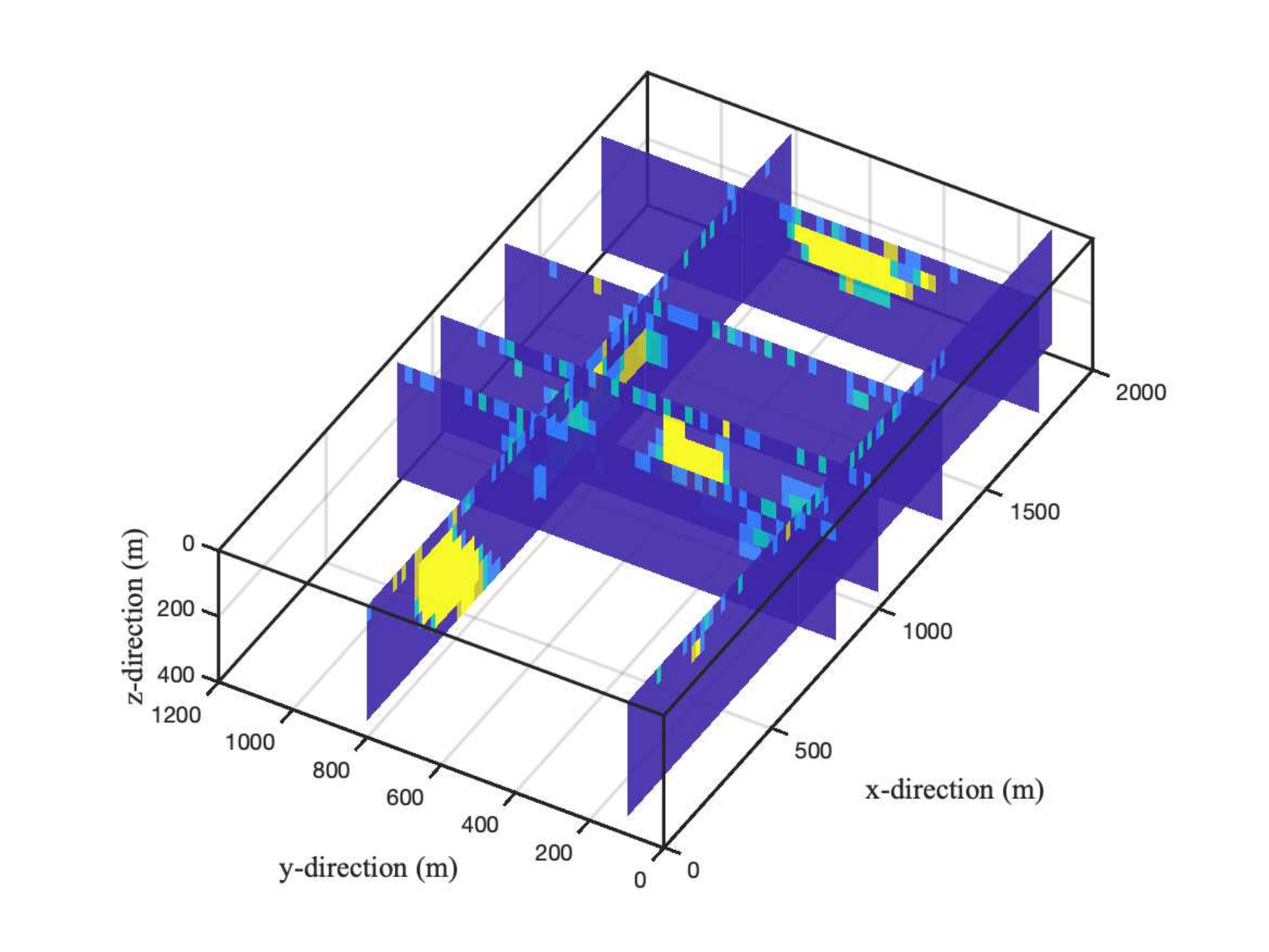}}
\subfigure[$(.64, .96)$.\label{figure9j}]{\includegraphics[width=0.24\textwidth]{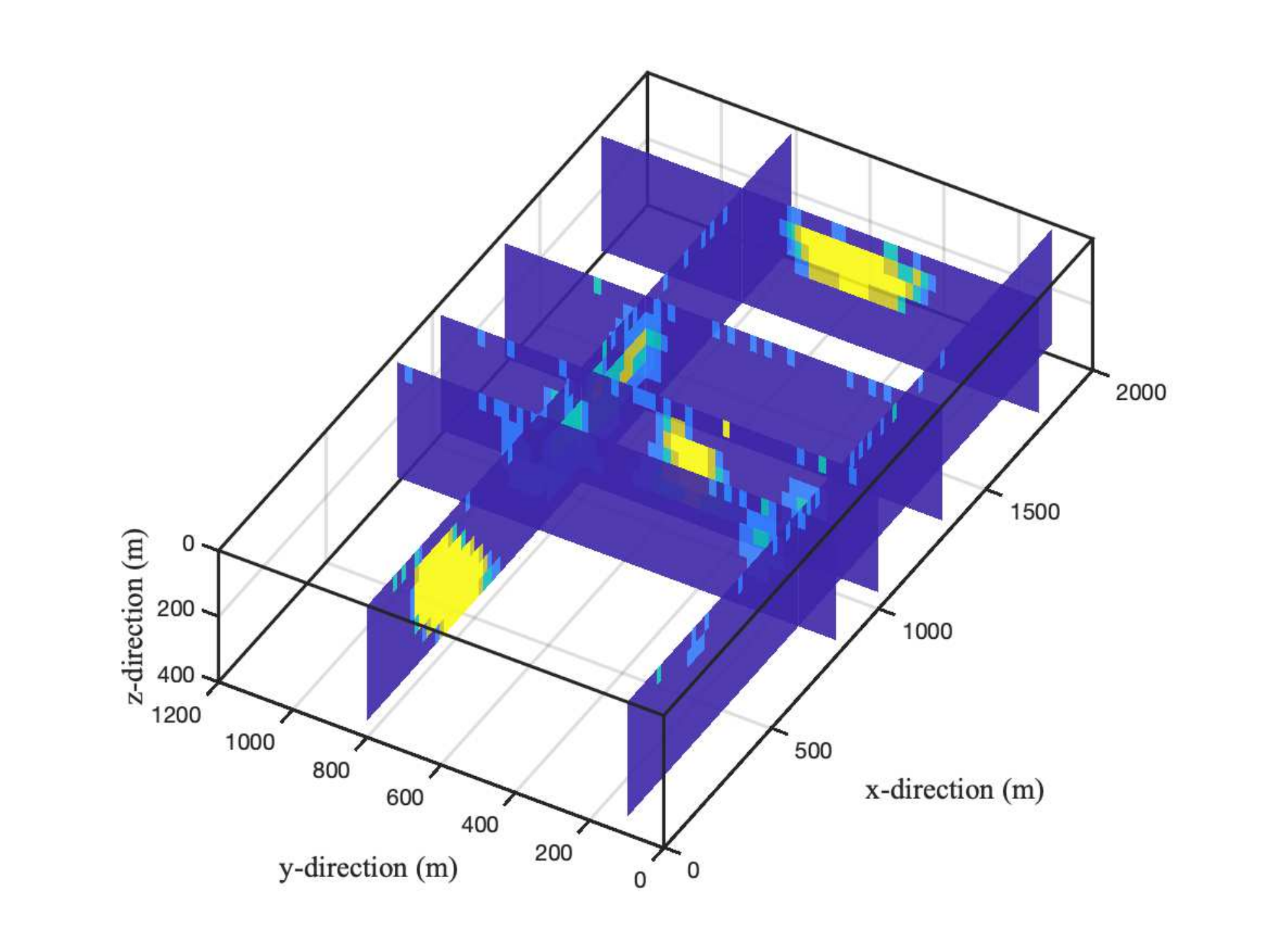}}
\subfigure[$(.75,.81)$.\label{figure9k}]{\includegraphics[width=0.24\textwidth]{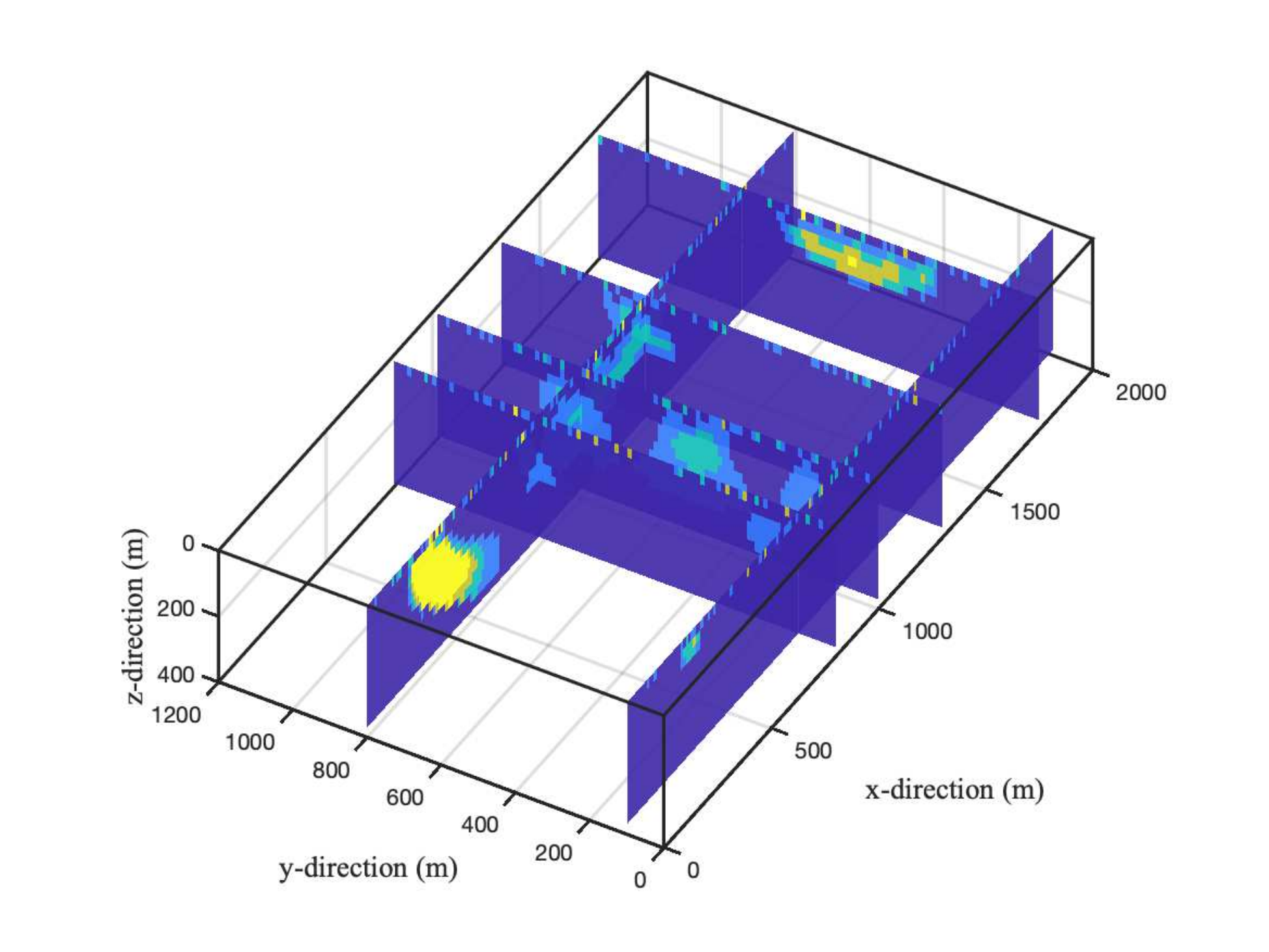}}
\subfigure[$(.70,1.00)$.\label{figure9l}]{\includegraphics[width=0.24\textwidth]{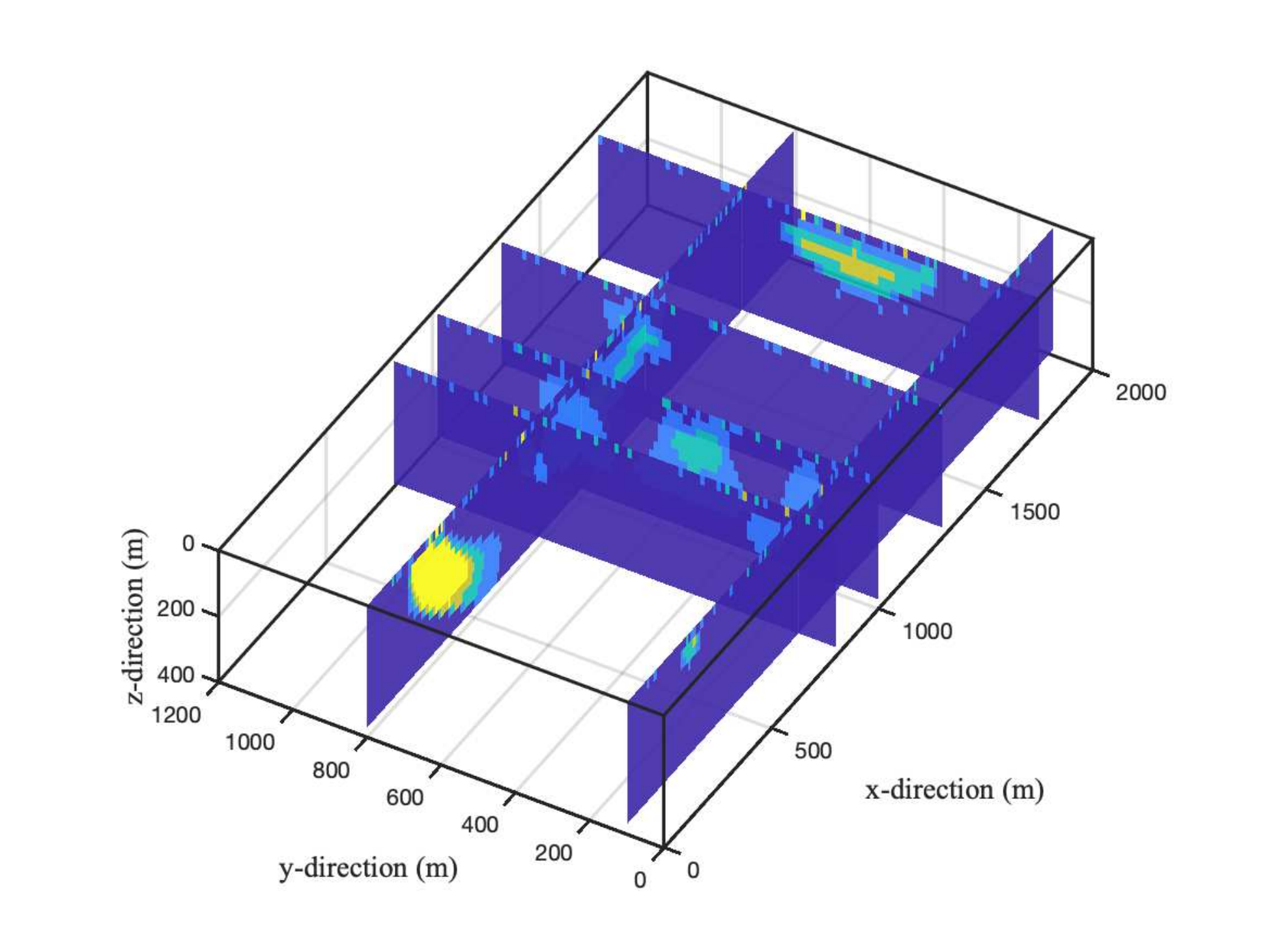}}
\subfigure[\texttt{RSVD}: $(.57,.96)$.\label{figure9m}]{\includegraphics[width=0.24\textwidth]{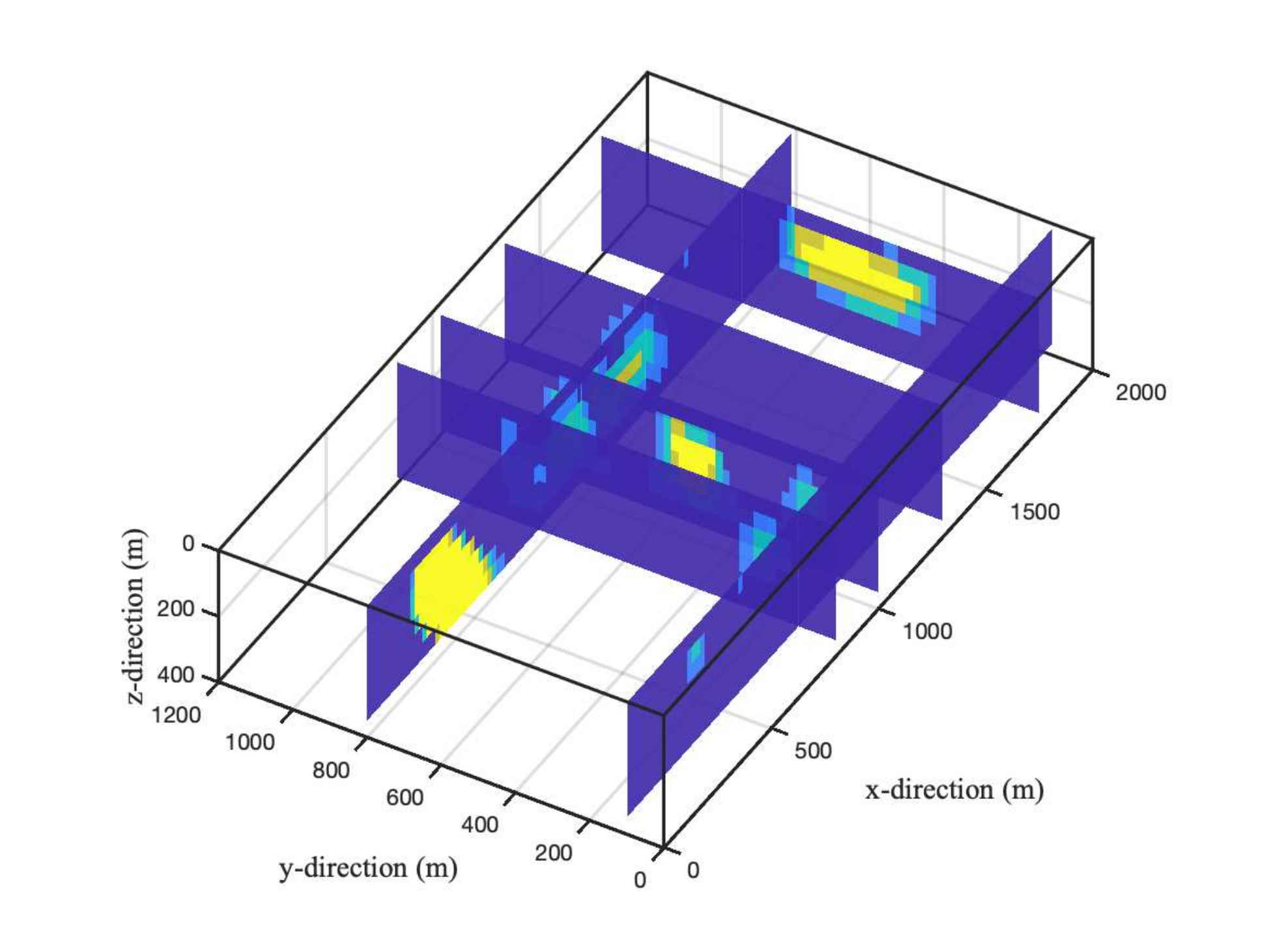}}
\subfigure[$(.57,.93)$.\label{figure9n}]{\includegraphics[width=0.24\textwidth]{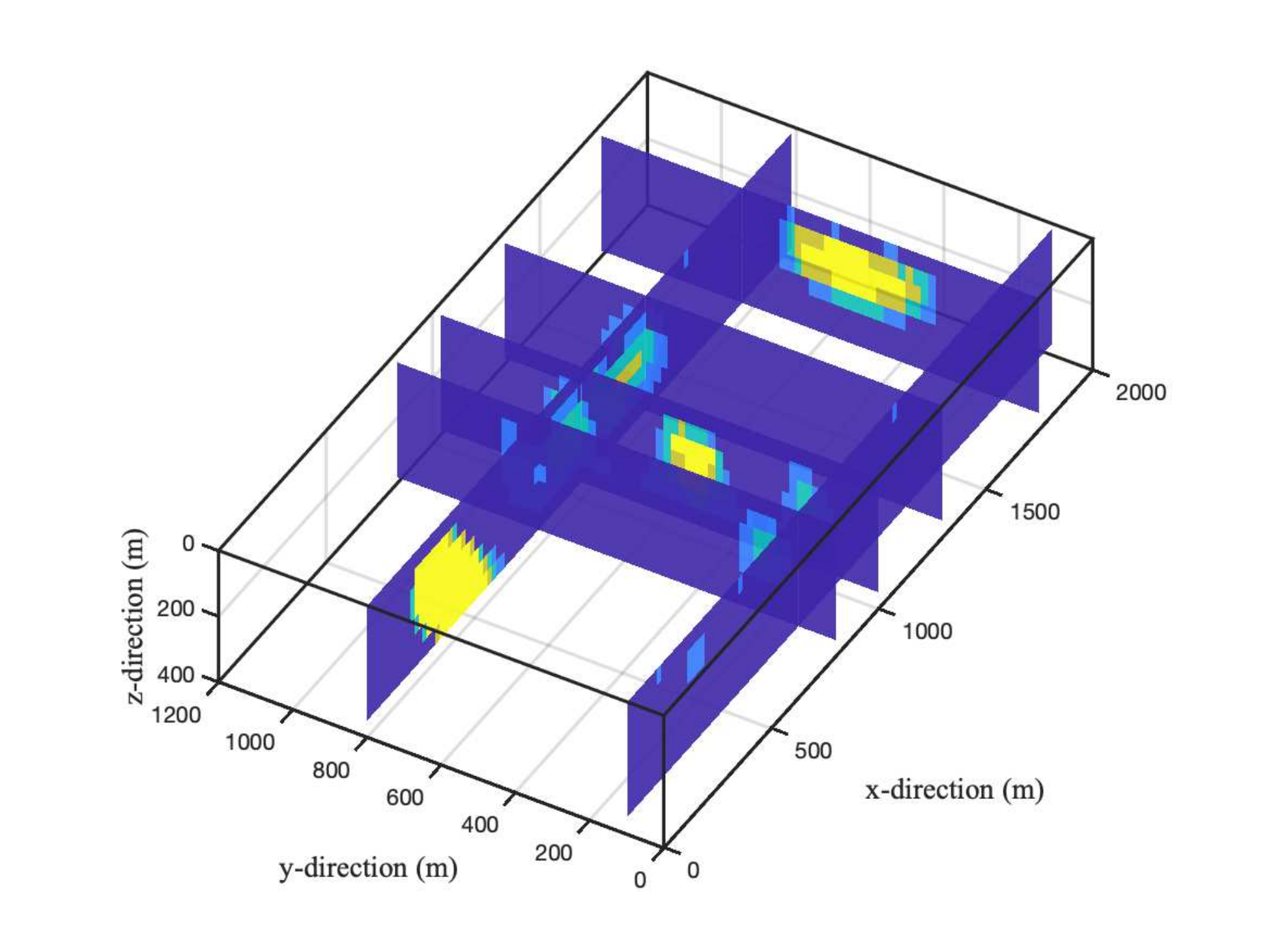}}
\subfigure[$(.60,.95)$.\label{figure9o}]{\includegraphics[width=0.24\textwidth]{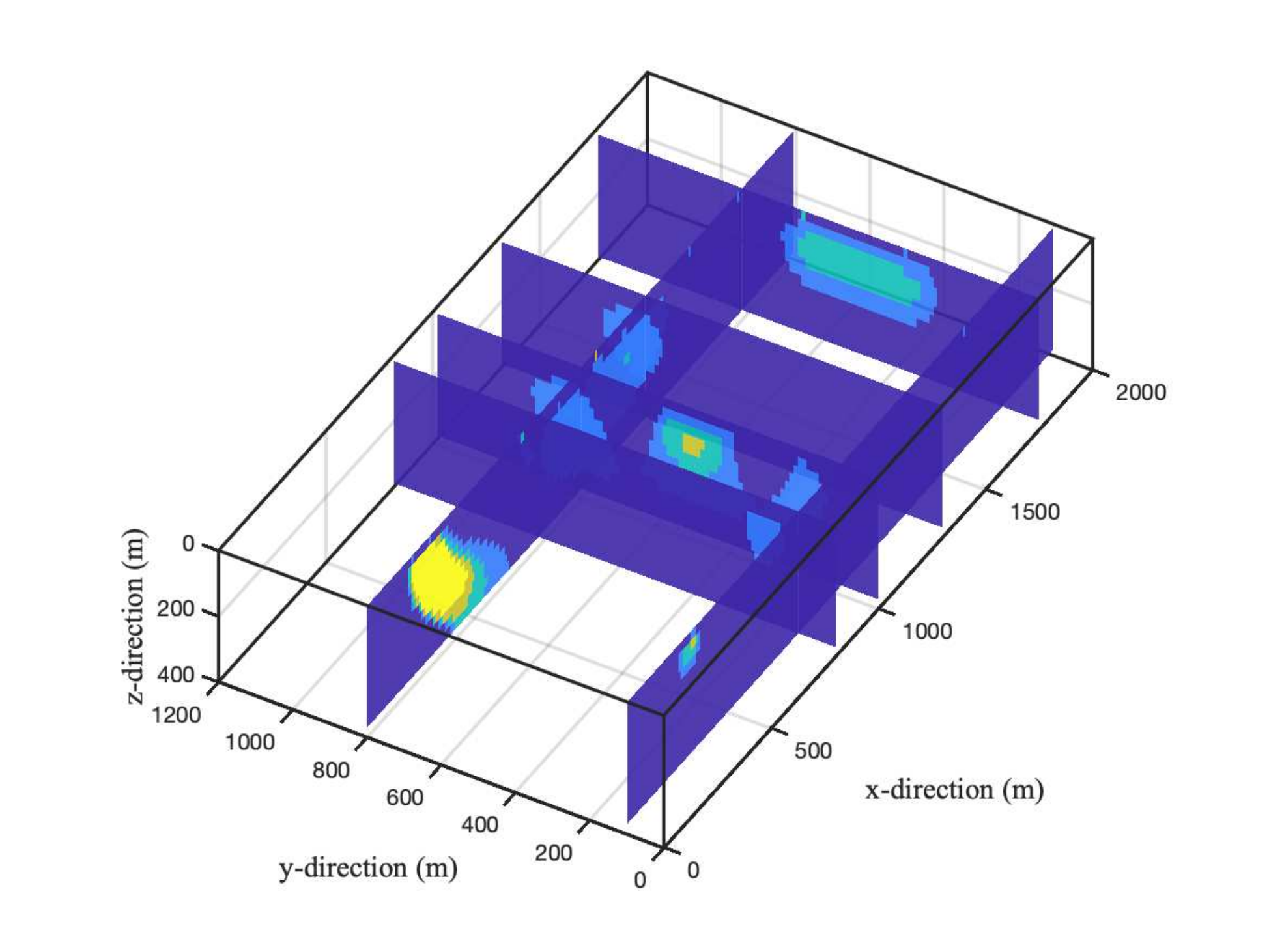}}
\subfigure[$(.61,.91)$.\label{figure9p}]{\includegraphics[width=0.24\textwidth]{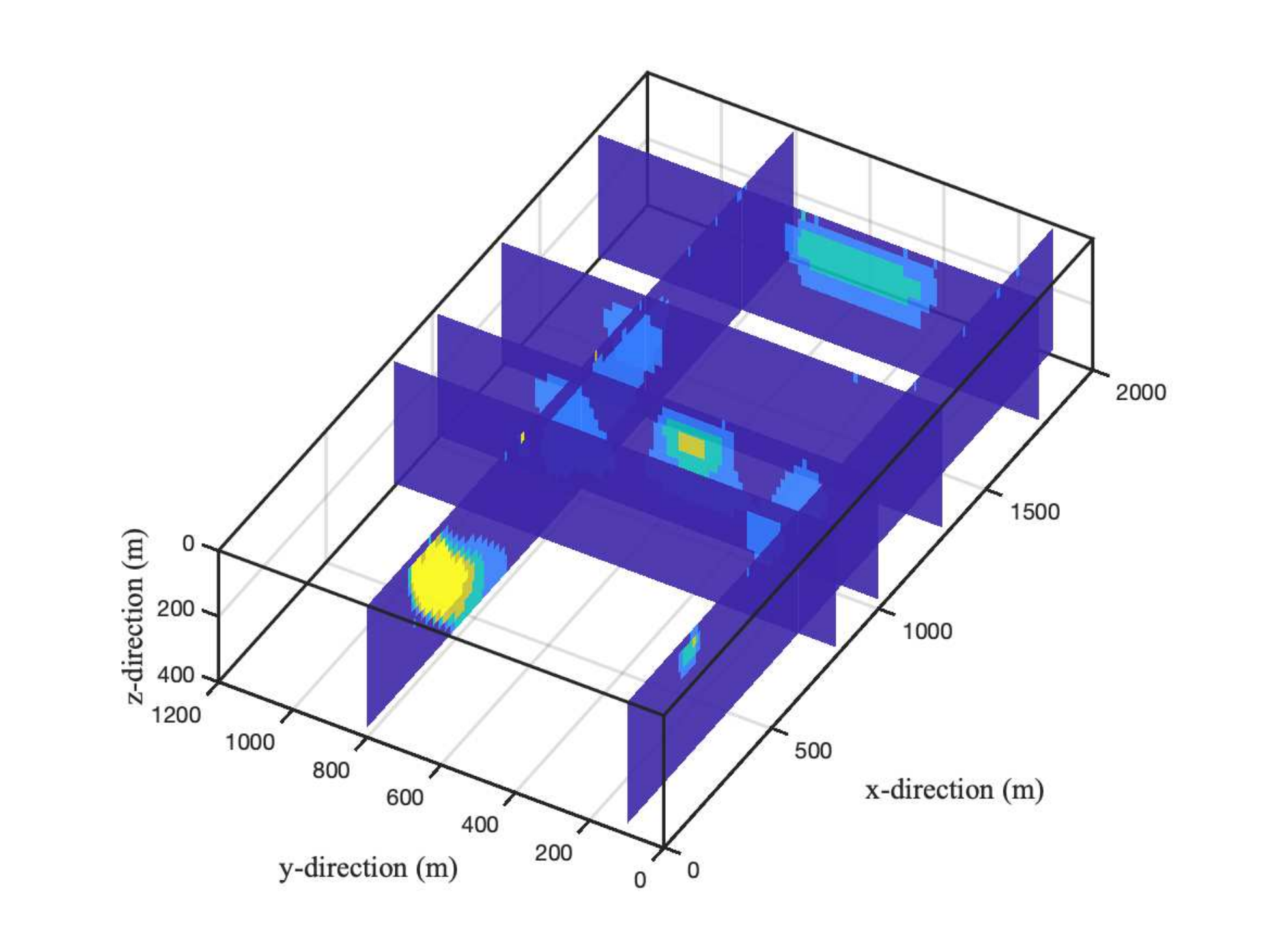}}
\caption{For \gravity~data the predicted anomalies obtained using \texttt{GKB} in Figures~\ref{figure9a}-\ref{figure9d} and \texttt{RSVD} in Figures~\ref{figure9e}-\ref{figure9h}, with the corresponding reconstructed volumes in Figures~\ref{figure9i}-\ref{figure9l} and Figures~\ref{figure9m}-\ref{figure9p}, respectively.  In each case the first row for \texttt{GKB} indicates the choices of  $\ell$ and $t$ in each column. The choices $t=750$ and $t=1500$ for $\ell=4$,  and with $t=2296$ and $t=4593$  for $\ell=7$, correspond to $t=\floor(m/8)$ and $t=\floor(m/4)$ for  $(m,n) = (6000, 48000)$ and 
$(18375, 257250)$, respectively.  Given are the pairs $(K, \texttt{Cost}\textrm{s})$,  (number of iterations to convergence and computational cost in seconds) in the captions of the anomalies, and $(\texttt{RE}, \chi^2/(m+\sqrt{2m}))$ in the captions of the reconstructions. 
Results for all cases are summarized in Table~\ref{tableB.6} with timings in Table~\ref{tableB.4}. The units for the anomalies are mGal.\label{figure9}}
\end{center}
\end{figure}

We first compare a set of solutions for which the timing results were compared in Section~\ref{sec:methodconvergence}. Figure~\ref{figure9} illustrates the predicted anomalies and reconstructed volumes for gravity data inverted by both algorithms, with resolutions given by $\ell=4$ and $\ell=7$ with $t=\floor(m/8)$ and $t=\floor(m/4)$. For the cases using $\ell=4$ it can be seen that the predicted anomalies are generally less accurate than with $\ell=7$. Moreover, there is little deterioration in the anomaly predictions when using $t=\floor(m/8)$ instead of $t=\floor(m/4)$, except that the results with the \texttt{GKB} show more residual noise. On the other hand, it is more apparent from consideration of the reconstructed volumes shown in Figures~\ref{figure9i}-\ref{figure9p} that the \texttt{RSVD} algorithm does yield better results in all cases, and specifically the high resolution $\ell=7$ results are very good, even using $t=\floor(m/8)$. When including the consideration of the computational cost, it is clear that if using $\ell=7$ it is sufficient to use $t=\floor(m/8)$ and the \texttt{RSVD} algorithm, but that a reasonable result may even be obtained using the same algorithm but with $\ell=4$ and requiring less than $5$ minutes of compute time. 

\begin{figure}[ht!]\begin{center}
\subfigure[\texttt{GKB}: $\ell=4, t=750$,        $(5,136\textrm{s})$.   \label{figure10a}]{\includegraphics[width=0.24\textwidth]{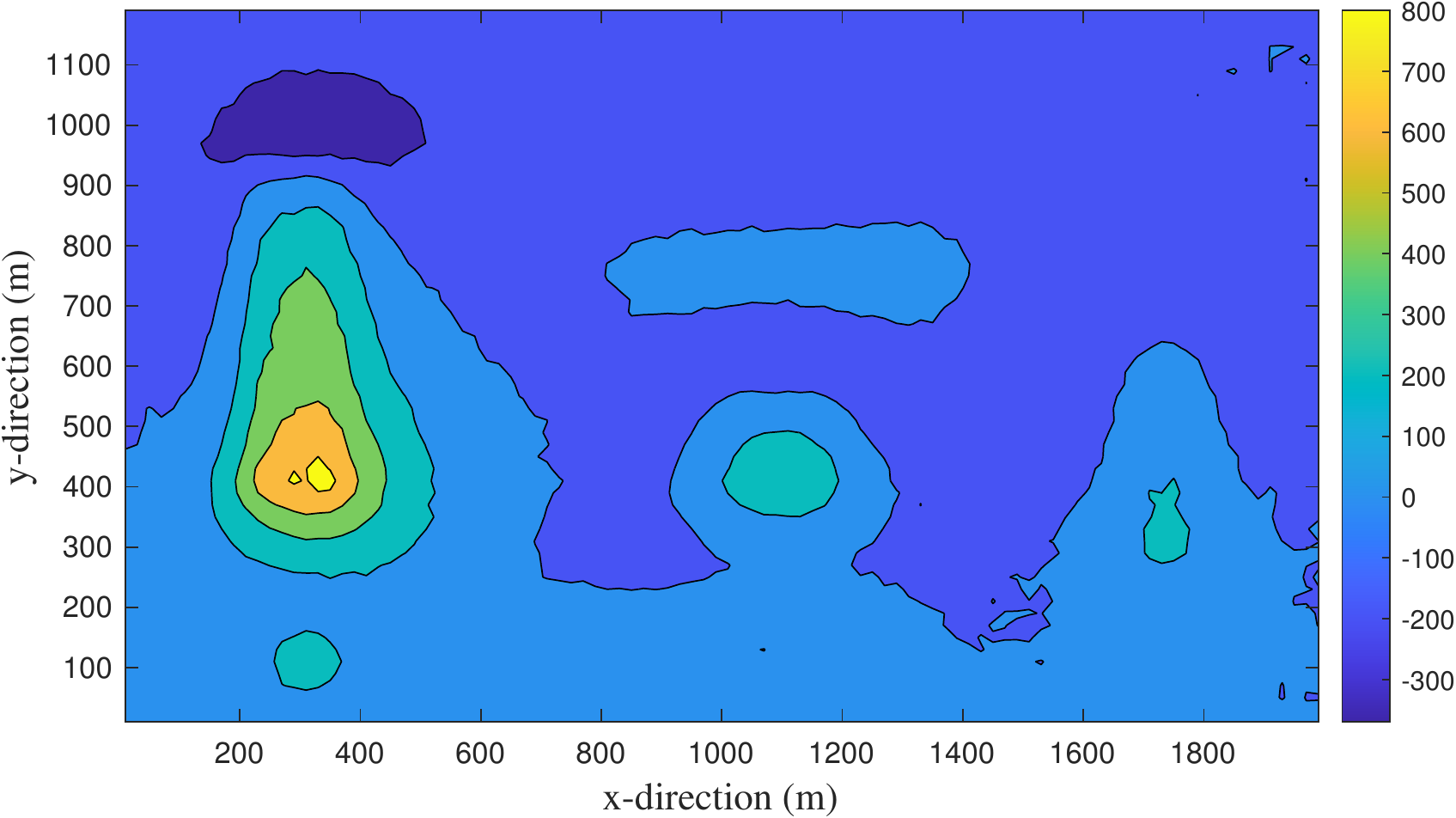}}
\subfigure[$\ell=4, t=1500$,    $(5, 343\textrm{s})$.   \label{figure10b}]{\includegraphics[width=0.24\textwidth]{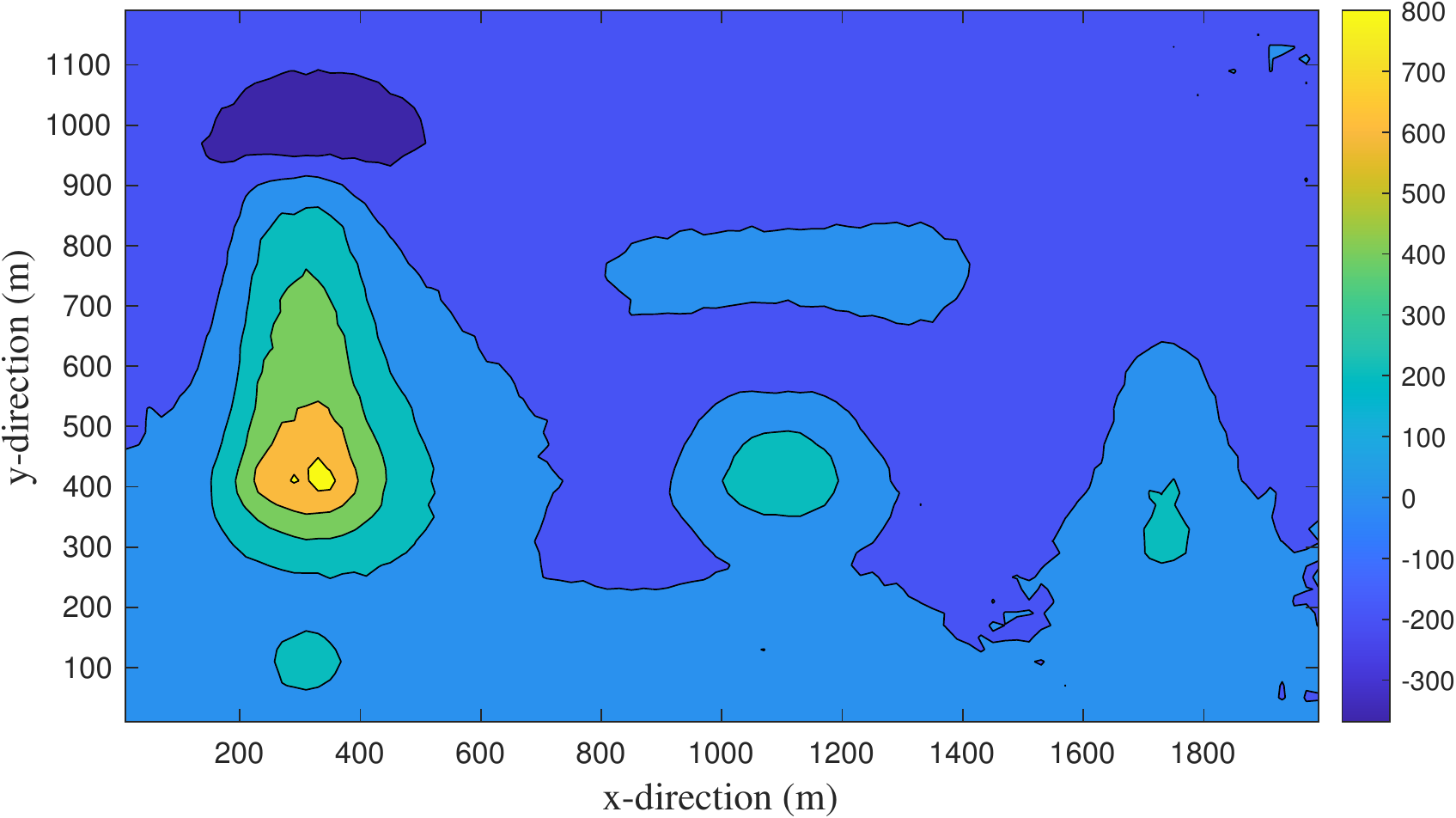}}
\subfigure[$\ell=7, t=2296$,     $(7, 3809\textrm{s})$. \label{figure10c}]{\includegraphics[width=0.24\textwidth]{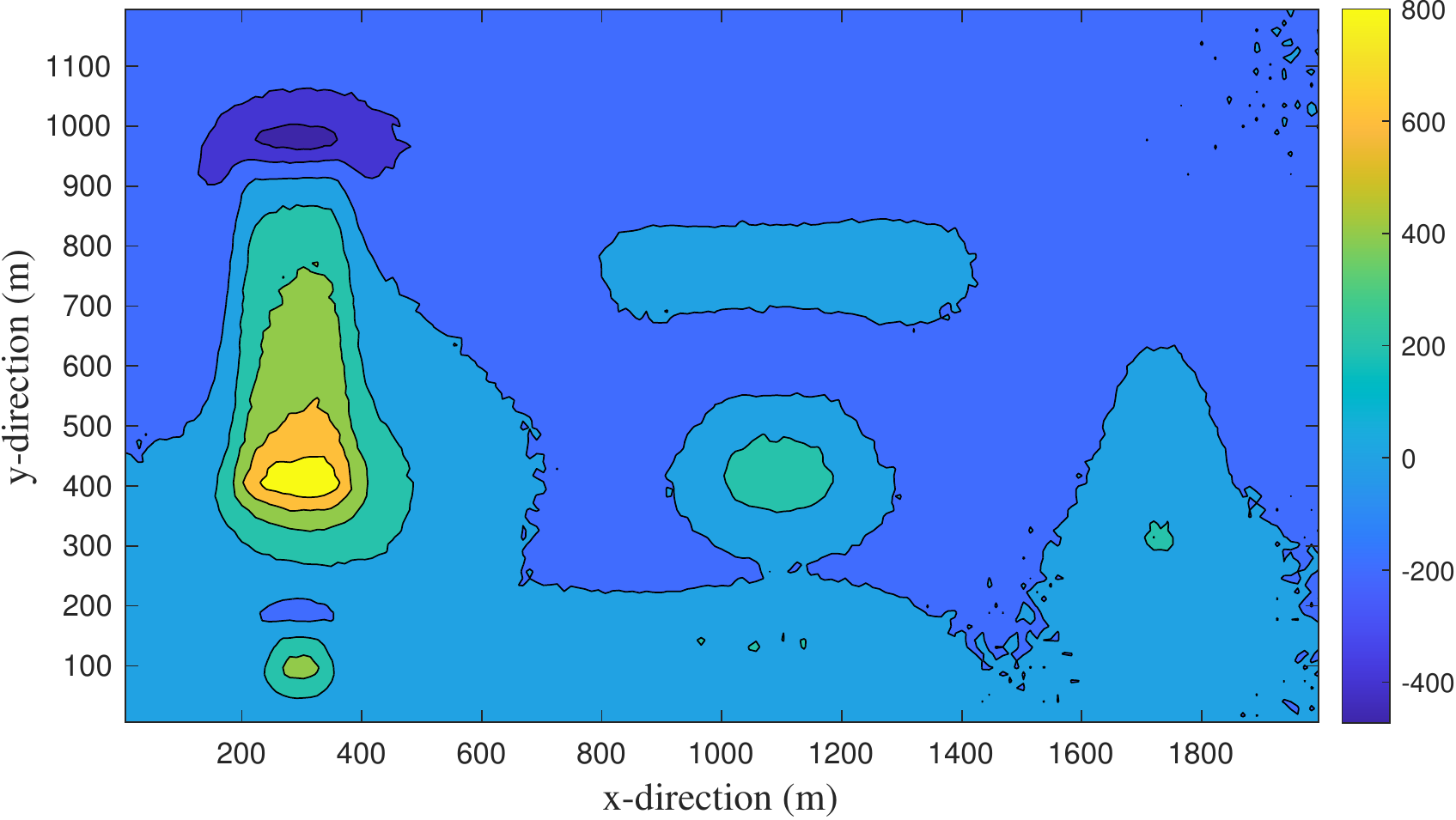}}
\subfigure[$\ell=7, t=4593$,     $(7, 10979\textrm{s})$.\label{figure10d}]{\includegraphics[width=0.24\textwidth]{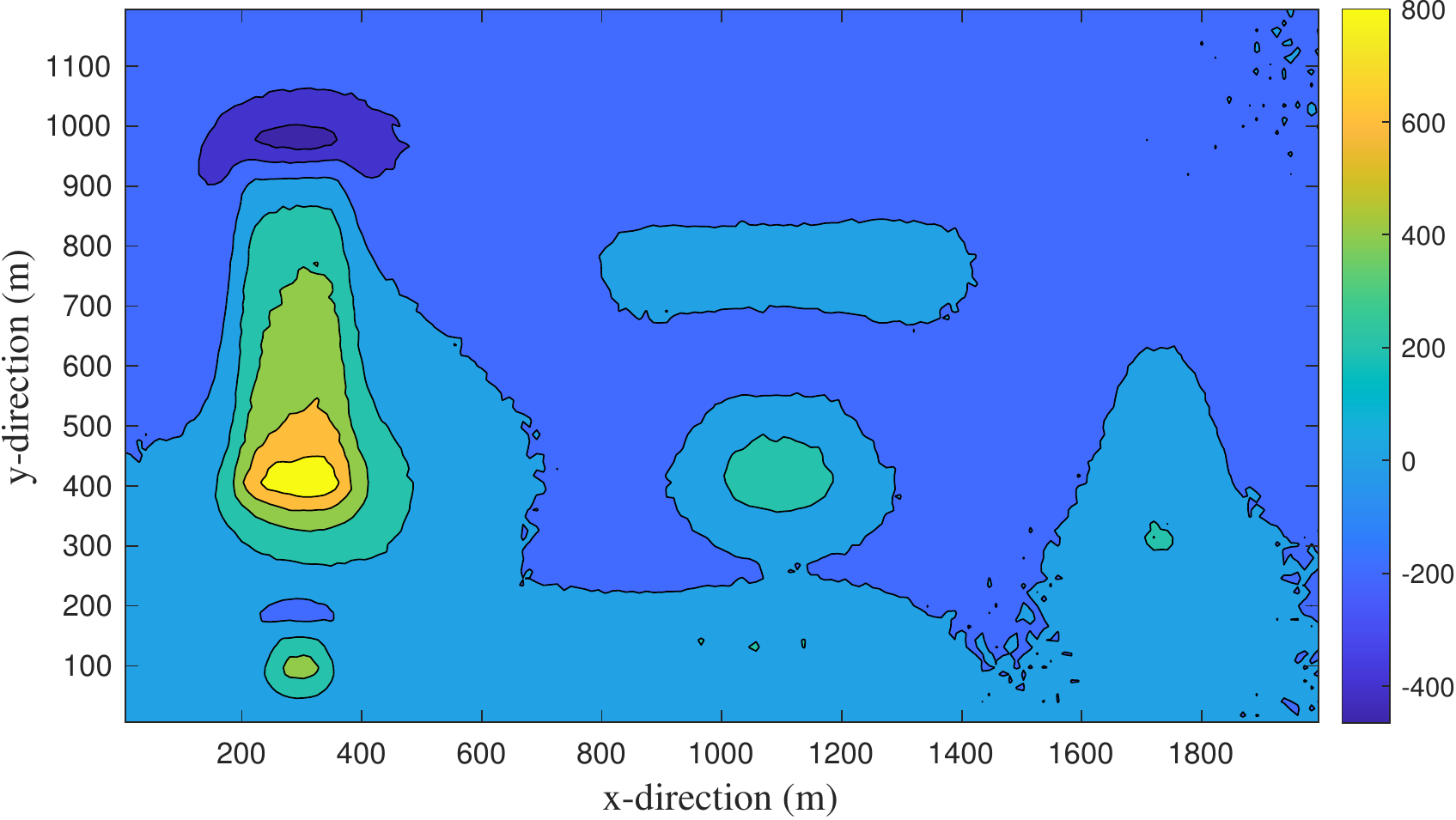}}
\subfigure[\texttt{RSVD}:    $(25, 883\textrm{s})$.   \label{figure10e}]{\includegraphics[width=0.24\textwidth]{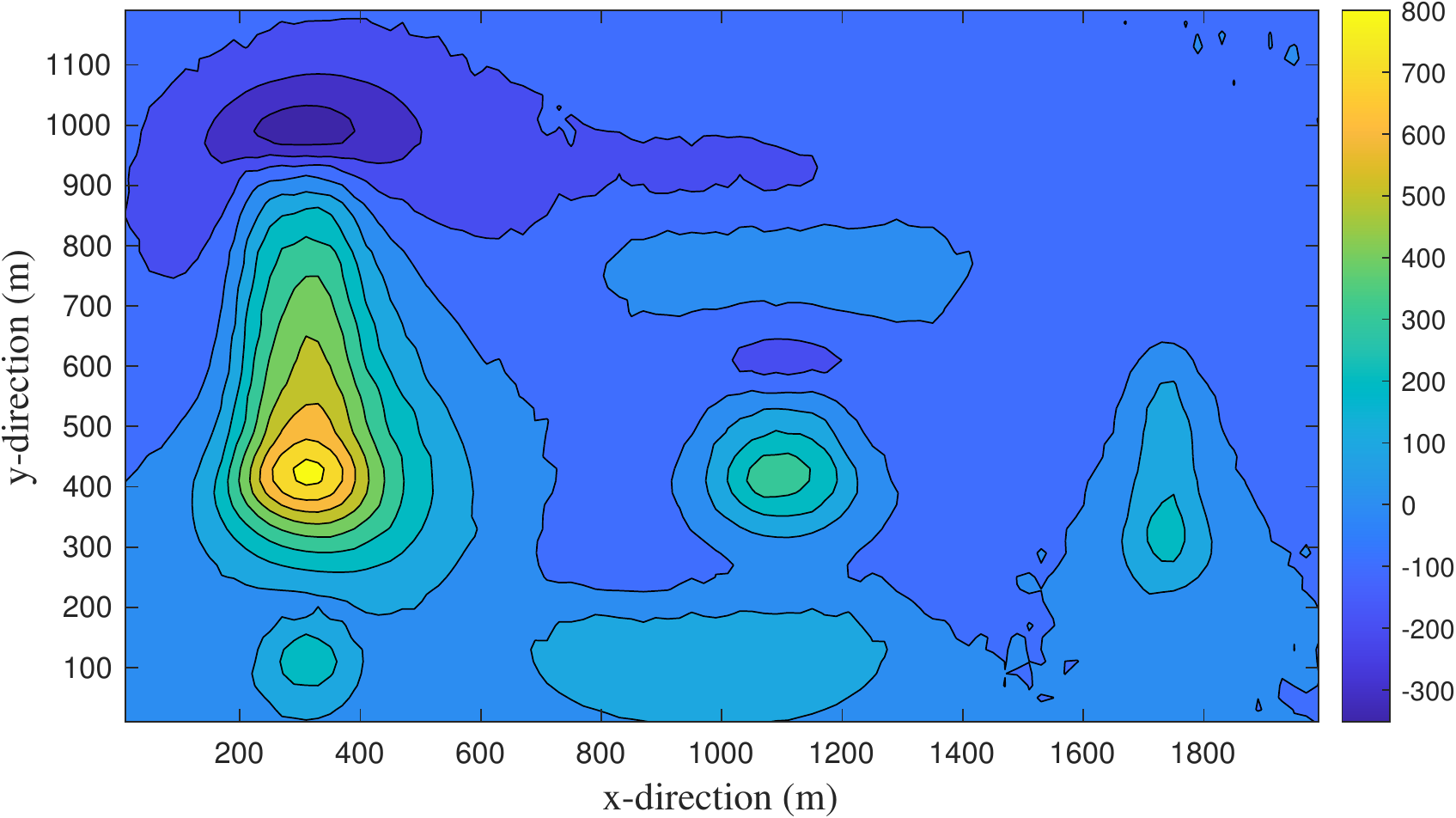}}
\subfigure[   $(9, 650\textrm{s})$.   \label{figure10f}]{\includegraphics[width=0.24\textwidth]{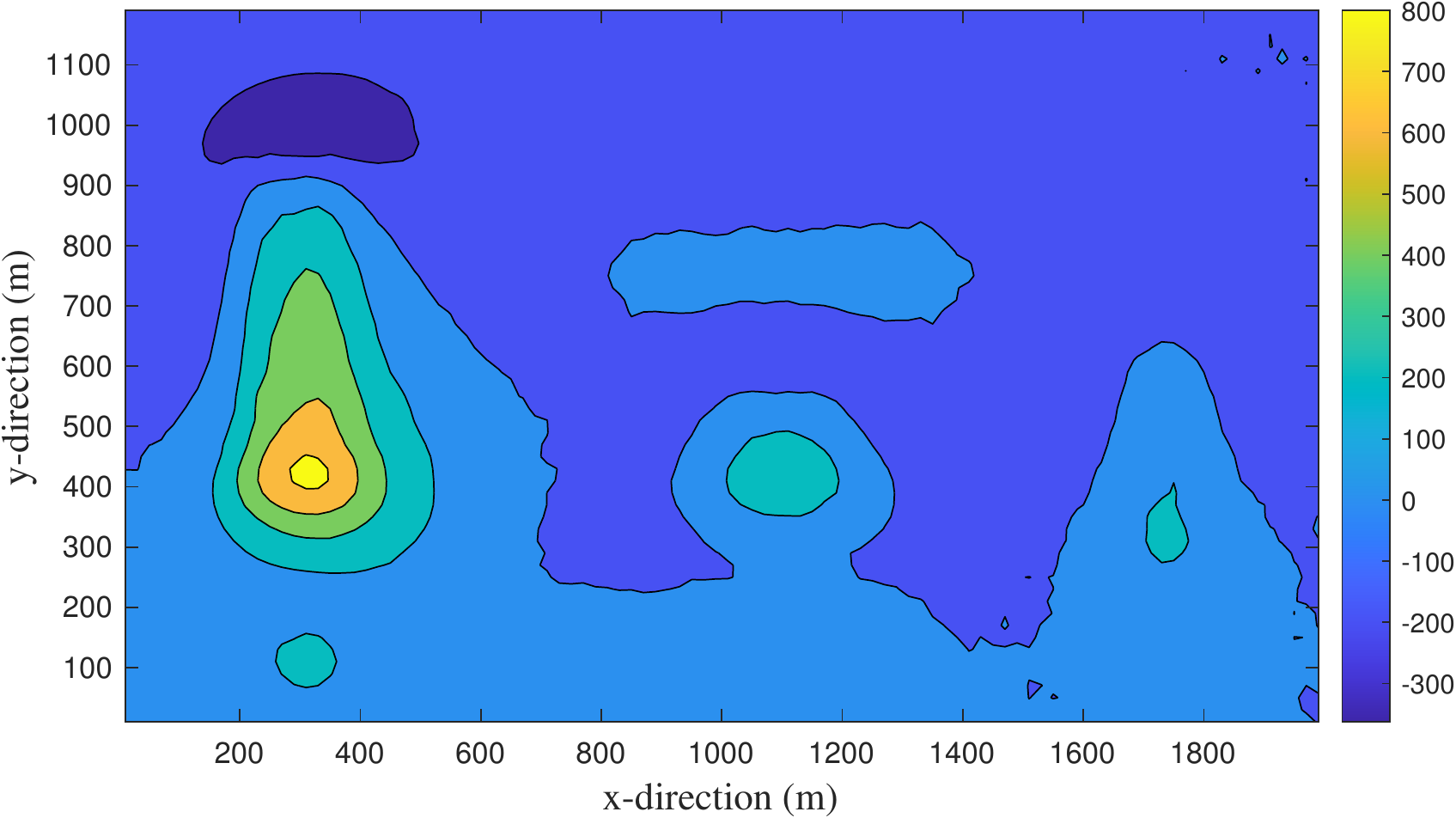}}
\subfigure[   $(14, 6618\textrm{s})$. \label{figure10g}]{\includegraphics[width=0.24\textwidth]{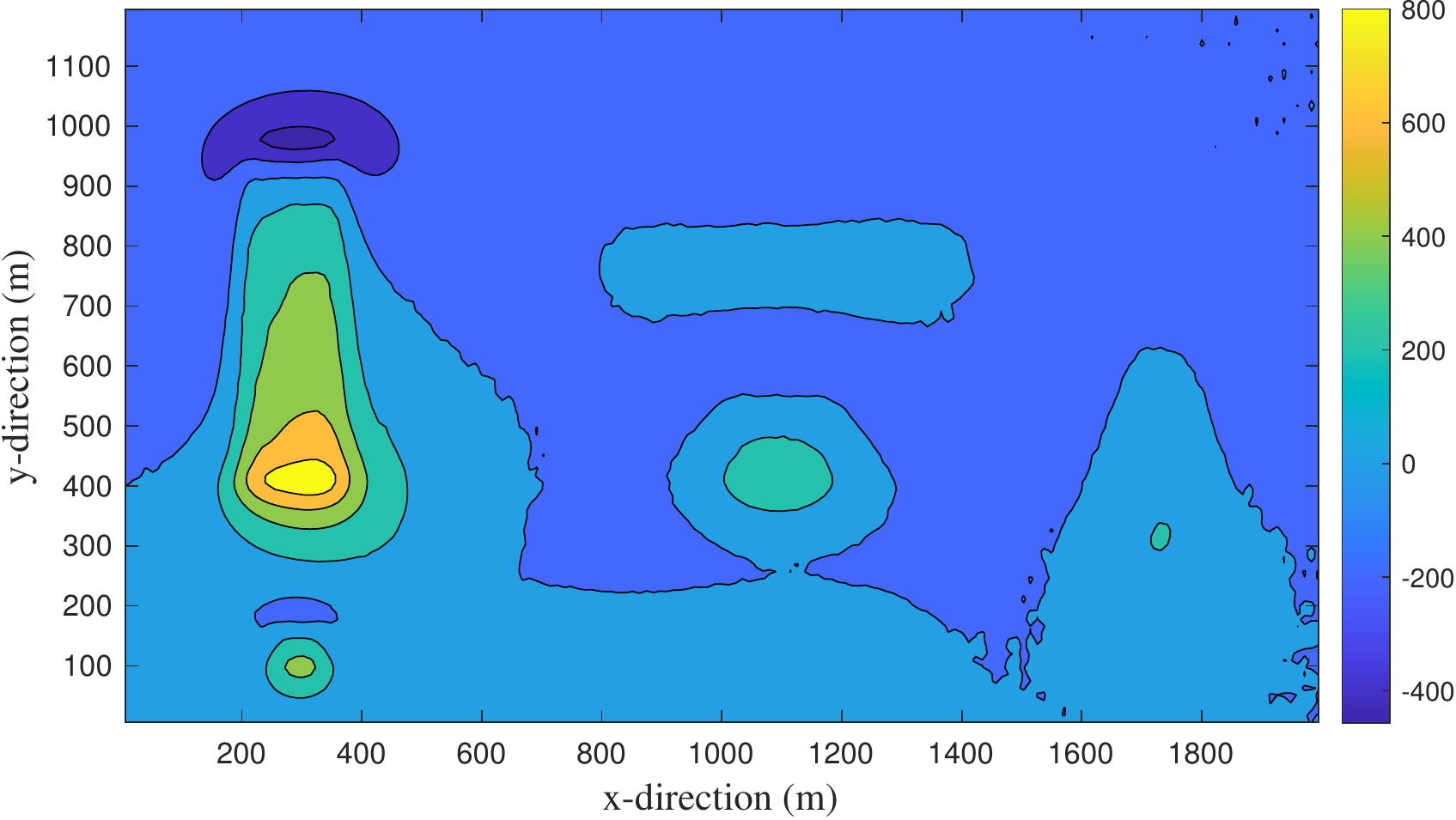}}
\subfigure[ $(13, 12949\textrm{s})$.\label{figure10h}]{\includegraphics[width=0.24\textwidth]{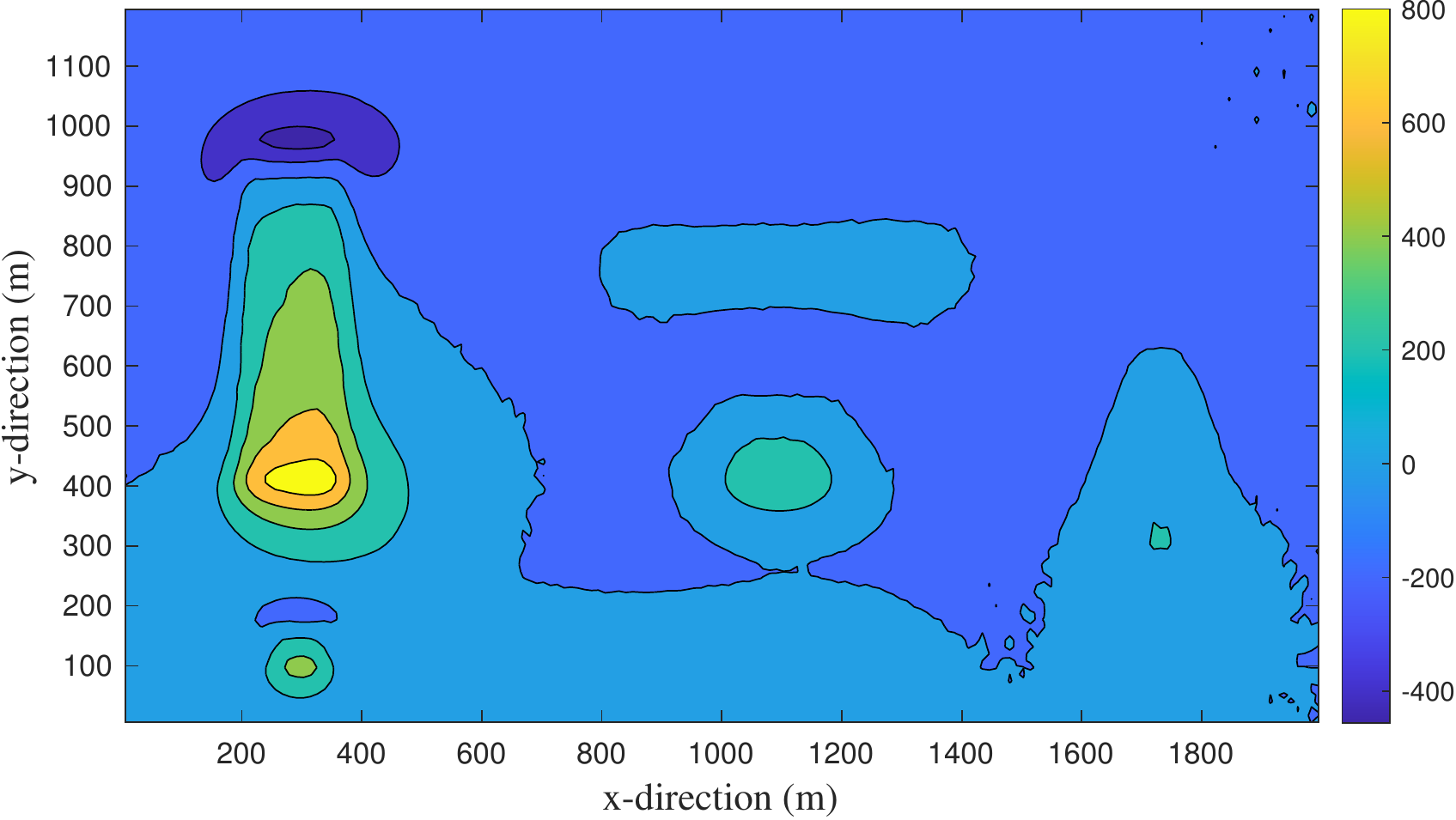}}
\subfigure[\texttt{GKB}:         $(.63,.90)$.\label{figure10i}]{\includegraphics[width=0.25\textwidth]{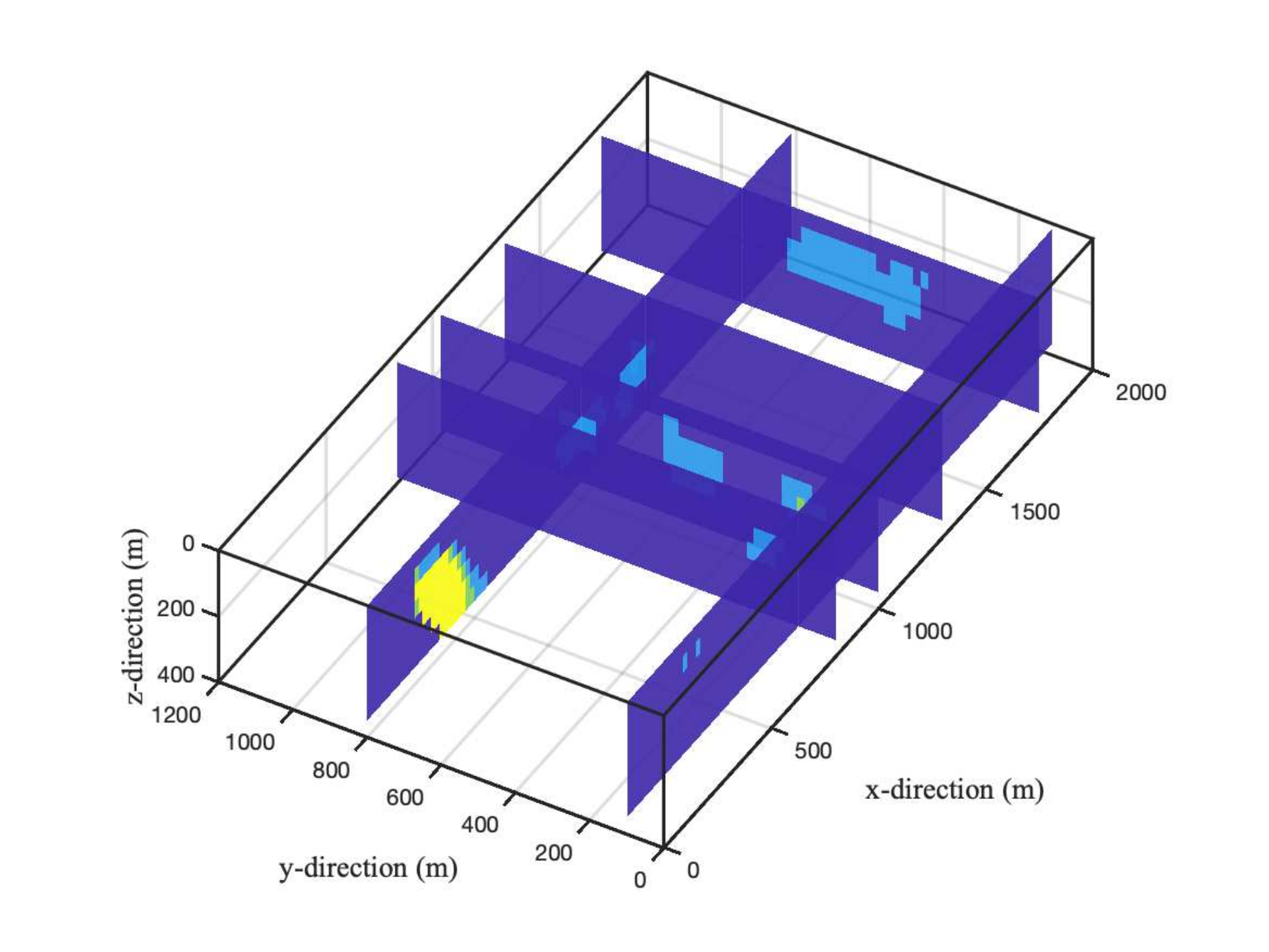}}
\subfigure[    $(.63,.91)$.\label{figure10j}]{\includegraphics[width=0.24\textwidth]{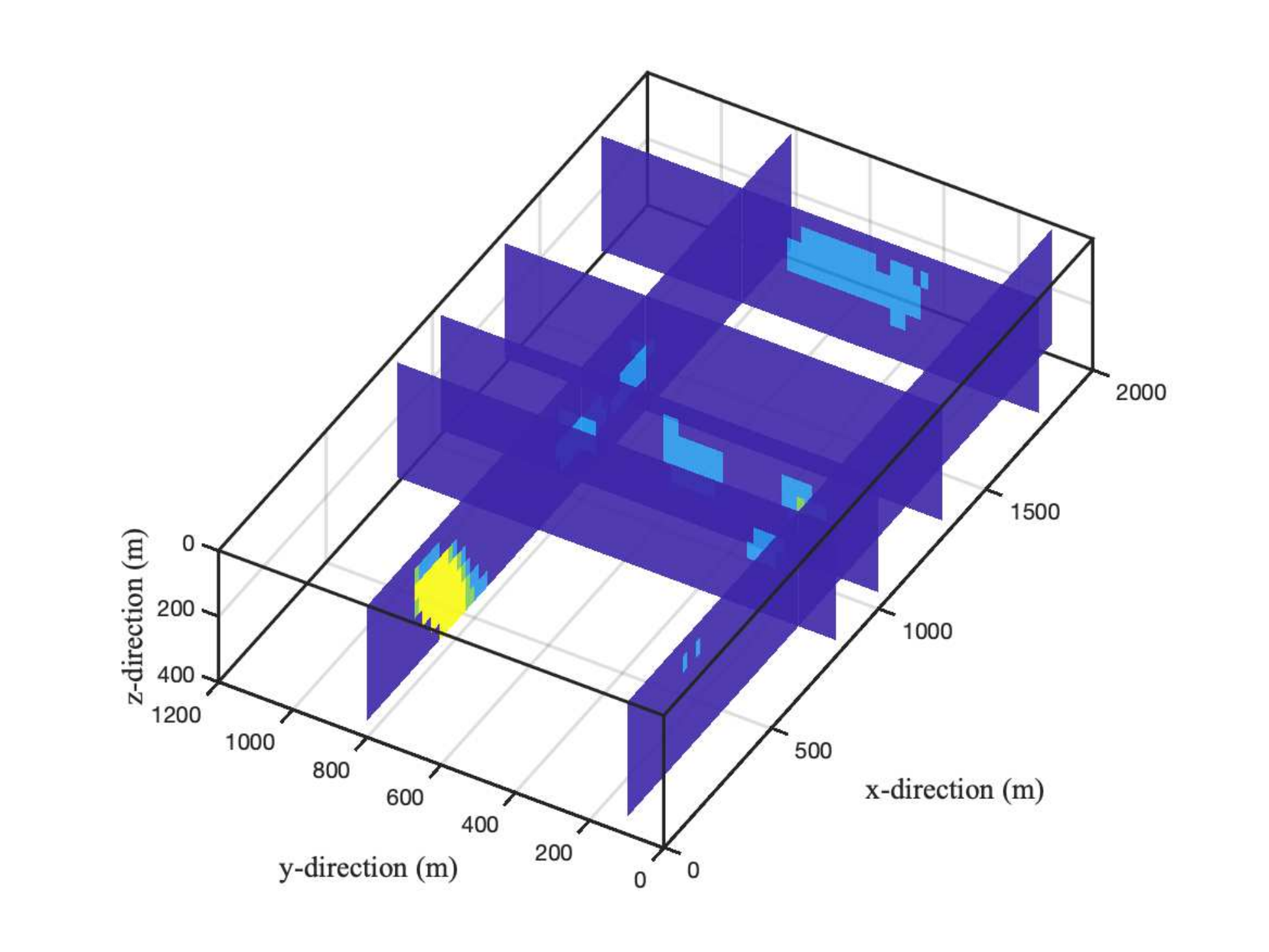}}
\subfigure[     $(.67,.90)$.\label{figure10k}]{\includegraphics[width=0.24\textwidth]{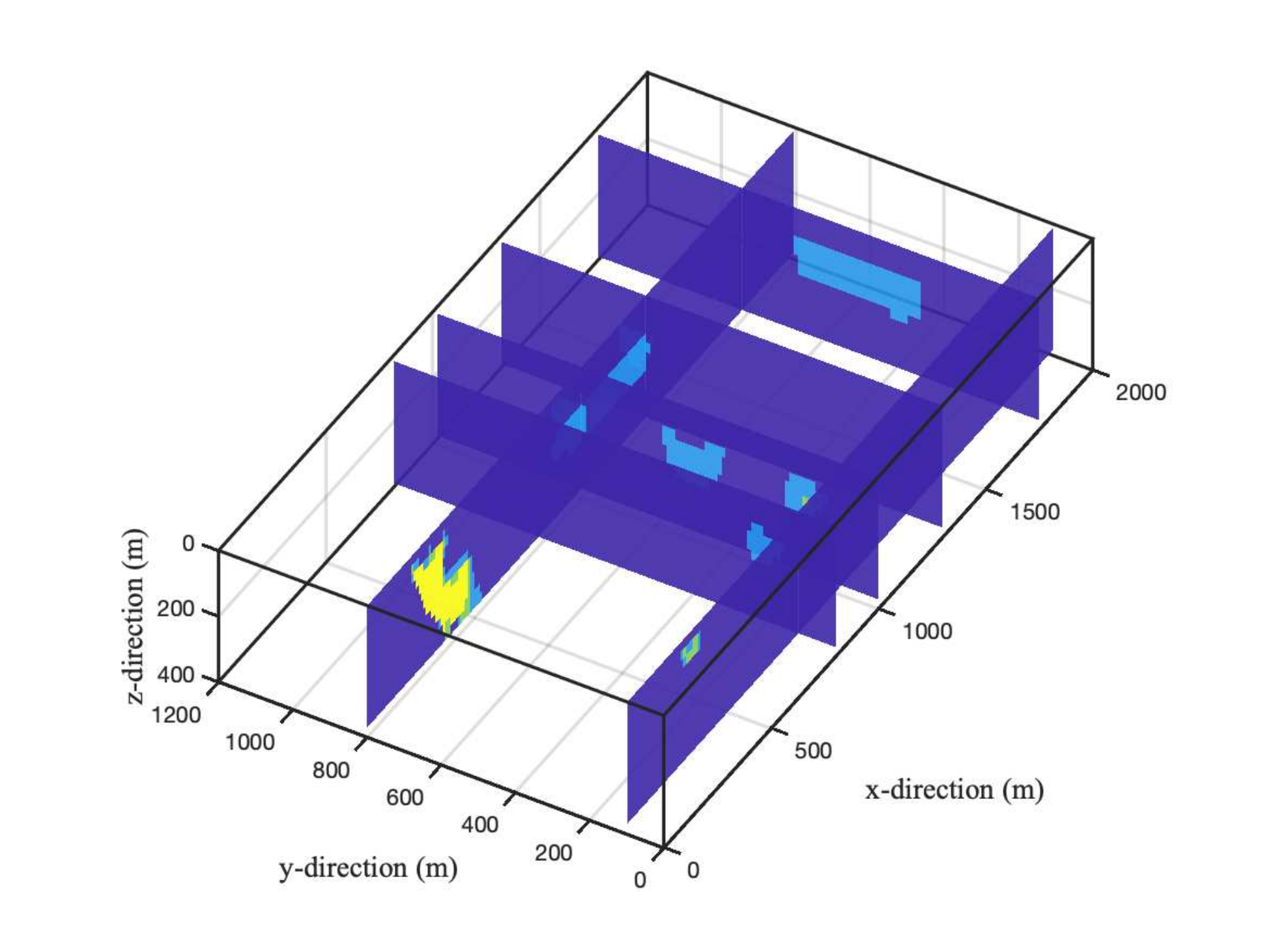}}
\subfigure[    $(.68,.92)$.\label{figure10l}]{\includegraphics[width=0.24\textwidth]{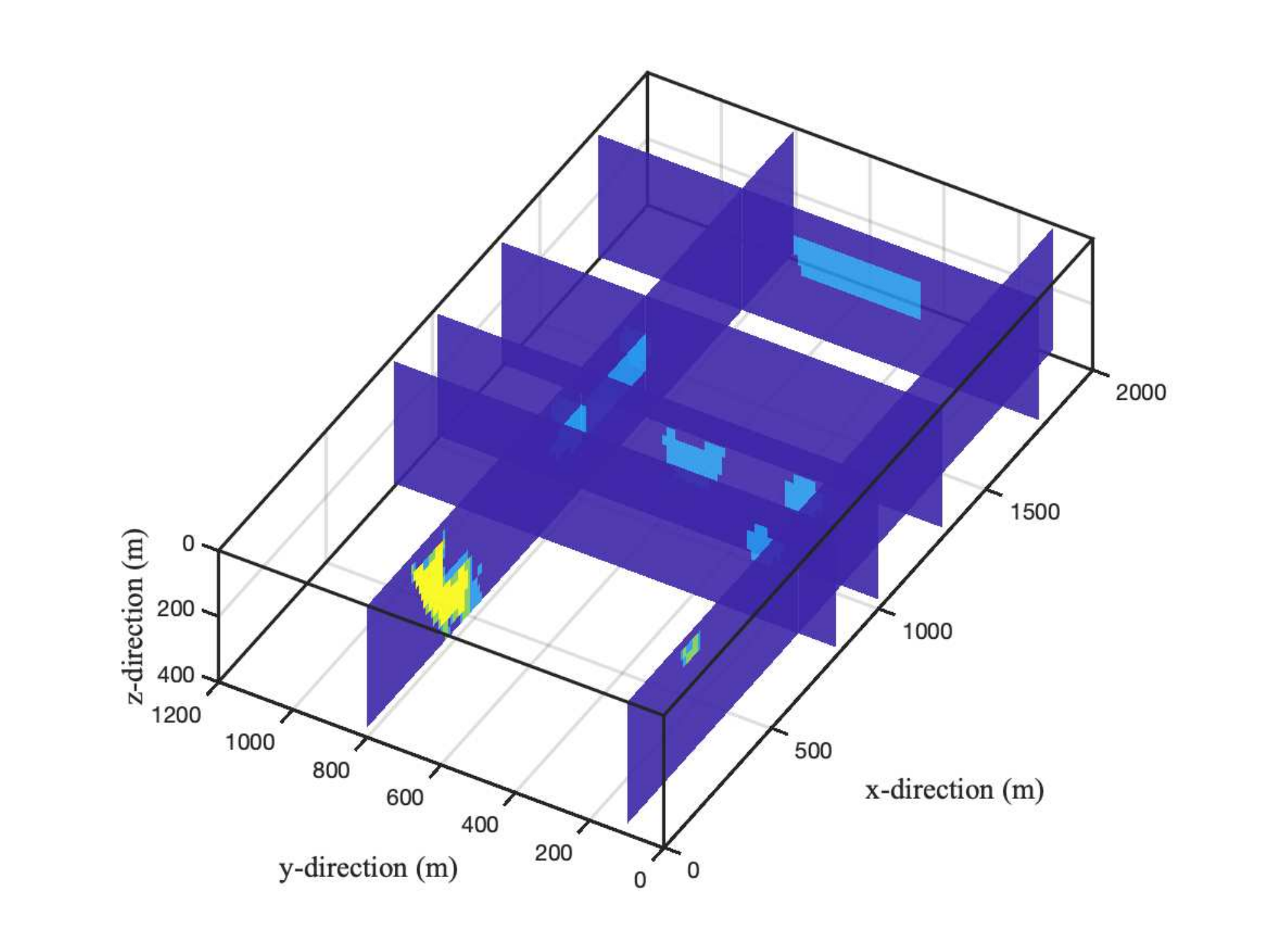}}
\subfigure[\texttt{RSVD}:  $(.64,1.11)$.\label{figure10m}]{\includegraphics[width=0.25\textwidth]{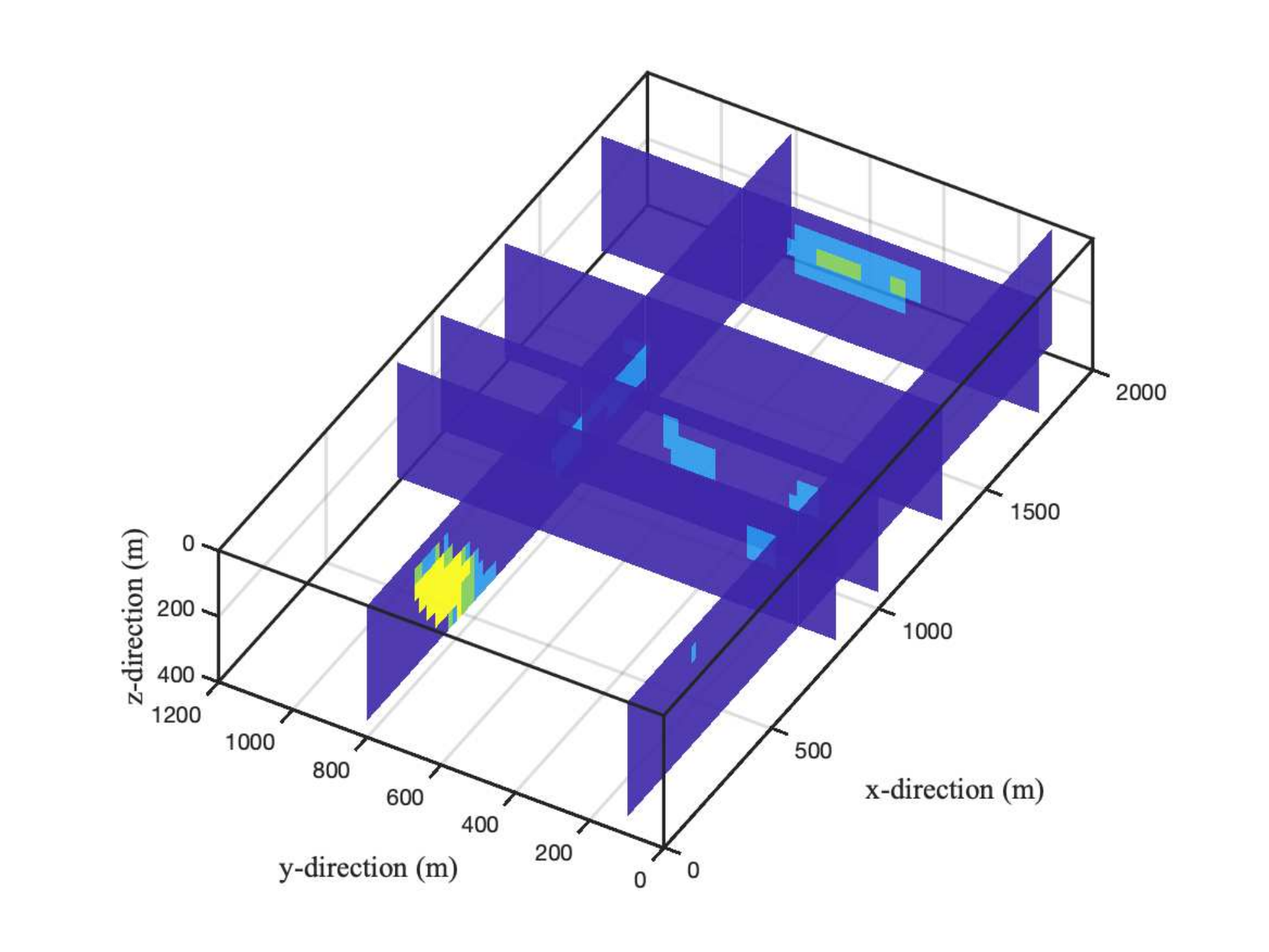}}
\subfigure[    $(.63,.93)$.\label{figure10n}]{\includegraphics[width=0.24\textwidth]{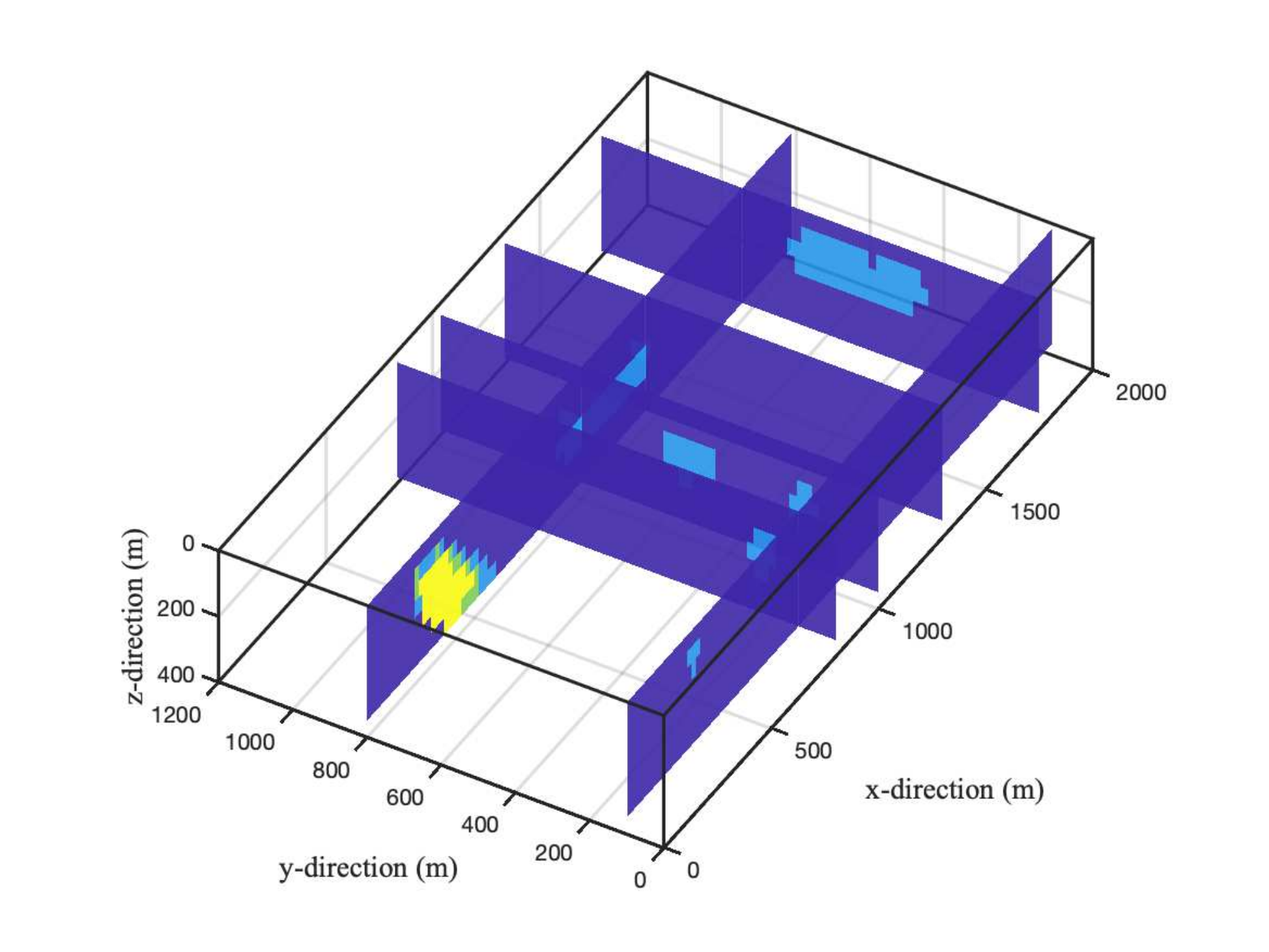}}
\subfigure[   $(.70,.99)$.\label{figure10o}]{\includegraphics[width=0.24\textwidth]{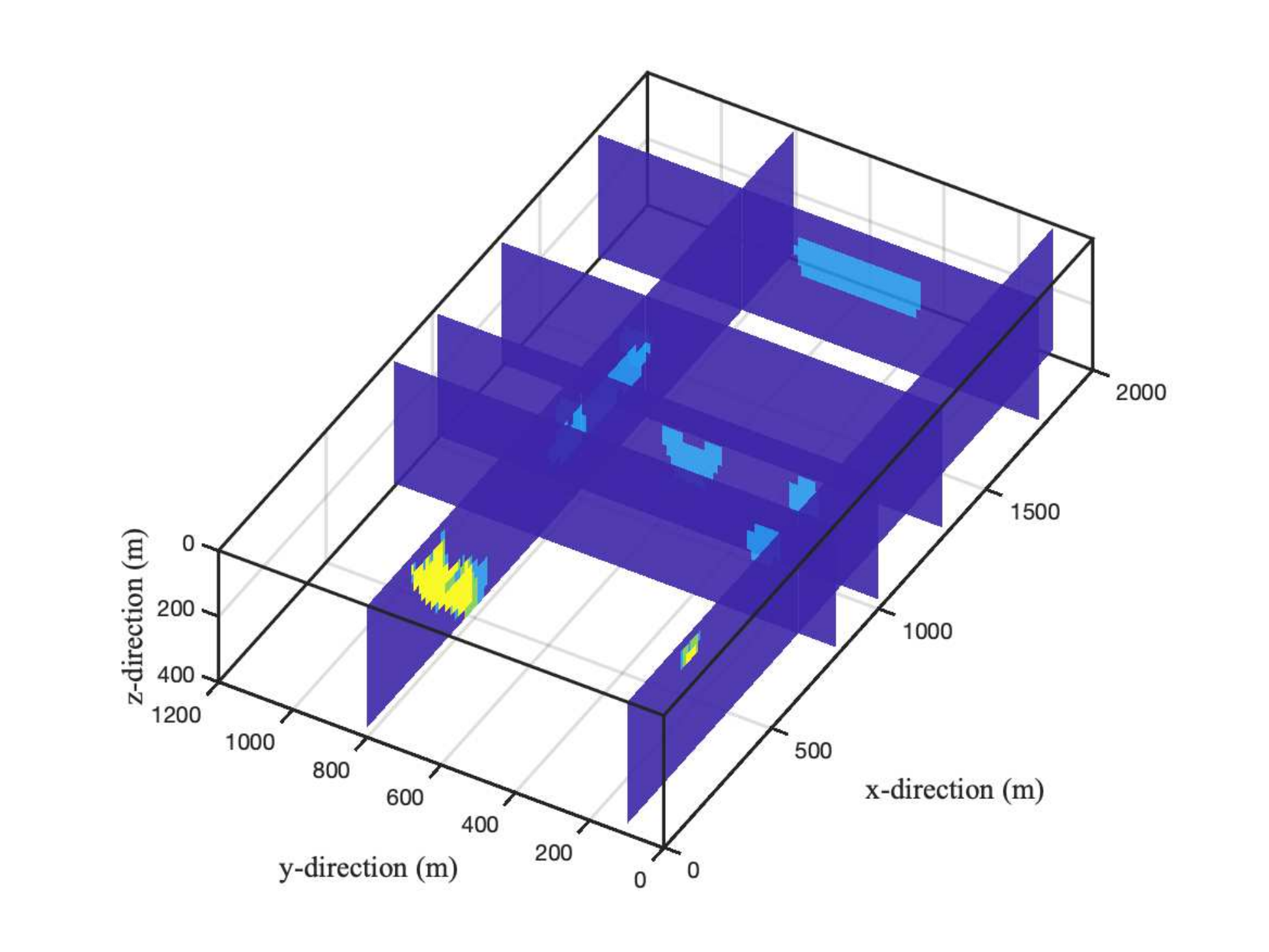}}
\subfigure[   $(.69,.90)$.\label{figure10p}]{\includegraphics[width=0.24\textwidth]{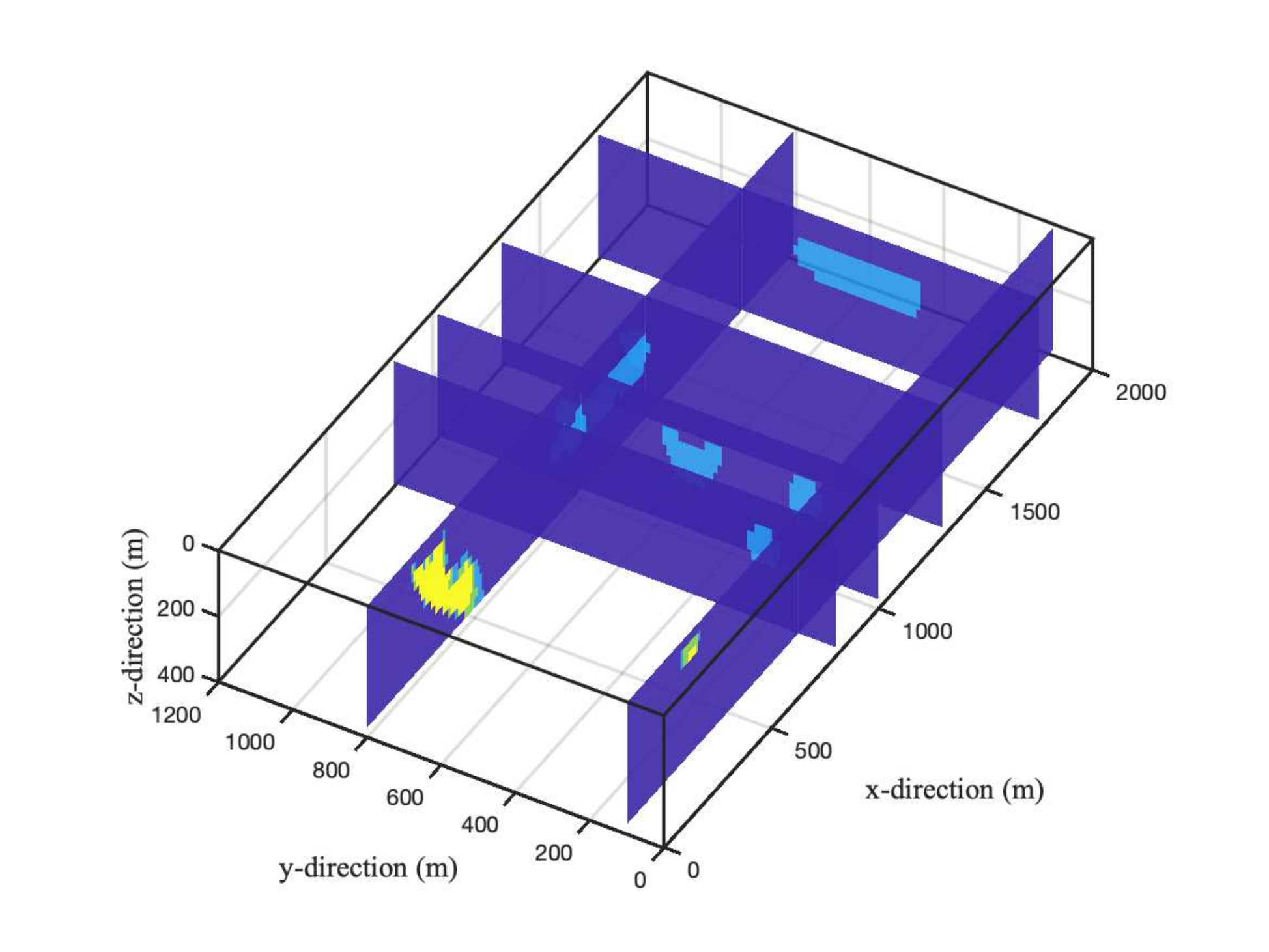}}
\caption{For \magnetic~data  the predicted anomalies obtained using \texttt{GKB} in Figures~\ref{figure10a}-\ref{figure10d} and \texttt{RSVD} in Figures~\ref{figure10e}-\ref{figure10h}, with the corresponding reconstructed volumes in Figures~\ref{figure10i}-\ref{figure10l} and Figures~\ref{figure10m}-\ref{figure10p}, respectively.  In each case the first row for \texttt{GKB} indicates the choices of  $\ell$ and $t$ in each column. The choices $t=750$ and $t=1500$ for $\ell=4$,  and with $t=2296$ and $t=4593$  for $\ell=7$, correspond to $t=\floor(m/8)$ and $t=\floor(m/4)$ for  $(m,n) = (6000, 48000)$ and 
$(18375, 257250)$, respectively.  Given are the pairs $(K, \texttt{Cost}\textrm{s})$,  (number of iterations to convergence and computational cost in seconds) in the captions of the anomalies, and $(\texttt{RE}, \chi^2/(m+\sqrt{2m}))$ in the captions of the reconstructions. 
Results for all cases are summarized in Table~\ref{tableB.5} with timings in Table~\ref{tableB.4}. The units for the anomalies are nT.
\label{figure10}}
\end{center}
\end{figure}

The results for the inversion of the magnetic data are illustrated in Figure~\ref{figure10} for the same cases as for the inversion of gravity data in illustrated in Figure~\ref{figure9}. Now, in contrast to the gravity results, the predicted anomalies are in good agreement with the true data for the results obtained using the \texttt{GKB} algorithm, with apparently greater accuracy for the lower resolution solutions, $\ell=4$ for both choices of $t$. On the other hand, the predicted magnetic anomalies are less satisfactory for small $\ell$ and $t$ but acceptable for large $\ell$. Then, considering the reconstructed volumes, there is a lack of resolution for $\ell=4$ which is evidenced by  the loss of the small well near the surface, which is seen when $\ell=7$ for both cases of $t$, when using the \texttt{GKB}. The other structures in the domain are also resolved better with $\ell=7$, but there is little gain from using $t=\floor(m/4)$ over $t=\floor(m/8)$. Then, considering the reconstructions obtained using the \texttt{RSVD} algorithm, while it is clear that the result with $\ell=4$ and small $t$ is unacceptable, the anomaly and reconstructed volume with $\ell=4$ and $t=\floor(m/4)$ is acceptable and achieved in reasonable time, approximately $11$ minutes, far faster than using  $\ell=7$ with \texttt{GKB}. Thus, this may contradict the conclusion that one should use the \texttt{GKB} algorithm within the magnetic data inversion algorithm. If there is a large amount of data and a high resolution volume is required, then it is important to use \texttt{GKB} in order to limit computational cost. Otherwise, it can be sufficient to use the \texttt{RSVD} provided $t\ge \floor(m/8)$ for a coarser resolution solution obtained at reasonable computational cost. 

\begin{figure}[ht!]\begin{center}
\subfigure[$11: (9, 13595\textrm{s})$. \label{figure11a}]{\includegraphics[width=0.24\textwidth]{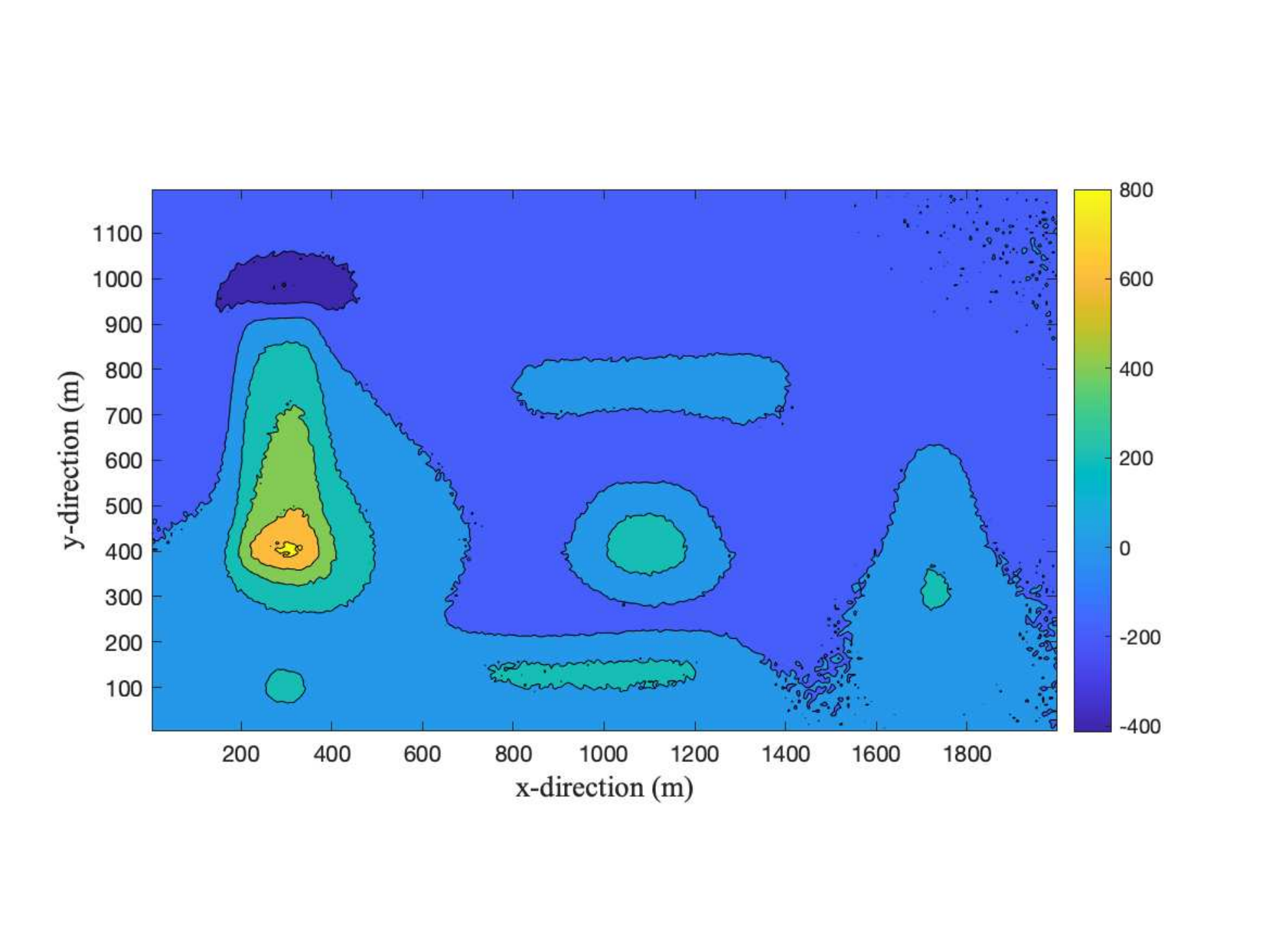}}
\subfigure[$12: (8, 21649\textrm{s})$. \label{figure11b}]{\includegraphics[width=0.24\textwidth]{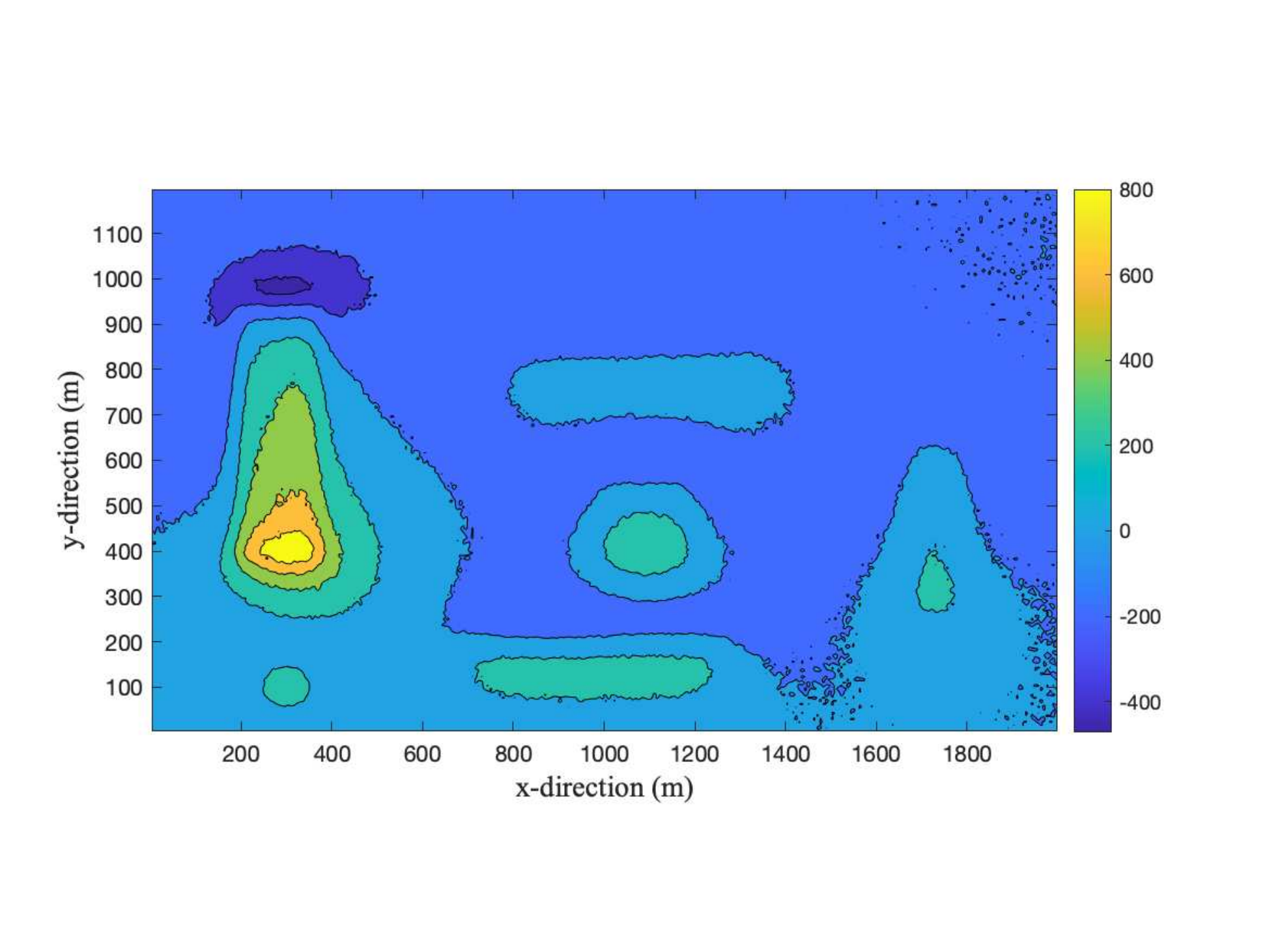}}
\subfigure[$11: (.74,.98)$.\label{figure11c}]{\includegraphics[width=0.24\textwidth]{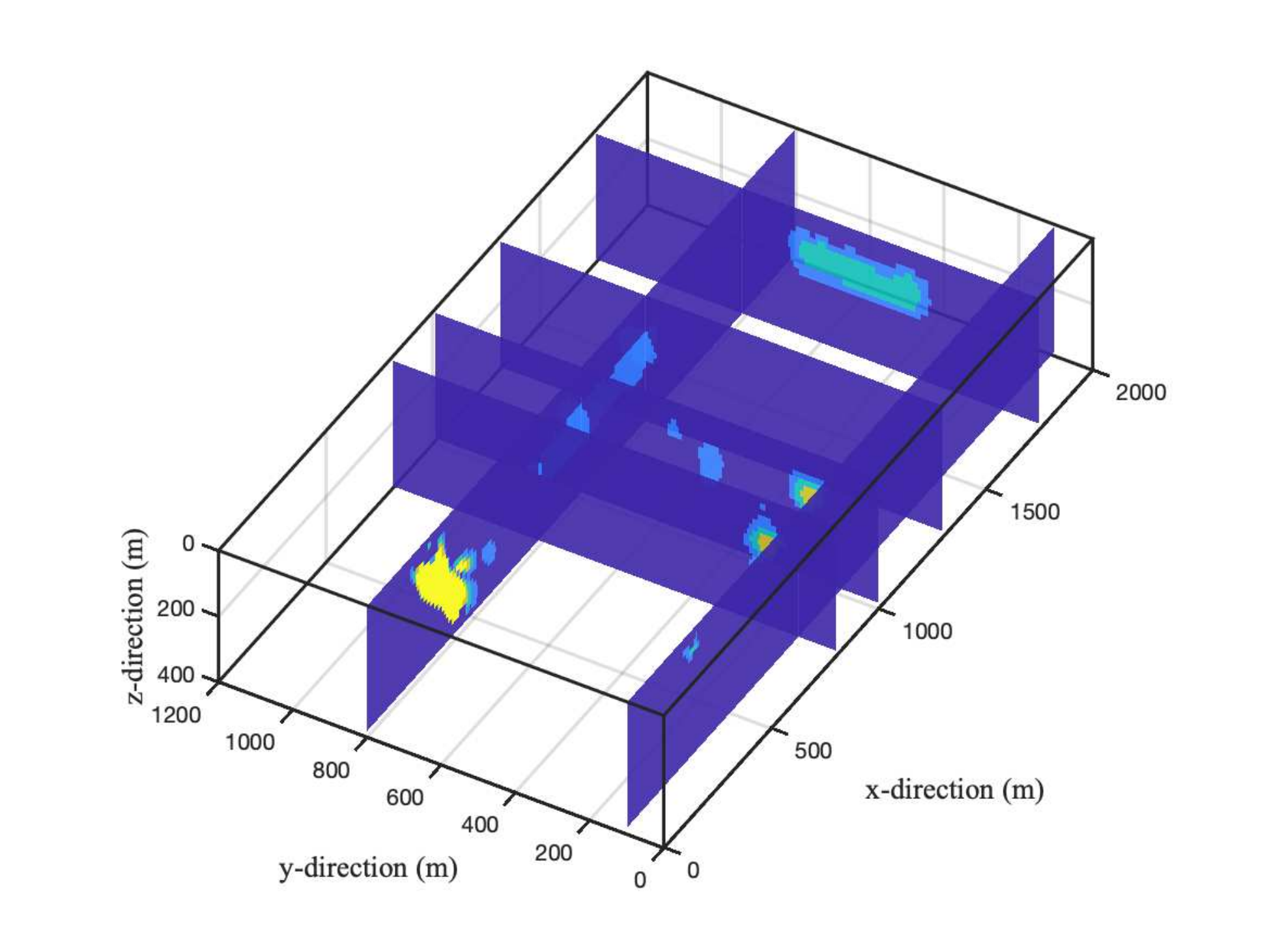}}
\subfigure[$12: (.74,.92)$.\label{figure11d}]{\includegraphics[width=0.24\textwidth]{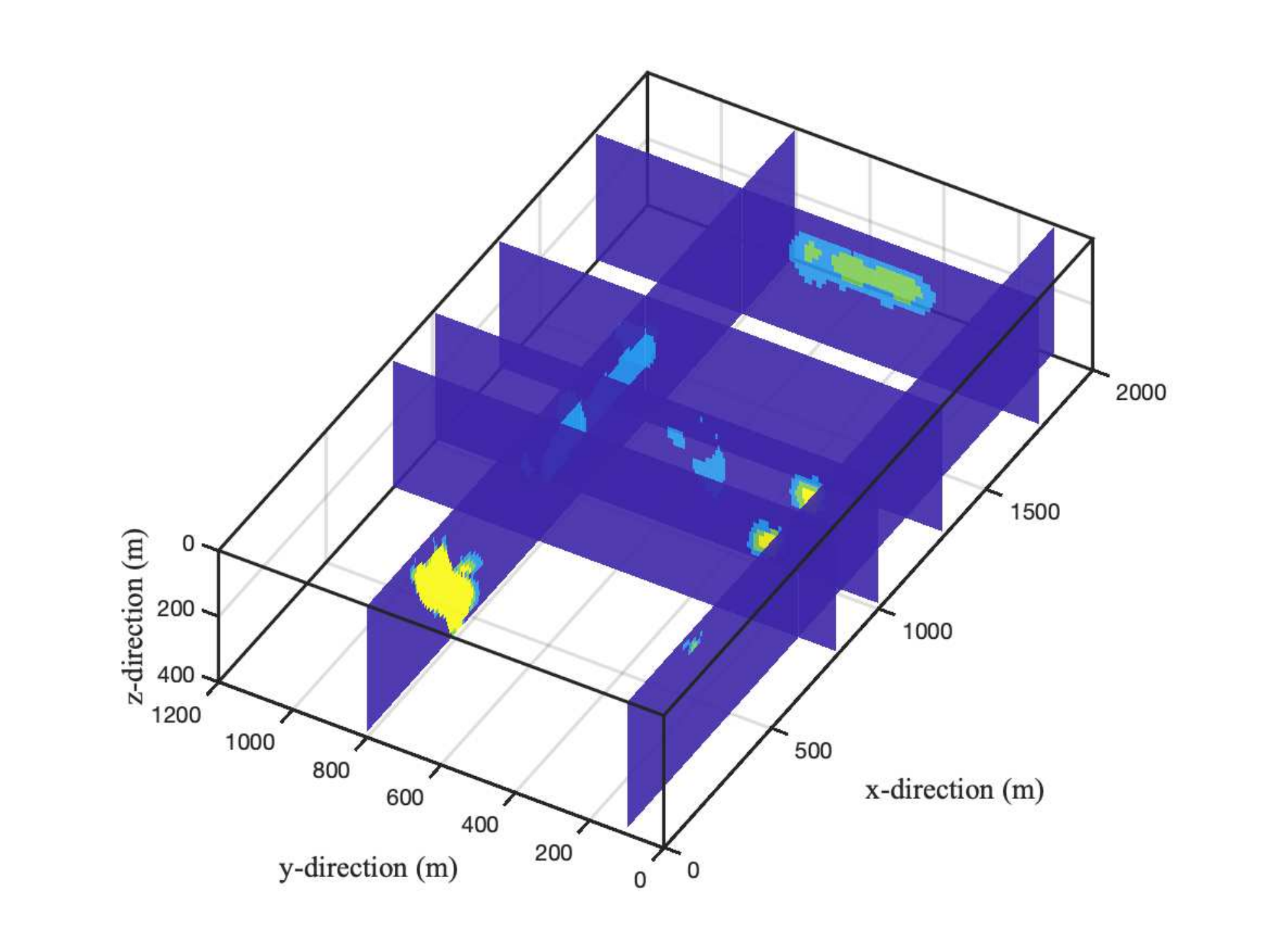}}
\subfigure[$(15, 21266\textrm{s})$.\label{figure11e}]{\includegraphics[width=0.24\textwidth]{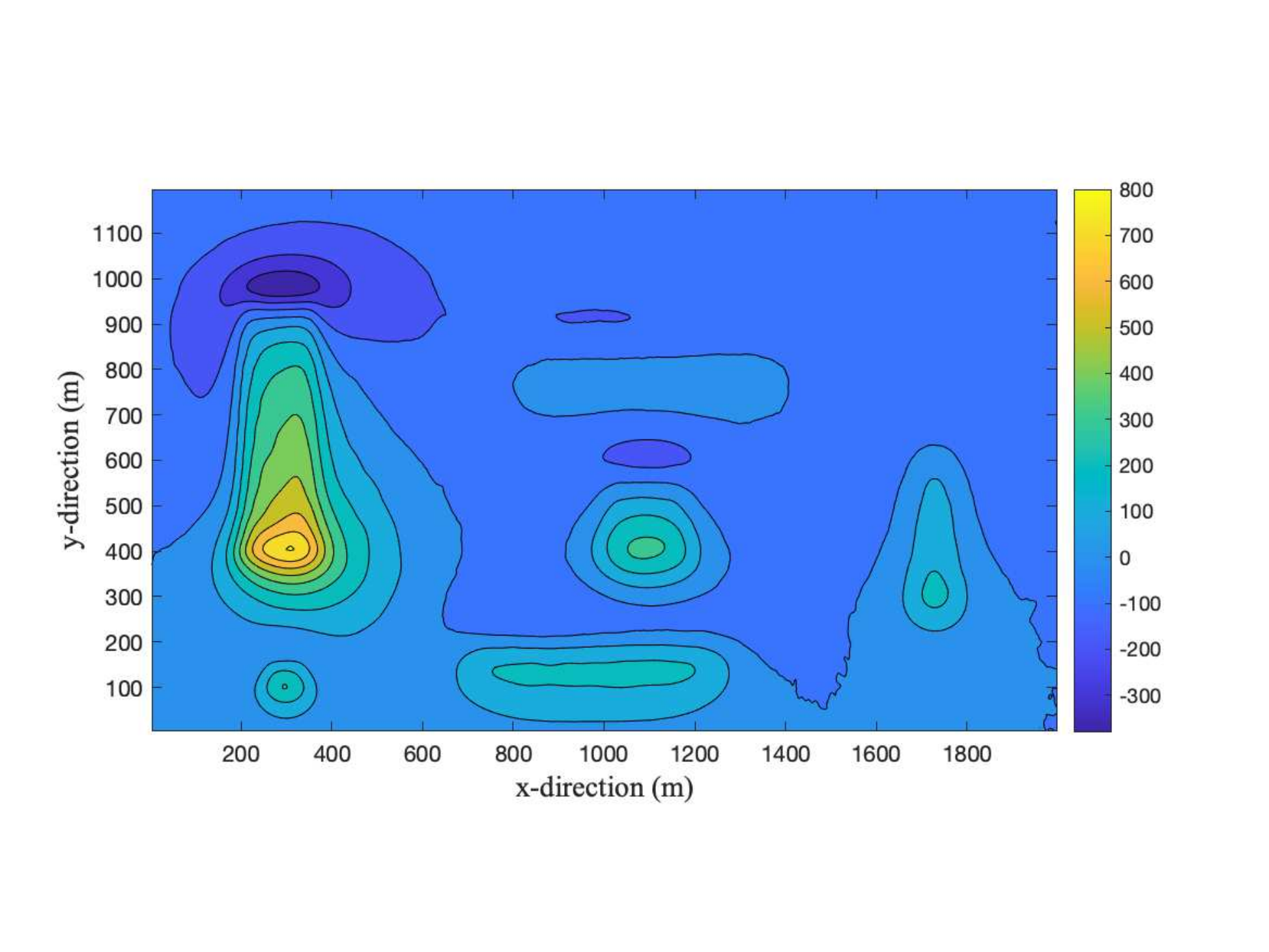}}
\subfigure[$(16, 41981\textrm{s})$.\label{figure11f}]{\includegraphics[width=0.24\textwidth]{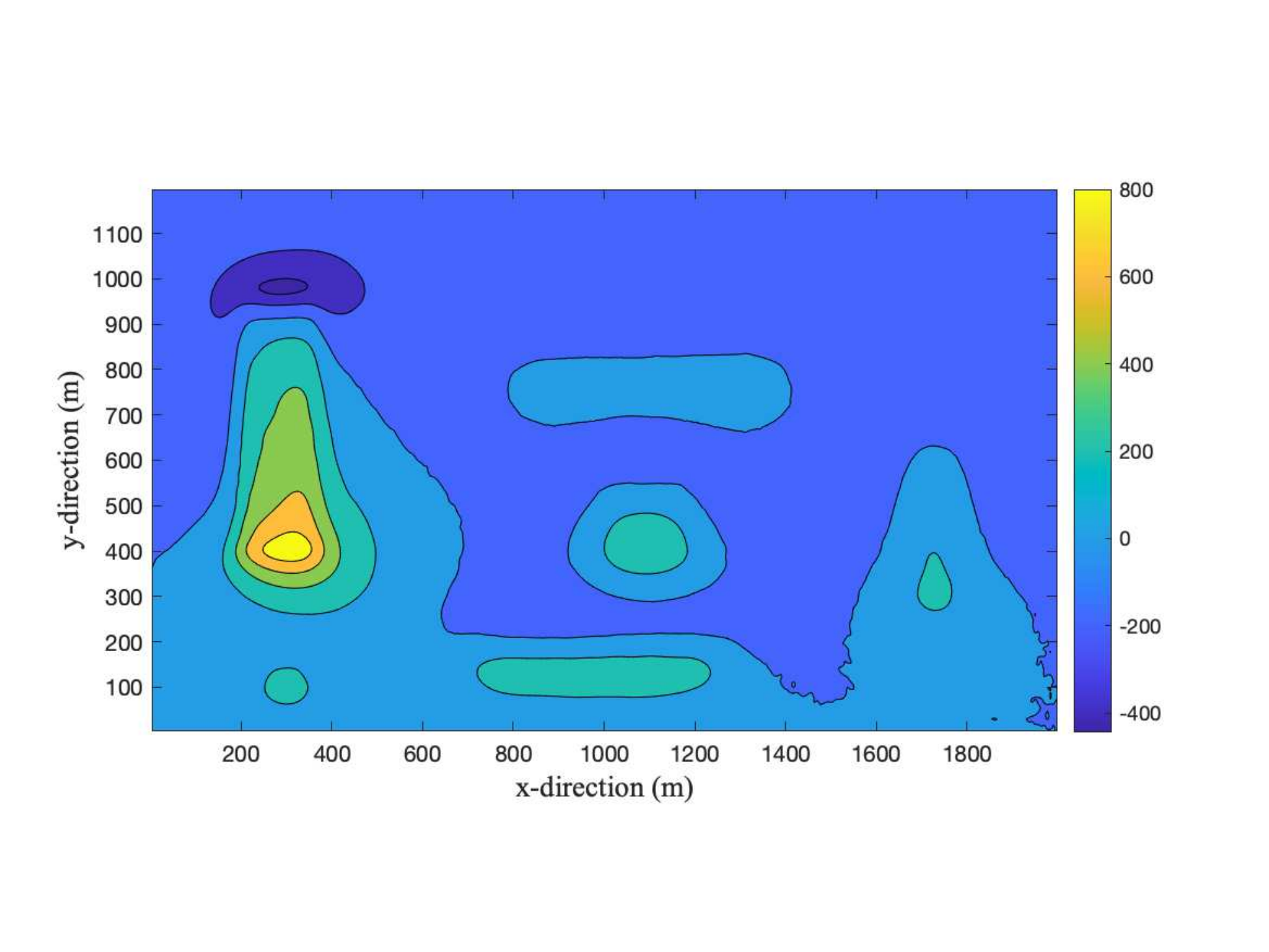}}
\subfigure[$(.74,.99)$.\label{figure11g}]{\includegraphics[width=0.24\textwidth]{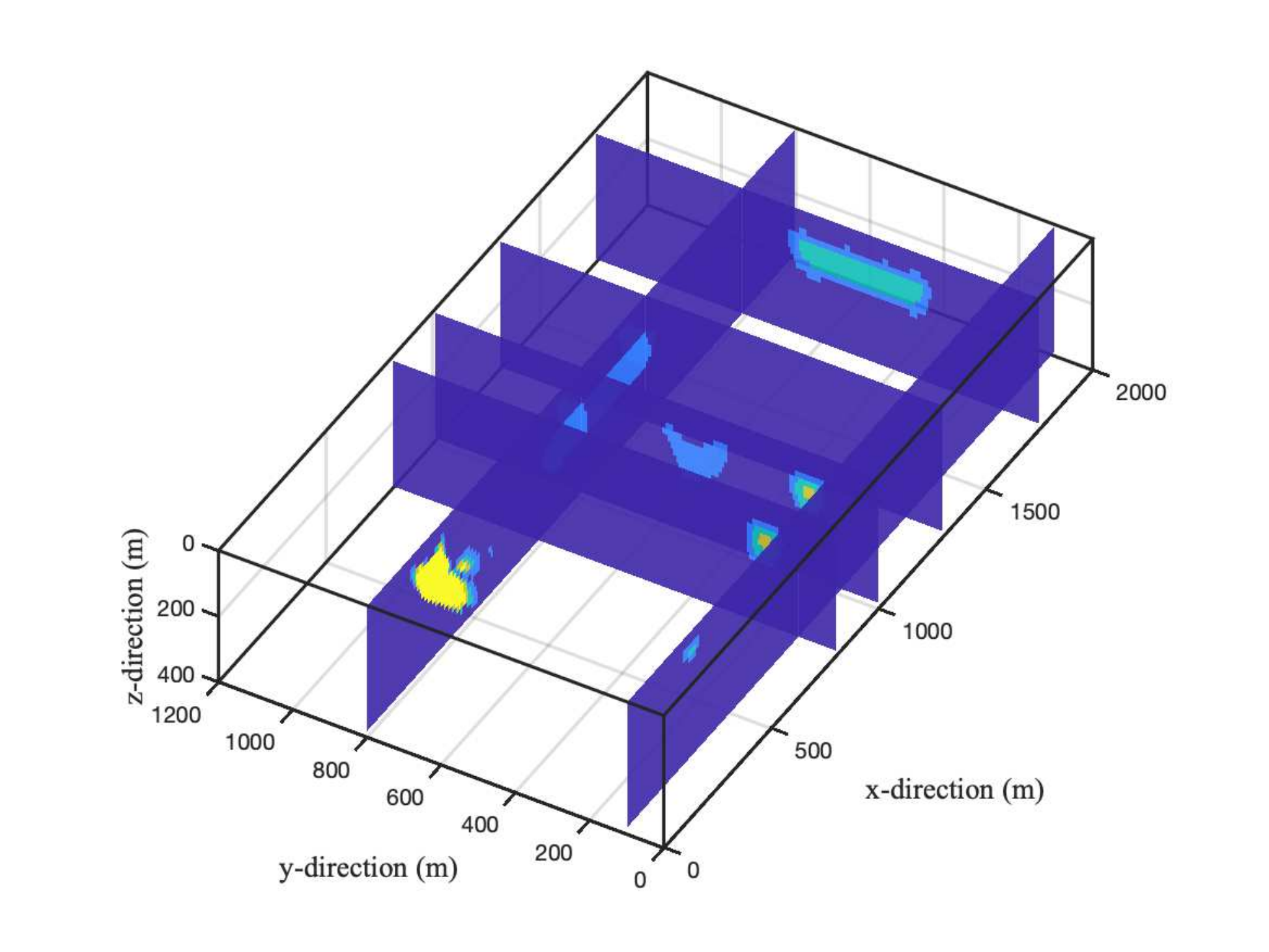}}
\subfigure[$(.73,.98)$.\label{figure11h}]{\includegraphics[width=0.24\textwidth]{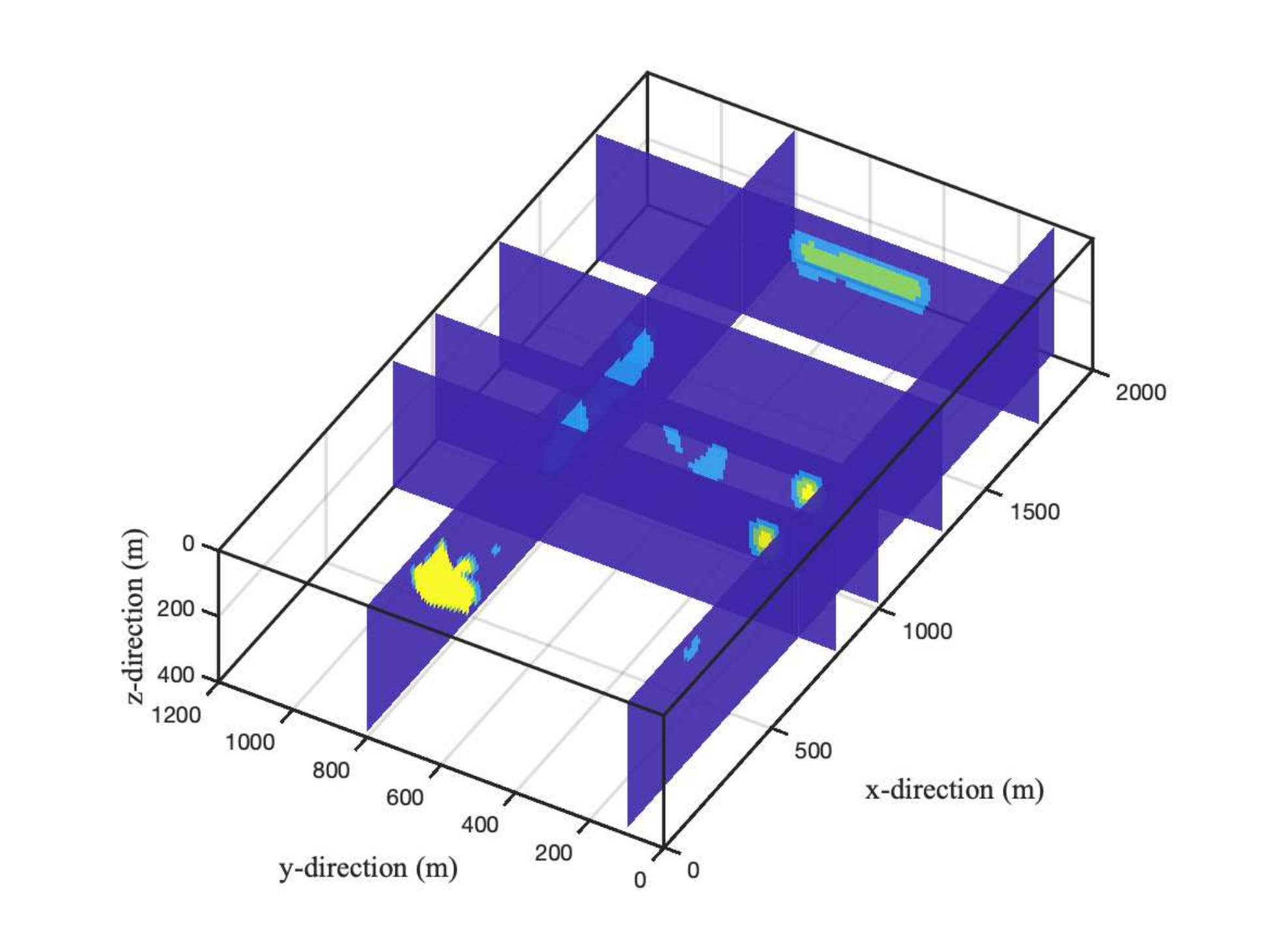}}
\caption{The magnetic anomalies and reconstructed volumes using the \texttt{GKB} and \texttt{RSVD} algorithms in Figures~\ref{figure11a}-\ref{figure11d} and \ref{figure11e}-\ref{figure11h}, respectively. The first row indicates the choice of $\ell=11$, for which  $t=2268=\floor(m/20)$, $m=45375$ and $n=998250$ 
 and oversampled projected problem of size $2381$, or $\ell=12$, with $t=2700=\floor(m/20)$, $m=54000$ and $n= 1296000$,  and oversampled projected problem of size $2835$. 
 Given are the pairs $(K, \texttt{Cost}\textrm{s})$,  (number of iterations to convergence and computational cost in seconds) in the captions of the anomalies, and $(\texttt{RE}, \chi^2/(m+\sqrt{2m}))$ in the captions of the reconstructions. The units for the anomalies are nT.
 \label{figure11}}
\end{center}
\end{figure}

We now investigate the quality of solutions obtained for magnetic data using higher resolution data sets, and both \texttt{GKB} and \texttt{RSVD} algorithms to assess which algorithm is best suited for such larger problems. In these cases we pick $t=\floor(m/20)$, to assess quality with a necessarily restricted subspace size as compared to the size of the given data set. Results using $\ell=11$ with $t=2268$ and $\ell=12$ with $t=2700$, corresponding to $m=45375$ and $n=998250$, and $m=54000$ and $n=1296000$, respectively, are illustrated in Figure~\ref{figure11}. For these large scale problems, the memory becomes too large for implementation on the  environment with just $32$GB RAM. Thus, these timings are for an implementation using a desktop computer with  the
Intel(R) Xeon (R) Gold 6138 CPU 
2.00GHz chip and with \textsc{Matlab} release 2019b. Comparing the results between $m=45375$ and $m=54000$ ($\ell=11$ and $\ell=12$) it can be seen that the predicted anomalies are always better for the larger problem, and in particular the result shown in Figure~\ref{figure11e} shows greater artifacts when using \texttt{RSVD}. The obtained reconstruction for this case, shown in Figure~\ref{figure11g} is, however, acceptable. Overall, trading off between computational cost and solution quality, there seems little gain in using $\ell=12$ and the results with $\ell=11$ obtained with the \texttt{GKB} algorithm in $227$ minutes (nearly $4$ hours)  are suitable. These results also show that it is sufficient to use a relatively smaller projected space, $t=\floor(m/20)$ when $m$ is larger. Indeed, notice that even in these cases the largest matrix required by both algorithms is of size $n \times t_p$ and requires $17.7$GB and $27.4$GB, for $\ell=11$ and $\ell=12$, respectively. Effectively, it is this large memory requirement that limits the given implementation using either \texttt{GKB} or \texttt{RSVD} for larger size problems.

\begin{figure}[ht!]\begin{center}
\subfigure[$11: (21, 30833\textrm{s})$. \label{figure12a}]{\includegraphics[width=0.24\textwidth]{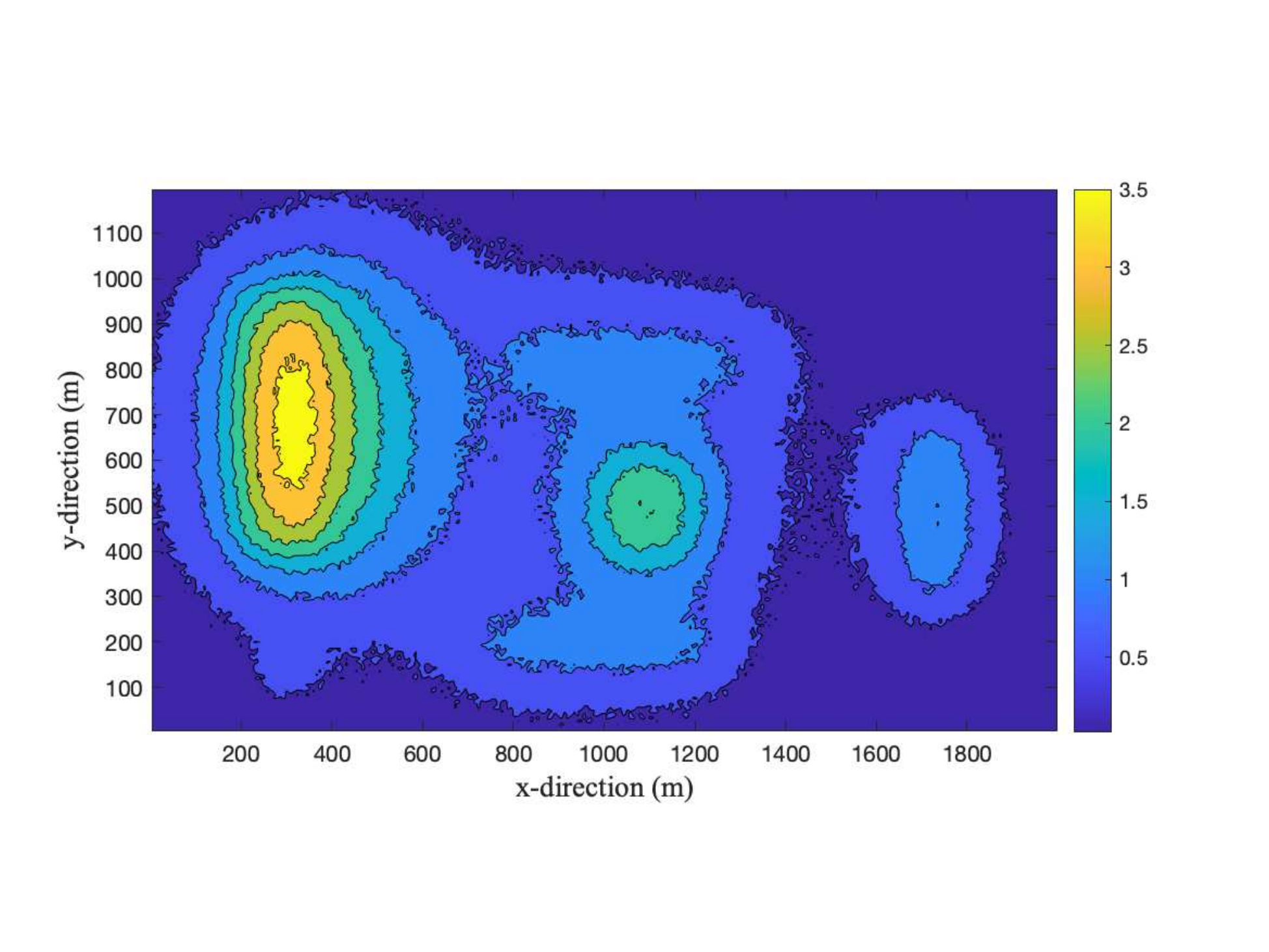}}
\subfigure[$12: (21, 59837\textrm{s})$. \label{figure12b}]{\includegraphics[width=0.24\textwidth]{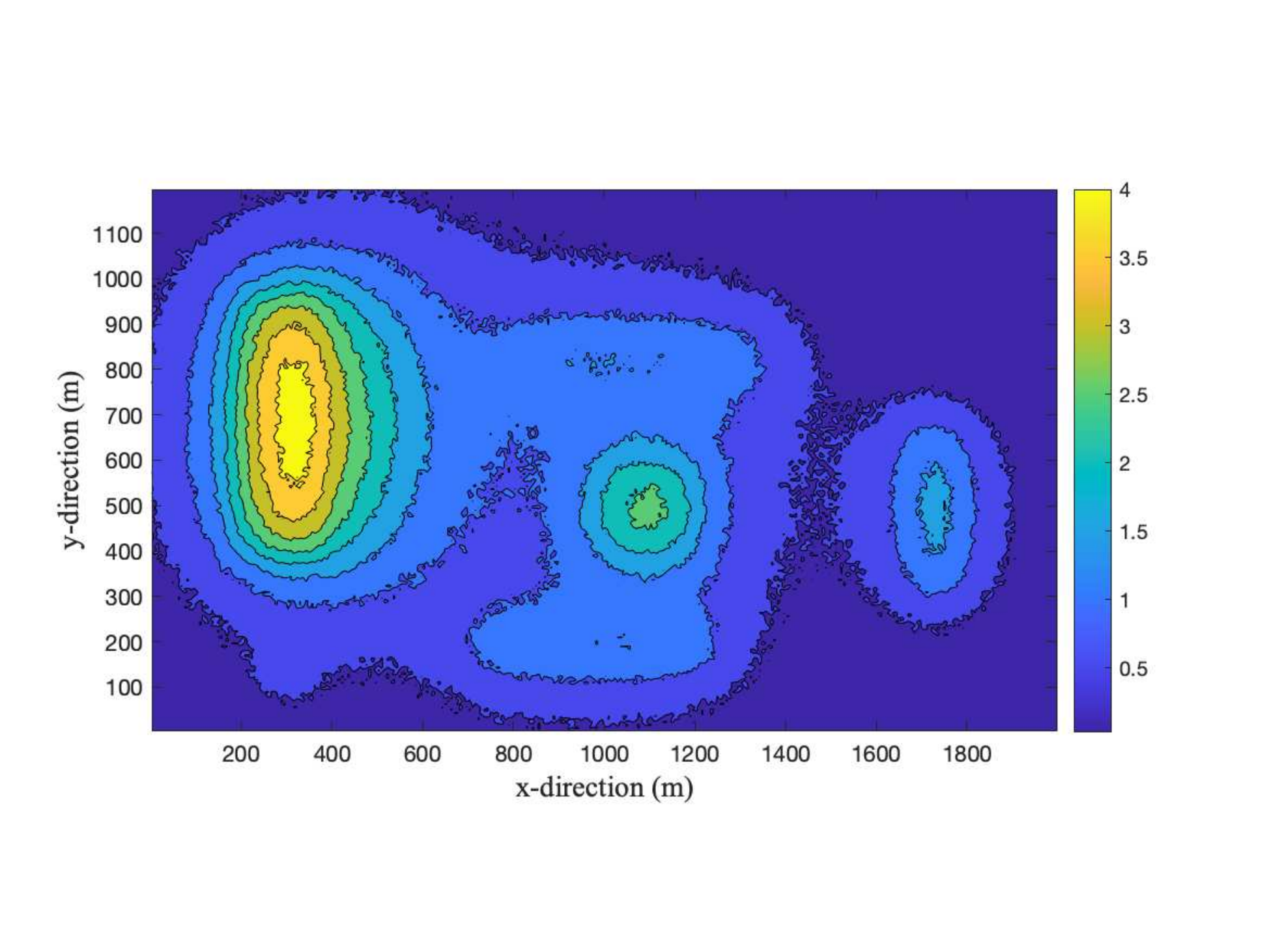}}
\subfigure[$11:  (1.02, .76)$.\label{figure12c}]{\includegraphics[width=0.23\textwidth]{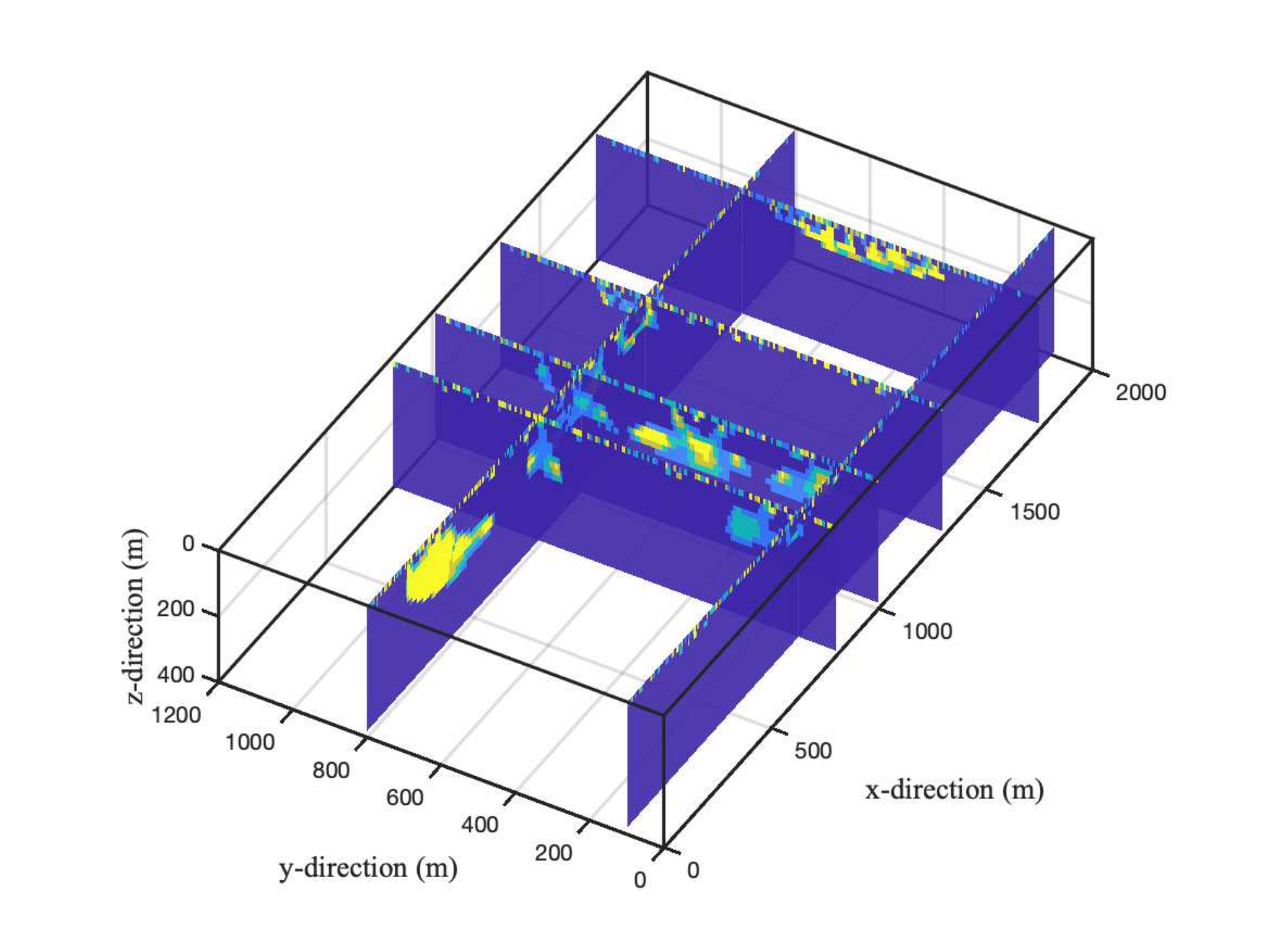}}
\subfigure[$12: (1.00, .78)$.\label{figure12d}]{\includegraphics[width=0.23\textwidth]{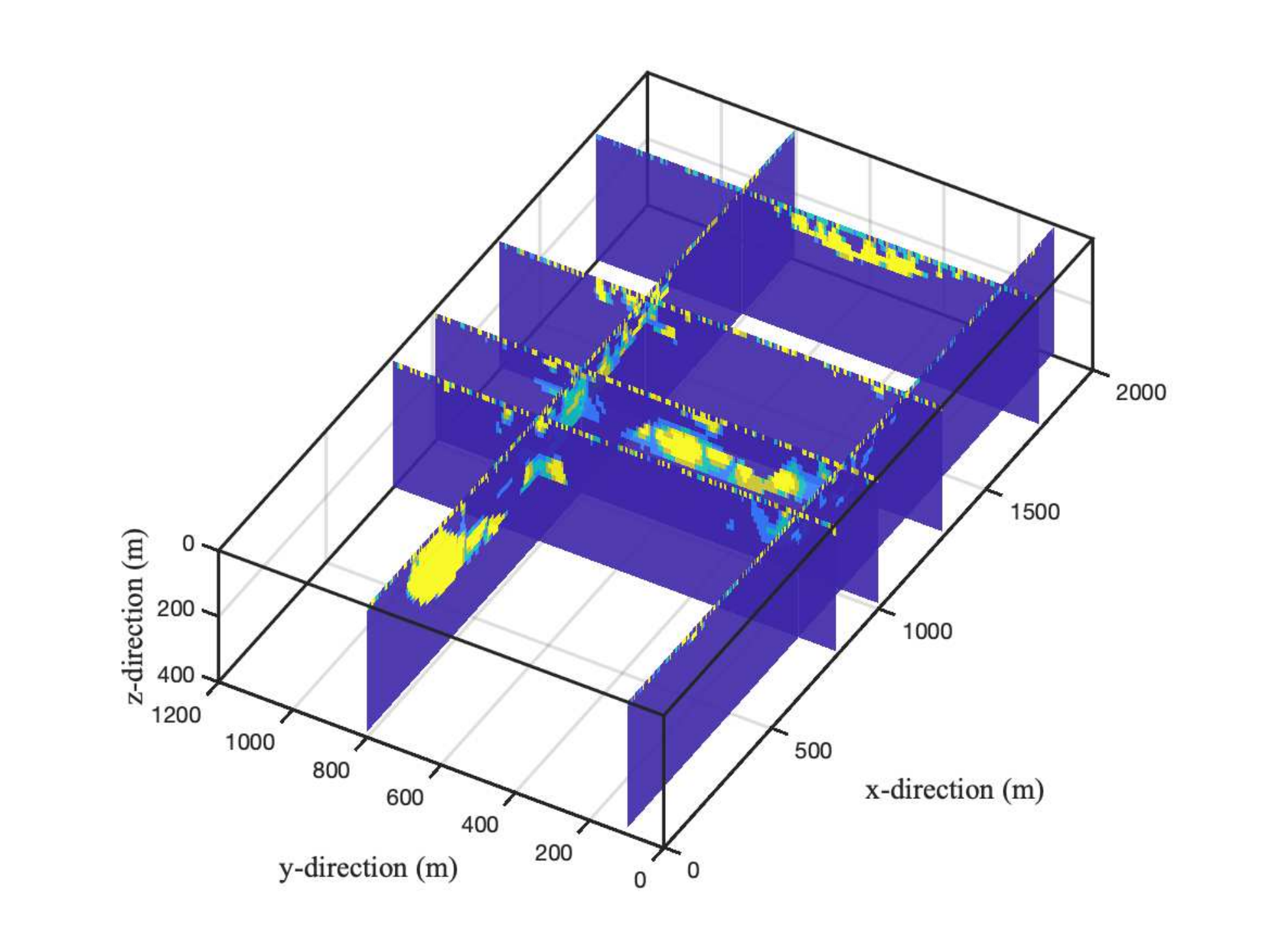}}
\subfigure[ $ (8, 11779\textrm{s})$.\label{figure12e}]{\includegraphics[width=0.24\textwidth]{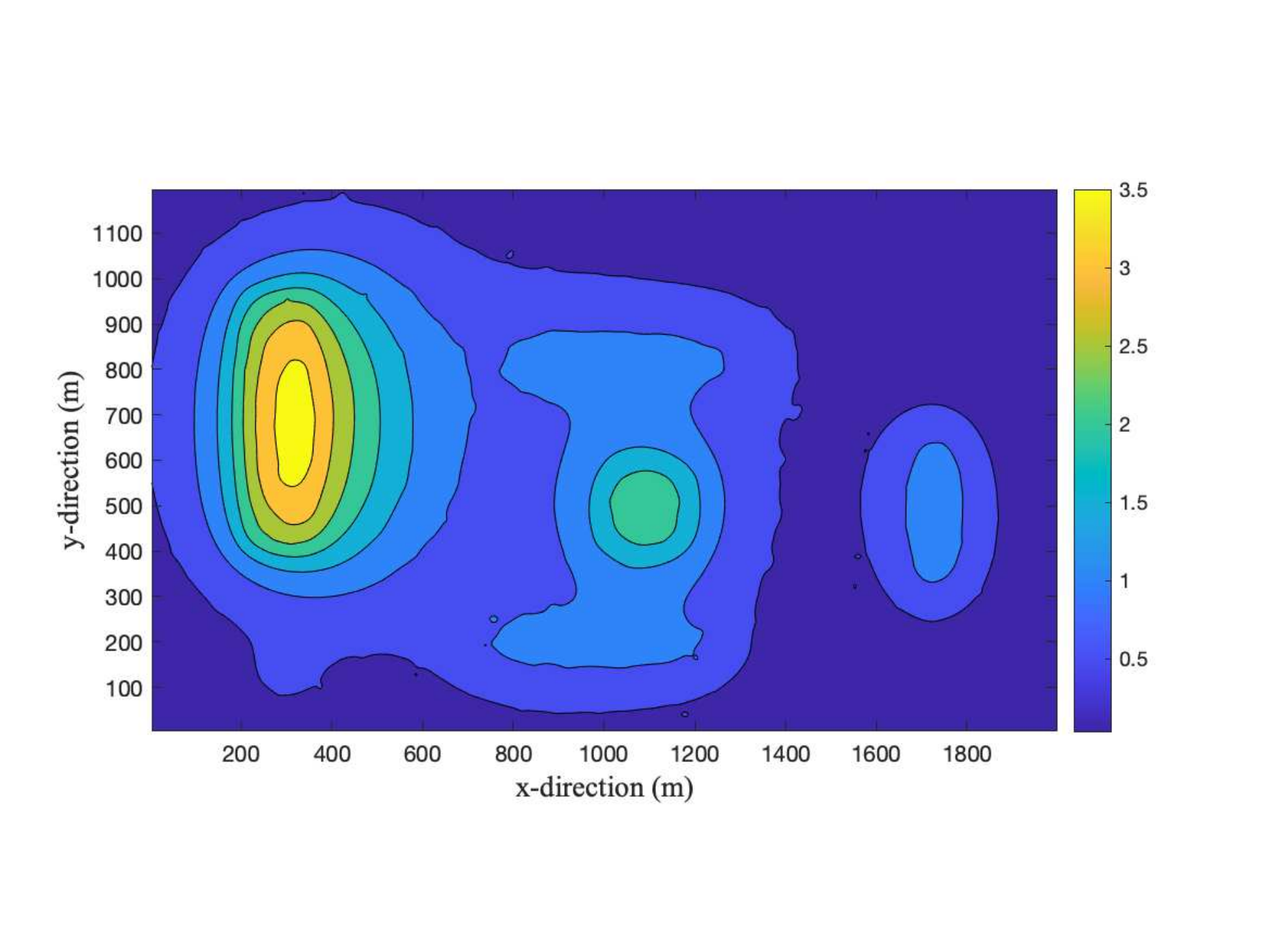}}
\subfigure[$(8, 30212\textrm{s})$.\label{figure12f}]{\includegraphics[width=0.24\textwidth]{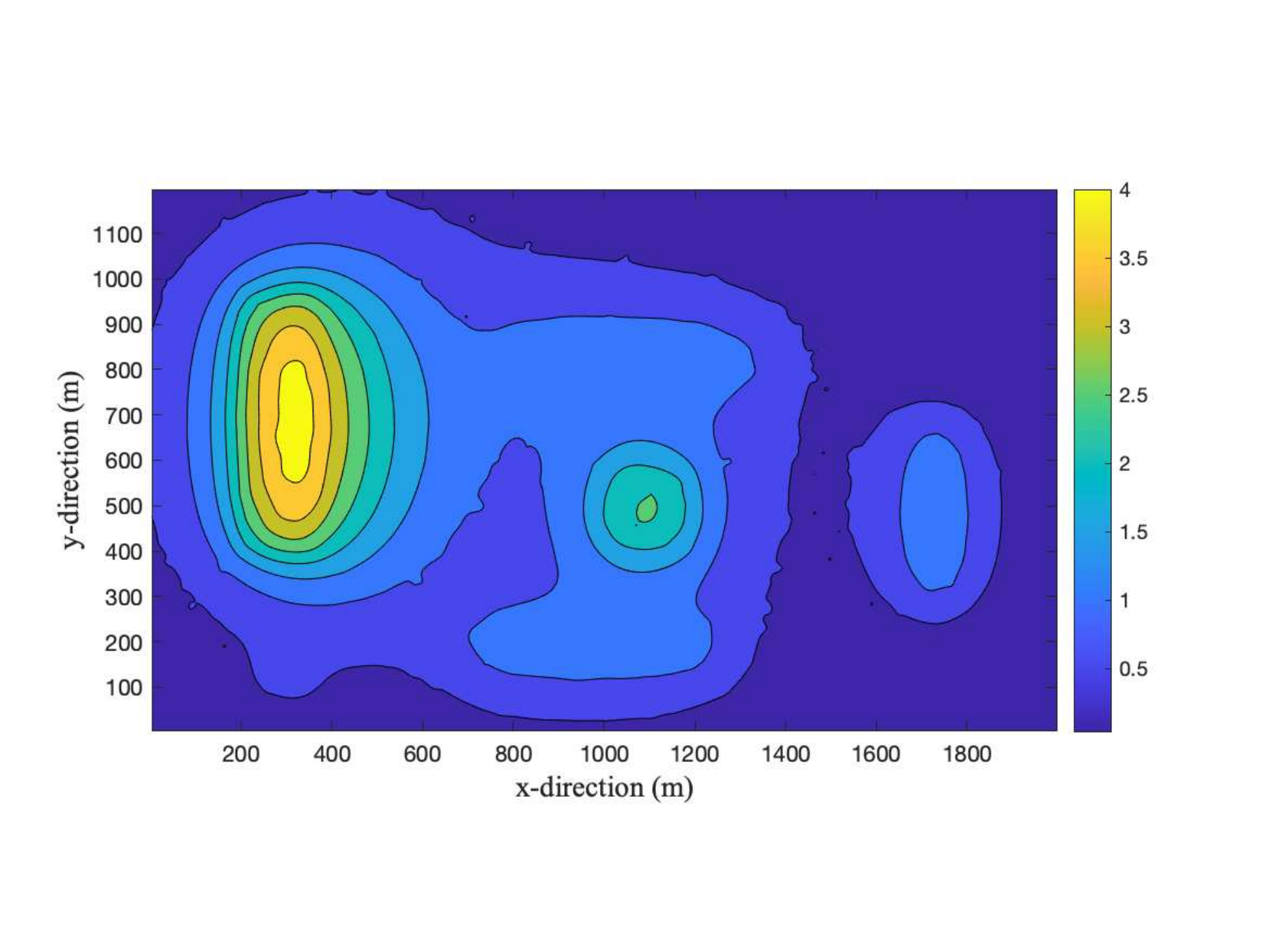}}
\subfigure[$(.61, 1.00)$.\label{figure12g}]{\includegraphics[width=0.24\textwidth]{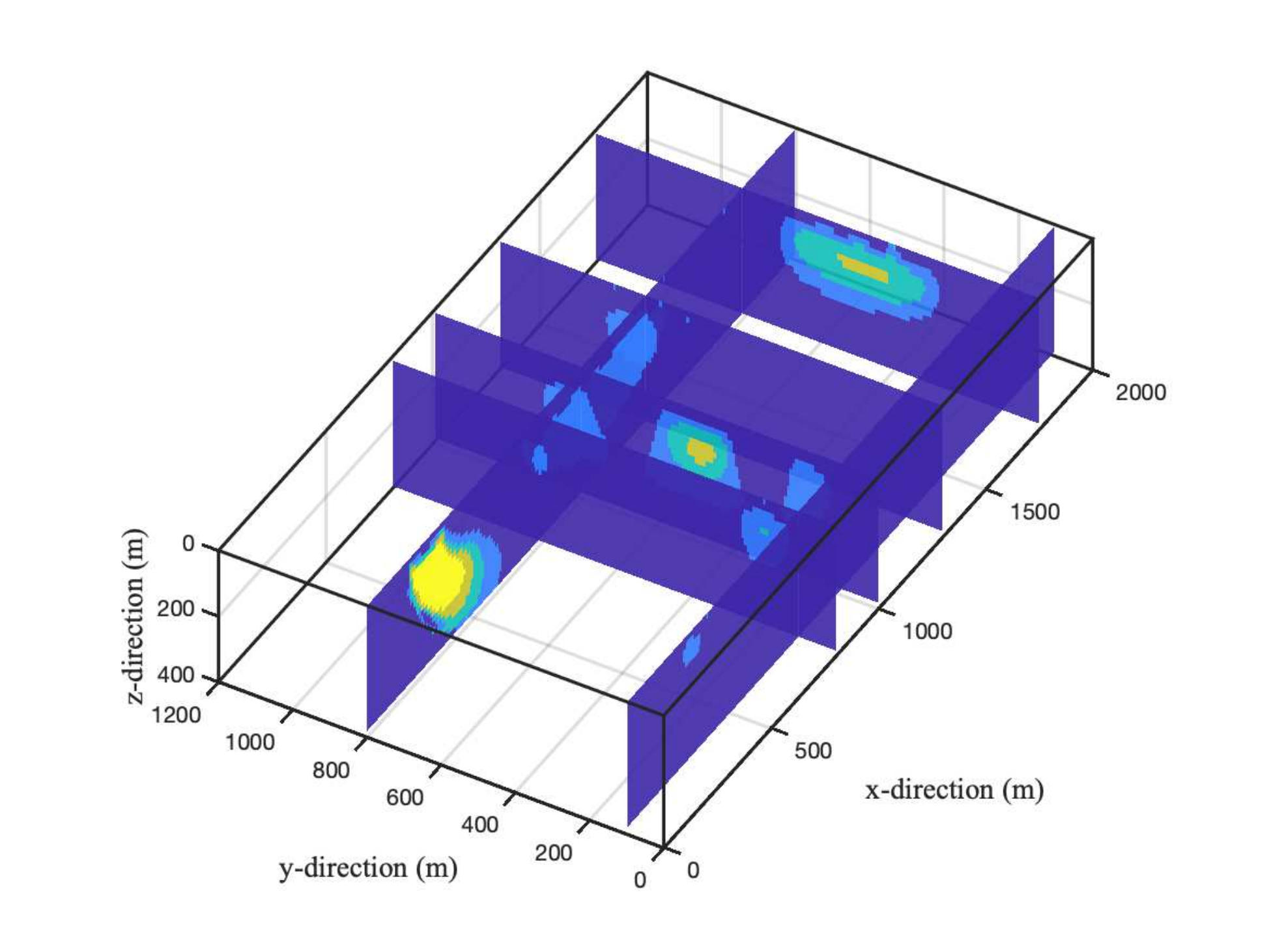}}
\subfigure[$(.59, .99)$.\label{figure12h}]{\includegraphics[width=0.24\textwidth]{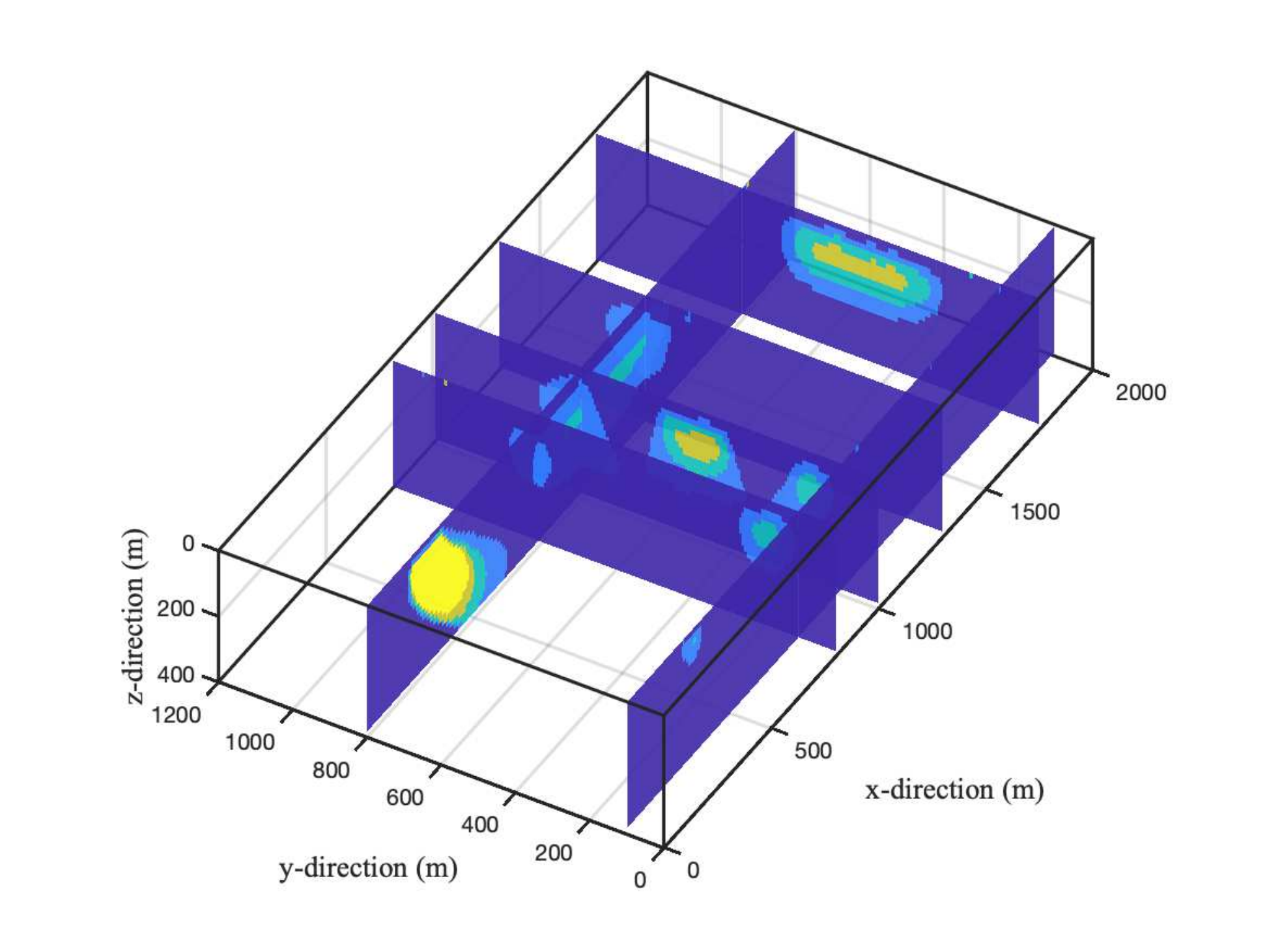}}
\caption{The gravity anomalies and reconstructed volumes using the \texttt{GKB} and \texttt{RSVD} algorithms in Figures~\ref{figure12a}-\ref{figure12d} and \ref{figure12e}-\ref{figure12h}, respectively. The first row indicates the choice of $\ell=11$, for which  $t=2268=\floor(m/20)$, $m=45375$ and $n=998250$ 
 and oversampled projected problem of size $2381$, or $\ell=12$, with $t=2700=\floor(m/20)$, $m=54000$ and $n= 1296000$,  and oversampled projected problem of size $2835$. 
 Given are the pairs $(K, \texttt{Cost}\textrm{s})$,  (number of iterations to convergence and computational cost in seconds) in the captions of the anomalies, and $(\texttt{RE}, \chi^2/(m+\sqrt{2m}))$ in the captions of the reconstructions. The units for the anomalies are mGal.
 \label{figure12}}
\end{center}
\end{figure}

Numerical experiments for the inversion of gravity data, similar to the testing for the magnetic data, demonstrates that indeed the \texttt{RSVD} algorithm with power iteration is to be preferred for the inversion of gravity data, yielding acceptable solutions at lower cost than when using the \texttt{GKB} algorithm. Representative results are detailed in Figure~\ref{figure12} for the same parameter settings as given in Figure~\ref{figure11} for the magnetic problem.  

\subsection{Real Data}\label{sec:real}
\begin{figure}[ht!]\begin{center}
\subfigure[Given anomaly. \label{figure13a}]{\includegraphics[width=0.32\textwidth]{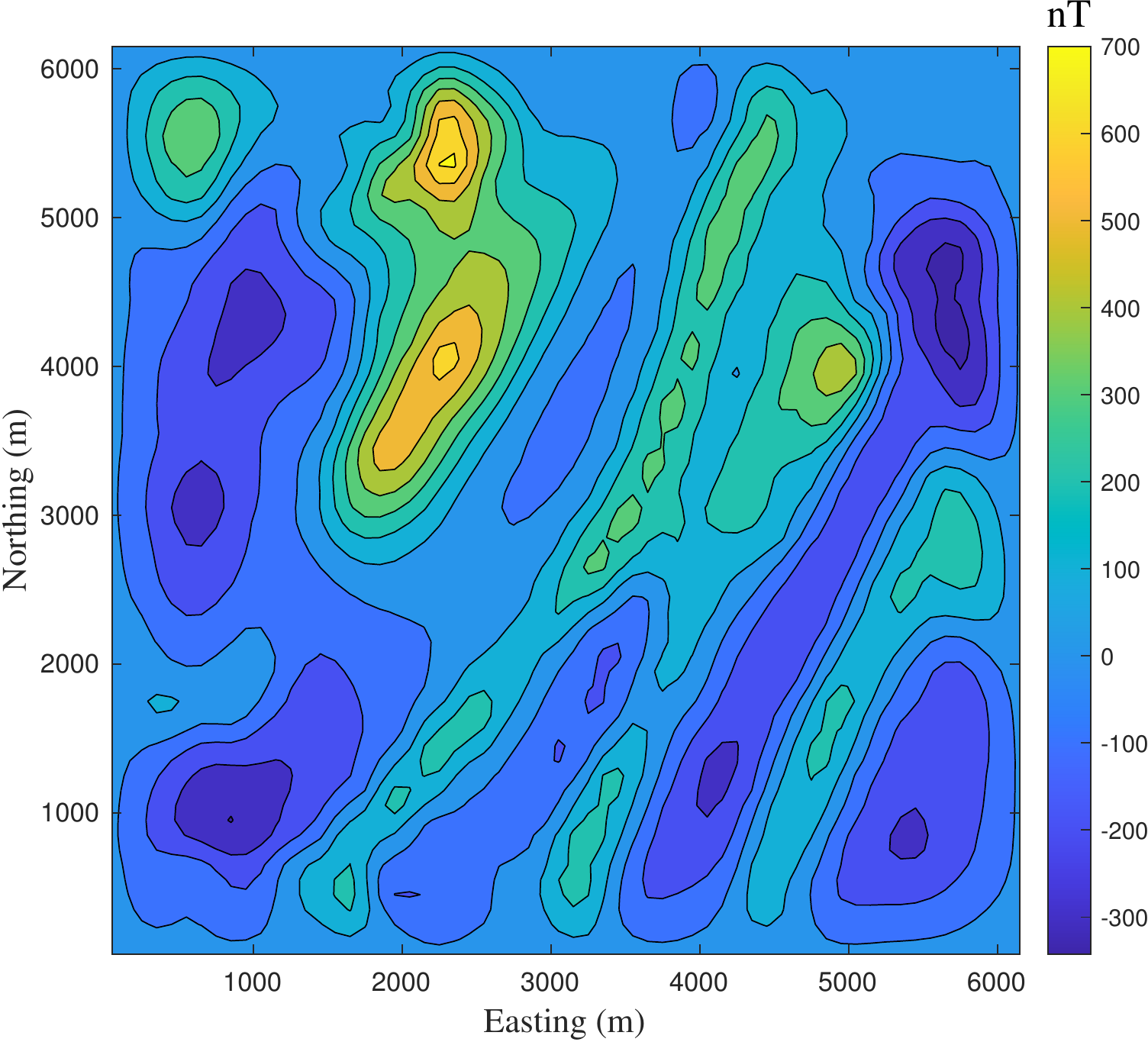}}
\subfigure[Predicted: $n=103680$.\label{figure13b}]{\includegraphics[width=0.32\textwidth]{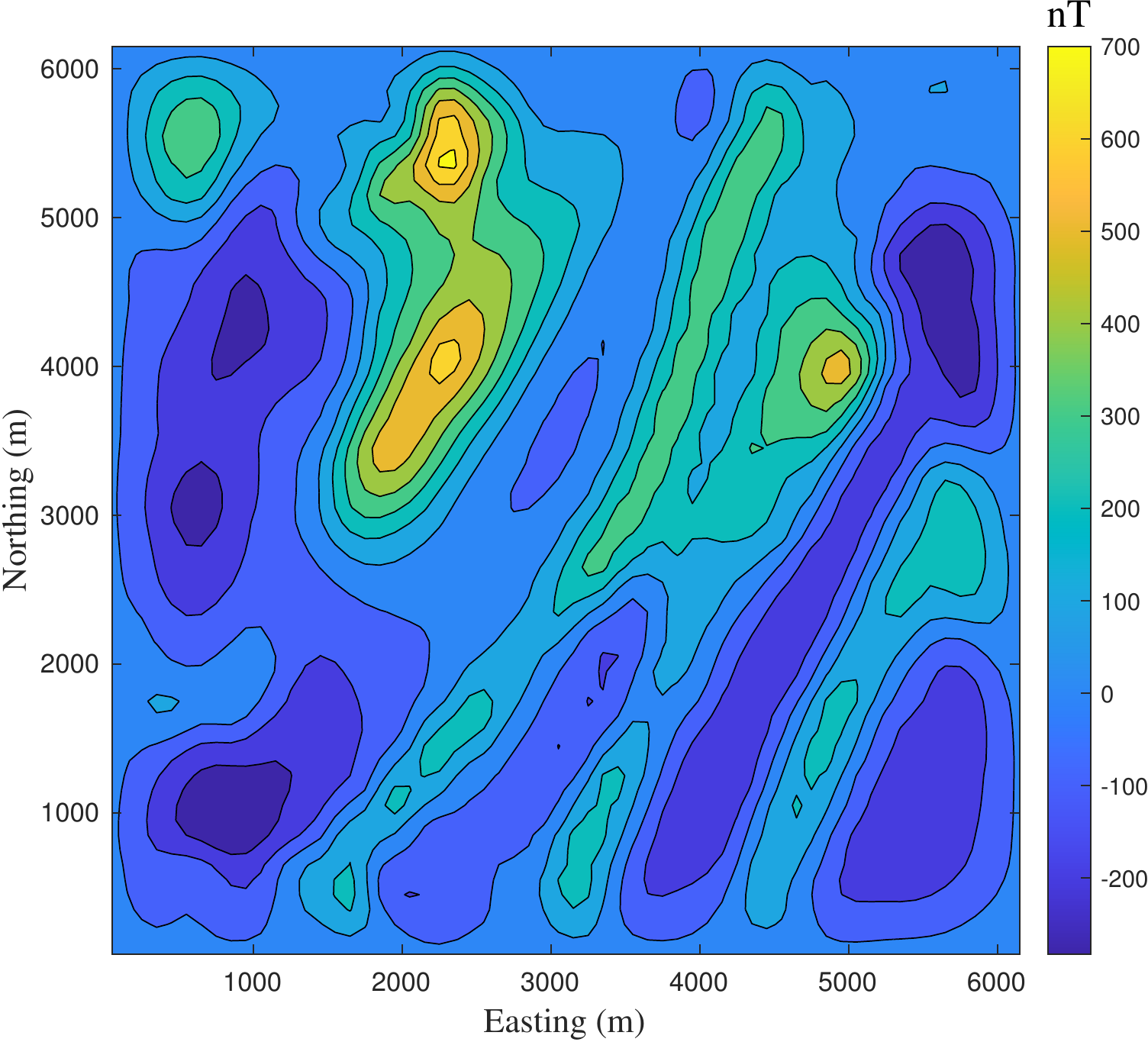}}
\subfigure[Predicted: $n=1238976$.\label{figure13c}]{\includegraphics[width=0.32\textwidth]{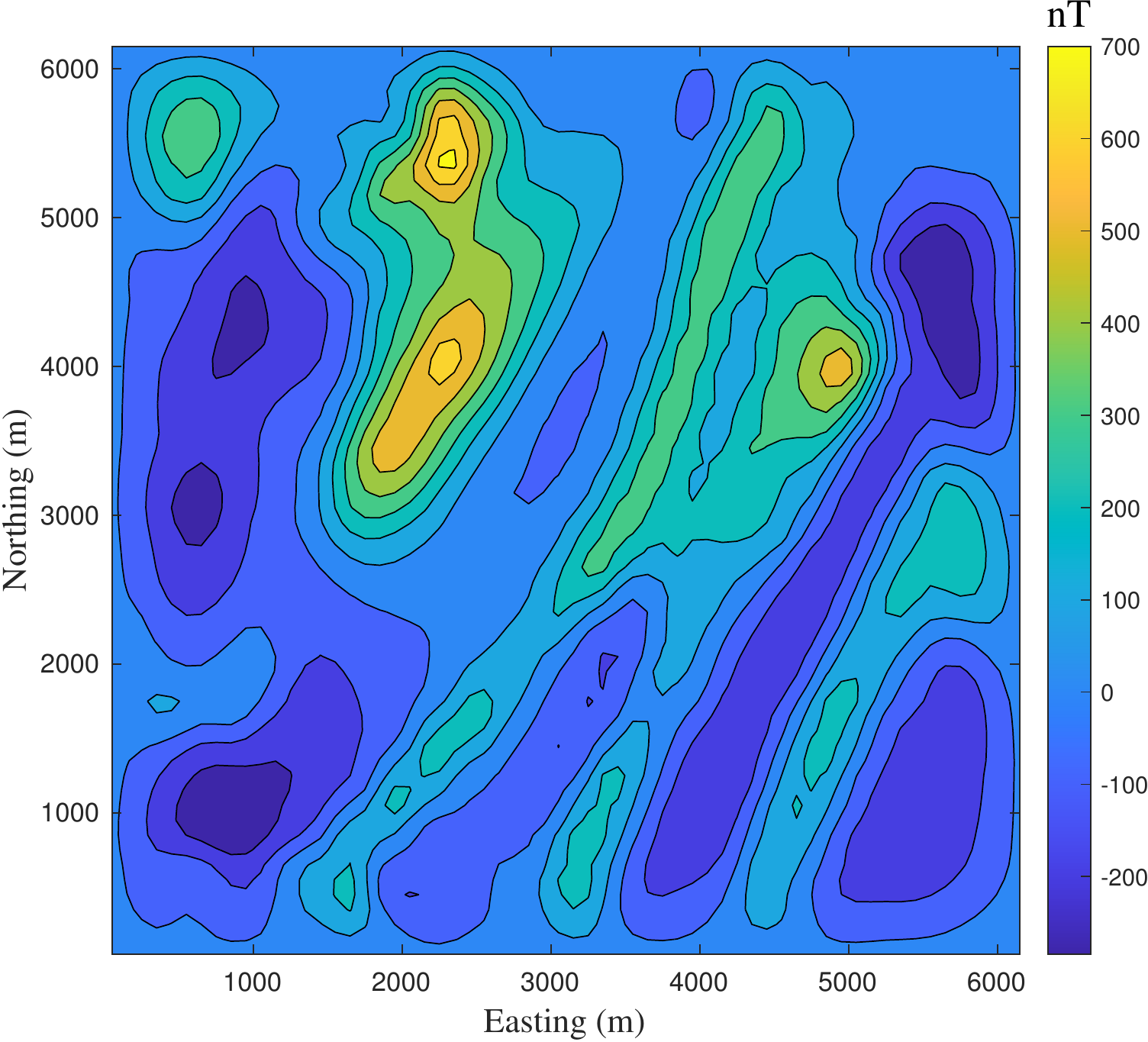}}
\caption{The given magnetic anomaly in Figure~\ref{figure13a} and the obtained predicted anomalies for the inversion using the parameters for the first and last lines of data in Table~\ref{table3} in Figures~\ref{figure13b}-\ref{figure13c}, respectively.\label{figure13}}
\end{center}
\end{figure}

For validation of the simulated results on a practical data set we apply the \texttt{GKB} algorithm for the inversion of a 
magnetic field anomaly that was collected over a portion of the Wuskwatim Lake region in Manitoba,
Canada. This data set was discussed in \cite{Pilkington:2009} and also used in \cite{vatankhah2019improving} for inversion using the \texttt{RSVD} algorithm with a single power iteration. Further details of the geological relevance of this data set is given in these references. Moreover, its use makes for direct comparison with these existing results. Here we use a grid of $62 \times 62=3184$ measurements at $100$m intervals in the East-North direction with padding of $5$ cells yielding a horizontal cross section of size $72 \times 72$ in the East-North directions. The depth dimension is discretized with  $\Delta z=100$m, yielding a regular cube, to $\Delta z=8$m for rectangular prisms with a smaller edge length in the depth dimension for a total depth of $2000$m, and providing increasing values of $n$ from $103680$ to $1238976$ as detailed in Table~\ref{table3}. The given magnetic anomaly is illustrated in Figure~\ref{figure13a}.

\begin{figure}[ht!]\begin{center}
\subfigure[Regularization function for $K$. \label{figure14a}]{\includegraphics[width=0.48\textwidth]{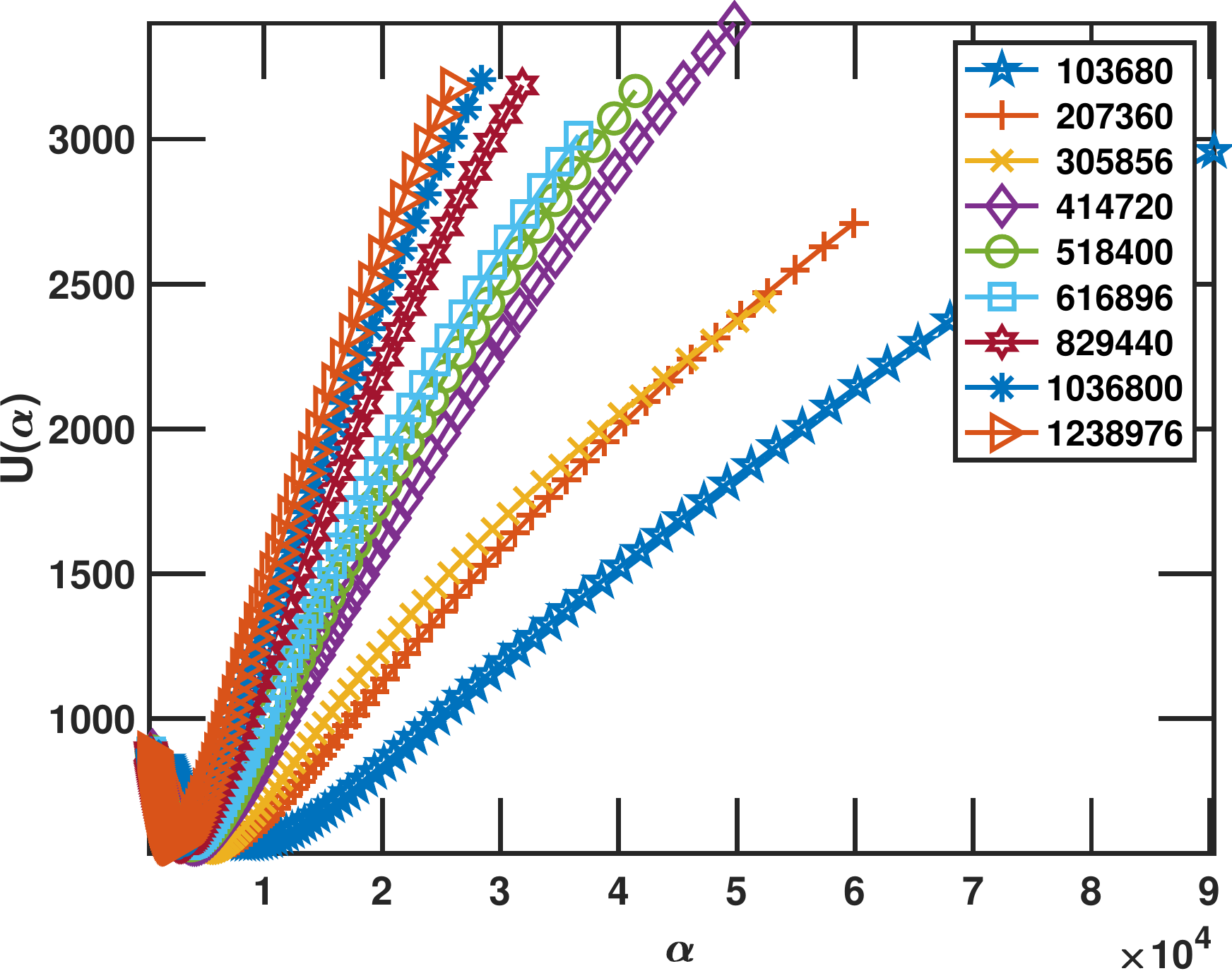}}
\subfigure[$\chi^2$ with $k$. \label{figure14b}]{\includegraphics[width=0.48\textwidth]{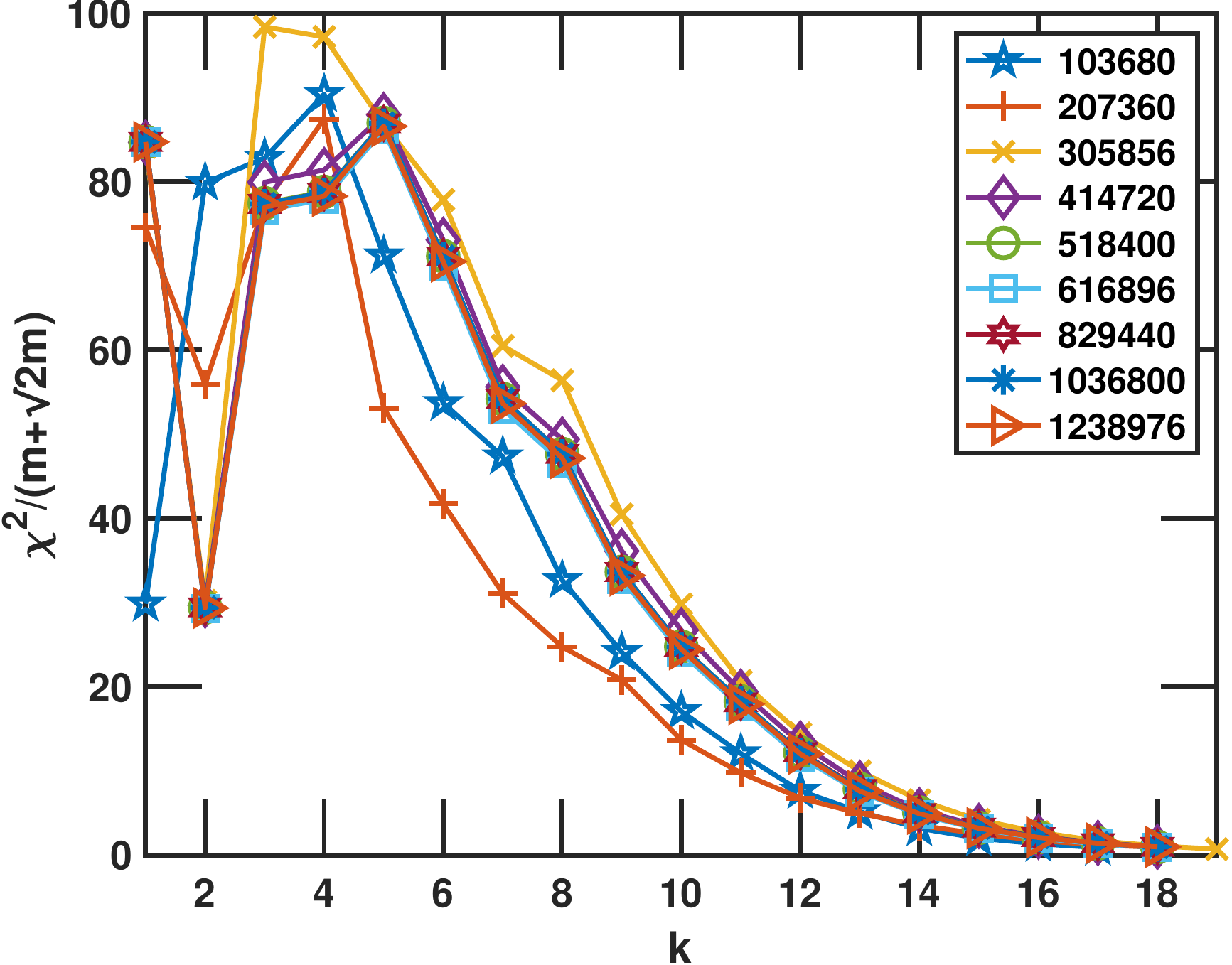}}
\caption{The plot of the regularization function $U(\alpha)$ for the \texttt{UPRE} algorithm, at the final iteration $K$ for increasing values of $n$ as indicated in Table~\ref{table3} in Figure~\ref{figure14a} and the progression of the scaled $\chi^2$ estimate as a function of iteration $k$ and for increasing $n$ in Figure~\ref{figure14b}.\label{figure14}}
\end{center}
\end{figure}

In each inversion the \texttt{GKB} algorithm is run with $t=480$, corresponding to $t=\floor(m/8)$, where $m=3184$ and oversampled projected space of size $504$, and a noise distribution based on \eqref{noisydata} is employed using $\tau_1=.02$ and $\tau_2=.018$. All inversions converge to the tolerance  $ \chi^2/(m+\sqrt{2m}))<1$ in no more than $19$ iterations for all problem sizes, as given in Table~\ref{table3}. The computational cost measured in seconds is also given in Table~\ref{table3} and demonstrates that it is feasible to invert for large parameter volumes, in times ranging from just under $5$ minutes for the coarsest resolution, to just over $73$ minutes for the volume with the highest resolution. Here the computations are performed on a MacBook Pro laptop with 2.5 GHz Dual-Core Intel Core i7 chip and $16$GB memory. In Figure~\ref{figure14a} we show that the \texttt{UPRE} function has a well-defined minimum at the final iteration for all resolutions, and in Figure~\ref{figure14b} that the convergence of the scaled $\chi^2$ value is largely independent of $n$. The final regularization parameter $\alpha^{(K)}$ decreases with increasing $n$, while the initial $\alpha$ found using \eqref{eq:alpha1} increases with $n$, as reported in Table~\ref{table3}. 

\begin{figure}[ht!]\begin{center}
\subfigure[Iso-surface using $n=103680$.\label{figure15a}]{\includegraphics[width=0.45\textwidth]{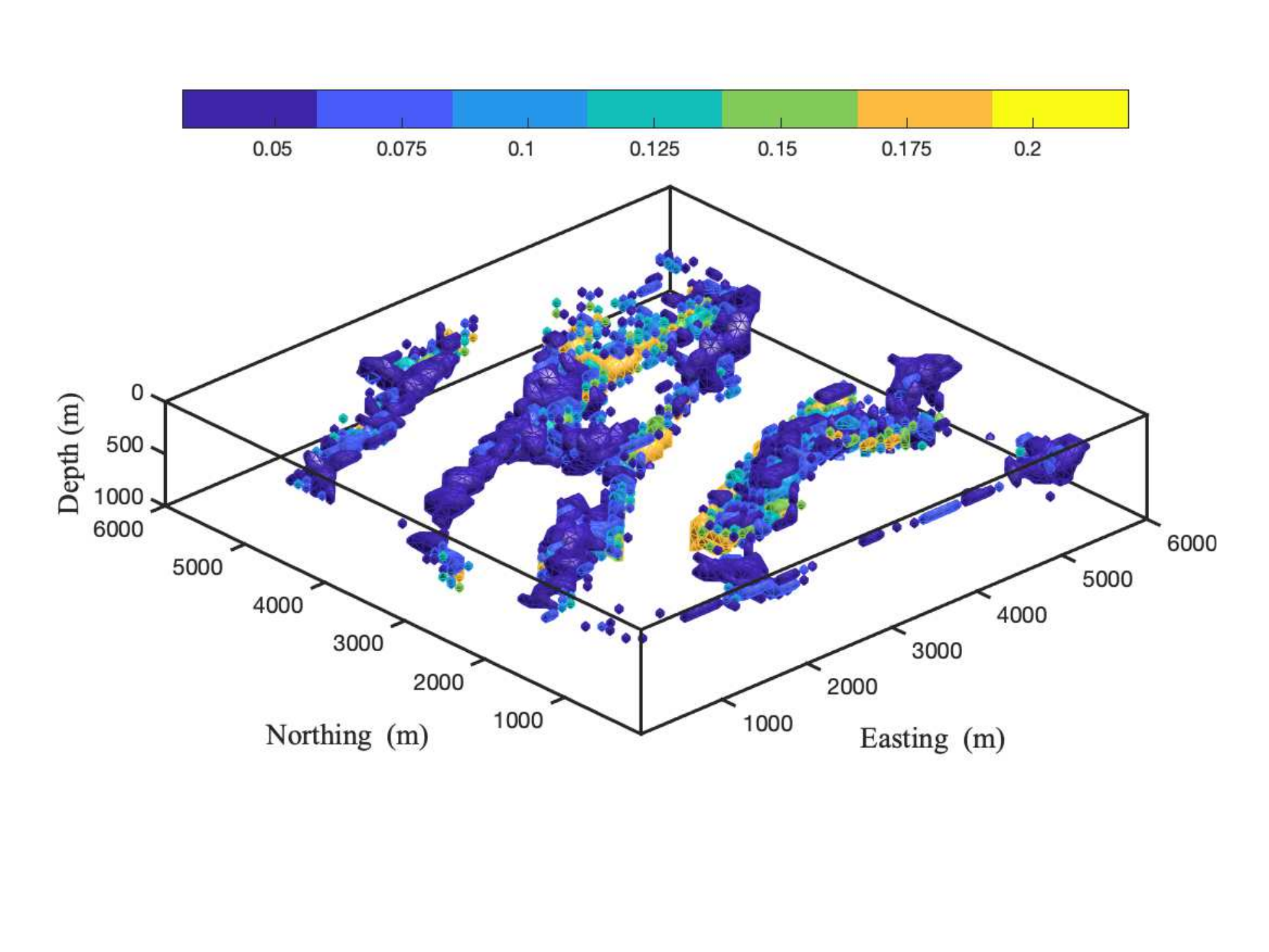}}
\subfigure[Iso-surface  using $n=1238976$.\label{figure15b}]{\includegraphics[width=0.45\textwidth]{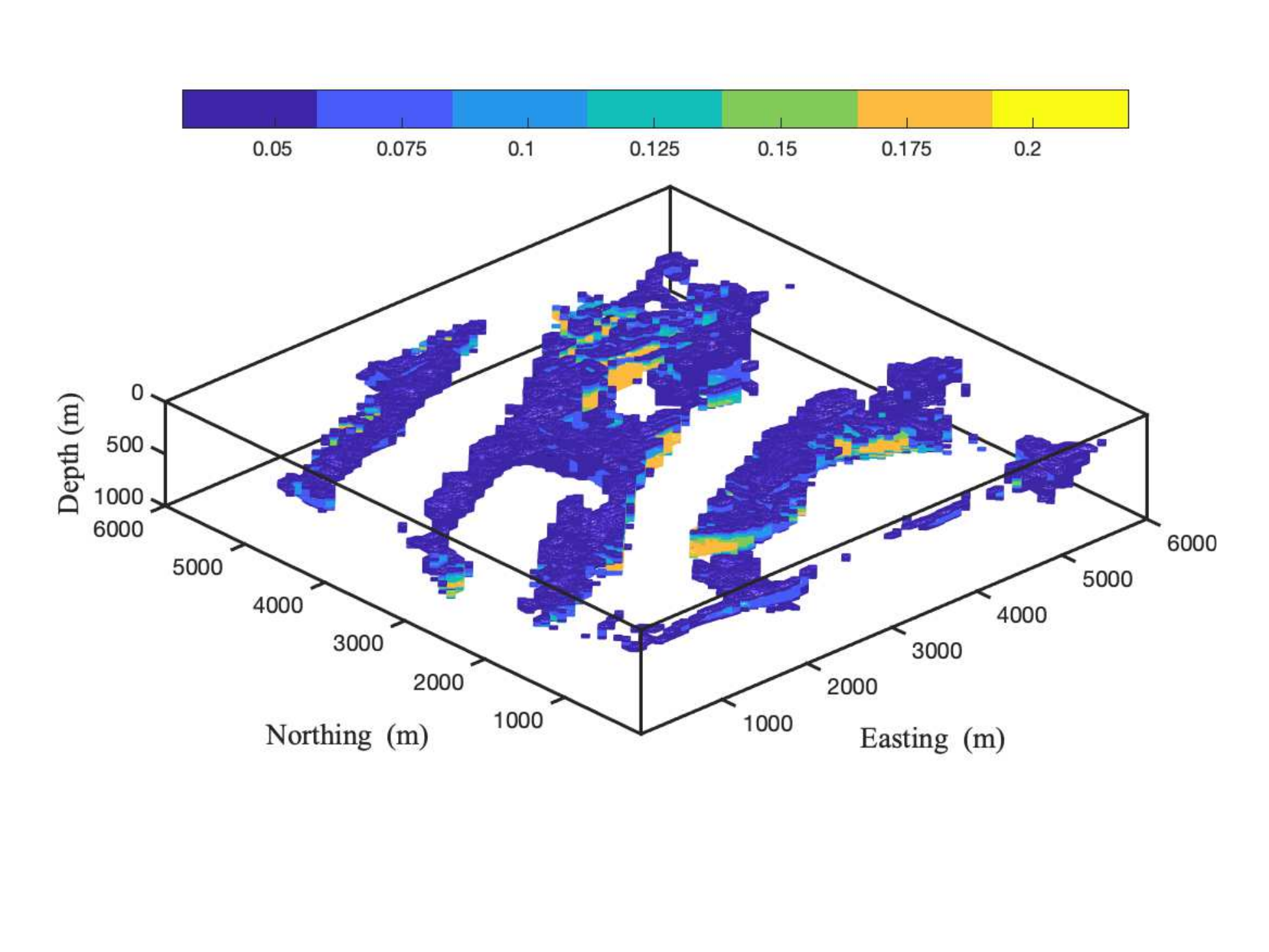}}
\caption{The reconstructed volumes showing parameters $\kappa>0.05$ and depth from $0$ to $1000$,  corresponding to the predicted anomalies in Figure~\ref{figure13}. \label{figure15}}
\end{center}
\end{figure}

\begin{figure}[ht!]\begin{center}
\subfigure[$300$m\label{figure16a}]{\includegraphics[width=0.19\textwidth]{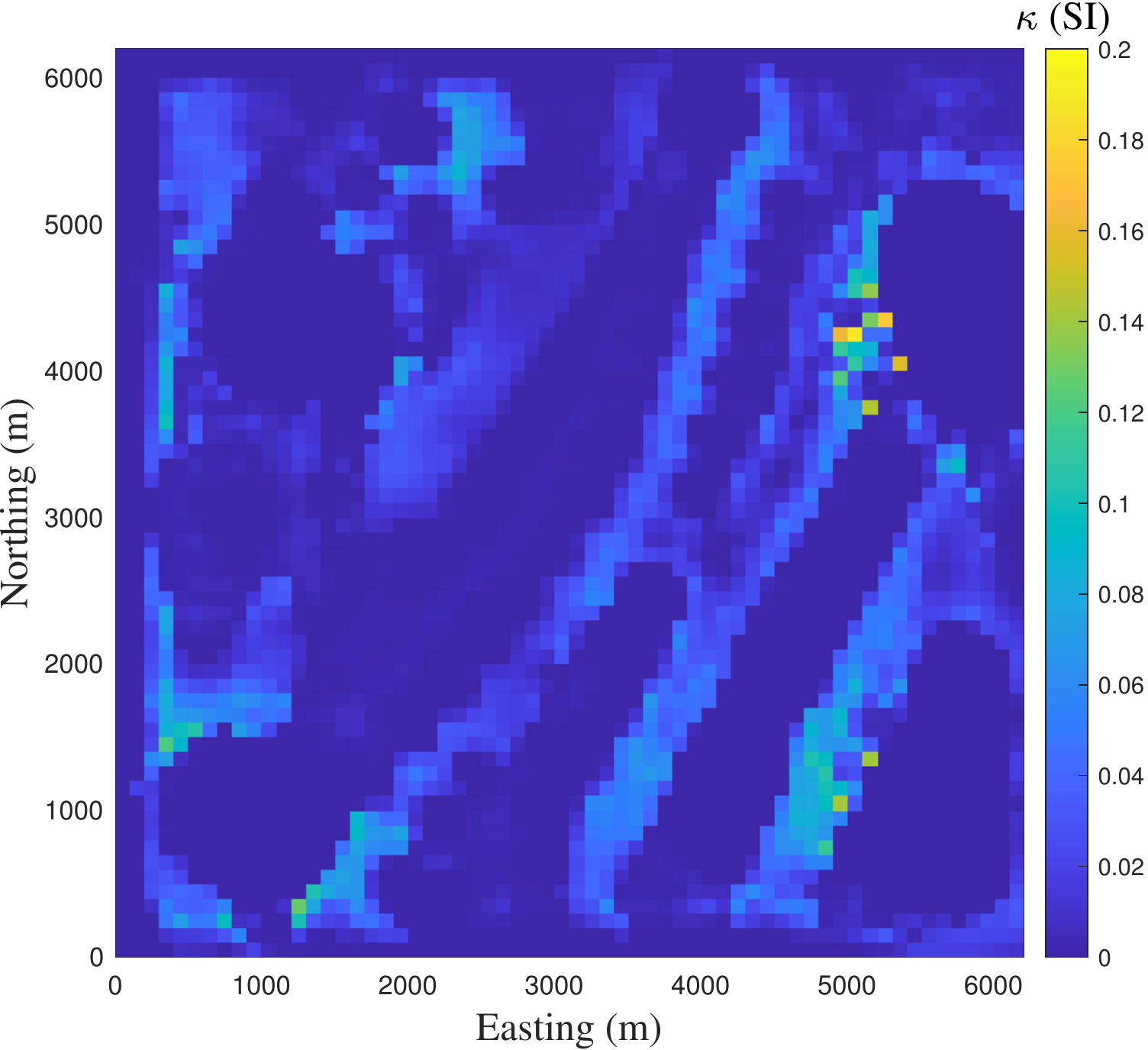}}
\subfigure[$500$m\label{figure16b}]{\includegraphics[width=0.19\textwidth]{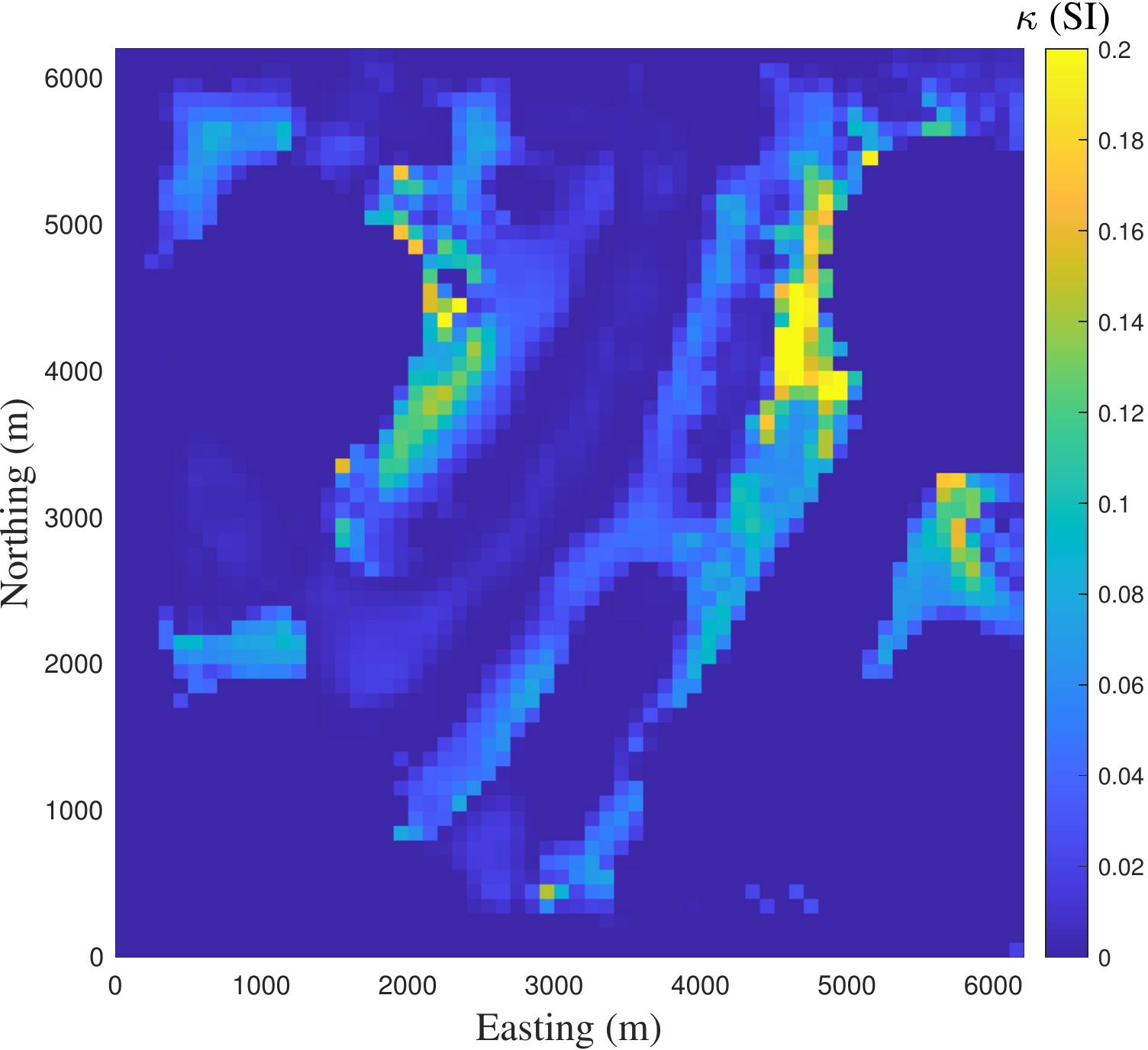}}
\subfigure[$700$m\label{figure16c}]{\includegraphics[width=0.19\textwidth]{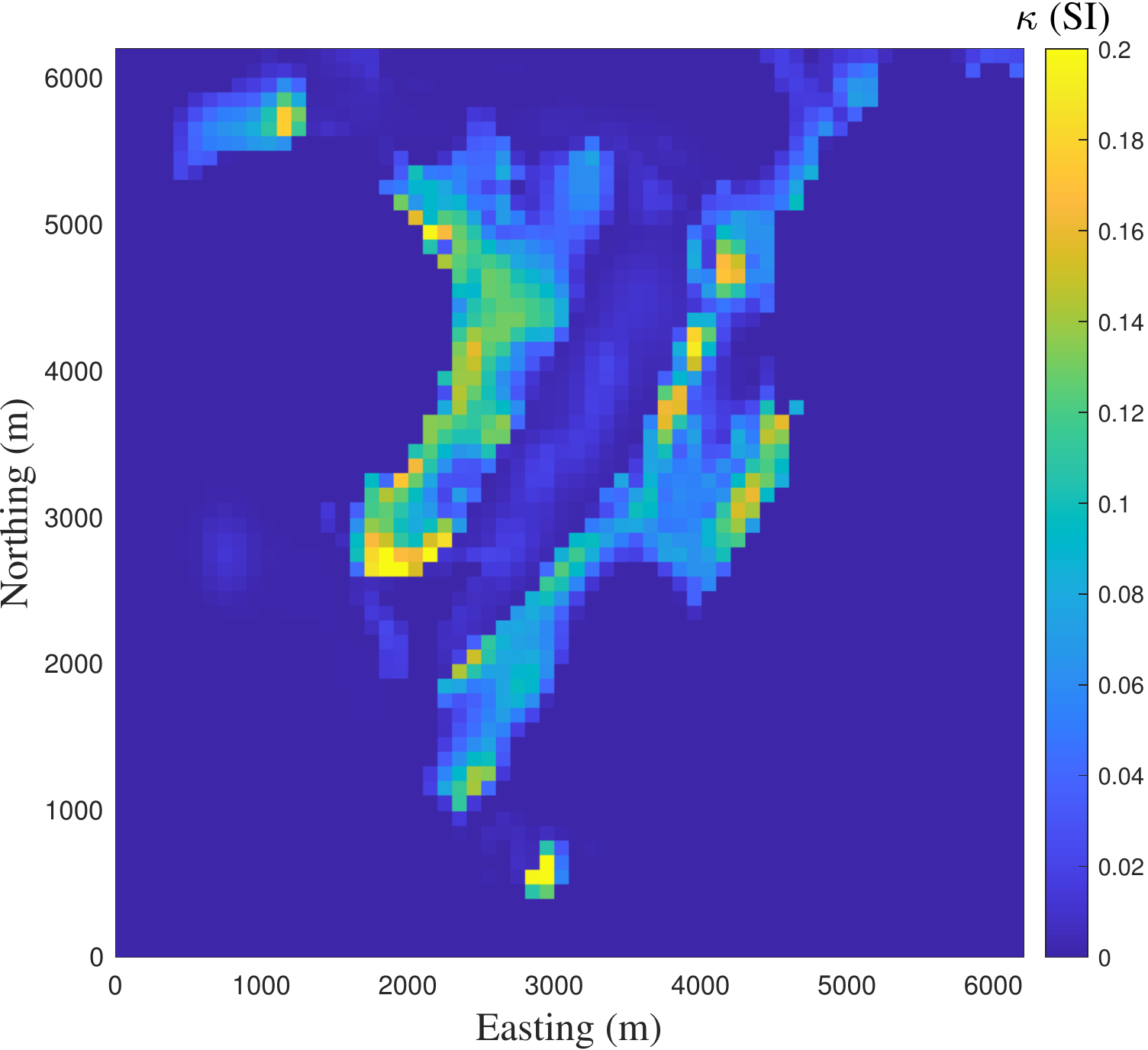}}
\subfigure[$900$m\label{figure16d}]{\includegraphics[width=0.19\textwidth]{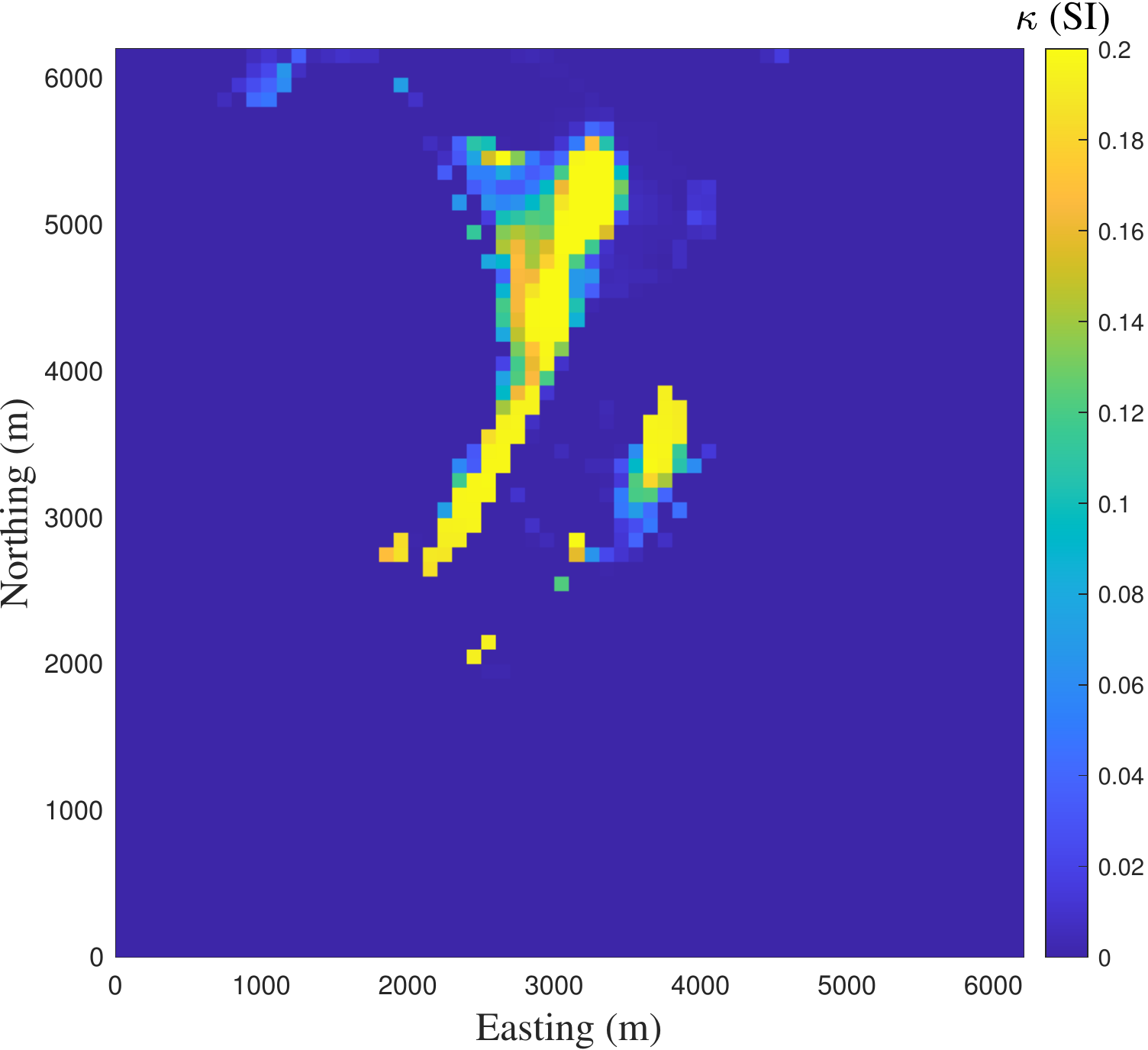}}
\subfigure[$1100$m\label{figure16e}]{\includegraphics[width=0.19\textwidth]{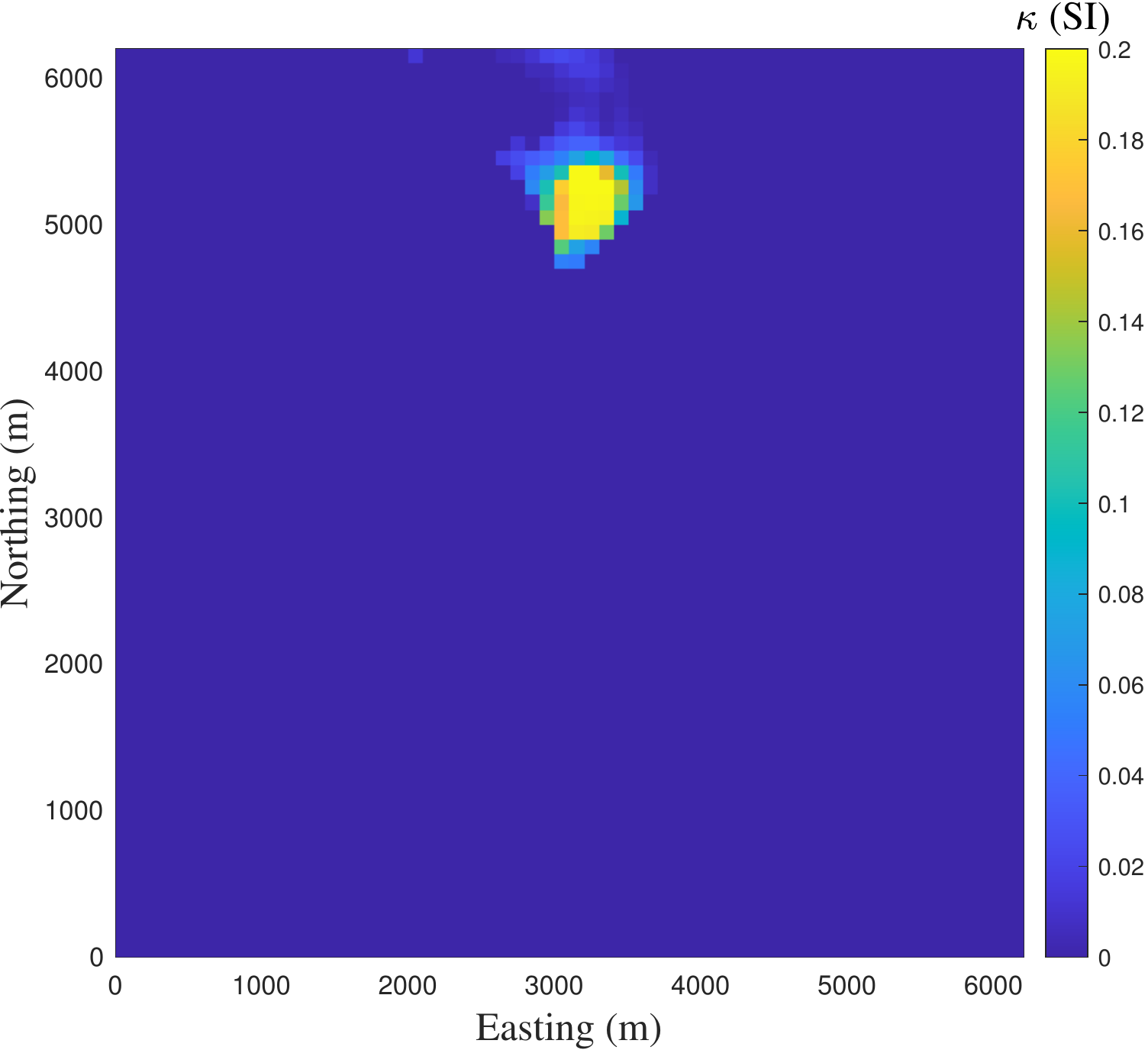}}
\subfigure[$300$m\label{figure16f}]{\includegraphics[width=0.19\textwidth]{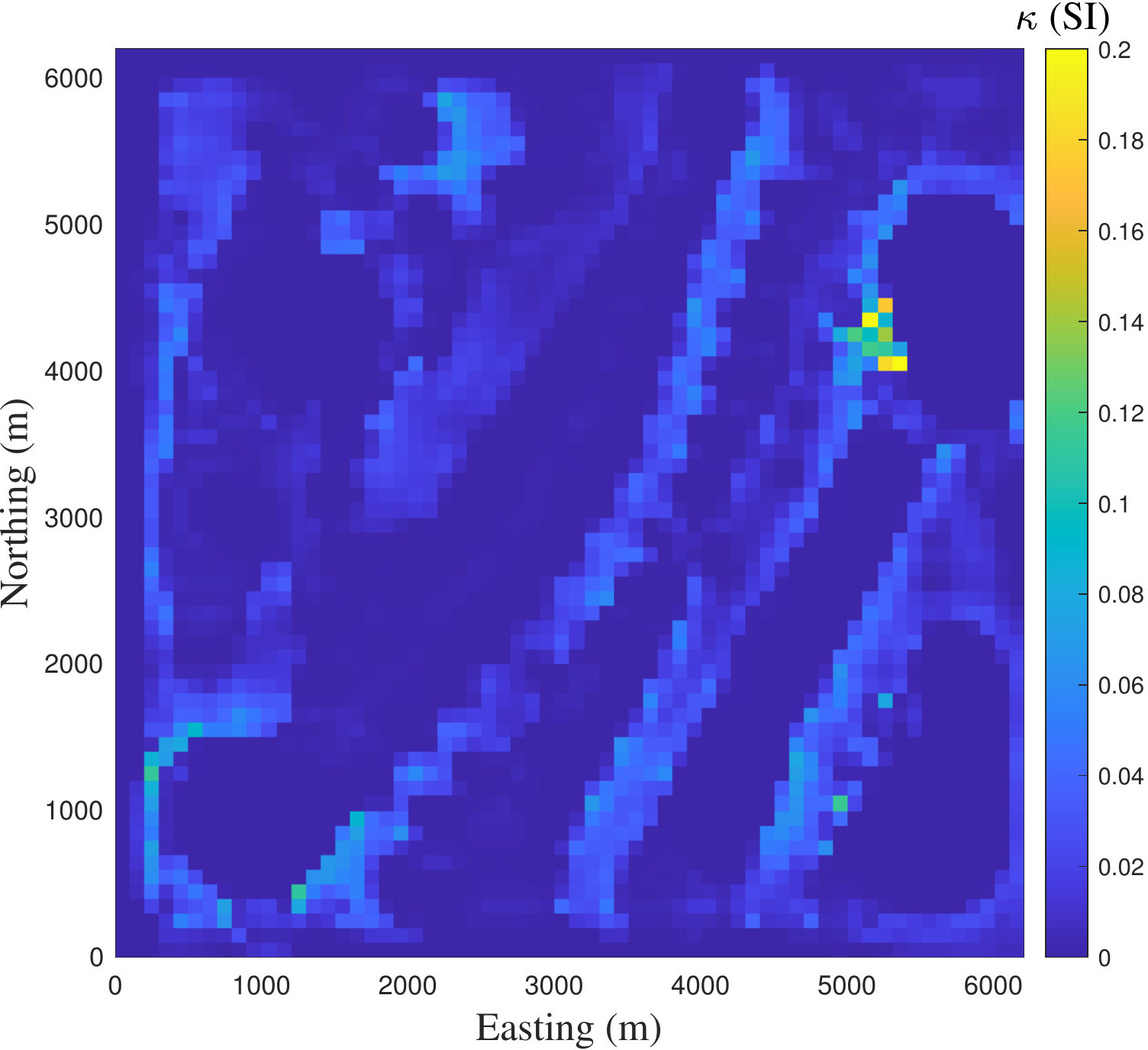}}
\subfigure[$500$m\label{figure16g}]{\includegraphics[width=0.19\textwidth]{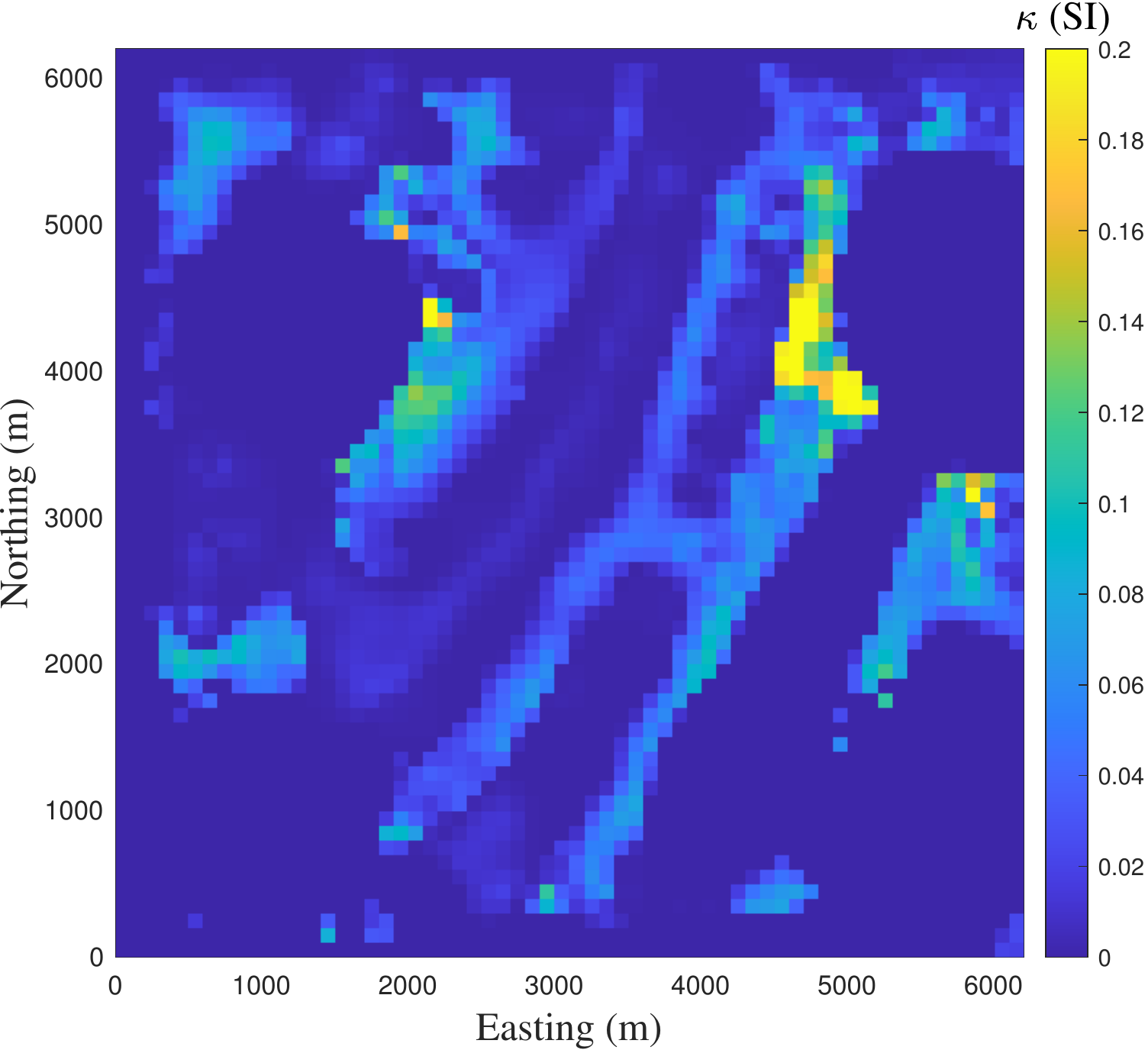}}
\subfigure[$700$m\label{figure16h}]{\includegraphics[width=0.19\textwidth]{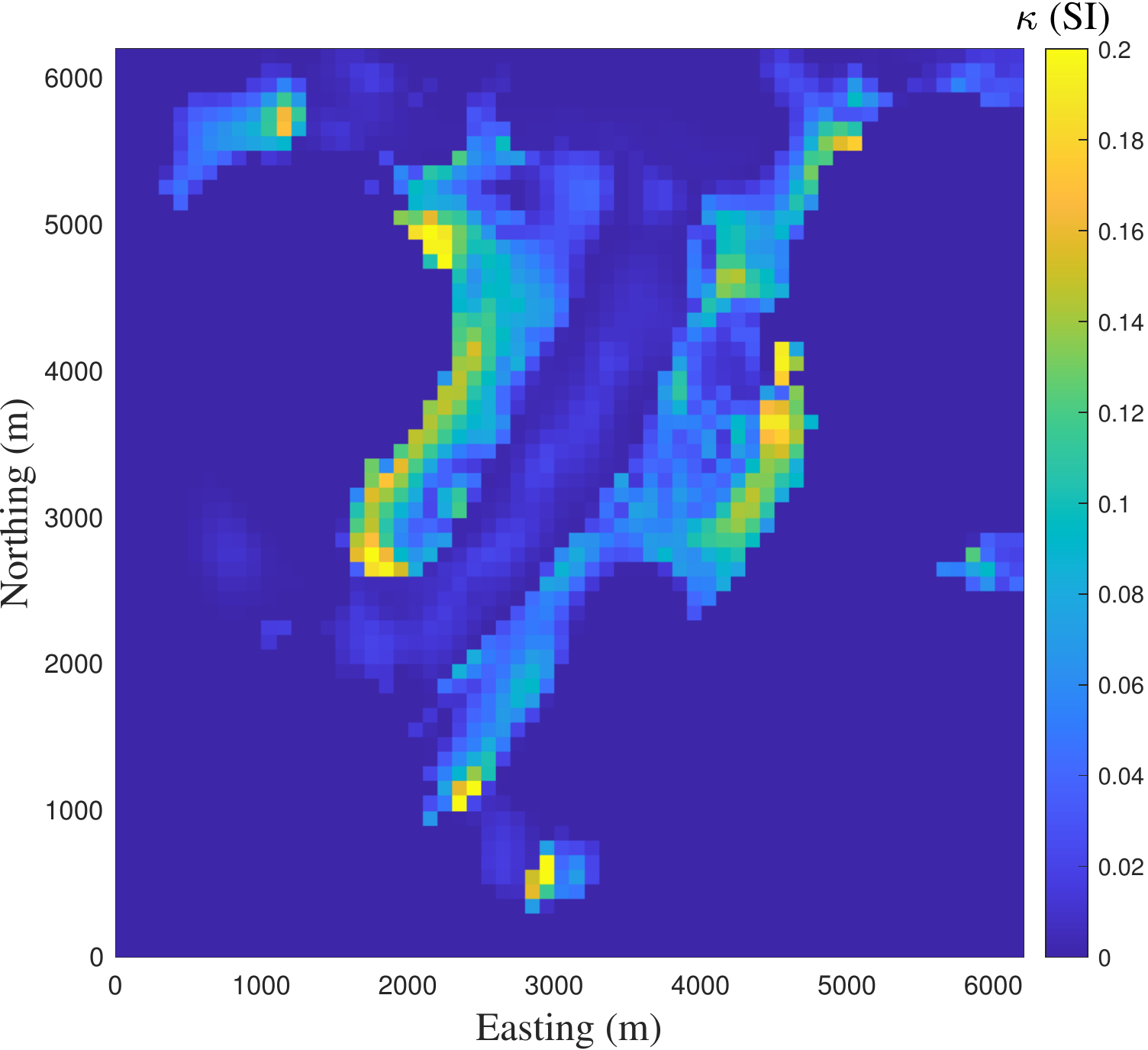}}
\subfigure[$900$m\label{figure16i}]{\includegraphics[width=0.19\textwidth]{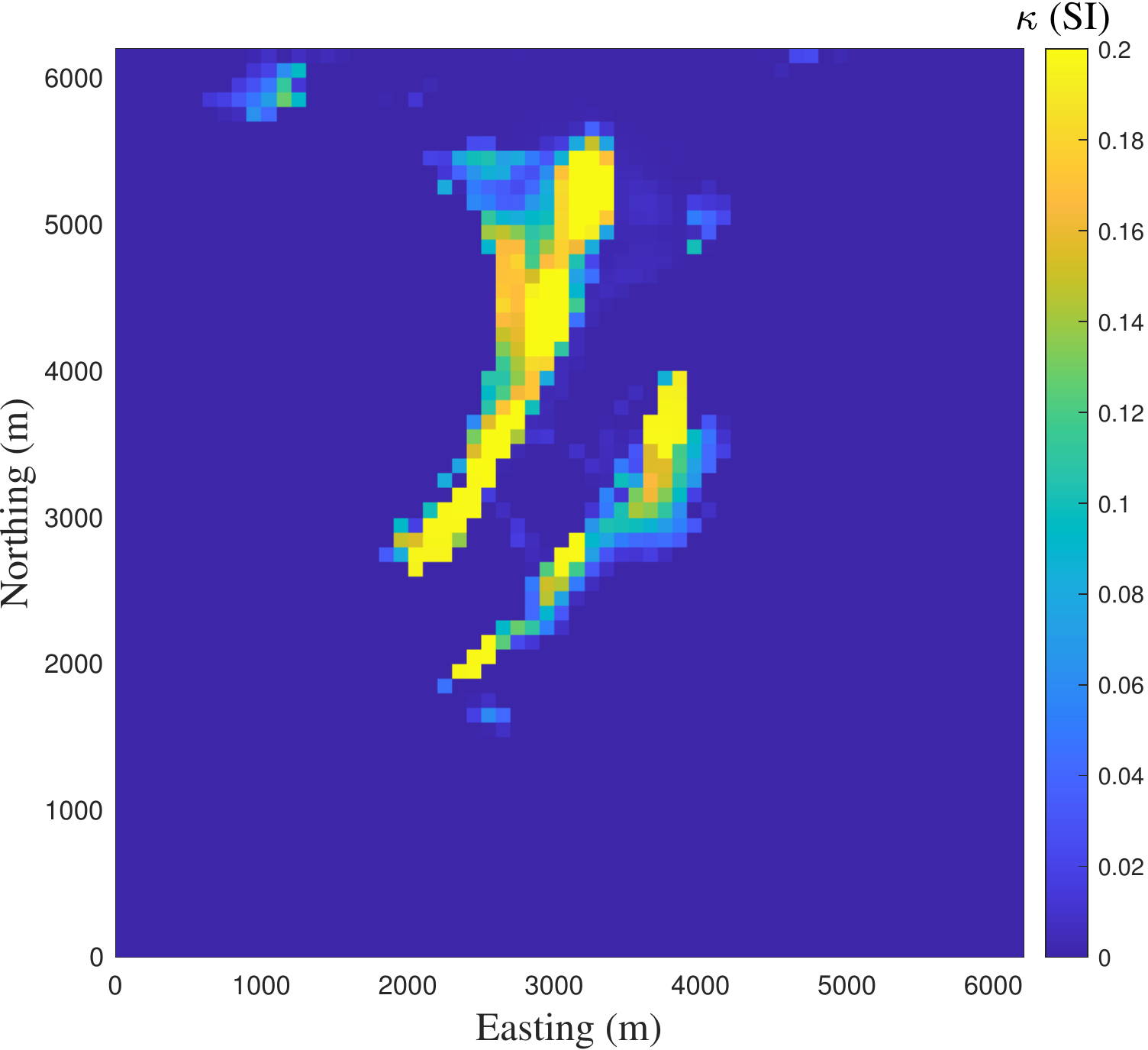}}
\subfigure[$1100$m\label{figure16j}]{\includegraphics[width=0.19\textwidth]{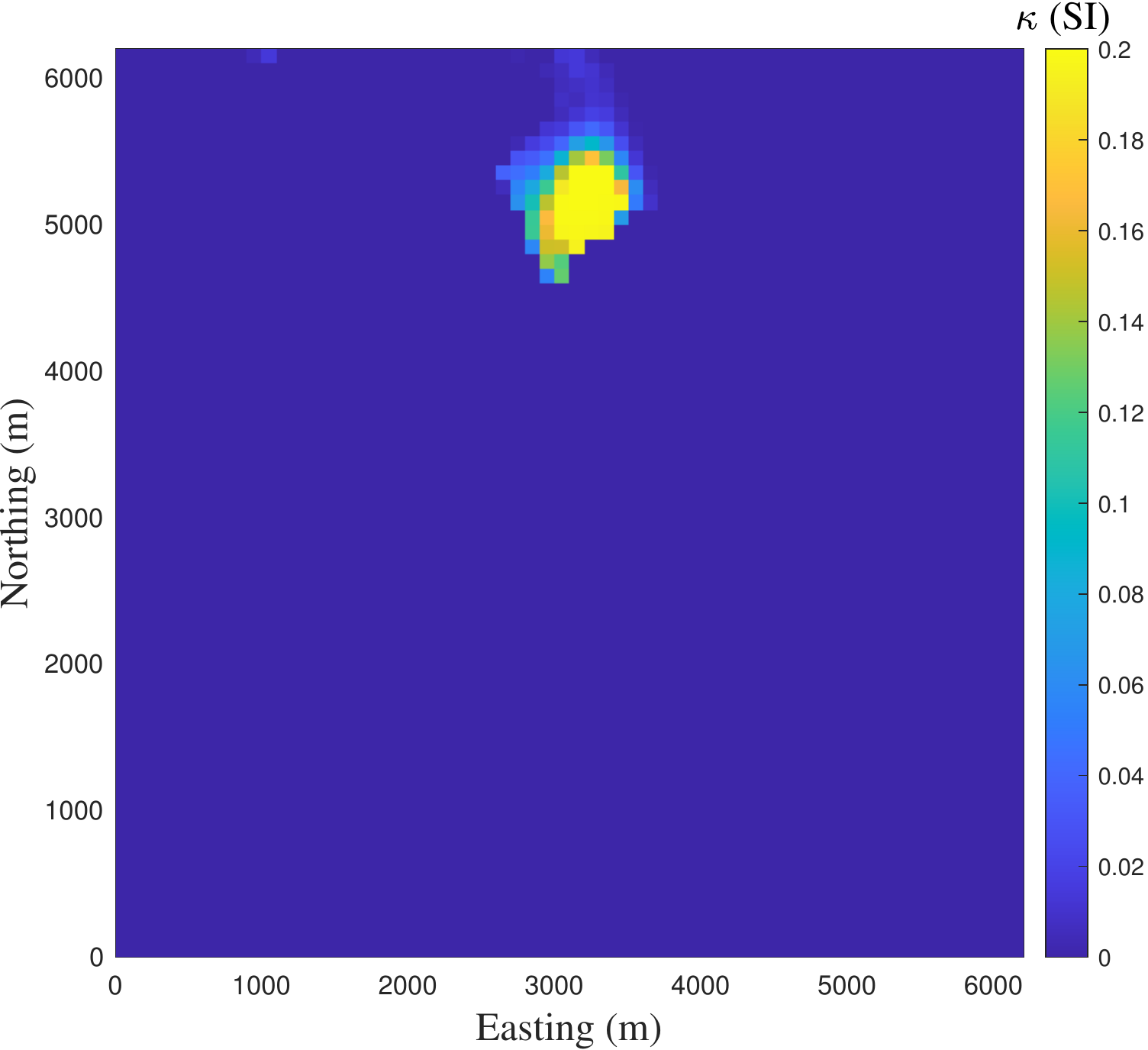}}
\caption{Slices through the volumes illustrated in Figure~\ref{figure15} for depths $300$, $500$, $700$, $900$ and $1100$, for $n=103680$ in Figures~\ref{figure16a}-\ref{figure16e} and for $n=1238976$ in Figures~\ref{figure16f}-\ref{figure16j}. \label{figure16}}
\end{center}
\end{figure}

Results of the inversion, for the coarsest and finest resolutions are presented in Figures~\ref{figure13}, \ref{figure15} and \ref{figure16}, for anomalies, reconstructed volumes, and depth slices through the volume domain, respectively. First, from Figures~\ref{figure13b}-\ref{figure13c}, as compared to Figure 16b in \cite{vatankhah2019improving}, we see that the predicted anomalies provide better agreement to the measured anomaly, with respect to structure and the given values. Moreover, more structure is seen in the volumes presented in Figures~\ref{figure15a}-\ref{figure15b} as compared to Figure19  \cite{vatankhah2019improving}, and the increased resolution provides greater detail in Figure~\ref{figure15b}as compared to Figure~\ref{figure15a}. Here the volumes are presented for the depth from $0$ to $1000$m only, but it is seen in Figures~\ref{figure16e} and \ref{figure16j}, which are the slices at depth $1100$m, that there is little structure evident at greater depth. Comparing the depth slices for increasing depth, we see that the use of the higher resolution leads to more structure at increased depth. Moreover,  the  results are consistent with those presented in \cite{vatankhah2019improving} for the use of the \texttt{RSVD} for a projected size $t=1100$  as compared to $t=480$ used here. It should also be noted that the \texttt{RSVD} algorithm with one power iteration does not converge within $50$ steps, under the same configurations for $m$, $n$ and $t$. 

\begin{table}[ht]\begin{center}\begin{tabular}{|*{8}{c|}}
\hline
$n$& $\nbz$&$\Delta z$&$K$ & $\alpha^{(1)}$ & $\alpha^{(K)}$   &${\chi^2}/{(m+\sqrt{2m})}$&$\textrm{Cost} (s)$\\
\hline
\hline$ 103680$&$     20$&$    100$&$     17$&$4.60e+05$&$  8558$&$  0.87$&$   334$\\
\hline$ 207360$&$     40$&$     50$&$     18$&$5.36e+06$&$  5887$&$  0.90$&$   754$\\
\hline$ 305856$&$     59$&$     33$&$     19$&$2.07e+07$&$  4930$&$  0.70$&$  1126$\\
\hline$ 414720$&$     80$&$     25$&$     18$&$6.09e+07$&$  4116$&$  0.95$&$  1513$\\
\hline$ 518400$&$    100$&$     20$&$     18$&$1.33e+08$&$  3701$&$  0.94$&$  2018$\\
\hline$ 616896$&$    119$&$     16$&$     18$&$2.43e+08$&$  3386$&$  0.90$&$  2095$\\
\hline$ 829440$&$    160$&$     12$&$     18$&$6.90e+08$&$  2933$&$  0.95$&$  3091$\\
\hline$1036800$&$    200$&$     10$&$     18$&$1.51e+09$&$  2627$&$  0.95$&$  3690$\\
\hline$1238976$&$    239$&$      8$&$     18$&$2.80e+09$&$  2396$&$  0.96$&$  4389$\\
\hline
\end{tabular}\caption{
Inversion of magnetic data as illustrated in Figure~\ref{figure13} for $m=3844$ on a grid of $62\times62$ stations, with $\Delta x=\Delta y=100$m and padding of $5$ cells in both $x$ and $y$-directions, yielding blocks of size $\nbr=5184$. The inversion uses the \texttt{GKB} algorithm with $t=480$ ($\floor(m/8)$) and $t_p=504$. The noise in the algorithm uses \eqref{noisydata} as  given for the simulations    with $\tau_1=.02$ and $\tau_2=.018$. These results are obtained using a MacBook Pro laptop with 2.5 GHz Dual-Core Intel Core i7 chip and $16$GB memory.} \label{table3}\end{center}\end{table}

\section{Conclusions and Future Work}\label{sec:conclusions}
Two algorithms,  \texttt{GKB} and \texttt{RSVD},  for the focused inversion of potential field data with all operations for the sensitivity matrix $G$ implemented using a fast \texttt{2DFFT} algorithm have been developed and validated for the inversion of both gravity and magnetic data sets.  The results show first that it is distinctly more efficient to use the \texttt{2DFFT} for operations with matrix $G$ rather than direct multiplication. This is independent of algorithm and data set, for all large scale implementations considered. Moreover, the implementation using the \texttt{2DFFT} makes it feasible to solve these large scale problems on a standard desktop computer without any code modifications to handle multiple cores or GPUs, which is not possible due to memory constraints when $m$ and $n$ increase. While both algorithms are improved with this implementation, the results show that the impact on the \texttt{GKB} efficiency is greater than that on the \texttt{RSVD} efficiency. A theoretical analysis of the computational cost of each algorithm for a single iterative step demonstrates that the \texttt{GKB} should be faster, but this is not always realized in practice as the problem size increases, with commensurate increase in the size of the projected space. Then, the efficiency of \texttt{GKB} deteriorates, and the advantage of using \texttt{builtin} routines from \textsc{Matlab} for the \texttt{RSVD} algorithm is crucial. 

When considering the computational cost to convergence for both algorithms, which also then includes the cost due to the requiring  projected spaces that are of reasonable size relative to $m$, the results confirm earlier published results that it is more efficient to use \texttt{RSVD}, with $t\ge \floor(m/8)$ for inversion of gravity data. Moreover, generally larger projected spaces are required when using \texttt{RSVD} for the inversion of magnetic data. On the other hand, prior published work did not contrast \texttt{GKB} with \texttt{RSVD} for the inversion of magnetic data. Here, our results contribute a new conclusion to the literature, namely that \texttt{GKB} is more efficient for these large-scale problems and can use also $t \approx  \floor(m/8)$  rather than larger spaces for use with \texttt{RSVD}. Critically, which algorithm to use is determined by the spectral space for the underlying problem-specific sensitivity matrix $G$, as discussed in \cite{vatankhah2019improving}. Moreover, we can relax the restriction $t \approx  \floor(m/8)$, indeed satisfactory results are achieved using $t\approx m/20$ for large problems, for the inversion of magnetic data. 

It should be noted that equivalent conclusions can be made when the implementations use padding, only that generally fewer iterations to convergence are required. Furthermore, all the implementations use the automatic determination of the regularization parameter using the \texttt{UPRE} function. The suitability of the \texttt{UPRE} function was demonstrated in earlier references, and is thus not reproduced here, but results that are not reported here demonstrated that the earlier results still hold for these large scale problems and algorithms. 

Overall, it has been shown that the use of the \texttt{BTTB} structure inherent in the sensitivity matrices leads to fast algorithms that make it feasible to solve large-scale focusing inversion problems using standard \texttt{GKB} and \texttt{RSVD} algorithms on desktop environments, without modifications to handle either multiple cores or GPUs. It is clear that yet greater efficiency could be achieved with such modifications, that may then be architecture specific and thus less flexible. Moreover, these results suggest that the development of alternative algorithms that avoid the need to use storage of matrices of size $n \times t$, is desirable and is a topic for future study.

\section*{Acknowledgments}
The authors would like to thank Dr. Mark Pilkington for providing data from the Wuskwatim Lake area. Rosemary Renaut acknowledges the support of NSF grant  DMS 1913136:   ``Approximate Singular Value Expansions and Solutions of
Ill-Posed Problems". 

\bibliography{../../../../cvstuff/UPRE}
\bibliographystyle{plainnat}

\appendix
\section{Multiplication using BTTB structure}\label{Appendix:BTTB}
We first consider the multiplication $G\bfx$ where $\bfx\in\Rm{n}$ and use the column block structure of $G$ which was given in \eqref{blockG} to see that $ G\bfx=\sum_{r=1}^{\nbz}\Gr \bfx^{(r)}$ where $\bfx$ is blocked consistently with $G$. 
Now each $\Gr$ has $\bttb$ structure and can be embedded in a circulant matrix in order to evaluate $\Gr \bfx^{(r)}$ using the \texttt{2DFFT} as described in \cite{Vogel:2002}. Specifically the first column of the circulant extension is reshaped into $T\in\Rmn{(\nsx+\nbx-1)}{(\nsy+\nby-1)}$, and $\bfx^{(r)}$ is reshaped and embedded into $W\in\Rmn{(\nsx+\nbx-1)}{(\nsy+\nby-1)}$, see \cite{hogue2019tutorial}. Now we assume that the \texttt{2DFFT} of $T$ is precomputed and that $\dotstar$ represents element-wise multiplication.  Then,  $\Gr \bfx^{(r)}$ is  extracted from  $\ifftt(\fftt(T)\dotstar \fftt(W))$, with cost
\begin{equation}\label{costblockfft}\texttt{Cost}_{\Gr \bfx^{(r)}} = \texttt{Cost}_{\fftt(W)} + \texttt{Cost}_{\dotstar} + \texttt{Cost}_{\ifftt()}. \end{equation}
Here, the \texttt{2DFFT} of $W$ is  computed as $\oneDFFT((\oneDFFT(W))^T)^T$, where the $\oneDFFT$ is applied to  each column of the array independently.  Using the cost of a $\oneDFFT$ as $(n/2)\log_2(n)$ for an $n$-length vector,  \cite{Vogel:2002}, this gives, using $\nbr\approx m$ except when the padding is large, 
\begin{eqnarray*}
\texttt{Cost}_{\fftt(W)} &\approx &  2\nby(\nbx\log_2(2\nbx)) + 2\nbx(\nby\log_2(2\nby)) \\
&=&2m(\log_2(2\nbx)) + \log_2(2\nby)) = 2m\log_2(4m).
\end{eqnarray*} 
The element-wise complex multiplication in \eqref{costblockfft} is for a reshaped vector of size $(\nsx+\nbx-1)(\nsy+\nby-1)  \approx 4m$, and each complex multiplication requires $6$ \texttt{flops}.  
Furthermore,  the  inverse \texttt{2DFFT} requires approximately the same number of operations as the forward \texttt{2DFFT}.  Hence 
$$ \texttt{Cost}_{\Gr \bfx^{(r)}} \approx 4m\log_2(4m)+24m,$$
and 
\begin{equation}\label{costGx} \texttt{Cost}_{G \bfx} \approx  4m\nbz\log_2(4m)+24m\nbz+(m-1)\nbz \approx  4n\log_2(4m)+25n+\mathit{LOT},\end{equation}
where the first term is for the multiplication and the second for the summation over the $\nbz$ vectors of length $m$. It is then immediate that the dominant cost for obtaining $GX$, for $X \in \Rmn{n}{t_p}$, ignoring all but third order terms  is 
$$ \texttt{Cost}_{G X} \approx 4t_pn\log_2(4m) +\mathit{LOT}.$$ 

The derivation of the computation, and the cost, for obtaining $G^T\bfy$ for $\bfy\in\Rm{m}$ follows similarly, noting that 
$G^T \bfy = [G^{(1)}, G^{(2)}, \dots G^{(\nbz)}]^T\bfy$,   requires the computation of $(\Gr)^T\bfy$ for each $r$ and that no summation is required as in \eqref{costGx}. Hence $ \texttt{Cost}_{G^T \bfy} \approx 4n\log_2(4m)$ and $ \texttt{Cost}_{G Y} \approx 4t_pn\log_2(4m)$. Furthermore, we note that  $X^TG^T = (GX)^T$ and $Y^TG = (G^TY)^T$. Thus, the computations and computational costs are immediately obtained from those of $GX$ and $G^TY$, respectively.

\section{Supporting Numerical Results of Simulations}\label{app:table}
Supporting results illustrated as figures in Sections~\ref{sec:costperiteration}-\ref{sec:methodsolutions} are reported in a set of tables, with captions describing the details. Table~\ref{tableB.3} reports the timing for one iteration of the inversion algorithm using both \texttt{GKB} and \texttt{RSVD} algorithms for \texttt{magnetic} data inversion, comparing timings using matrix $G$ directly and the \texttt{2DFFT}.  The time to convergence for the algorithms is given in Table~\ref{tableB.3} for both \texttt{magnetic} and \texttt{gravity} data sets for domains without padding. Tables~\ref{tableB.5}-\ref{tableB.6} give the details of the number of iteration steps to convergence $K$ and the resulting relative errors, \texttt{RE}, for the timing results of Table~\ref{tableB.4}.
\begin{table}[htb!]\begin{center}\begin{tabular}{|*{12}{c|}}\hline
\multicolumn{3}{|c|}{ \texttt{magnetic}} &\multicolumn{4}{|c|}{\texttt{WITH \texttt{2DFFT}}}&\multicolumn{4}{|c|}{\texttt{Direct use of $G$}}\\
\hline$\ell$&$t$ &$t_p$&\texttt{GKB}&\texttt{RSVD}&\texttt{P}\texttt{GKB}&\texttt{P}\texttt{RSVD}&\texttt{GKB}&\texttt{RSVD}&\texttt{P}\texttt{GKB}&\texttt{P}\texttt{RSVD}\\  
\hline
\hline$      4$&$    150$&$    157$&$     2$&$     3$&$     2$&$     2$&$    25$&$     3$&$    31$&$     3$\\
\hline$      4$&$    240$&$    252$&$     3$&$     4$&$     3$&$     4$&$    40$&$     4$&$    49$&$     5$\\
\hline$      4$&$    300$&$    315$&$     4$&$     6$&$     4$&$     5$&$    51$&$     5$&$    62$&$     6$\\
\hline$      4$&$    750$&$    787$&$    16$&$    16$&$    17$&$    14$&$   132$&$    12$&$   161$&$    15$\\
\hline$      4$&$   1000$&$   1050$&$    26$&$    22$&$    28$&$    20$&$   180$&$    16$&$   220$&$    19$\\
\hline$      4$&$   1500$&$   1575$&$    52$&$    35$&$    56$&$    32$&$   283$&$    26$&$   344$&$    31$\\
\hline$      4$&$   2000$&$   2100$&$    87$&$    49$&$    95$&$    46$&$   393$&$    36$&$   477$&$    45$\\ \hline
\hline$      5$&$    234$&$    245$&$     8$&$    13$&$     7$&$    11$&$   120$&$    12$&$   143$&$    14$\\
\hline$      5$&$    375$&$    393$&$    14$&$    20$&$    14$&$    18$&$   193$&$    18$&$   232$&$    22$\\
\hline$      5$&$    468$&$    491$&$    18$&$    26$&$    18$&$    24$&$   244$&$    22$&$   294$&$    27$\\
\hline$      5$&$   1171$&$   1229$&$    72$&$    70$&$    77$&$    68$&$   633$&$    55$&$   765$&$    66$\\
\hline$      5$&$   1562$&$   1640$&$   115$&$    96$&$   125$&$    90$&$   862$&$    74$&$  1044$&$    89$\\
\hline$      5$&$   2343$&$   2460$&$   230$&$   151$&$   257$&$   144$&$  1347$&$   118$&$  1633$&$   142$\\
\hline$      5$&$   3125$&$   3281$&$   389$&$   215$&$   435$&$   208$&$  1869$&$   169$&$  2278$&$   211$\\ \hline
\hline$      6$&$    337$&$    353$&$    19$&$    29$&$    16$&$    20$&$   430$&$   440$&$   532$&$  1597$\\
\hline$      6$&$    540$&$    567$&$    36$&$    48$&$    32$&$    35$&$   689$&$  1996$&$   831$&$  2985$\\
\hline$      6$&$    675$&$    708$&$    49$&$    60$&$    46$&$    45$&$   867$&$   977$&$  1050$&$  2821$\\
\hline$      6$&$   1687$&$   1771$&$   213$&$   164$&$   224$&$   127$&$  2255$&$   465$&$  2739$&$  1301$\\
\hline$      6$&$   2250$&$   2362$&$   351$&$   227$&$   382$&$   182$&$  3068$&$  1235$&$  3738$&$  2425$\\
\hline$      6$&$   3375$&$   3543$&$   733$&$   376$&$   818$&$   315$&$  4798$&$  1279$&$  5890$&$  2834$\\
\hline$      6$&$   4500$&$   4725$&$  1259$&$   542$&$  1413$&$   475$&$  6666$&$  2108$&$ 61661$&$  3487$\\ \hline
\hline$      7$&$    459$&$    481$&$    41$&$    56$&$    54$&$    72$&NA&NA&NA&NA\\
\hline$      7$&$    735$&$    771$&$    84$&$    94$&$   104$&$   117$&NA&NA&NA&NA\\
\hline$      7$&$    918$&$    963$&$   117$&$   121$&$   145$&$   150$&NA&NA&NA&NA\\
\hline$      7$&$   2296$&$   2410$&$   554$&$   346$&$   674$&$   433$&NA&NA&NA&NA\\
\hline$      7$&$   3062$&$   3215$&$   944$&$   496$&$  1136$&$   601$&NA&NA&NA&NA\\
\hline$      7$&$   4593$&$   4822$&$  1999$&$   854$&$  2409$&$  1061$&NA&NA&NA&NA\\
\hline$      7$&$   5000$&$   5250$&$  2317$&$   949$&$  2868$&$  1192$&NA&NA&NA&NA\\
\hline
\end{tabular}
\caption{Timing results in seconds for one step of the  inversion algorithm applied to \texttt{magnetic} potential field data for the simulations described in Table~\ref{table2} without padding and with padding (indicated by \texttt{P}),  and for problem sizes up to $\ell=7$.  $t_p=\floor(1.05t)$ is the size of the oversampled projected space for  \texttt{GKB} and \texttt{RSVD} implementations.  The columns under \texttt{Direct use of $G$} do not use the \texttt{2DFFT}. These results are illustrated in Figures~\ref{figure4}-\ref{figure7}, along with the equivalent set of results for the inversion of \texttt{gravity} data. \label{tableB.3}}
\end{center}\end{table}

\begin{table}[htb!]\begin{center}\begin{tabular}{|*{9}{c|}}\hline
&&&\multicolumn{2}{c|}{\texttt{gravity}}
&\multicolumn{2}{c|}{\texttt{magnetic}}
&\multicolumn{2}{c|}{$\texttt{Cost}_{\texttt{GKB}} /\texttt{Cost}_{\texttt{RSVD}}$}\\ \hline
$\ell$&$t$ &$t_p$
&\multicolumn{1}{c|}{\texttt{GKB}}&\multicolumn{1}{c|}{\texttt{RSVD}}
&\multicolumn{1}{c|}{\texttt{GKB}}&\multicolumn{1}{c|}{\texttt{RSVD}}
&\multicolumn{1}{c|}{\texttt{gravity}}&\multicolumn{1}{c|}{\texttt{magnetic}}\\
\hline
\hline$      4$&$    150$&$    157$&$    78$&$    56$&$    40$&$   172^*$&$  1.40$&$  0.23$\\
\hline$      4$&$    240$&$    252$&$   111$&$    79$&$    60$&$   282^*$&$  1.40$&$  0.21$\\
\hline$      4$&$    300$&$    315$&$   153$&$   100$&$    70$&$   351^*$&$  1.53$&$  0.20$\\
\hline$      4$&$    750$&$    787$&$   248$&$   216$&$   136$&$   883^*$&$  1.15$&$  0.15$\\
\hline$      4$&$   1000$&$   1050$&$   323$&$   285$&$   197$&$  1182^*$&$  1.13$&$  0.17$\\
\hline$      4$&$   1500$&$   1575$&$   494$&$   436$&$   343$&$   650$&$  1.13$&$  0.53$\\
\hline$      4$&$   2000$&$   2100$&$   641$&$   588$&$   739$&$   771$&$  1.09$&$  0.96$\\ \hline
\hline$      5$&$    234$&$    245$&$   265$&$   166$&$   152$&$   509^*$&$  1.60$&$  0.30$\\
\hline$      5$&$    375$&$    393$&$   411$&$   259$&$   174$&$   811^*$&$  1.59$&$  0.21$\\
\hline$      5$&$    468$&$    491$&$   342$&$   325$&$   199$&$  1014^*$&$  1.05$&$  0.20$\\
\hline$      5$&$   1171$&$   1229$&$  1064$&$   835$&$   626$&$  2582^*$&$  1.27$&$  0.24$\\
\hline$      5$&$   1562$&$   1640$&$  1235$&$   997$&$   948$&$  2121$&$  1.24$&$  0.45$\\
\hline$      5$&$   2343$&$   2460$&$  1899$&$  1492$&$  1728$&$  2126$&$  1.27$&$  0.81$\\
\hline$      5$&$   3125$&$   3281$&$  2918$&$  2052$&$  2915$&$  2971$&$  1.42$&$  0.98$\\ \hline
\hline$      6$&$    337$&$    353$&$   595$&$   296$&$   246$&$   923^*$&$  2.01$&$  0.27$\\
\hline$      6$&$    540$&$    567$&$   671$&$   424$&$   347$&$  1514^*$&$  1.58$&$  0.23$\\
\hline$      6$&$    675$&$    708$&$   802$&$   527$&$   413$&$  1877^*$&$  1.52$&$  0.22$\\
\hline$      6$&$   1687$&$   1771$&$  2704$&$  1385$&$  1077$&$  2581$&$  1.95$&$  0.42$\\
\hline$      6$&$   2250$&$   2362$&$  2518$&$  1597$&$  1608$&$  2937$&$  1.58$&$  0.55$\\
\hline$      6$&$   3375$&$   3543$&$  4308$&$  2483$&$  3071$&$  4142$&$  1.73$&$  0.74$\\ 
\hline$      6$&$   4500$&$   4725$&$  6925$&$  3429$&$  6699$&$  5109$&$  2.02$&$  1.31$\\ \hline
\hline$      7$&$    459$&$    481$&$  1427$&$   679$&$   594$&$  2157^*$&$  2.10$&$  0.28$\\
\hline$      7$&$    735$&$    771$&$  1642$&$  1104$&$  1070$&$  3524^*$&$  1.49$&$  0.30$\\
\hline$      7$&$    918$&$    963$&$  2084$&$  1218$&$  1026$&$  4413^*$&$  1.71$&$  0.23$\\
\hline$      7$&$   2296$&$   2410$&$  5732$&$  3311$&$  3809$&$  6618$&$  1.73$&$  0.58$\\
\hline$      7$&$   3062$&$   3215$&$  6959$&$  4490$&$  5639$&$  8469$&$  1.55$&$  0.67$\\
\hline$      7$&$   4593$&$   4822$&$ 12347$&$  6979$&$ 10979$&$ 12949$&$  1.77$&$  0.85$\\
\hline$      7$&$   5000$&$   5250$&$ 13975$&$  7711$&$ 12544$&$ 13239$&$  1.81$&$  0.95$\\
\hline
\end{tabular}
\caption{Timing results to convergence for inversion of \texttt{gravity}  and \texttt{magnetic} potential field data for the simulations described in Table~\ref{table2} without padding, for problem sizes up to $\ell=7$. Entries with $*$ indicate that the algorithm did not converge.   In the last two columns the relative costs of \texttt{GKB} as compared to \texttt{RSVD}. Values greater than $1$, less than $1$, indicate that the \texttt{RSVD} is overall faster, slower, respectively. In general  \texttt{RSVD} is faster for inversion of \texttt{gravity} data but slower for inversion of  \texttt{magnetic} data. Still, as problem size increases, the relative efficiency of \texttt{GKB} for the \texttt{magnetic} data decreases, $\texttt{Cost}_{\texttt{GKB}} /\texttt{Cost}_{\texttt{RSVD}}$  increases towards $1$.  Results for relative errors and number of iterations are presented in Tables~\ref{tableB.5}-\ref{tableB.6}, for \texttt{magnetic} and \texttt{gravity} data, respectively. \label{tableB.4}}
\end{center}\end{table}

\begin{table}[htb!]\begin{center}\begin{tabular}{|*{12}{c|}}\hline
\multicolumn{3}{|c|}{ \texttt{magnetic}} 
&\multicolumn{2}{c|}{\texttt{GKB}}&\multicolumn{2}{c|}{\texttt{RSVD}}
&\multicolumn{2}{c|}{\texttt{P}\texttt{GKB}}&\multicolumn{2}{c|}{\texttt{P}\texttt{RSVD}}\\
\hline$\ell$&$t$ &$t_p$&$K~$&$\texttt{RE}$&$K~$&$\texttt{RE}$&$~~K~$&$\texttt{RE}$&$~~K~$&$RE$  \\  

\hline$      4$&$    150$&$    157$&$     10$&$  0.71$&$     25$&$  0.72$&$      5$&$  0.63$&$     25$&$  0.79$\\
\hline$      4$&$    240$&$    252$&$      9$&$  0.68$&$     25$&$  0.69$&$      5$&$  0.63$&$     25$&$  0.72$\\
\hline$      4$&$    300$&$    315$&$      8$&$  0.66$&$     25$&$  0.69$&$      5$&$  0.63$&$     25$&$  0.74$\\
\hline$      4$&$    750$&$    787$&$      5$&$  0.63$&$     25$&$  0.64$&$      4$&$  0.63$&$     25$&$  0.65$\\
\hline$      4$&$   1000$&$   1050$&$      5$&$  0.63$&$     25$&$  0.63$&$      4$&$  0.63$&$     19$&$  0.65$\\
\hline$      4$&$   1500$&$   1575$&$      5$&$  0.63$&$      9$&$  0.63$&$      5$&$  0.62$&$      8$&$  0.64$\\
\hline$      4$&$   2000$&$   2100$&$      7$&$  0.61$&$      8$&$  0.63$&$      6$&$  0.60$&$      7$&$  0.63$\\ \hline
\hline$      5$&$    234$&$    245$&$     12$&$  0.81$&$     25$&$  0.82$&$      6$&$  0.71$&$     25$&$  0.89$\\
\hline$      5$&$    375$&$    393$&$      8$&$  0.72$&$     25$&$  0.78$&$      6$&$  0.69$&$     25$&$  0.82$\\
\hline$      5$&$    468$&$    491$&$      7$&$  0.70$&$     25$&$  0.77$&$      6$&$  0.69$&$     25$&$  0.80$\\
\hline$      5$&$   1171$&$   1229$&$      7$&$  0.66$&$     25$&$  0.70$&$      6$&$  0.66$&$     25$&$  0.71$\\
\hline$      5$&$   1562$&$   1640$&$      7$&$  0.66$&$     15$&$  0.70$&$      6$&$  0.66$&$     12$&$  0.71$\\
\hline$      5$&$   2343$&$   2460$&$      7$&$  0.65$&$     10$&$  0.67$&$      6$&$  0.66$&$      9$&$  0.68$\\
\hline$      5$&$   3125$&$   3281$&$      8$&$  0.65$&$     10$&$  0.67$&$      8$&$  0.66$&$      9$&$  0.68$\\ \hline
\hline$      6$&$    337$&$    353$&$     10$&$  0.74$&$     25$&$  0.73$&$      5$&$  0.67$&$     25$&$  0.77$\\
\hline$      6$&$    540$&$    567$&$      8$&$  0.69$&$     25$&$  0.69$&$      5$&$  0.67$&$     25$&$  0.73$\\
\hline$      6$&$    675$&$    708$&$      7$&$  0.67$&$     25$&$  0.68$&$      5$&$  0.67$&$     25$&$  0.71$\\
\hline$      6$&$   1687$&$   1771$&$      5$&$  0.64$&$     13$&$  0.66$&$      5$&$  0.65$&$     10$&$  0.68$\\
\hline$      6$&$   2250$&$   2362$&$      5$&$  0.64$&$     11$&$  0.65$&$      5$&$  0.66$&$     10$&$  0.68$\\
\hline$      6$&$   3375$&$   3543$&$      5$&$  0.64$&$     10$&$  0.64$&$      5$&$  0.67$&$      9$&$  0.66$\\
\hline$      6$&$   4500$&$   4725$&$      7$&$  0.62$&$      9$&$  0.63$&$      5$&$  0.67$&$      9$&$  0.66$\\ \hline
\hline$      7$&$    459$&$    481$&$     11$&$  0.78$&$     25$&$  0.80$&$      6$&$  0.69$&$     25$&$  0.80$\\
\hline$      7$&$    735$&$    771$&$     10$&$  0.74$&$     25$&$  0.75$&$      6$&$  0.69$&$     25$&$  0.76$\\
\hline$      7$&$    918$&$    963$&$      7$&$  0.69$&$     25$&$  0.73$&$      6$&$  0.69$&$     25$&$  0.75$\\
\hline$      7$&$   2296$&$   2410$&$      7$&$  0.67$&$     14$&$  0.70$&$      5$&$  0.72$&$     12$&$  0.72$\\
\hline$      7$&$   3062$&$   3215$&$      7$&$  0.68$&$     13$&$  0.70$&$      6$&$  0.69$&$     11$&$  0.71$\\
\hline$      7$&$   4593$&$   4822$&$      7$&$  0.68$&$     13$&$  0.69$&$      6$&$  0.70$&$     11$&$  0.72$\\
\hline$      7$&$   5000$&$   5250$&$      7$&$  0.68$&$     12$&$  0.68$&$      6$&$  0.70$&$     11$&$  0.72$\\
\hline
\end{tabular}
\caption{Results for inversion of \texttt{magnetic} potential field data for the simulations described in Table~\ref{table2} without padding and with padding and for problem sizes up to $\ell=7$. The maximum number of iterations is set to $25$  in all cases.  $t_p$ is the size of the projected space  for  \texttt{GKB} and   \texttt{RSVD} implementations.  Reported are the number of iterations to convergence, $K$, for convergence as defined by \eqref{scaledchi2}, with $K=25$ indicating that the simulation did not converge to the given tolerance.  The calculated relative error $\texttt{RE}$ for the given $K$ are also given, for both unpadded and padded cases respectively. \label{tableB.5}}
\end{center}\end{table}

\begin{table}[htb!]\begin{center}\begin{tabular}{|*{12}{c|}}\hline
\multicolumn{3}{|c|}{\texttt{gravity} } 
&\multicolumn{2}{c|}{\texttt{GKB}}&\multicolumn{2}{c|}{\texttt{RSVD}}
&\multicolumn{2}{c|}{\texttt{P}\texttt{GKB}}&\multicolumn{2}{c|}{\texttt{P}\texttt{RSVD}}\\
\hline$\ell$&$t$ &$t_p$&$K~$&$\texttt{RE}$&$K~$&$\texttt{RE}$&$~~K~$&$\texttt{RE}$&$~~K~$&$RE$  \\  
\hline$      4$&$    150$&$    157$&$     19$&$  1.00$&$      8$&$  0.56$&$     20$&$  1.00$&$      8$&$  0.56$\\
\hline$      4$&$    240$&$    252$&$     16$&$  0.97$&$      7$&$  0.56$&$     17$&$  0.97$&$      7$&$  0.56$\\
\hline$      4$&$    300$&$    315$&$     17$&$  0.97$&$      7$&$  0.56$&$     17$&$  0.96$&$      7$&$  0.56$\\
\hline$      4$&$    750$&$    787$&$      9$&$  0.76$&$      6$&$  0.57$&$     10$&$  0.76$&$      6$&$  0.57$\\
\hline$      4$&$   1000$&$   1050$&$      8$&$  0.66$&$      6$&$  0.57$&$      8$&$  0.66$&$      6$&$  0.57$\\
\hline$      4$&$   1500$&$   1575$&$      7$&$  0.64$&$      6$&$  0.57$&$      7$&$  0.64$&$      6$&$  0.57$\\
\hline$      4$&$   2000$&$   2100$&$      6$&$  0.62$&$      6$&$  0.57$&$      7$&$  0.64$&$      7$&$  0.58$\\ \hline
\hline$      5$&$    234$&$    245$&$     21$&$  1.05$&$      8$&$  0.49$&$     21$&$  1.05$&$      9$&$  0.51$\\
\hline$      5$&$    375$&$    393$&$     19$&$  1.01$&$      8$&$  0.50$&$     21$&$  1.03$&$      9$&$  0.53$\\
\hline$      5$&$    468$&$    491$&$     12$&$  0.82$&$      8$&$  0.50$&$     13$&$  0.84$&$      8$&$  0.53$\\
\hline$      5$&$   1171$&$   1229$&$     12$&$  0.78$&$      8$&$  0.53$&$     11$&$  0.79$&$      8$&$  0.57$\\
\hline$      5$&$   1562$&$   1640$&$      9$&$  0.66$&$      7$&$  0.53$&$      9$&$  0.68$&$      8$&$  0.58$\\
\hline$      5$&$   2343$&$   2460$&$      8$&$  0.65$&$      7$&$  0.55$&$      8$&$  0.67$&$      8$&$  0.59$\\ 
\hline$      5$&$   3125$&$   3281$&$      8$&$  0.64$&$      7$&$  0.57$&$      8$&$  0.66$&$      7$&$  0.60$\\ \hline
\hline$      6$&$    337$&$    353$&$     24$&$  1.03$&$      8$&$  0.56$&$     23$&$  1.03$&$      8$&$  0.60$\\
\hline$      6$&$    540$&$    567$&$     15$&$  0.90$&$      7$&$  0.58$&$     15$&$  0.93$&$      7$&$  0.61$\\
\hline$      6$&$    675$&$    708$&$     14$&$  0.89$&$      7$&$  0.58$&$     14$&$  0.91$&$      7$&$  0.62$\\
\hline$      6$&$   1687$&$   1771$&$     13$&$  0.85$&$      7$&$  0.61$&$     12$&$  0.85$&$      6$&$  0.64$\\
\hline$      6$&$   2250$&$   2362$&$      8$&$  0.70$&$      6$&$  0.62$&$      8$&$  0.71$&$      6$&$  0.64$\\
\hline$      6$&$   3375$&$   3543$&$      7$&$  0.69$&$      6$&$  0.63$&$      8$&$  0.71$&$      6$&$  0.64$\\
\hline$      6$&$   4500$&$   4725$&$      7$&$  0.69$&$      6$&$  0.63$&$      8$&$  0.70$&$      6$&$  0.64$\\ \hline
\hline$      7$&$    459$&$    481$&$     24$&$  1.07$&$      8$&$  0.56$&$     25$&$  1.08$&$      8$&$  0.59$\\
\hline$      7$&$    735$&$    771$&$     16$&$  0.95$&$      8$&$  0.57$&$     16$&$  0.95$&$      8$&$  0.59$\\
\hline$      7$&$    918$&$    963$&$     15$&$  0.93$&$      7$&$  0.58$&$     15$&$  0.94$&$      7$&$  0.60$\\
\hline$      7$&$   2296$&$   2410$&$     11$&$  0.75$&$      7$&$  0.60$&$     11$&$  0.75$&$      7$&$  0.60$\\
\hline$      7$&$   3062$&$   3215$&$      9$&$  0.72$&$      7$&$  0.60$&$     10$&$  0.73$&$      7$&$  0.61$\\
\hline$      7$&$   4593$&$   4822$&$      8$&$  0.70$&$      7$&$  0.61$&$      9$&$  0.71$&$      7$&$  0.61$\\
\hline$      7$&$   5000$&$   5250$&$      8$&$  0.70$&$      7$&$  0.61$&$      9$&$  0.71$&$      7$&$  0.61$\\
\hline
\end{tabular}
\caption{Results for inversion of \texttt{gravity} potential field data for the simulations described in Table~\ref{table2} without padding and with padding and for problem sizes up to $\ell=7$. The maximum number of iterations is set to $25$  in all cases.  $t_p$ is the size of the projected space  for  \texttt{GKB} and   \texttt{RSVD} implementations.  Reported are the number of iterations to convergence, $K$, for convergence as defined by \eqref{scaledchi2}, with $K=25$ indicating that the simulation did not converge to the given tolerance.  The calculated relative error $\texttt{RE}$ for the given $K$ are also given, for both unpadded and padded cases respectively. \label{tableB.6}}
\end{center}\end{table}




\end{document}